\RequirePackage{ifpdf}
\documentclass[12pt,letterpaper]{JHEP3}         
\usepackage{amssymb,amsfonts}
\usepackage{amsmath}
\usepackage{epsfig}

\def\be{\begin{eqnarray}}
\def\ee{\end{eqnarray}}
\newcommand{\nn}{\nonumber}
\newcommand\para{\paragraph{}}
\newcommand{\ft}[2]{{\textstyle\frac{#1}{#2}}}
\newcommand{\eqn}[1]{(\ref{#1})}
\newcommand{\ppp}[2]{\frac{\partial {#1}}{\partial {#2}}}

\newcommand{\up}{|\!\uparrow\,\rangle}
\newcommand{\down}{|\!\downarrow\,\rangle}

\newcommand\bpi{\mbox{\boldmath $\pi$}}

\def\Dslash{\,\,{\raise.15ex\hbox{/}\mkern-13mu D}}
\def\Dbarslash{\,\,{\raise.15ex\hbox{/}\mkern-12mu {\bar D}}}
\def\delslash{\,\,{\raise.15ex\hbox{/}\mkern-10mu \partial}}
\def\delbarslash{\,\,{\raise.15ex\hbox{/}\mkern-9mu {\bar\partial}}}
\def\pslash{\,\,{\raise.15ex\hbox{/}\mkern-11mu p}}
\def\qslash{\,\,{\raise.15ex\hbox{/}\mkern-9mu q}}
\def\kslash{\,\,{\raise.15ex\hbox{/}\mkern-11mu k}}
\def\eslash{\,\,{\raise.15ex\hbox{/}\mkern-9mu \epsilon}}

\newcommand{\slsh}[1]{\,\,{\raise.15ex\hbox{/}\mkern-12mu {#1}}}

\newcommand{\sign}{{\rm sign}}
\newcommand{\Tr}{{\rm Tr}}
\newcommand{\tr}{{\rm tr}}

\newcommand{\oket}{|0\rangle}

\newcommand{\ep}{\epsilon^{\mu\nu\rho}}

\title{{\Huge The Quantum Hall Effect} \\
{\large TIFR Infosys Lectures}}

\author{David Tong\\
Department of Applied Mathematics and Theoretical Physics, \\
Centre for Mathematical Sciences, \\
Wilberforce Road, \\
Cambridge, CB3 OBA, UK \\ {}\\
http://www.damtp.cam.ac.uk/user/tong/qhe.html\\
\email{d.tong@damtp.cam.ac.uk}
\\{}\\{}\\{}\\{}\\{}\\{}\\{}\\{}\\{}\\{}\\{}\\{}\\{}\\{}
\\{}\\{}\\{}\\{}\\{}\\{}
}

\preprint{January 2016}

\abstract{ \\ {}

There are surprisingly few dedicated books on the quantum Hall effect. Two prominent ones are
\begin{itemize}
\item Prange and Girvin, ``{\it The Quantum Hall Effect}"
\end{itemize}
This is a collection of articles by most of the main players circa 1990. The basics are described well but there's nothing about Chern-Simons theories or the importance of the edge modes.
\begin{itemize}
\item J. K. Jain, ``{\it Composite Fermions}"
\end{itemize}
As the title suggests, this book focuses on the composite fermion approach as a lens through which to view all aspects of the quantum Hall effect. It has many good explanations but doesn't cover the more field theoretic aspects of the subject.

\para
There are also a number of good multi-purpose condensed matter textbooks which contain extensive descriptions of the quantum Hall effect. Two, in particular, stand out:

\begin{itemize}
\item  Eduardo Fradkin, {\it Field Theories of Condensed Matter Physics}
\end{itemize}
\begin{itemize}
\item Xiao-Gang Wen, {\it Quantum Field Theory of Many-Body Systems: From the Origin of Sound to an Origin of Light and Electrons}
\end{itemize}

\para
Several    excellent lecture notes covering the various topics discussed in these lectures are available on the web. Links can be found on the course webpage: \href{http://www.damtp.cam.ac.uk/user/tong/qhe.html}{http://www.damtp.cam.ac.uk/user/tong/qhe.html}.}

\begin{document}

\newpage

\subsection*{\center{Acknowledgements}}

{\ }\\
These lectures were given in TIFR, Mumbai. I'm grateful to the students, postdocs, faculty and director for their excellent questions and comments which helped me a lot in understanding what I was saying. 

\para
To first approximation, these lecture notes contain no references to original work.  I've included some footnotes with pointers to review articles and a handful of  key papers. More extensive references can be found in the review articles mentioned earlier, or in the book of reprints,  ``{\it Quantum Hall Effect}", edited by Michael Stone.


\para My thanks to everyone in TIFR for their warm hospitality. Thanks also to Bart Andrews for comments and typo-spotting.  
These lecture notes were written as preparation for research funded by the European Research Council under the European UnionÕs Seventh
Framework Programme (FP7/2007-2013), ERC grant agreement STG 279943, ``Strongly Coupled Systems".

\para
\subsection*{\center{Magnetic Scales}}

\be \mbox{Cyclotron Frequency:}\ \ \omega_B = \frac{eB}{m}\nn\ee
\be \mbox{Magnetic Length:}\ \ l_B=\sqrt{\frac{\hbar}{eB}}\nn\ee
\be \mbox{Quantum of Flux:}\ \ \Phi_0 = \frac{2\pi\hbar}{e}\nn\ee
\be \mbox{Hall Resistivity:}\ \ \rho_{xy} = \frac{2\pi\hbar}{e^2}\frac{1}{\nu}\nn\ee


\newcommand{\upket}{|\uparrow\,\rangle}
\newcommand{\downket}{|\downarrow\,\rangle}
\newcommand{\rightket}{|\rightarrow\,\rangle}
\newcommand{\leftket}{|\leftarrow\,\rangle}
\newcommand{\upbra}{\langle\,\uparrow|}
\newcommand{\downbra}{\langle\,\downarrow|}
\newcommand{\rightbra}{\langle\,\rightarrow|}
\newcommand{\leftbra}{\langle\,\leftarrow|}
\newcommand{\zeroket}{|\,0\,\rangle}
\newcommand{\oneket}{|\,1\,\rangle}
\newcommand{\zerobra}{\langle\,0|}
\newcommand{\onebra}{\langle\,1|}

\newpage

\section{The Basics}\label{basicsec}

\subsection{Introduction}

Take a bunch of electrons, restrict them to move in a two-dimensional plane and turn on a strong magnetic field. This simple set-up provides the setting for some of the most wonderful and surprising results in physics. These phenomena are known collectively as the {\it quantum Hall effect}.

\para
The name comes from the most experimentally visible of these surprises. The Hall conductivity (which we will define below) takes quantised values
\be \sigma_{xy} = \frac{e^2}{2\pi\hbar} \,\nu\nn\ee%
Originally it was found that $\nu$ is, to extraordinary precision, integer valued. Of course, we're very used to things being quantised at the microscopic, atomic level. But this is something different: it's the quantisation of an emergent, macroscopic property in a dirty system involving many many particles and its explanation requires something new. It turns out that this something new is the role that topology can play in quantum many-body systems. Indeed, ideas of topology and geometry will be a constant theme throughout these lectures.

\para
Subsequently, it was found that $\nu$ is not only restricted to take integer values, but can also take very specific rational values. The most  prominent fractions experimentally are $\nu=1/3$ and $\nu=2/5$ but there are many dozens of different fractions that have been seen.
This needs yet another  ingredient. This time, it is the interactions between electrons which result in a highly correlated quantum state that is now recognised as a new state of matter. It is here that the most remarkable things happen. The charged particles that roam around these systems carry a fraction of the charge of the electron, as if the electron has split itself into several pieces. Yet this occurs despite the fact that the electron is (and remains!) an indivisible constituent of matter.

\para
In fact, it is not just the charge of the electron that fractionalises: this happens to the ``statistics" of the electron as well. Recall that the electron is a fermion, which means that the distribution of many electrons is governed by the Fermi-Dirac distribution function. When the electron splits, so too does its fermionic nature. The individual constituents are no longer fermions, but neither are they bosons. Instead they are new entities known as {\it anyons} which, in the simplest cases, lie somewhere between bosons and fermions. In more complicated examples even this description breaks down: the resulting objects are called {\it non-Abelian anyons} and provide physical embodiment of the kind of non-local entanglement famous in quantum mechanics.

\para
Because of this kind of striking behaviour,  the quantum Hall effect has been a constant source of new ideas, providing hints of where to look for  interesting  and novel phenomena, most of them related to the ways in which the mathematics of topology impinges on quantum physics. Important examples include the subject of topological insulators, topological order and topological quantum computing. All of them have their genesis in the quantum Hall effect. 

\para
Underlying all of these phenomena is an impressive theoretical edifice, which involves a tour through some of the most beautiful and important developments in theoretical and mathematical physics over the past decades. The first attack on the problem focussed on the microscopic details of the electron wavefunctions. Subsequent approaches looked at the system from a more coarse-grained, field-theoretic perspective where a subtle construction known as Chern-Simons theory plays the key role. Yet another perspective comes from the edge of the sample where certain excitations live  that know more about what's happening inside than you might think. The main purpose of these lectures is to describe these different approaches and the intricate and surprising links between them.

\subsection{The Classical Hall Effect}\label{classicalhallsec}

The original, classical Hall effect was discovered in 1879 by Edwin Hall. It is a simple consequence of the motion of charged particles in a magnetic field. We'll start these lectures by reviewing the underlying physics of the Hall effect. This will provide a useful background for our discussion of the quantum Hall effect. 

\para
\EPSFIGURE{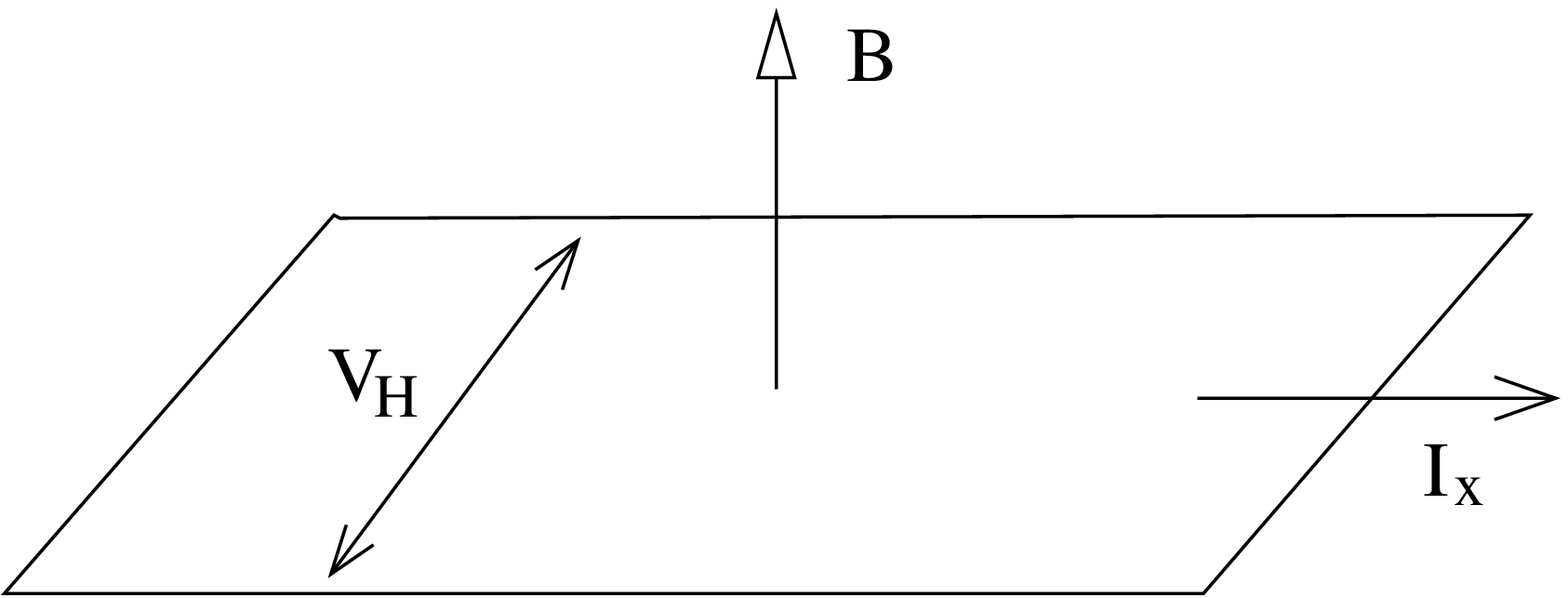,height=60pt}{The classical Hall effect}
Here's the set-up. We turn on a constant magnetic field, ${\bf B}$ pointing in the $z$-direction. Meanwhile, the electrons are restricted to move only in the $(x,y)$-plane. A constant current $I$ is made to flow in the $x$-direction. The {\it Hall effect} is the statement that this induces a voltage $V_H$ ($H$ is for ``Hall") in the $y$-direction. This is shown in the figure to the right. 

\subsubsection{Classical Motion in a Magnetic Field}
\label{clasmotionsec}

The Hall effect arises from the fact that a magnetic field causes charged particles to move in circles. Let's recall the basics. The equation of motion for a particle of mass $m$ and charge $-e$ in a magnetic field is
\be m\frac{d{\bf v}}{dt} = -e{\bf v}\times {\bf B}\nn\ee
When the magnetic field points in the $z$-direction, so that ${\bf B} = (0,0,B)$, and the particle moves only in the transverse plane, so ${\bf v} = (\dot{x},\dot{y},0)$, the equations of motion become two, coupled differential equations
\be m\ddot{x} = -eB\dot{y}\ \ \ {\rm and}\ \ \ m\ddot{y} = eB\dot{x}\label{eom1}\ee
The general solution is
\be x(t) = X - R\sin(\omega_B t + \phi)\ \ \ {\rm and}\ \ \ y(t) = Y + R\cos(\omega_B t+\phi)\label{classicalmotion}\ee
\EPSFIGURE{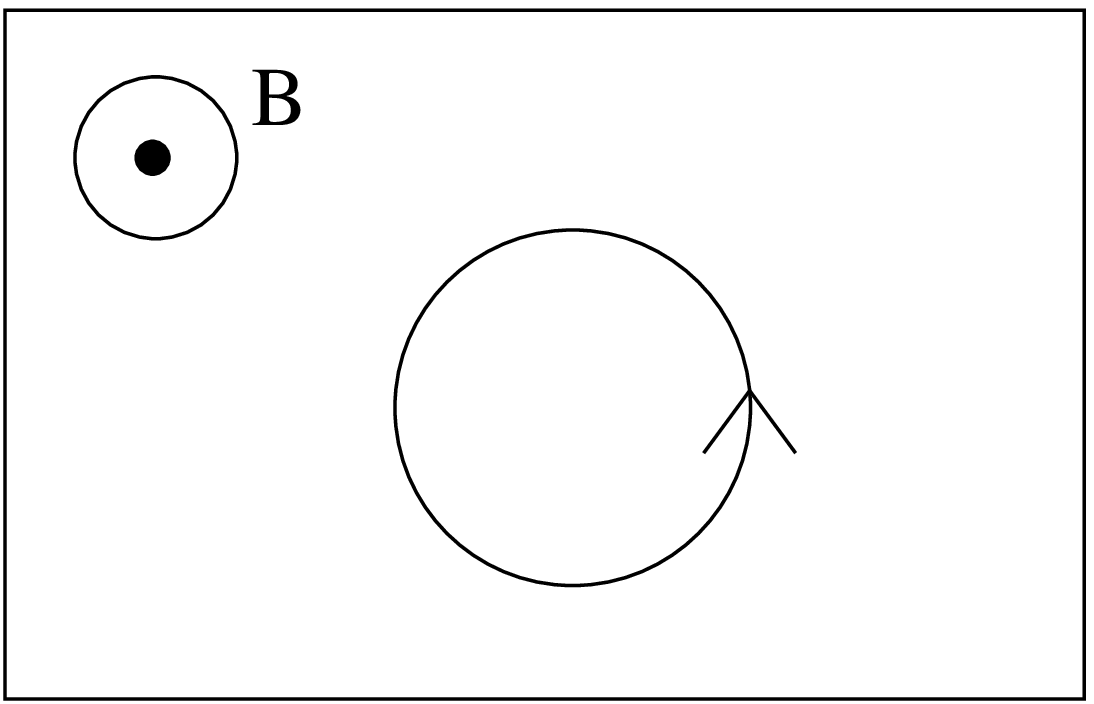,height=60pt}{}
\noindent
We see that the particle moves in a circle which, for $B>0$, is in an anti-clockwise direction. The centre of the circle, $(X,Y)$, the radius of the circle $R$ and the phase $\phi$ are all arbitrary. These are the four integration constants from solving the two second order differential equations. However, the frequency with which the particle goes around the circle is fixed, and given by
\be \omega_B = \frac{eB}{m}\label{cyclotron}\ee
This is called the {\it cyclotron frequency}.   

\subsubsection{The Drude Model}\label{drudesec}

Let's now repeat this calculation with two further ingredients. The first is an electric field, ${\bf E}$. This will accelerate the charges and, in the absence of a magnetic field, would result in a current in the direction of ${\bf E}$. The second ingredient is a linear friction term, which is supposed to capture the effect of the electron bouncing off whatever impedes its progress, whether impurities, the underlying lattice or other electrons. The resulting equation of motion is
\be m\frac{d{\bf v}}{dt} = -e{\bf E} -e {\bf v} \times {\bf B} - \frac{m{\bf v}}{\tau}\label{drude}\ee
The coefficient $\tau$ in the friction term is called the {\it scattering time}. It can be thought of as the average time between collisions.

\para
The equation of motion \eqn{drude} is the simplest model of charge transport, treating the mobile electrons as if they were classical billiard balls. It is called the {\it Drude model} and we met it already in the lectures on \href{http://www.damtp.cam.ac.uk/user/tong/em.html}{\it Electromagnetism}.

\para
We're interested in equilibrium solutions  of  \eqn{drude} which have  
$d{\bf v}/dt=0$. The velocity of the particle must then solve
\be {\bf v} + \frac{e\tau}{m}{\bf v}\times {\bf B} = -\frac{e\tau}{m}{\bf E} \label{vase}\ee
The current density ${\bf J}$ is related to the velocity by
\be {\bf J} = -ne{\bf v}\nn\ee
where $n$ is the density of charge carriers. In matrix notation, \eqn{vase} then becomes
\be \left(\begin{array}{cc} 1 & \omega_B \tau \\ -\omega_B\tau & 1\end{array}\right) {\bf J} = \frac{e^2n\tau}{m}{\bf E}\nn\ee
We can invert this matrix to get an equation of the form
\be {\bf J} = \sigma {\bf E}\nn\ee
This equation is known as {\it Ohm's law}: it tells us how the current flows in response to an electric field. The proportionality constant $\sigma$ is the {\it conductivity}. The slight novelty is that, in the presence of a magnetic field,  $\sigma$ is not a single number: it is a matrix. It is sometimes called the {\it conductivity tensor}. We write it as
\be \sigma=\left(\begin{array}{cc}\sigma_{xx} & \sigma_{xy} \\ -\sigma_{xy} & \sigma_{xx}\end{array}\right)\label{conductivity}\ee
The structure of the matrix, with identical diagonal components, and equal but opposite off-diagonal components, follows from rotational invariance. From the Drude model, we get the explicit expression for the conductivity,
\be \sigma = \frac{\sigma_{DC}}{1+\omega_B^2\tau^2}\left(\begin{array}{cc}1 & -\omega_B\tau\\ \omega_B\tau & 1\end{array}\right) \ \ {\rm with}\ \ \ \sigma_{DC} = \frac{ne^2\tau}{m}\nn\ee
Here $\sigma_{DC}$ is the DC conductivity in the absence of a magnetic field. (This is the same result that we derived in the \href{http://www.damtp.cam.ac.uk/user/tong/em.html}{\it Electromagnetism} lectures). The off-diagonal terms in the matrix are responsible for the Hall effect: in equilibrium, a current in the $x$-direction requires an  electric field with a component in the $y$-direction.

\para
Although it's not directly relevant for our story, it's worth pausing to think about how we actually approach equilibrium in the Hall effect. We start by putting an electric field in the $x$-direction. This gives rise to a current density $J_x$, but this current is deflected due to the magnetic field and bends towards the $y$-direction. In a finite material, this results in a build up of charge along the edge and an associated electric field $E_y$. This continues until the electric field $E_y$ cancels the bending of due to the magnetic field, and the electrons then travel only in the $x$-direction. It's this induced electric field $E_y$ which is responsible for the Hall voltage $V_H$.

\subsubsection*{Resistivity vs Resistance}

The {\it resistivity} is defined as the inverse of the conductivity. This remains true when both are matrices, 
\be \rho = \sigma^{-1} = \left(\begin{array}{cc} \rho_{xx} & \rho_{xy} \\ -\rho_{xy} & \rho_{yy}\end{array}\right)\label{inversecond}\ee
From the Drude model, we have
\be \rho = \frac{1}{\sigma_{DC}}\left(\begin{array}{cc} 1 & \omega_B \tau \\ -\omega_B\tau & 1\end{array}\right)\label{drho}\ee
The off-diagonal components of the resistivity tensor, $\rho_{xy}= \omega_B\tau/\sigma_{DC}$, have  a couple of rather nice properties. First, they are independent of the scattering time $\tau$. This means that they capture something fundamental about the material itself as opposed to the dirty messy stuff that's responsible for scattering. 

\para
 The second nice property is to do with what we measure. 
Usually we measure the {\it resistance} $R$, which differs from the resistivity $\rho$ by geometric factors. However, for $\rho_{xy}$,  these two things coincide. 
To see this, consider a sample of material of length $L$ in the $y$-direction. We drop a voltage $V_y$ in the $y$-direction and measure the resulting current $I_x$ in the $x$-direction. The transverse resistance is 
\be R_{xy} = \frac{V_y}{I_x} = \frac{LE_y}{L J_x} = \frac{E_y}{J_x} = -\rho_{xy}\nn\ee
This has the happy consequence that what we calculate, $\rho_{xy}$, and what we measure, $R_{xy}$, are, in this case, the same. In contrast, if we measure the longitudinal resistance $R_{xx}$ then we'll have to divide by the appropriate lengths to extract the resistivity $\rho_{xx}$. Of course, these lectures are about as theoretical as they come. We're not actually going to measure anything. Just pretend.

%

\para
While we're throwing different definitions around, here's one more. For a current $I_x$ flowing in the $x$-direction, and the associated electric field $E_y$ in the $y$-direction, the  {\it Hall coefficient} is defined by 
\be R_H =  -\frac{E_y}{J_xB}=\frac{\rho_{xy}}{B}\nn\ee
So in the Drude model, we have
\be R_H = \frac{\omega_B}{B \sigma_{DC}} =\frac{1}{ne}\nn\ee
As promised, we see that the Hall coefficient depends only on microscopic information about the material: the charge and density of the conducting particles. The Hall coefficient does not depend on the scattering time $\tau$; it is insensitive to whatever friction processes are at play in the material.

\para
\EPSFIGURE{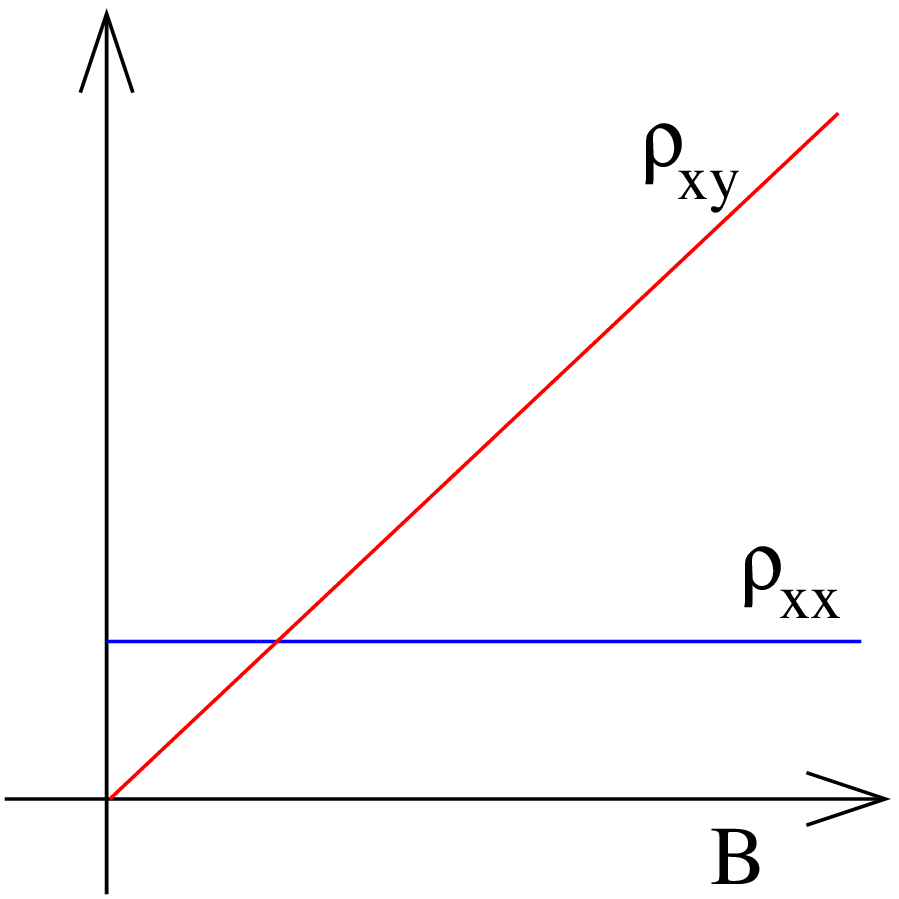,height=100pt}{}
We now have all we need to make an experimental prediction! The two resistivities should be
\be \rho_{xx} = \frac{m}{ne^2\tau} \ \ \ {\rm and}\ \ \ \rho_{xy} = \frac{B}{ne}\nn\ee
Note that only $\rho_{xx}$ depends on the scattering time $\tau$, and $\rho_{xx}\rightarrow 0$ as scattering processes become less important and $\tau\rightarrow \infty$. 
If we plot the two resistivities  as a function of the magnetic field, then our classical expectation is that they  should look the figure on the right. 
%
%

\subsection{Quantum Hall Effects}

Now we understand the classical expectation. And, of course, this expectation is borne out whenever we can trust classical mechanics. But the world is governed by quantum mechanics. This becomes important at low temperatures and strong magnetic fields where more interesting things can happen. 

\para
It's useful to distinguish between two different quantum Hall effects which are associated to two related phenomena. These are called the {\it integer} and {\it fractional} quantum Hall effects. Both were first discovered experimentally and only subsequently understood theoretically. Here we summarise the basic facts about these effects. The goal of these lectures is to understand in more detail what's going on.

\subsubsection{Integer Quantum Hall Effect}

The first experiments exploring the quantum regime of the Hall effect were performed in 1980 by von Klitzing, using samples prepared by Dorda and Pepper\footnote{K. v Klitzing, G. Dorda, M. Pepper, ``{\it New Method for High-Accuracy Determination of the Fine-Structure Constant Based on Quantized Hall Resistance}", \href{http://journals.aps.org/prl/abstract/10.1103/PhysRevLett.45.494}{Phys. Rev. Lett. {\bf 45} 494}.}. The resistivities look like this:
\be \raisebox{-1.1ex}{\epsfxsize=3.2in\epsfbox{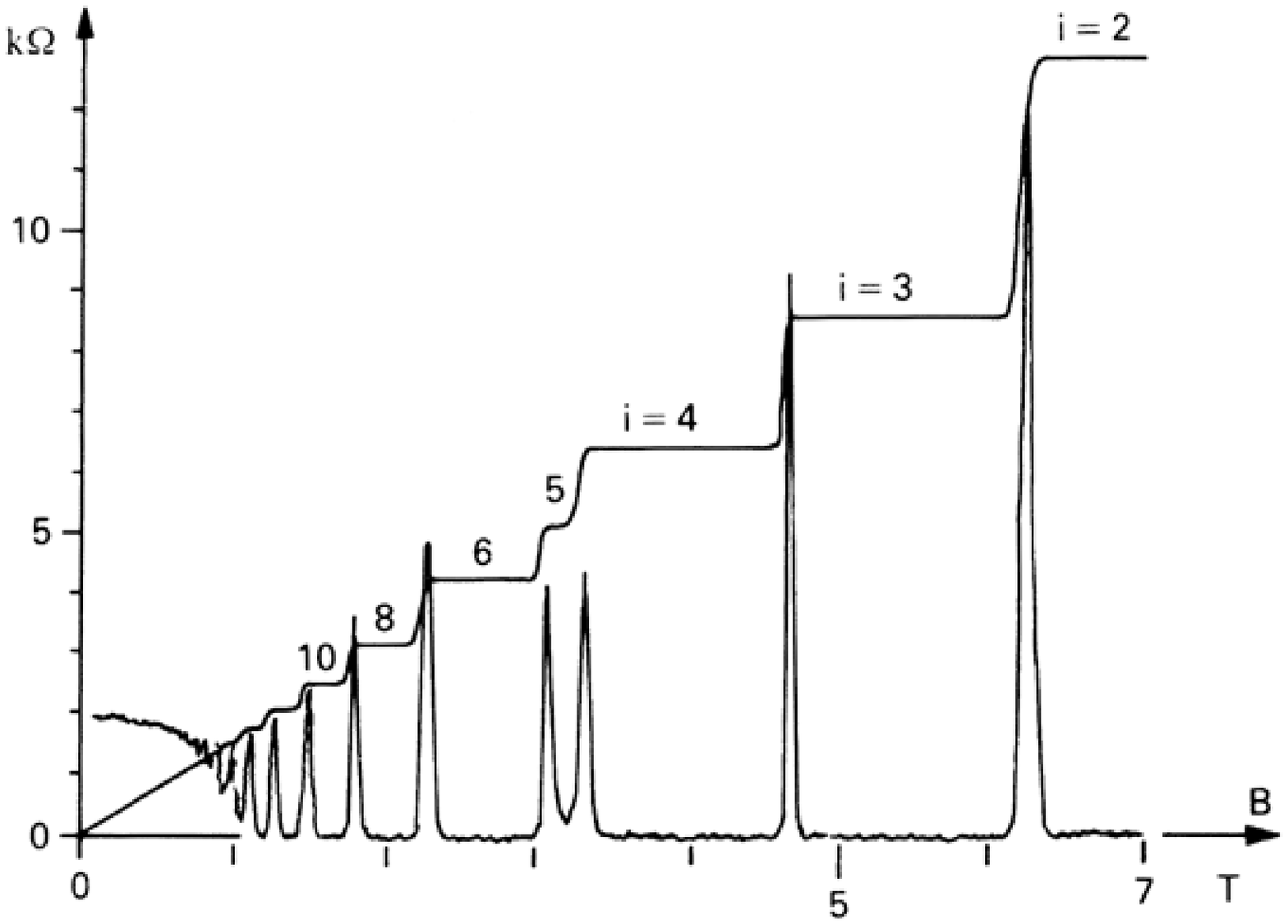}}\nn\ee
This is the {\it integer quantum Hall effect}. For this, von Klitzing was awarded the 1985 Nobel prize. 


\para
%

Both the Hall resistivity $\rho_{xy}$ and the longitudinal resistivity $\rho_{xx}$ exhibit interesting behaviour. Perhaps the most striking feature in the data is the that the Hall resistivity $\rho_{xy}$ sits on a plateau for a range of magnetic field, before jumping suddenly to the next plateau. On these plateau, the resistivity takes the value
\be \rho_{xy} = \frac{2\pi \hbar}{e^2}\, \frac{1}{\nu}\ \ \ \nu\in{\bf Z}\label{rhoxy}\ee
The value of $\nu$ is measured to be an integer to an extraordinary accuracy --- something like  one part in $10^{9}$. The quantity $2\pi\hbar/e^2$  is called the {\it quantum of resistivity} (with $-e$, the electron charge). It is now used as the standard for measuring of resistivity. Moreover, the integer quantum Hall effect is now used as the basis for measuring the ratio of fundamental constants $2\pi\hbar/e^2$ sometimes referred to as the von Klitzing constant. This means that, by definition, the $\nu=1$ state in \eqn{rhoxy} is exactly integer!

\para
The centre of each of these plateaux occurs when the magnetic field takes the value
\be B = \frac{2\pi \hbar n}{\nu e} = \frac{n}{\nu} \Phi_0 \nn\ee
where $n$ is the electron density and $\Phi_0 = {2\pi\hbar}/{e}$ is known as the {\it flux quantum}. As we will review in Section \ref{iqhesec}, these are the values of the magnetic field at which the first $\nu\in {\bf Z}$ Landau levels are filled. In fact, as we will see, it is very easy to argue that the Hall resistivity should take value \eqn{rhoxy} when $\nu$ Landau levels are filled. The surprise is that the plateau exists, with the quantisation  persisting over a range of magnetic fields. 

\para
There is a clue in the experimental data about the origin of the plateaux. Experimental systems are typically dirty, filled with impurities. The technical name for this is {\it disorder}. Usually one wants to remove this dirt to get at the underlying physics. Yet, in the quantum Hall effect, as  you increase the amount of disorder (within reason) the plateaux become more prominent, not less. In fact, in the absence of disorder, the plateaux are expected to vanish completely. That sounds odd: how can the presence of dirt give rise to something as exact and pure as an integer? This is something we will explain in  Section \ref{iqhesec}.

\para
The longitudinal resistivity $\rho_{xx}$ also exhibits a surprise. When $\rho_{xy}$ sits on a plateau, the longitudinal resistivity vanishes: $\rho_{xx}=0$. It spikes only when $\rho_{xy}$ jumps to the next plateau. 

\para
Usually we would think of a system with $\rho_{xx}=0$ as a perfect conductor. But there's something a little counter-intuitive about vanishing resistivity in the presence of a magnetic field. To see this, we can return to the simple definition \eqn{inversecond} which, in components, reads
\be \sigma_{xx} = \frac{\rho_{xx}}{\rho_{xx}^2+\rho_{xy}^2}\ \ \ {\rm and}\ \ \ \sigma_{xy} = \frac{-\rho_{xy}}{\rho_{xx}^2+\rho_{xy}^2}\label{rescond}\ee
%
%
If  $\rho_{xy} = 0$ then we get the familiar relation between conductivity and  resistivity: $\sigma_{xx}= 1/\rho_{xx}$. But if $\rho_{xy}\neq 0$, then we have the more interesting relation above. In particular, we see 
\be \rho_{xx} = 0\ \ \ \Rightarrow \ \ \ \sigma_{xx} =0\ \ \ \ \ \ ({\rm if}\ \rho_{xy}\neq 0)\nn\ee
While we would usually call a system with $\rho_{xx}=0$ a perfect conductor, we would usually call a system with $\sigma_{xx}=0$ a perfect insulator! What's going on?

\para
This particular surprise has more to do with the words we use to describe the phenomena than the underlying physics. In particular, it has nothing to do with quantum mechanics: this behaviour occurs in the Drude model in the limit $\tau\rightarrow \infty$ where there is no scattering. 
In this situation, the current is flowing perpendicular to the applied electric field, so  ${\bf E}\cdot {\bf J} =0$. But recall that ${\bf E}\cdot {\bf J}$ has the interpretation as the work done in accelerating charges. The fact that this vanishes means that we have a steady current flowing without doing any work and, correspondingly, without any dissipation. The fact that $\sigma_{xx}=0$ is telling us that no current is flowing in the longitudinal direction (like an insulator) while the fact that $\rho_{xx}=0$ is telling us that there is no dissipation of energy (like in a perfect conductor).

\subsubsection{Fractional Quantum Hall Effect}

As the disorder is decreased, the integer Hall plateaux become less prominent. But other plateaux emerge at fractional values. This was discovered in 1982 by Tsui and St\"ormer using samples prepared by Gossard\footnote{D. C. Tsui, H. L. Stormer, and A. C. Gossard, ``{\it Two-Dimensional Magnetotransport in the Extreme Quantum Limit}", \href{http://journals.aps.org/prl/abstract/10.1103/PhysRevLett.48.1559}{Phys. Rev. Lett. {\bf 48} (1982)1559}.}. The resistivities look like this:
\be \raisebox{-1.1ex}{\epsfxsize=3.3in\epsfbox{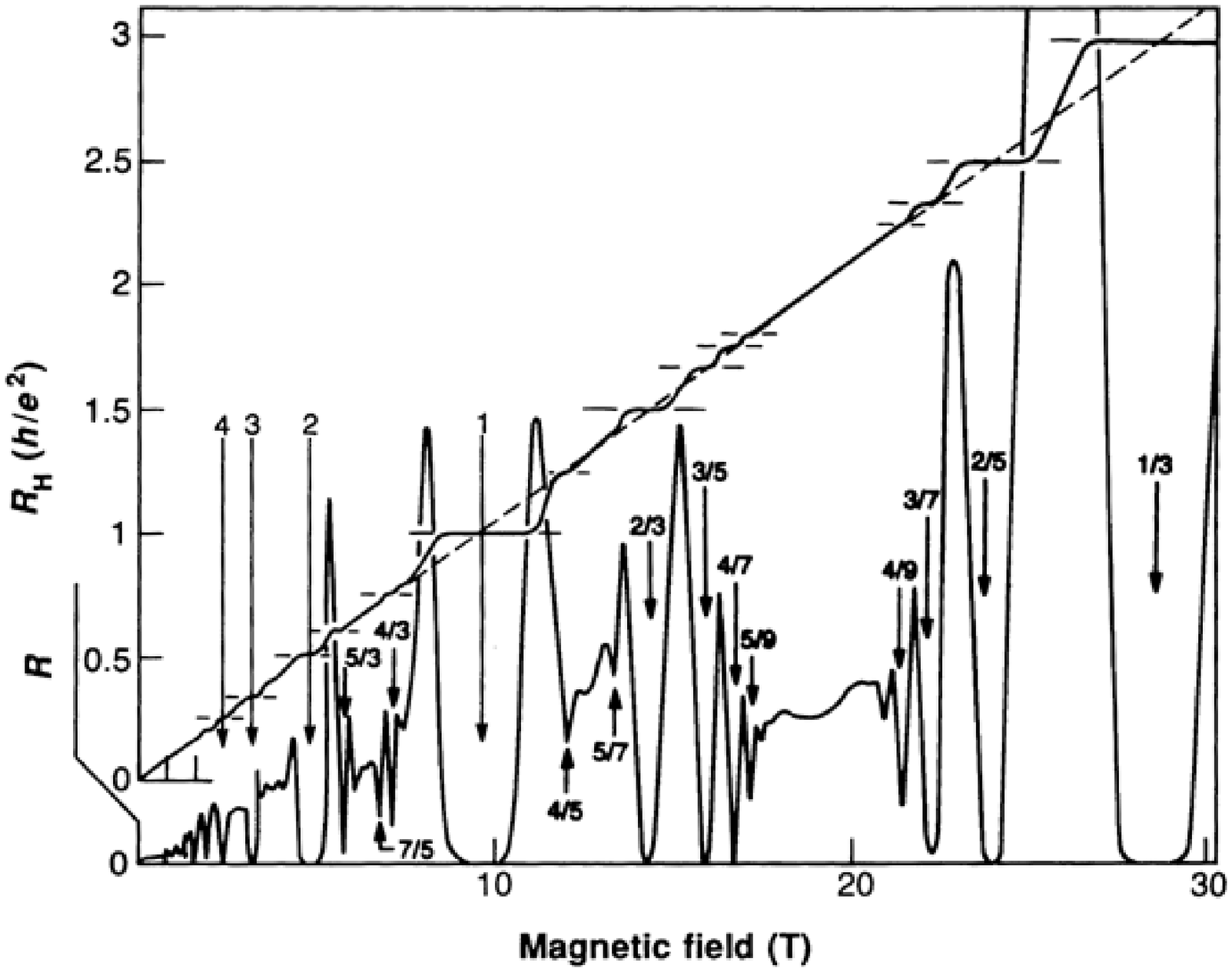}}\nn\ee
The is the {\it fractional quantum Hall effect}. 
On the plateaux, the Hall resistivity again takes the simple form \eqn{rhoxy}, but now with $\nu$ a rational number
\be \nu \in {\bf Q}\nn\ee
Not all fractions appear. The most prominent plateaux sit at $\nu = 1/3, 1/5$ (not shown above) and $2/5$ but there are many more. The vast majority of these have denominators which are odd. But there are exceptions: in particular a clear plateaux has been observed at $\nu=5/2$. As the disorder is decreased, more and more plateaux emerge. It seems plausible that in the limit of a perfectly clean sample, we would get an infinite number of plateaux which brings us back to the classical picture of a straight line for $\rho_{xy}$!

\para
The integer quantum Hall effect can be understood using free electrons. In contrast, to explain the fractional quantum Hall effect we need to take interactions between electrons into account. This makes the problem much harder and much richer. The basics of the theory were first laid down by Laughlin\footnote{R. B. Laughlin, ``{\it The Anomalous Quantum Hall Effect: An Incompressible Quantum Fluid with Fractionally Charged Excitations}," \href{http://journals.aps.org/prl/abstract/10.1103/PhysRevLett.50.1395}{Phys. Rev. Lett. 50, 1395 (1983)}.}, but the subject has since expanded in a myriad of different directions. The 1998 Nobel prize was awarded to Tsui, St\"ormer and Laughlin. Sections \ref{fqhesec} onwards will be devoted to aspects of the fractional quantum Hall effect. 

\subsubsection*{Materials}

These lectures are unabashedly theoretical. We'll have nothing to say about how one actually constructs these phases of matter in the lab. Here I want to merely throw out a few  technical words in an attempt to breed familiarity.

\para
The integer quantum Hall effect was originally discovered in a $Si$ MOSFET (this stands for ``metal-oxide-semiconductor field-effect transistor"). This is a metal-insulator-semiconductor sandwich, with electrons trapped in the ``inversion band" of width $\sim 30 \AA$ between the insulator and semi-conductor. 
Meanwhile the fractional quantum Hall effect was discovered in a $GaAs$-$GaAlAs$ heterostructure. A lot of the subsequent work was done on this system, and it usually goes by the name $GaAs$ (Gallium Arsenide if your chemistry is rusty). In both these systems, the density of electrons is around $n\sim 10^{11} - 10^{12}\ cm^{-2}$.

\para
More recently, both quantum Hall effects have been discovered in graphene, which is a two dimensional material with relativistic electrons. The physics here is similar in spirit, but differs in details.

\subsection{Landau Levels}\label{llevelsec}

 It won't come as a surprise to learn that the physics of the quantum Hall effect involves quantum mechanics. In this section, we will review the quantum mechanics of free particles moving in a background magnetic field and the resulting phenomenon of Landau levels. We will look at these Landau levels in a number of different ways. Each is useful to highlight different aspects of the physics and they will all be important for describing the quantum Hall effects.

\para
Throughout this discussion, we will neglect the spin of the electron. This is more or less appropriate for most physically realised quantum Hall systems. The reason is that in the presence of a magnetic field $B$ there is a Zeeman splitting between the energies of the up and down spins given by $\Delta = 2\mu_B B$ where $\mu_B = e\hbar/2m$ is the Bohr magneton. We will be interested in large magnetic fields where large energies are needed to flip the spin. This means that, if we restrict to low energies, the electrons act as if they are effectively spinless. (We will, however,  add a caveat to this argument below.)

\para
Before we get to the quantum theory, we first need to briefly review some of the structure of classical mechanics in the presence of a magnetic field. The Lagrangian for a particle of charge $-e$ and mass $m$ moving in a background magnetic field ${\bf B} = \nabla \times {\bf A}$ is
\be L = \frac{1}{2}m\dot{\bf x}^2 - e\dot{\bf x}\cdot {\bf A}\nn\ee
Under a gauge transformation, ${\bf A} \rightarrow {\bf A} + \nabla \alpha$, the Lagrangian changes by a total derivative: $L\rightarrow L - e\dot{\alpha}$. This is enough to ensure that the equations of motion \eqn{eom1} remain unchanged under a gauge transformation.

\para
The {\it canonical momentum} arising from this Lagrangian is 
\be {\bf p} = \ppp{L}{\dot{\bf x}}=m\dot{\bf x} - e{\bf A}\nn\ee
This differs from what we called momentum when we were in high school, namely $m\dot{\bf x}$. We will refer to $m\dot{\bf x}$ as the {\it mechanical momentum}. 

\para
We can compute the Hamiltonian
\be H = \dot{\bf x}\cdot{\bf p} - L = \frac{1}{2m}({\bf p}+ e{\bf A})^2\nn\ee
If we write the Hamiltonian in terms of the mechanical momentum then it looks the same as it would in the absence of a magnetic field: $H = \ft12 m \dot{\bf x}^2$. This is the statement that a magnetic field does no work and so doesn't change the energy of the system. However, there's more to the Hamiltonian framework than just the value of $H$. We need to remember which variables are canonical. This information is encoded in the Poisson bracket structure of the theory (or, in fancy language, the symplectic structure on phase space) and, in the quantum theory, is transferred onto commutation relations between operators. The fact that ${\bf x}$ and ${\bf p}$ are canonical means that
\be \{x_i,p_j\} = \delta_{ij}\ \ \ \ {\rm with}\ \ \ \ \{x_i,x_j\} = \{p_i,p_j\} = 0\label{poissonb}\ee
Importantly, ${\bf p}$ is not gauge invariant. This means that the numerical value of ${\bf p}$ can't have any physical meaning since it depends on our choice of gauge. In contrast, the mechanical momentum $m\dot{\bf x}$ is gauge invariant; it measures what you would physically call ``momentum".  But it doesn't have canonical Poisson structure. Specifically, the  Poisson bracket of the mechanical momentum with itself is non-vanishing,
\be \{m\dot{x}_i,m\dot{x}_j\} = \{ p_i+eA_i, p_j + eA_j\} = -e\left(\ppp{A_j}{x^i}-\ppp{A_i}{x^j}\right)=-e\epsilon_{ijk}B_k\label{mechmombracket}\ee

\subsubsection*{Quantisation}

Our task is to solve for the spectrum and wavefunctions of the quantum Hamiltonian,
\be H = \frac{1}{2m} ({\bf p} + e{\bf A})^2\label{ham}\ee
Note that we're not going to put hats on operators in this course; you'll just have to remember that they're quantum operators. Since the particle is restricted to lie in the plane, we write ${\bf x}=(x,y)$. Meanwhile, we take the magnetic field to be constant and perpendicular to this plane,  $\nabla\times {\bf A}= B\hat{\bf z}$. The canonical commutation relations that follow from \eqn{poissonb} are
\be [x_i,p_j] = i\hbar\delta_{ij}\ \ \ \ {\rm with}\ \ \ \ [x_i,x_j] = [p_i,p_j] = 0\nn\ee
We will first derive the energy spectrum using a purely algebraic method. This is very similar to the algebraic solution of the harmonic oscillator and has the advantage that we don't need to specify a choice of gauge potential ${\bf A}$. The disadvantage is that we don't get to write down specific wavefunctions in terms of the positions of the electrons. We will rectify this  in Sections \ref{landaugaugesec} and \ref{symgaugesec}.

\para
To proceed, we work with the commutation relations for the mechanical momentum. We'll give it  a new name (because the time derivative in $\dot{\bf x}$ suggests that we're working in the Heisenberg picture which is not necessarily true). We write
\be\bpi ={\bf p} + e{\bf A} =  m\dot{\bf x}\label{mmom}\ee
Then the commutation relations following from the Poisson bracket \eqn{mechmombracket} are
\be [\pi_x,\pi_y] = -i e\hbar B\label{picoms}\ee
At this point we introduce new variables. These are raising and lowering operators, entirely analogous to those that we use in the harmonic oscillator. They are defined by
\be a = \frac{1}{\sqrt{2e\hbar B}}\left(\pi_x-i\pi_y\right)\ \ \ {\rm and}\ \ \ a^\dagger =  \frac{1}{\sqrt{2e\hbar B}}\left(\pi_x+i\pi_y\right)\nn\ee
The commutation relations for $\bpi$ then tell us that $a$ and $a^\dagger$ obey
\be[a,a^\dagger] = 1\nn\ee
which are precisely the commutation relations obeyed by the raising and lowering operators of the harmonic oscillator.  Written in terms of these operators, the Hamiltonian \eqn{ham} even takes the same form as that of the harmonic oscillator
\be H = \frac{1}{2m} \bpi\cdot\bpi = \hbar\omega_B\left(a^\dagger a + \frac{1}{2}\right)\nn\ee
where $\omega_B = eB/m$ is the cyclotron frequency that we met previously \eqn{cyclotron}.

\para
Now it's simple to finish things off. We can construct the Hilbert space in the same way as the harmonic oscillator: we first introduce a ground state $|0\rangle$ obeying $a|0\rangle=0$ and build the rest of the Hilbert space by acting with $a^\dagger$,
\be a^\dagger |n\rangle = \sqrt{n+1}|n+1\rangle\ \ \ {\rm and}\ \ \ a|n\rangle = \sqrt{n}|n-1\rangle\nn\ee
The state $|n\rangle$ has energy
\be E_n = \hbar\omega_B \left(n+\frac{1}{2}\right)\ \ \ \ \ \ \ n\in {\bf N}\label{llevel}\ee
We learn that in the presence of a magnetic field, the energy levels of a particle become equally spaced, with the gap between each level proportional to the magnetic field $B$. The energy levels  are called {\it Landau levels}. Notice that this is not a small change: the spectrum looks very very different from that of a free particle in the absence of a magnetic field.

\para
There's something a little disconcerting about the above calculation. We started with a particle moving in a plane. This has two degrees of freedom. But we ended up writing this in terms of the harmonic oscillator which has just a single degree of freedom. It seems like we lost something along the way! And, in fact, we did. The energy levels \eqn{llevel} are the correct spectrum of the theory but, unlike for the harmonic oscillator, it turns out that each energy level does not have a unique state associated to it. Instead there is a degeneracy of states. A wild degeneracy. We will return to the algebraic approach in Section \ref{symgaugesec} and demonstrate this degeneracy. But  it's simplest to first turn to a specific choice of the gauge potential ${\bf A}$, which we do shortly. 

\subsubsection*{A Quick Aside:  The role of spin}

The splitting between Landau levels is $\Delta = \hbar \omega_B = e\hbar B/m$. But, for free electrons,  this precisely coincides with the Zeeman splitting $\Delta = g\mu_B\ B$ between spins, where $\mu_B = e\hbar/2m$  is the Bohr magneton and, famously, $g=2$ . It looks as if the spin up particles in Landau level $n$ have exactly the same energy as the spin down particles in level $n+1$. In fact, in real materials, this does not happen. The reason is twofold. First,  the true value of the cyclotron frequency is $\omega_B = eB/m_{\rm eff}$, where $m_{\rm eff}$ is the effective mass of the electron moving in its environment. Second, the $g$ factor can also vary due to effects of band structure. For $GaAs$, the result is that the Zeeman energy is typically about 70 times smaller than the cyclotron energy. This means that first the $n=0$ spin-up Landau level fills, then the $n=0$ spin-down, then the $n=1$ spin-up and so on. 
For other materials (such as the interface between $ZnO$ and $MnZnO$) the relative size of the energies can be flipped and you can fill levels in a different order. This results in different fractional quantum Hall states. In these notes, we will mostly ignore these issues to do with spin. (One exception is Section \ref{multiwfsec} where we discuss wavefunctions for particles with spin).

\subsubsection{Landau Gauge}\label{landaugaugesec}

To find wavefunctions corresponding to the energy eigenstates, we first need to specify a gauge potential ${\bf A}$ such that 
\be \nabla\times {\bf A} = B\hat{\bf z}\nn\ee 
There is, of course, not a unique choice. In this section and the next we will describe two different choices of ${\bf A}$. 

\para
In this section, we work with the choice
\be {\bf A} = xB\hat{\bf y}\label{lgauge}\ee
This is called {\it Landau gauge}. Note that the magnetic field ${\bf B}$ is invariant under both translational symmetry and rotational symmetry in the $(x,y)$-plane. However, the choice of ${\bf A}$ is not; it breaks translational symmetry in the $x$ direction (but not in the $y$ direction) and rotational symmetry. This means that, while the physics will be invariant under all symmetries, the intermediate calculations will not be manifestly invariant. This kind of compromise is typical when dealing with magnetic field. 

\para
The Hamiltonian \eqn{ham} becomes
\be H = \frac{1}{2m}\left(p_x^2 + (p_y+eBx)^2\right)\nn\ee
Because we have manifest translational invariance in the $y$ direction, we can look for energy eigenstates which are also eigenstates of $p_y$. These, of course, are just plane waves in the $y$ direction. This motivates an ansatz using the separation of variables,
\be \psi_k(x,y) = e^{iky} f_k(x)\label{sepansatz}\ee
Acting on this wavefunction with the Hamiltonian, we see that the operator $p_y$ just gets replaced by its eigenvalue $\hbar k$,
\be H\psi_k(x,y) = \frac{1}{2m}\left(p_x^2 + (\hbar k + eBx)^2\right) \psi_x(x,y) \equiv H_k \psi_k(x,y)\nn\ee
But this is now something very familiar: it's the Hamiltonian for a harmonic oscillator in the $x$ direction, with the centre displaced from the origin, 
\be H_k = \frac{1}{2m} p_x^2 + \frac{m\omega_B^2}{2} (x+kl_B^2)^2\label{shiftedho}\ee
The frequency of the harmonic oscillator is again the cyloctron frequency $\omega_B = eB/m$, and we've also introduced  a length scale $l_B$. This is a characteristic length scale which governs any quantum phenomena in a magnetic field. It is  called the {\it magnetic length}. 
\be l_B = \sqrt\frac{\hbar}{eB}\nn\ee
To give you some sense for this, in a magnetic field of $B=1$ {\it Tesla}, the magnetic length for an electron is $l_B\approx 2.5 \times 10^{-8}\ m$.

\para
Something rather strange has happened in the Hamiltonian \eqn{shiftedho}: the momentum in the $y$ direction, $\hbar k$, has turned into the position of the harmonic oscillator in the $x$ direction, which is now centred at $x=-kl_B^2$.

\para
Just as in the algebraic approach above, we've reduced the problem to that of the harmonic oscillator. The energy eigenvalues are
\be E_n = \hbar \omega_B \left(n+\frac{1}{2}\right)\nn\ee
But now we can also write down the explicit wavefunctions. They depend on two quantum numbers, $n\in {\bf N}$ and $k\in {\bf R}$,
\be \psi_{n,k}(x,y) \sim e^{iky} H_n(x+kl_B^2)e^{-(x+kl_B^2)^2/2l_B^2}\label{landaupsi}\ee
with $H_n$ the usual Hermite polynomial wavefunctions of the harmonic oscillator. The $\sim$ reflects the fact that we have made no attempt to normalise these these wavefunctions. 

\para
The wavefunctions look like strips, extended in the $y$ direction but exponentially localised around $x=-kl_B^2$ in the $x$ direction. 
However, the large degeneracy means that by taking linear combinations of these states, we can cook up wavefunctions that have pretty much any shape you like. Indeed, in the next section we will choose a different ${\bf A}$ and see very different profiles for the wavefunctions.

\subsubsection*{Degeneracy}

One advantage of this approach is that we can immediately see the degeneracy in each Landau level. The wavefunction  \eqn{landaupsi} depends on two quantum numbers, $n$ and $k$ but the energy levels depend only on $n$. Let's now see how large this degeneracy is.

\para
To do this, we need to restrict ourselves to a finite region of the $(x,y)$-plane. We pick a rectangle with sides of lengths $L_x$ and $L_y$. We want to know how many states fit inside this rectangle.

\para
Having a finite size $L_y$ is like putting the system in a box in the $y$-direction. We know that the effect of this is to quantise the momentum $k$ in units of $2\pi/L_y$.

\para
Having a finite size $L_x$ is somewhat more subtle. The reason is that, as we mentioned above, the gauge choice \eqn{lgauge} does not have manifest translational invariance in the $x$-direction. This means that our argument will be a little heuristic. Because the wavefunctions \eqn{landaupsi} are exponentially localised around $x=-kl_B^2$, for a finite sample restricted to $0\leq x\leq L_x$ we would  expect the allowed $k$ values to range between $-L_x/l_B^2 \leq k \leq 0$. The end result is that the number of states is
\be {\cal N} = \frac{L_y}{2\pi} \int_{-L_x/l_B^2}^0 dk = \frac{L_xL_y}{2\pi l_B^2} = \frac{eBA}{2\pi\hbar}\label{degeneracy}\ee
where $A=L_xL_y$ is the area of the sample. Despite the slight approximation used above, this turns out to be the exact answer for the number of states on a torus. (One can do better taking the wavefunctions on a torus to be elliptic theta functions).

\EPSFIGURE{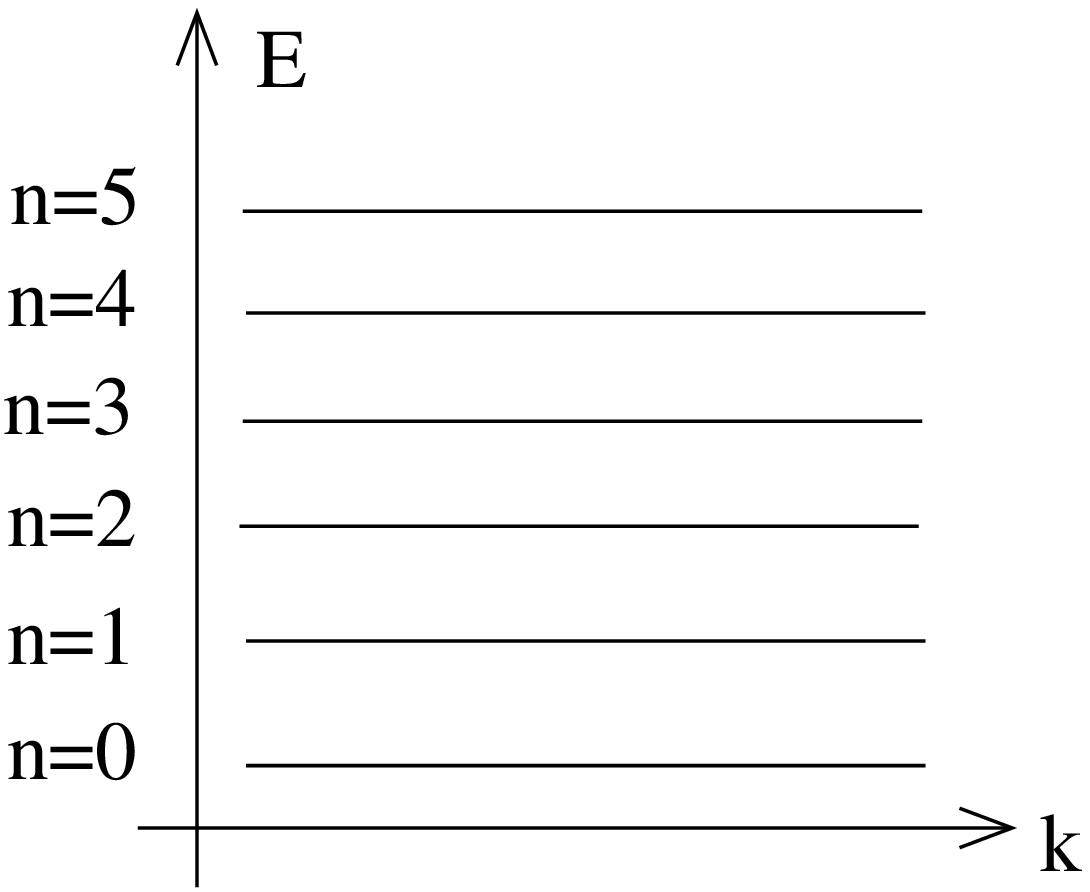,height=100pt}{Landau Levels}
\para
The degeneracy \eqn{degeneracy} is very very large. There are a macroscopic number of states in each Landau level. The resulting spectrum looks like the figure on the right, with $n\in {\bf N}$ labelling the Landau levels and the energy independent of $k$. This degeneracy will be responsible for much of the interesting physics of the fractional quantum Hall effect that we will meet in Section \ref{fqhesec}.

\para
It is common to introduce some new notation to describe the degeneracy \eqn{degeneracy}. We write
\be {\cal N} = \frac{AB}{\Phi_0}\ \ \ \ {\rm with}\ \ \ \ \ \Phi_0=\frac{2\pi\hbar}{e}\label{deggy}\ee
$\Phi_0$ is called the {\it quantum of flux}. It can be thought of as the magnetic flux contained within the area $2\pi l_B^2$. It plays an important role in a number of quantum phenomena in the presence of magnetic fields.

\subsubsection{Turning on an Electric Field}\label{turningesec}

The Landau gauge is useful for working in rectangular geometries. One of the things that is particularly easy in this gauge is the addition of an electric field $E$ in the $x$ direction. We can implement this by the addition of an electric potential $\phi = -Ex$. The resulting Hamiltonian is
\be H = \frac{1}{2m}\left(p_x^2 + (p_y+eBx)^2\right) - eEx\label{eham}\ee
We can again use the ansatz \eqn{sepansatz}. We simply have to complete the square to again write the Hamiltonian as that of a displaced harmonic oscillator. The states are related to those that we had previously, but with a  shifted argument
\be \psi(x,y) = \psi_{n,k}(x-mE/eB^2,y)\label{shiftedstate}\ee
and the energies are now given by
\be E_{n,k} = \hbar \omega_B \left(n+\frac{1}{2}\right) + eE \left(kl_B^2  - \frac{eE}{m\omega_B^2}\right) + \frac{m}{2}\frac{E^2}{B^2}\label{movingenergy}\ee
%
\EPSFIGURE{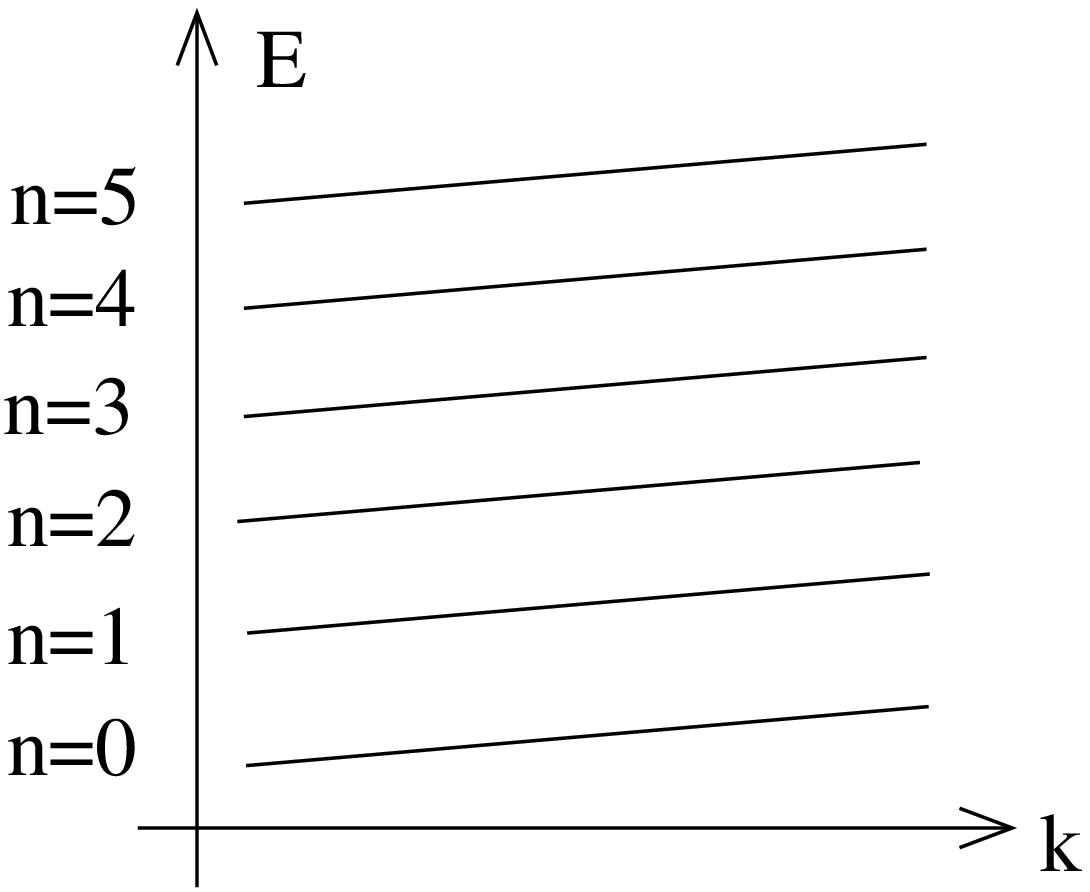,height=100pt}{Landau Levels in an electric field}
\noindent
This is interesting. The degeneracy in each Landau level has now been lifted. The energy in each level now depends linearly on $k$, as shown in the figure.

\para
Because the energy now depends on the momentum, it means that states now drift in the $y$ direction. The group velocity is
\be v_y = \frac{1}{\hbar}\frac{\partial E_{n,k}}{\partial k} = e\hbar El_B^2 = \frac{E}{B}\label{drifting}\ee
This result is one of the surprising joys of classical physics: if you put an electric field ${\bf E}$ perpendicular to a magnetic field ${\bf B}$  then the cyclotron orbits of the electron drift. But they don't drift in the direction of the electric field! Instead they drift in the direction ${\bf E}\times {\bf B}$. Here we see the quantum version of this statement.

\para
The fact that the particles are now moving also provides a natural interpretation of the energy \eqn{movingenergy}.  A wavepacket with momentum $k$ is now localised at position $x = - kl_B^2 - eE/m\omega_B^2$; the middle term above 
can be thought of as the potential energy of this wavepacket. The final term can be thought of as the kinetic energy for the particle: $\frac{1}{2}mv_y^2$.

\subsubsection{Symmetric Gauge}\label{symgaugesec}

Having understood the basics of Landau levels, we're now going to do it all again. This time we'll work in {\it symmetric gauge}, with 
\be {\bf A} = -\frac{1}{2}{\bf r}\times {\bf B} = -\frac{yB}{2}\hat{\bf x} + \frac{xB}{2}\hat{\bf y}\label{sgauge}\ee
This choice of gauge breaks translational symmetry in both the $x$ and the $y$ directions. However, it does preserve rotational symmetry about the origin. This means that angular momentum is a good quantum number.

\para
The main reason for studying Landau levels in symmetric gauge is that this is most convenient language 
for describing  the fractional quantum Hall effect. We shall look at this in Section \ref{fqhesec}. However, as we now see, there are also a number of pretty things that happen in symmetric gauge.

\subsubsection*{The Algebraic Approach Revisited}

At the beginning of this section, we provided a simple algebraic derivation of the energy spectrum \eqn{llevel} of a particle in a magnetic field. But we didn't provide an algebraic derivation of the degeneracies of these Landau levels. Here we rectify this. As we will see, this derivation only really works in the symmetric gauge.

\para
Recall that the algebraic approach uses the mechanical momenta $\bpi = {\bf p} + e{\bf A}$. This is gauge invariant, but non-canonical. We can use this to build ladder operators $a = (\pi_x-i\pi_y)/\sqrt{2e\hbar B}$ which obey $[a,a^\dagger ]=1$. In terms of these creation operators, the Hamiltonian takes the harmonic oscillator form,
\be H = \frac{1}{2m}\bpi\cdot\bpi = \hbar \omega_B\left(a^\dagger a+ \frac{1}{2}\right)\nn\ee
To see the degeneracy in this language, we start by introducing yet another kind of ``momentum", 
\be \tilde{\bpi} = {\bf p} - e{\bf A}\label{tildepi}\ee
This differs from the mechanical momentum \eqn{mmom} by the minus sign. This means that, in contrast to $\bpi$, this new momentum is {\it not} gauge invariant. We should be careful when interpreting the value of $\tilde{\bpi}$ since it can change depending on choice of gauge potential ${\bf A}$.

\para
The commutators of this new momenta differ from \eqn{picoms} only by a minus sign
\be [\tilde{\pi}_x,\tilde{\pi}_y] = i e\hbar B\label{tildepicom}\ee
However, the lack of gauge invariance shows up when we take the commutators of $\bpi$ and $\tilde{\bpi}$. We find
\be [\pi_x,\tilde{\pi}_x] = 2ie\hbar\ppp{A_x}{x}\ \ ,\ \ [\pi_y,\tilde{\pi}_y]= 2ie\hbar\ppp{A_y}{y}\ \  , \ \ [\pi_x,\tilde{\pi}_y] = [\pi_y,\tilde{\pi}_x]=ie\hbar\left(\ppp{A_x}{y}+\ppp{A_y}{x}\right)\nn\ee
This is unfortunate. It means that we cannot, in general, simultaneously diagonalise $\tilde{\bpi}$ and the Hamiltonian $H$ which, in turn, means that we can't use  $\tilde{\pi}$ to tell us about other quantum numbers in the problem.

\para
There is, however, a happy exception to this. In symmetric gauge \eqn{sgauge} all these commutators vanish and we have
\be [\pi_i,\tilde{\pi}_j]=0\nn\ee
We can now define a second pair of raising and lowering operators, 
\be b = \frac{1}{\sqrt{2e\hbar B}}\left(\tilde{\pi}_x+i\tilde{\pi}_y\right)\ \ \ {\rm and}\ \ \ b^\dagger =  \frac{1}{\sqrt{2e\hbar B}}\left(\tilde{\pi}_x-i\tilde{\pi}_y\right)\nn\ee
These too obey
\be[b,b^\dagger] = 1\nn\ee
It is this second pair of creation operators that provide the degeneracy of the Landau levels. We define the ground state  $|0,0\rangle$ to be annihilated by both lowering operators, so that $a|0,0\rangle=b|0,0\rangle=0$. Then the general state in the Hilbert space is $|n,m\rangle$ defined by
\be |n,m\rangle = \frac{a^{\dagger\,n}b^{\dagger\,m}}{\sqrt{n!m!}}|0,0\rangle\nn\ee
The energy of this state is given by the usual Landau level expression \eqn{llevel}; it depends on $n$ but not on $m$.

\subsubsection*{The Lowest Landau Level}

Let's now construct the wavefunctions in the symmetric gauge. We're going to focus attention on the lowest Landau level, $n=0$, since this will be of primary interest when we come to discuss the fractional quantum Hall effect. The states in the lowest Landau level are annihilated by $a$, meaning $a|0,m\rangle = 0$
The trick is to convert this into a differential equation. The lowering operator is 
\be a &=& \frac{1}{\sqrt{2e\hbar B}}\left(\pi_x-i\pi_y\right)\nn\\ &=& \frac{1}{\sqrt{2e\hbar B}}\left(p_x-ip_y + e(A_x -i A_y)\right)
\nn\\ &=&  \frac{1}{\sqrt{2e\hbar B}}\left(-i\hbar\left(\frac{\partial}{\partial x} - i\frac{\partial}{\partial y}\right) + \frac{eB}{2}(-y - i x)\right) 
\nn\ee
At this stage, it's useful to work in complex coordinates on the plane. We introduce 
\be z = x-iy  \ \ \ {\rm and} \ \ \ \bar{z} = x+iy\nn\ee
Note that this is the opposite to how we would normally define these variables! It's annoying but it's because we want the wavefunctions below to be holomorphic rather than anti-holomorphic. (An alternative would be to work with magnetic fields $B<0$ in which case we get to use the usual definition of holomorphic. However, we'll stick with our choice above throughout these lectures). We also introduce the corresponding holomorphic and anti-holomorphic derivatives
\be \partial  = \frac{1}{2}\left(\frac{\partial}{\partial x} + i\frac{\partial}{\partial y}\right) \ \ \ {\rm and}\ \ \ \bar{\partial}  = \frac{1}{2}\left(\frac{\partial}{\partial x} - i\frac{\partial}{\partial y}\right) \nn\ee
which obey $\partial z= \bar{\partial} \bar{z} = 1$ and $\partial \bar{z} = \bar{\partial} z=0$. In terms of these holomorphic coordinates, $a$ takes the simple form 
\be a = -i\sqrt{2}\left( l_B \bar{\partial} + \frac{z}{4l_B} \right)\nn\ee
and, correspondingly, 
\be  a^\dagger = -i\sqrt{2}\left( l_B {\partial} - \frac{\bar{z}}{4l_B} \right)\nn\ee
which we've chosen to write in terms of the magnetic length $l_B=\sqrt{\hbar/eB}$. The lowest Landau level wavefunctions $\psi_{LLL}(z,\bar{z})$ are then those which are annihilated by this differential operator. But this is easily solved: they are
\be \psi_{LLL}(z,\bar{z})= f(z) \, e^{-|z|^2/4l_B^2}\nn\ee
for any holomorphic function $f(z)$. 

\para
We can construct the specific states $|0,m\rangle$ in the lowest Landau level by similarly writing $b$ and $b^\dagger$ as differential operators. We find
\be b = -i\sqrt{2}\left(l_B\partial + \frac{\bar{z}}{4l_B}\right)\ \ \ {\rm and}\ \ \ b^\dagger = -i\sqrt{2}\left(l_B\bar{\partial} - \frac{{z}}{4l_B}\right)\nn\ee
The lowest state $\psi_{LLL,m=0}$ is annihilated by both $a$ and $b$. There is a unique such state given by
\be \psi_{LLL,m=0} \sim e^{-|z|^2/4l_B^2}\nn\ee
We can now construct the higher states by acting with $b^\dagger$. Each time we do this, we pull down a factor of $z/2l_B$. This gives us a basis of lowest Landau level wavefunctions in terms of holomorphic monomials
\be \psi_{LLL,m}  \sim \left(\frac{z}{l_B}\right)^m e^{-|z|^2/4l_B^2}\label{lllwf}\ee
This particular basis of states has another advantage: these are eigenstates of angular momentum. To see this, we define angular momentum operator, 
\be J = i\hbar\left(x\ppp{}{y} - y\ppp{}{x}\right)
= \hbar(z\partial -\bar{z}\bar{\partial}) \label{angmomop}\ee
Then, acting on these lowest Landau level states we have
\be
J\psi_{LLL,m} = \hbar m\, \psi_{LLL,m}\nn\ee
The wavefunctions \eqn{lllwf} provide a basis  for the lowest Landau level. But it is a simple matter to extend this to write down wavefunctions for all high Landau levels; we simply need to act with the raising operator $a^\dagger = -i\sqrt{2}(l_B\partial - \bar{z}/4l_B)$. However, we won't have any need for the explicit forms of these higher Landau level wavefunctions in what follows. 
  
\subsubsection*{Degeneracy Revisited}

In symmetric gauge, the profiles of the wavefunctions \eqn{lllwf} form concentric rings around the origin. The higher the angular momentum $m$, the further out the ring. This, of course, is very different from the strip-like wavefunctions that we saw in Landau gauge \eqn{landaupsi}. You shouldn't read too much into this other than the fact that the profile of the wavefunctions is not telling us anything physical as it is not gauge invariant. 

\para
However, it's worth seeing how we can see the degeneracy of states in symmetric gauge. The wavefunction with angular momentum $m$ is peaked on a ring of radius $r = \sqrt{2m} l_B$. This means that in  a disc shaped region of area $A = \pi R^2$, the number of states is roughly (the integer part of) 
\be {\cal N} = R^2/2l_B^2= A/2\pi l_B^2 = \frac{eBA}{2\pi \hbar}\nn\ee
which agrees with our earlier result \eqn{degeneracy}.

\para
There is yet another way of seeing this degeneracy that makes contact with the classical physics. In Section \ref{classicalhallsec}, we reviewed the classical motion of particles in a magnetic field. They go in circles. The most general solution to the classical equations of motion is given by \eqn{classicalmotion}, 
\be x(t) = X - R\sin(\omega_B t + \phi)\ \ \ {\rm and}\ \ \ y(t) = Y + R\cos(\omega_B t+\phi)\label{motion2}\ee
Let's try to tally this with our understanding of the exact quantum states in terms of Landau levels. To do this, we'll think about the coordinates labelling the centre of the orbit $(X,Y)$ as quantum operators. We can rearrange \eqn{motion2} to give
\be X &=& x(t) + R\sin(\omega_B t+\phi) = x  - \frac{\dot{y}}{\omega_B} = x - \frac{\pi_y}{m\omega_B} \nn\\
Y &=& y(t) - R\cos(\omega_B t+\phi) = y  + \frac{\dot{x}}{\omega_B} = y  + \frac{\pi_x}{m\omega_B}
\label{guiding}\ee
This feels like something of a slight of hand, but the end result is what we wanted: we have the centre of mass coordinates in terms of familiar quantum operators. Indeed, one can check that under time evolution, we have
\be i\hbar\dot{X} = [X,H]  = 0 \ \ \ ,\ \ \ i\hbar\dot{Y}=[Y,H] = 0\label{guidingno}\ee
confirming the fact that these are constants of motion.

\para
The definition of the centre of the orbit $(X,Y)$ given above holds in any gauge. If we now return to  symmetric gauge we can replace the $x$ and $y$ coordinates appearing here with the gauge potential \eqn{sgauge}. We end up with
\be X = \frac{1}{eB}\left(2 eA_y- \pi_y\right) = -\frac{\tilde{\pi}_y}{eB} \ \ \ {\rm and}\ \ \  Y = \frac{1}{eB}( -2eA_x  + \pi_x) = \frac{\tilde{\pi}_x}{eB}\nn\ee

\DOUBLEFIGURE{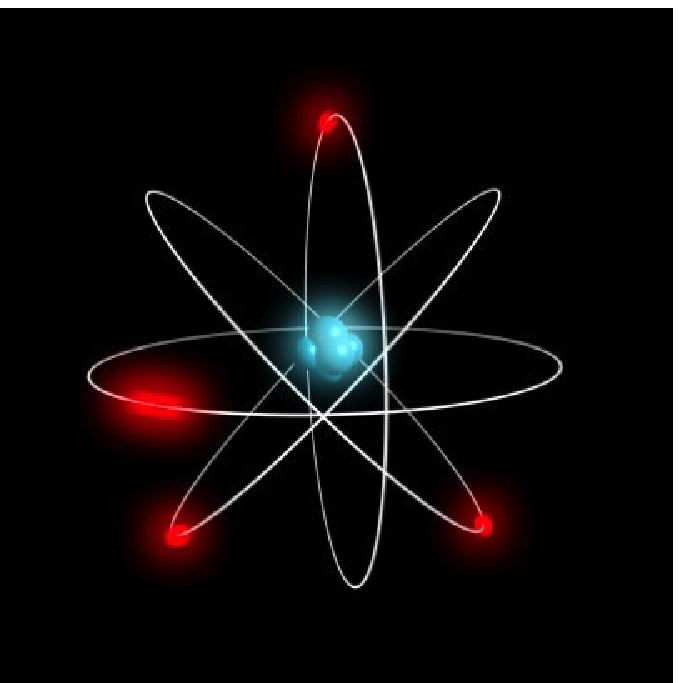,width=120pt}{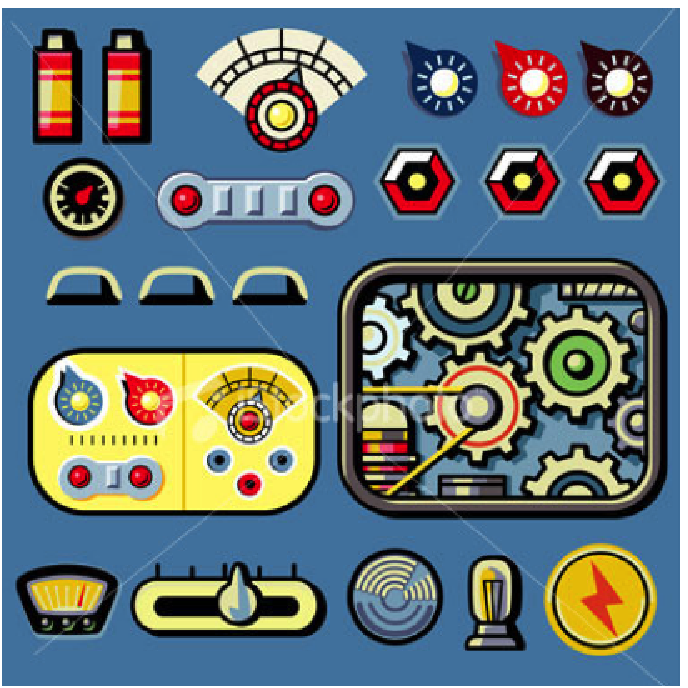,width=120pt}
{The  degrees of freedom $x$.}{The parameters $\lambda$.}
\noindent
where, finally, we've used the expression \eqn{tildepi} for the ``alternative momentum" $\tilde{\bpi}$. We see that, in symmetric gauge, this has the  alternative momentum has the nice interpretation of the centre of the orbit! The commutation relation \eqn{tildepicom} then tells us that the positions of the orbit in the $(X,Y)$ plane fail to commute with each other,
\be [X,Y] = i l_B^2\label{xyl}\ee
The lack of commutivity is precisely the magnetic length $l_B^2 = \hbar/eB$. The Heisenberg uncertainty principle now means that we can't localise states in both the $X$ coordinate and the $Y$ coordinate: we have to find a compromise. In general, the uncertainty is given by 
\be \Delta X \Delta Y = 2\pi l_B^2\nn\ee
A naive semi-classical count of the states then comes from taking the plane and parcelling it up into regions of area $2\pi l_B^2$. The number of states in an area $A$ is then
\be {\cal N} = \frac{A}{\Delta X\Delta Y}= \frac{A}{2\pi l_B^2} = \frac{eBA}{2\pi \hbar}\nn\ee
which is the counting that we've already seen above.

\subsection{Berry Phase}\label{berrysec}

There is one last topic that we need to review before we can start the story of the quantum Hall effect. This is the subject of Berry phase or, more precisely, the Berry holonomy\footnote{An excellent review of this subject can be found in the book {\it Geometric Phases in Physics} by Wilczek and Shapere}. This is not  a  topic which is relevant just in quantum Hall physics: it has  applications in many areas of  quantum mechanics and will arise over and over again in different guises in these lectures. Moreover, it is a topic which perhaps captures the spirit of the quantum Hall effect better than any other, for the Berry phase is the simplest demonstration of how geometry and topology can emerge from quantum mechanics. As we will see in these lectures, this is the heart of the quantum Hall effect.

\subsubsection{Abelian Berry Phase and Berry Connection}

We'll describe the Berry phase arising for a general Hamiltonian which we write as 
\be H(x^a;\lambda^i)\nn\ee
As we've illustrated, the Hamiltonian depends on two different kinds of variables. The $x^a$ are the degrees of freedom of the system. These are the things that evolve dynamically, the things that we want to solve for in any problem. They are typically things like the positions or spins of particles. 

\para
In contrast, the other variables $\lambda^i$ are the parameters of the Hamiltonian. They are fixed, with their values determined by some external apparatus that probably involves knobs and dials and flashing lights and things as shown above. We don't usually exhibit the dependence of $H$ on these variables\footnote{One exception is the classical subject of adiabatic invariants, where we also think about how $H$ depends on parameters $\lambda$. See section 4.6 of the notes on {\it Classical Dynamics}.}. 

\para
Here's the game. We pick some values for the parameters $\lambda$ and place the system in a specific  energy eigenstate $|\psi\rangle$ which,  for simplicity, we will take to be the ground state. We assume this ground state is unique (an assumption which we will later relax in Section \ref{nonabberrysec}). Now we very slowly vary the parameters $\lambda$. The Hamiltonian changes so, of course, the ground state also changes; it is $|\psi(\lambda(t))\rangle$. 
\para
There is a theorem in quantum mechanics called the {\it adiabatic theorem}. This states that if we place a system in a non-degenerate energy eigenstate and vary parameters sufficiently slowly, then the system will cling to that energy eigenstate. It won't be excited to any higher or lower states. 

\para
There is one caveat to the adiabatic theorem. 
How slow you have to be in changing the parameters depends on the energy gap from the state you're in to the nearest other state. This means that if you get {\it level crossing}, where another state becomes degenerate with the one you're in, then all bets are off. When the states separate again, there's no simple way to tell which linear combinations of the state you now sit in. However, level crossings are rare in quantum mechanics. In general, you have to tune three parameters to specific values in order to get two states to have the same energy. This follows by thinking about the a general Hermitian $2\times 2$ matrix which can be viewed as the Hamiltonian for the two states of interest. The general Hermitian $2\times 2$ matrix depends on 4 parameters, but its eigenvalues only coincide if it is proportional to the identity matrix. This means that three of those parameters have to be set to zero.

\para
The idea of the Berry phase arises in the following situation: we vary the parameters $\lambda$ but, ultimately, we put them back to their starting values. This means that we trace out a closed path in the space of parameters. We will assume that this path did not go through a point with level crossing. The question is: what state are we now in?

\para
The adiabatic theorem tells us most of the answer. If we started in the ground state, we also end up in the ground state. The only thing left uncertain is the phase of this new state
\be |\psi\rangle \rightarrow e^{i\gamma}|\psi\rangle\nn\ee
We often think of the overall phase of a wavefunction as being unphysical. But that's not the case here because this is a {\it phase difference}. For example, we could have started with two states and  taken only one of them on this journey while leaving the other unchanged. We could then interfere these two states and the phase $e^{i\gamma}$ would have physical consequence.

\para
So what is the phase $e^{i\gamma}$? There are two contributions. The first is simply the dynamical phase $e^{-iEt/\hbar}$ that is there for any energy eigenstate, even if the parameters don't change. But there is  also another, less obvious contribution to the phase. This is the {\it Berry phase}.

\subsubsection*{Computing the Berry Phase}

The wavefunction of the system evolves through the  time-dependent Schr\"odinger equation 
\be i\hbar\ppp{|\psi\rangle}{t} = H(\lambda(t))|\psi\rangle \label{seqn}\ee
For every choice of the parameters $\lambda$, we introduce a ground state with some fixed choice of phase. We call these reference states $|n(\lambda)\rangle$. There is no canonical way to do this; we just make an arbitrary choice. We'll  soon see how this choice affects the final answer.
The adiabatic theorem means that the ground state $|\psi(t)\rangle$ obeying \eqn{seqn} can be written as
\be |\psi(t)\rangle = U(t)\, |n(\lambda(t))\rangle \label{adev}\ee
where $U(t)$ is some time dependent phase. If we pick the $|n(\lambda(t=0))\rangle = |\psi(t=0)\rangle$ then we have $U(t=0)=1$. 
 Our task is then to determine $U(t)$ after we've taken $\lambda$ around the closed path and back to where we started. 
 
 \para
 There's always the dynamical  contribution to the phase, given by $e^{-i\int dt\, E_0(t)/\hbar}$ where $E_0$ is the ground state energy. This is not what's interesting here and we will ignore it simply by setting $E_0(t)=0$. However, there is an extra contribution. This arises by  
 plugging the adiabatic ansatz into \eqn{seqn}, and taking the overlap with $\langle\psi|$. We have
\be \langle\psi|\dot{\psi}\rangle = \dot{U}U^\star + \langle n|\dot{n}\rangle =0\nn\ee
where we've used the fact that, instantaneously, $H(\lambda) |n(\lambda)\rangle = 0$ to get zero on the right-hand side. (Note: this calculation is actually a little more subtle than it looks. To do a better job we would have to look more closely at corrections to the adiabatic evolution \eqn{adev}). This gives us an expression for the time dependence of the phase $U$,
\be U^\star \dot{U} = -\langle n|\dot{n}\rangle = -\langle n|\ppp{}{\lambda^i}|n\rangle\,\dot{\lambda}^i\label{uudot}\ee
It is useful to define the {\it Berry connection}
\be {\cal A}_i(\lambda) = -i\langle n|\ppp{}{\lambda^i}|n\rangle\label{berrycon}\ee
so that \eqn{uudot} reads
\be \dot{U} = -i {\cal A}_i \,\dot{\lambda}^i U\nn\ee
But this is easily solved. We have
\be U(t) = \exp\left(-i\int {\cal A}_i(\lambda)\,\dot{\lambda}^i \, dt\right)\nn\ee
Our goal is to compute the phase $U(t)$ after we've taken a closed path $C$ in parameter space. This is simply
\be e^{i\gamma} = \exp\left(-i\oint_C {\cal A}_i(\lambda)\,d\lambda^i\right)\label{berryphase}\ee
This is the {\it Berry phase}. Note that it doesn't depend on the time taken to change the parameters. It does, however, depend on the path taken through parameter space.

\subsubsection*{The Berry Connection}

Above we introduced the idea of the Berry connection \eqn{berrycon}. This is an example of a kind of object that you've seen before: it is like the gauge potential in electromagnetism! Let's explore this analogy a little further.

\para
In the relativistic form of electromagnetism, we have a gauge potential $A_\mu(x)$ where $\mu=0,1,2,3$ and $x$ are coordinates over Minkowski spacetime. There is a redundancy in the description of the gauge potential: all physics remains invariant under the gauge transformation
\be A_\mu \rightarrow A'_\mu = A_\mu + \partial_\mu \omega\label{gt}\ee
for any function $\omega(x)$. In our course on electromagnetism, we were taught that if we want to extract the physical information contained in $A_\mu$, we should compute the field strength 
\be F_{\mu\nu} = \ppp{A_\mu}{x^\nu} - \ppp{A_\nu}{x^\mu}\nn\ee
This contains the electric and magnetic fields. It is invariant under gauge transformations. 

\para
Now let's compare this to the Berry connection ${\cal A}_i(\lambda)$. Of course, this no longer depends on the coordinates of Minkowski space; instead it depends on the parameters $\lambda^i$. The number of these parameters is arbitrary; let's suppose that we have $d$ of them. This means that $i=1,\ldots, d$. In the language of differential geometry ${\cal A}_i(\lambda)$ is said to be a one-form over the space of parameters, while $A_i(x)$ is said to be a one-form over Minkowski space.

\para
There is also a redundancy in the information contained in the Berry connection ${\cal A}_i(\lambda)$. This follows from the arbitrary choice we made in fixing the phase of the reference states $|n(\lambda)\rangle$. We could just as happily have chosen a different set of reference states which differ by a phase. Moreover, we could pick a different phase for every choice of parameters $\lambda$, 
\be |n'(\lambda)\rangle =  e^{i\omega(\lambda)} \,|n(\lambda)\rangle\nn\ee
for any function $\omega(\lambda)$. If we compute the Berry connection arising from this new choice, we have
\be {\cal A}'_i  = -i\langle n'|\ppp{}{\lambda^i}|n'\rangle = {\cal A}_i + \ppp{\omega}{\lambda^i}\label{berrygt}\ee
This takes the same form as the gauge transformation \eqn{gt}.

\para
Following the analogy with electromagnetism, we might expect that the physical information in the Berry connection can be found in the gauge invariant field strength which, mathematically, is known as the {\it curvature} of the connection,
\be {\cal F}_{ij}(\lambda) = \ppp{{\cal A}_i}{\lambda^j}- \ppp{{\cal A}_j}{\lambda^i}\nn\ee
It's certainly true that ${\cal F}$ contains some physical information about our quantum system and we'll have use of this in later sections. But it's not the only gauge invariant quantity of interest. In the present context, the most natural thing to compute is the Berry phase \eqn{berryphase}. Importantly, this too is independent of the arbitrariness arising from the gauge transformation \eqn{berrygt}. This is because $\oint \partial_i \omega\,d\lambda^i = 0$. In fact, it's possible to write the Berry phase in terms of the field strength using the higher-dimensional version of Stokes' theorem
\be e^{i\gamma} =  \exp\left(-i\oint_C {\cal A}_i(\lambda)\,d\lambda^i\right) = \exp\left(-i\int_S {\cal F}_{ij}\,dS^{ij}\right)\label{berrygauss}\ee
where $S$ is a two-dimensional surface in the parameter space bounded by the path $C$.

%
%

\subsubsection{An Example: A Spin in a Magnetic Field}\label{berryspinsec}

The standard example of the Berry phase is very simple. It is a spin, with a Hilbert space consisting of just two states. The spin is placed in a magnetic field $\vec{B}$, with Hamiltonian which we take to be
\be H = -\vec{B}\cdot\vec{\sigma} +B \nn\ee
with $\vec{\sigma}$ the triplet of Pauli matrices and $B=|\vec{B}|$. The offset ensures that the ground state always has vanishing energy. Indeed, this Hamiltonian has two eigenvalues: $0$ and $+2B$. We denote the ground state as $\down$ and the excited state as $\up$,
\be H\down = 0 \ \ \ {\rm and}\ \ \ H\up = 2B\up\nn\ee
Note that these two states are non-degenerate as long as $\vec{B}\neq 0$. 

\para
We are going to treat the magnetic field as the parameters, so that $\lambda^i \equiv \vec{B}$ in this example. Be warned: this means that things are about to get confusing because we'll be talking about Berry connections ${\cal A}_i$ and curvatures ${\cal F}_{ij}$ over the space of magnetic fields. (As opposed to electromagnetism where we talk about magnetic fields over actual space). 

\para
The specific form of $\up$ and $\down$ will depend on the orientation of $\vec{B}$. To provide more explicit forms for these states, we write the magnetic field $\vec{B}$ in spherical polar coordinates
\be \vec{B} = \left(\begin{array}{c} B \sin\theta\cos\phi \\ B \sin\theta\sin\phi \\  B \cos\theta \end{array}\right)\nn\ee
with $\theta\in [0,\pi]$ and $\phi\in [0,2\pi)$
The Hamiltonian then reads
\be H =   -B\left(\begin{array}{cc} \cos\theta-1 &  e^{-i\phi}\sin\theta \\ e^{+i\phi}\sin\theta &\ \ -\cos\theta-1
 \end{array}\right)\nn\ee
 In these coordinates,  two normalised eigenstates are given by
 \be\down = \left(\begin{array}{c} e^{-i\phi}\sin\theta/2 \\ -\cos\theta/2\end{array}\right)\ \ \ {\rm and}\ \ \ 
 \up = 
 \left(\begin{array}{c}\ e^{-i\phi}\cos\theta/2 \\ \sin\theta/2\end{array}\right)\nn\ee
 These states play the role of our $|n(\lambda)\rangle$ that we had in our general derivation. Note, however, that they are not well defined for all values of $\vec{B}$. When we have $\theta=\pi$, the angular coordinate $\phi$ is not well defined. This means that $\down$ and  $\up$ don't have well defined phases. This kind of behaviour is typical of systems with non-trivial Berry phase. 
 
 \para
  We can easily compute the Berry phase arising from these states (staying away from $\theta=\pi$ to be on the safe side). We have
 \be {\cal A}_\theta = -i\langle\downarrow|\ppp{}{\theta}\down = 0\ \ \ {\rm and}\  \ \ {\cal A}_\phi =-i\langle\downarrow|\ppp{}{\phi}\down = -\sin^2\left(\frac{\theta}{2}\right)\nn\ee
 The resulting Berry curvature in polar coordinates is
 \be {\cal F}_{\theta\phi} = \ppp{{\cal A}_\phi}{\theta} - \ppp{{\cal A}_\theta}{\phi} = -\sin\theta\nn\ee
 This is simpler if we translate it back to cartesian coordinates where  the rotational symmetry is more manifest. It becomes
 \be {\cal F}_{ij}(\vec{B}) = -\epsilon_{ijk} \frac{B^k}{2|\vec{B}|^3}\nn\ee
 But this is interesting. It is a magnetic monopole! Of course, it's not a real magnetic monopole of electromagnetism: those are forbidden by the Maxwell equation. Instead it is, rather confusingly, a magnetic monopole in the space of magnetic fields.

    \DOUBLEFIGURE{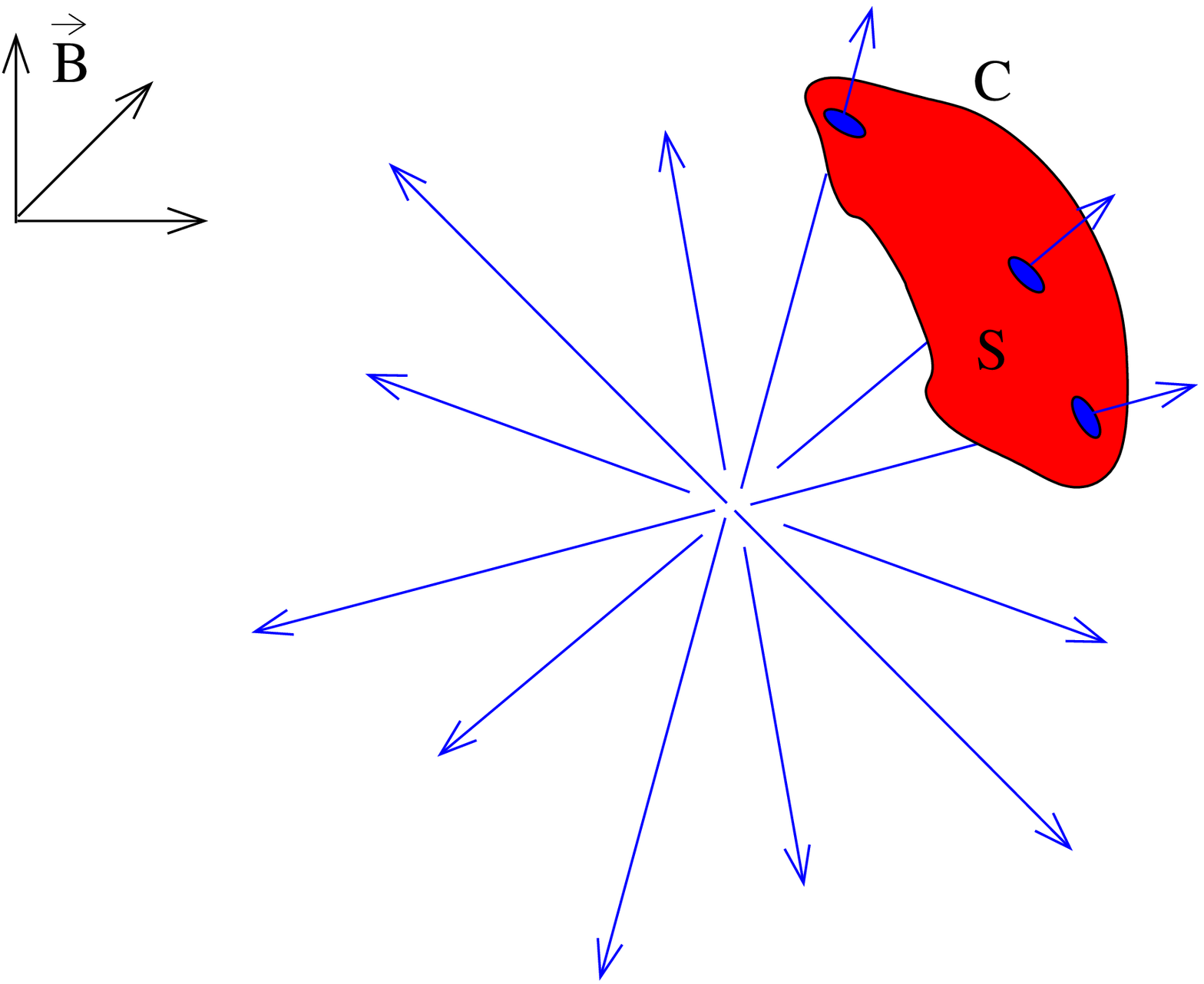,width=160pt}{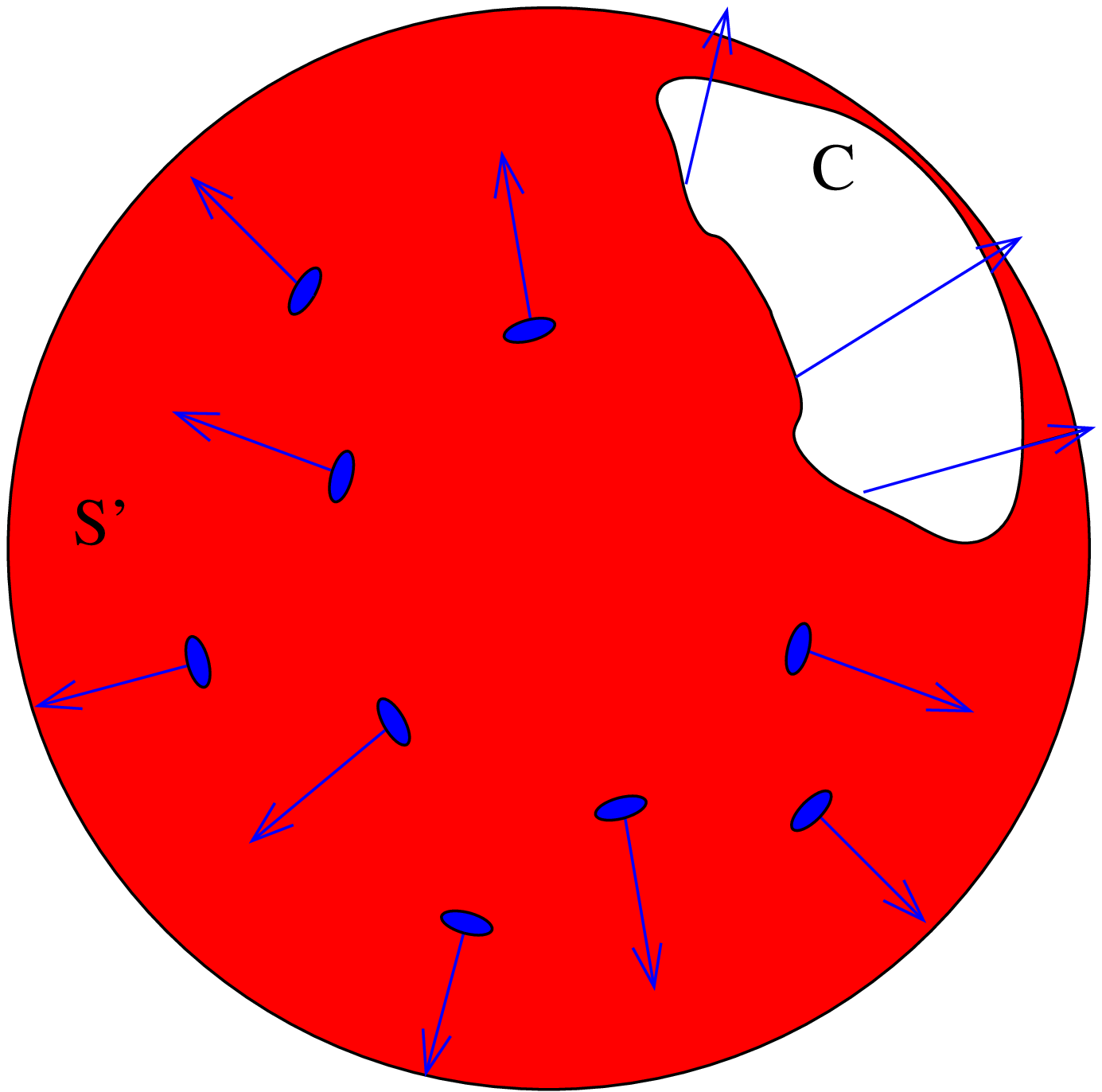,width=160pt}
{Integrating over $S$...}{...or over $S'$.}
 \para
 Note that the magnetic monopole sits at the point $\vec{B}=0$ where the two energy levels coincide. Here, the field strength is singular.  This is the point where we can no longer trust the Berry phase computation. Nonetheless, it is the presence of this level crossing and the resulting singularity which is dominating the physics of the Berry phase.

 \para
The magnetic monopole has charge $g=-1/2$, meaning that the integral of the Berry curvature  over any two-sphere ${\bf S}^2$ which surrounds the origin is
 \be \int_{{\bf S}^2} {\cal F}_{ij}\,dS^{ij} = 4\pi g = -2\pi\label{monopolecharge}\ee
Using this, we can easily compute the Berry phase for any path $C$ that we choose to take in the space of magnetic fields $\vec{B}$. We only insist that the path $C$ avoids the origin. 
 Suppose that the surface $S$, bounded by $C$, makes a solid angle $\Omega$. Then, using the form \eqn{berrygauss} of the Berry phase, we have
\be e^{i\gamma} =   \exp\left(-i\int_S {\cal F}_{ij}\,dS^{ij}\right) = \exp\left(\frac{i\Omega}{2}\right)\ee
 Note, however, that there is an ambiguity in this computation. We could choose to form $S$ as shown in the left hand figure. But we could equally well choose the surface $S'$ to go around the back of the sphere, as shown in the right-hand figure. In this case, the solid angle formed by $S'$ is $\Omega' = 4\pi - \Omega$. Computing the Berry phase using $S'$ gives
 \be e^{i\gamma'} =   \exp\left(-i\int_{S'} {\cal F}_{ij}\,dS^{ij}\right) = \exp\left(\frac{-i(4\pi - \Omega)}{2}\right)=e^{i\gamma}\ee
where the difference in sign in the second equality comes because the surface now has opposite orientation. 
So, happily, the two computations agree. Note, however, that this agreement requires that the charge of the monopole in \eqn{monopolecharge} is $2g\in {\bf Z}$. In the context of electromagnetism, this was Dirac's original argument for the quantisation of monopole charge. This quantisation extends to a general Berry curvature ${\cal F}_{ij}$ with an arbitrary number of parameters: the integral of the curvature over any closed surface must be quantised in units of $2\pi$, 
\be \int {\cal F}_{ij}\,dS^{ij} = 2\pi C\label{chernobyl}\ee
The integer $C\in {\bf Z}$ is called the {\it Chern number}.

\subsubsection{Particles Moving Around a Flux Tube}\label{absec}

 In our course on {\it Electromagentism}, we learned that the gauge potential $A_\mu$ is unphysical: the physical quantities that affect the motion of a particle are the electric and magnetic fields. 
 This statement is certainly true classically. Quantum mechanically, it requires some caveats. This is the subject of the Aharonov-Bohm effect. As we will show, aspects of the Aharonov-Bohm effect can be viewed as a special case of the Berry phase.

\para
The starting observation of the Aharonov-Bohm effect is that the gauge potential $\vec{A}$ appears in the Hamiltonian rather than the magnetic field $\vec{B}$. Of course, the Hamiltonian is invariant under gauge transformations so there's nothing wrong with this. Nonetheless, it does open up the possibility that the physics of a quantum particle can be sensitive to $\vec{A}$ in more subtle ways than a classical particle.

\subsubsection*{Spectral Flow} 

 \EPSFIGURE{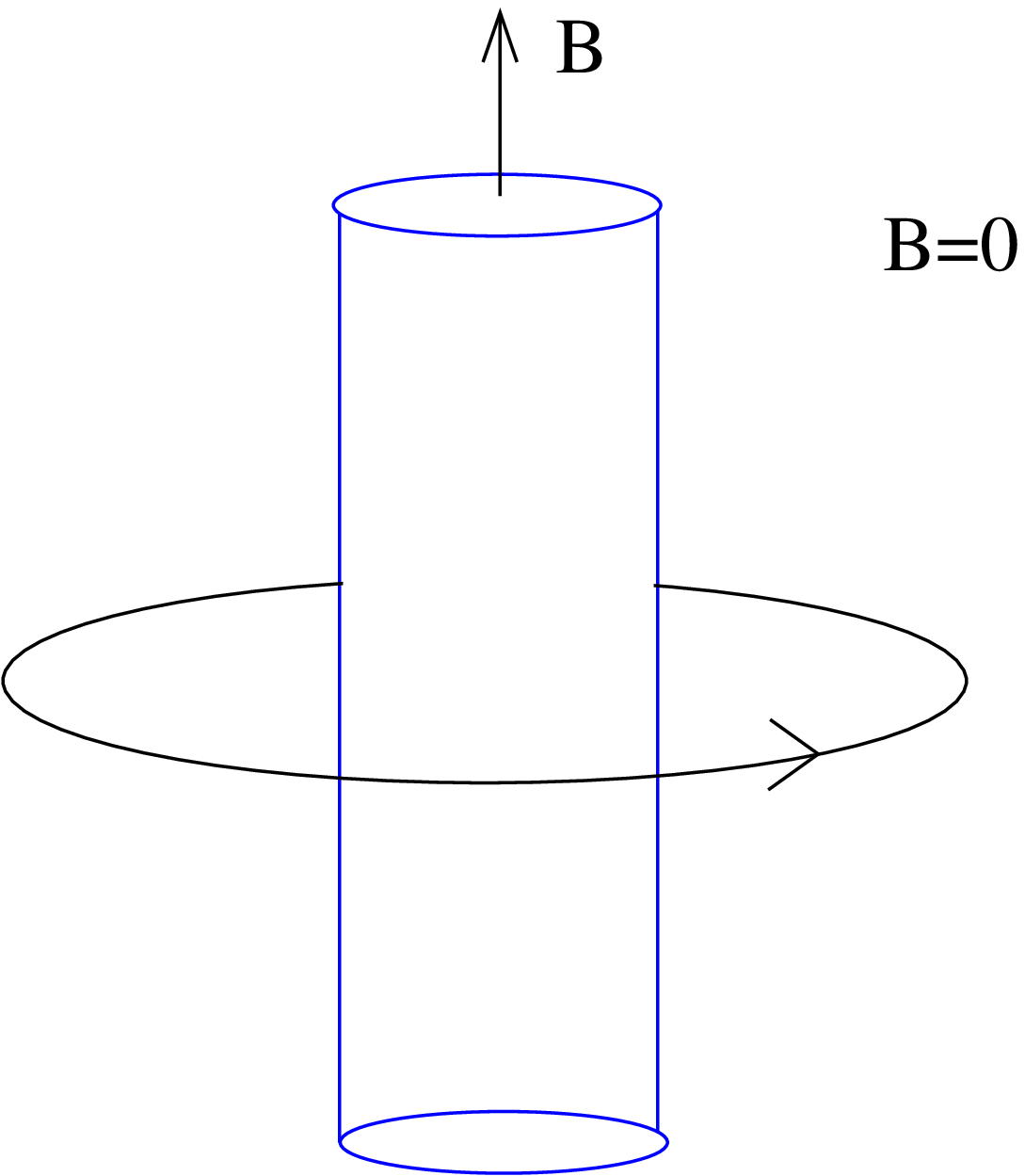,height=130pt}{A particle moving around a solenoid.}

To see how the gauge potential $\vec{A}$ can affect the physics, consider the set-up shown in the figure. We have a solenoid of area $A$, carrying magnetic field $\vec{B}$ and therefore magnetic flux $\Phi=BA$. Outside the solenoid the magnetic field is zero. However, the vector potential is not. This follows from Stokes' theorem which tells us that the line integral outside the solenoid is given by 
 \be \oint \vec{A}\cdot d\vec{r} = \int \vec{B}\cdot d\vec{S} = \Phi\nn\ee
 This is simply solved  in cylindrical polar coordinates by
 \be A_\phi = \frac{\Phi}{2\pi r} \nn\ee

    \begin{figure}[htb]
\begin{center}
\epsfxsize=3.2in\leavevmode\epsfbox{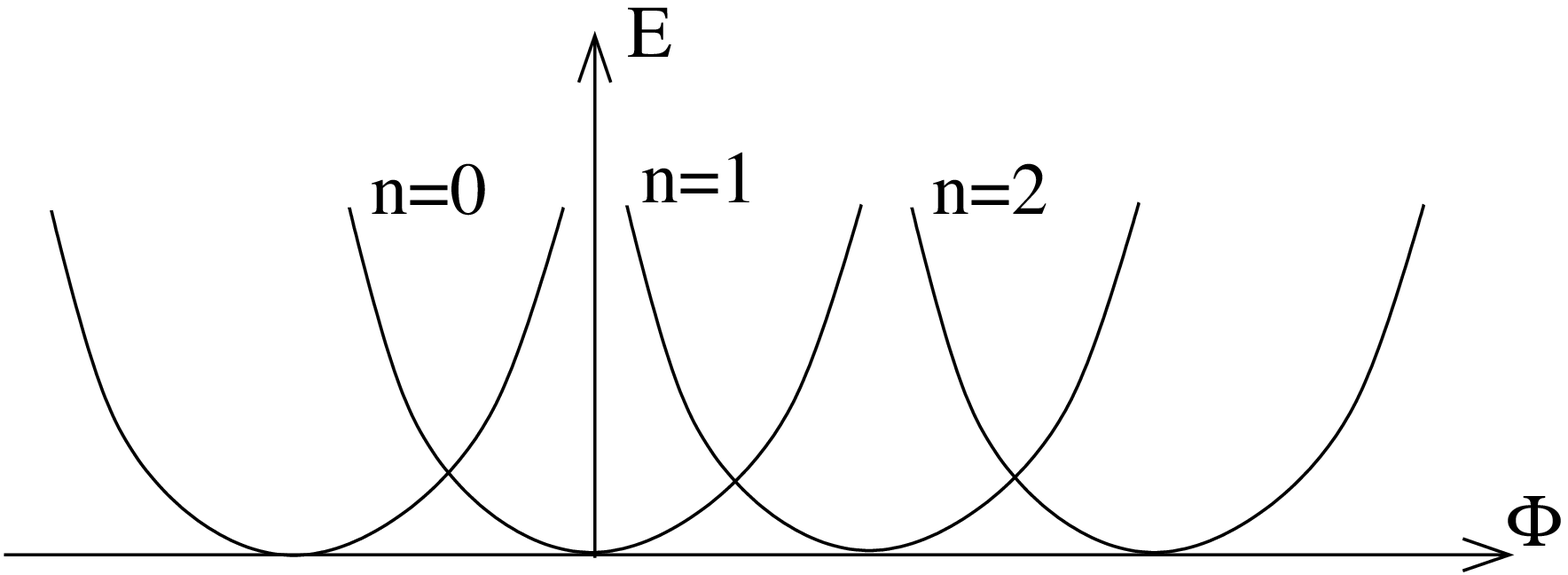}
\end{center}
\caption{The spectral flow for the energy states of a particle moving around a solenoid.}
\label{sflowfig}\end{figure}
\noindent
Now consider a charged quantum particle restricted to lie in a ring of radius $r$ outside the solenoid. The only dynamical degree of freedom is the angular coordinate $\phi\in [0,2\pi)$.  The Hamiltonian is
\be  H = \frac{1}{2m}\left(p_\phi + eA_\phi\right)^2 = \frac{1}{2mr^2}\left(-i\hbar\ppp{}{\phi} + \frac{e\Phi}{2\pi }\right)^2\nn\ee
We'd like to see how the presence of this solenoid affects the particle. The energy eigenstates are simply 
\be \psi = \frac{1}{\sqrt{2\pi r}} \,e^{in\phi}\ \ \ \ n\in {\bf Z}\nn\ee
where the requirement that $\psi$ is single valued around the circle means that we must take $n\in{\bf Z}$. Plugging this into the time independent Schr\"odinger equation $H\psi = E\psi$,  we find the spectrum
\be  E = \frac{1}{2mr^2}\left(\hbar n + \frac{e\Phi}{2\pi}\right)^2 = \frac{\hbar^2}{2mr^2}\left( n + \frac{\Phi}{\Phi_0}\right)^2\ \ \ \ n\in{\bf Z}
 \nn\ee
 Note that if $\Phi$ is an integer multiple of the quantum of flux $\Phi_0 = 2\pi\hbar/e$, then the spectrum is unaffected by the solenoid. But if the flux in the solenoid is not an integral multiple of $\Phi_0$ --- and there is no reason that it should be --- then the spectrum gets shifted. We see that the energy of the particle knows about the flux $\Phi$ even though the particle never goes near the region with magnetic field. The resulting energy spectrum is shown in Figure \ref{sflowfig}.

\para
Suppose now that we turn off the solenoid and place the particle in the $n=0$ ground state. Then we very slowly increase the flux. By the adiabatic theorem, the particle remains in the $n=0$ state. But,  by the time we have reached $\Phi=\Phi_0$, it is no longer in the ground state. It is now in the state that we previously labelled $n=1$. Similarly, each state $n$ is shifted to the next state, $n+1$. This is an example of a phenomenon is called {\it spectral flow}: under a change of parameter --- in this case $\Phi$ --- the spectrum of the Hamiltonian changes, or ``flows". As we change increase the flux by one unit $\Phi_0$ the spectrum returns to itself, but individual states have morphed into each other. We'll see related examples of spectral flow applied to the integer quantum Hall effect in Section \ref{pumpingsec}.

\subsubsection*{The Aharonov-Bohm Effect}

The situation described above smells like the  Berry phase story. We can cook up a very similar situation that demonstrates the relationship more clearly. Consider a set-up like the solenoid where the magnetic field is localised to some region of space. We again consider a particle which sits outside this region. However, this time we restrict the particle to lie in a small box. 
There can be some interesting physics going on inside the box; we'll capture this by including a potential $V({\vec x})$ in the Hamiltonian and, in order to trap the particle,  we take this potential to be infinite outside the box. 

\para
The fact that the box is ``small" means that the gauge potential  is approximately constant inside the box. If we place the centre of the box at position $\vec{x} = \vec{X}$, then the Hamiltonian of the system is then
\be H = \frac{1}{2m}(-i\hbar\nabla + e\vec{A}(\vec{X}))^2 + V(\vec{x}-\vec{X})\label{abmoving}\ee
We start by placing the centre of the box at position $\vec{x} = \vec{X}_0$ where we'll take the gauge potential to vanish: $\vec{A}(\vec{X}_0)=0$. 
 (We can always  do a gauge transformation to ensure that $\vec{A}$ vanishes at any point of our choosing). Now the Hamiltonian is of the kind that we solve in our first course on quantum mechanics. We will take the ground state to be
\be
\psi(\vec{x}-\vec{X}_0)\nn\ee
which is localised around $\vec{x}=\vec{X}_0$ as it should be. Note that  we have made a choice of phase in specifying this wavefunction. Now we slowly move the box in some path in space. In doing so, the gauge potential $\vec{A}(\vec{x}=\vec{X})$ experienced by the particle changes. It's simple to check that the Schr\"odinger equation for the Hamiltonian \eqn{abmoving} is solved by the state
\be \psi(\vec{x}-\vec{X}) = \exp\left(-\frac{ie}{\hbar} \int_{\vec{x}=\vec{X}_0}^{\vec{x}=\vec{X}} \vec{A}(\vec{x})\cdot d\vec{x}\right) \psi(\vec{x}-\vec{X}_0)\nn\ee
This works because when the $\nabla$ derivative hits the exponent, it brings down a factor which cancels the $e\vec{A}$ term in the Hamiltonian. We now play our standard Berry game: we take the box in a loop $C$ and bring it back to where we started. The wavefunction comes back to
\be
\psi(\vec{x}-\vec{X}_0) \ \rightarrow\  e^{i\gamma} \psi(\vec{x}-\vec{X}_0)\ \ \ {\rm with}\ \ \ 
e^{i\gamma} = 
\exp\left(-\frac{ie}{\hbar} \oint_C\vec{A}(\vec{x})\cdot d\vec{x}\right)\label{abphase1}\ee
 Comparing this to our general expression for the Berry phase, we see that in this particular context the Berry connection is actually identified with the electromagnetic potential,
 \be \vec{\cal A}(\vec{X}) = \frac{e}{\hbar}\, \vec{A}(\vec{x}=\vec{X})\nn\ee
 The electron has charge $q=-e$ but, in what follows, we'll have need to talk about particles with different charges. In general, if a particle of charge $q$ goes around a region containing flux $\Phi$, it will pick up an Aharonov-Bohm phase
 \be e^{iq\Phi/\hbar}\nn\ee
 This simple fact will play an important role in our discussion of the fractional quantum Hall effect.

\para
 There is an experiment which exhibits the Berry phase in the Aharonov-Bohm effect. It is a variant on the famous double slit experiment.  As usual, the particle can go through one of two slits. As usual, the wavefunction splits so the particle, in essence, travels through both. Except now, we hide a solenoid carrying magnetic flux $\Phi$ behind the wall. 
 The wavefunction of the particle is prohibited from entering the region of the solenoid, so the particle never experiences the magnetic field $\vec{B}$. Nonetheless, as we have seen, the presence of the solenoid induces  a phase different $e^{i\gamma}$   between particles that take the upper slit and those that take the lower slit. This phase difference  manifests itself as a change to the interference pattern seen on the screen. Note that  when $\Phi$ is an integer multiple of $\Phi_0$, the interference pattern remains unchanged; it is only sensitive to the fractional part of $\Phi/\Phi_0$.

\subsubsection{Non-Abelian Berry Connection}\label{nonabberrysec}

The Berry phase described above assumed that the ground state was unique. We now describe an important generalisation to the situation where the ground state is $N$-fold degenerate and remains so for all values of the parameter $\lambda$.

\para
 We should note from the outset that there's something rather special about this situation. If a Hamiltonian has an $N$-fold degeneracy then a generic perturbation will break this degeneracy. But here we want to change the Hamiltonian without breaking the degeneracy; for this to happen there usually has to be some symmetry protecting the states. We'll see a number of examples of how this can happen in these lectures. 

\para
We now play the same game that we saw in the Abelian case. We place the system in one of the $N$ degenerate ground states, vary the parameters in a closed path, and ask: what state does the system return to?

\para
This time the adiabatic theorem tells us only that the system clings to the particular energy eigenspace as the parameters are varied. But, now this eigenspace has $N$-fold degeneracy and the adiabatic theorem does not restrict how the state moves within this subspace. This means that, by the time we return the parameters to their original values, the state could lie anywhere within this $N$-dimensional eigenspace. We want to know how it's moved. This is no longer given just by a phase; instead we want to compute a unitary matrix $U\subset U(N)$. 

\para
We can compute this by following the same steps that we took for the Abelian Berry phase. To remove the boring, dynamical phase $e^{-iEt}$, we again assume that the ground state energy is $E=0$ for all values of $\lambda$. The time dependent Schr\"odinger equation is again
\be i\ppp{|\psi\rangle}{t} = H(\lambda(t))|\psi\rangle = 0\label{seqn2}\ee
This time, for every choice of parameters $\lambda$, we introduce an $N$-dimensional basis of ground states
\be |n^a(\lambda)\rangle\ \ \ \ a=1,\ldots,N\nn\ee
As in the non-degenerate case, there is no canonical way to do this. We could just as happily have picked any other choice of basis for each value of $\lambda$. We just pick one. We now think about how this basis evolves through the Schr\"odinger equation \eqn{seqn2}. We write
\be |\psi_a(t)\rangle = U_{ab}(t)\,|n_b(\lambda(t))\rangle\nn\ee
with $U_{ab}$ the components of a time-dependent unitary matrix $U(t)\subset U(N)$. Plugging this ansatz into \eqn{seqn2}, we have
\be |\dot{\psi}_a\rangle = \dot{U}_{ab}|n_b\rangle + U_{ab}|\dot{n}_b\rangle  =0\nn\ee
which, rearranging, now gives
\be U_{ac}^\dagger \dot{U}_{ab} = -\langle n_a|\dot{n}_b\rangle = -\langle n_a|\ppp{}{\lambda^i}|n_b\rangle\,\dot{\lambda}^i\label{uab}\ee
We again define a connection. This time it is a {\it non-Abelian Berry connection}, 
\be ({\cal A}_i)_{ba} = -i\langle n_a|\ppp{}{\lambda^i}|n_b\rangle\label{nonabconnection}\ee
We should think of ${\cal A}_i$ as an $N\times N$ matrix. It lives in the Lie algebra
 $u(N)$ and should be thought of as a $U(N)$ gauge connection over the space of parameters.

\para
The gauge connection ${\cal A}_i$ is the same kind of object that forms the building block of Yang-Mills theory. Just as in Yang-Mills theory, it suffers from an ambiguity in its definition. Here, the ambiguity arises from the arbitrary choice of basis vectors $|n_a(\lambda)\rangle$ for each value of the parameters $\lambda$. We could have quite happily picked a different basis at each point,
\be |n'_a(\lambda)\rangle = \Omega_{ab}(\lambda)\,|n_b(\lambda)\rangle\nn\ee
where $\Omega(\lambda) \subset U(N)$ is a unitary rotation of the basis elements. As the notation suggests, there is nothing to stop us picking different rotations for different values of the parameters so $\Omega$ can depend on $\lambda$. If we compute the Berry connection \eqn{nonabconnection} in this new basis, we find
\be {\cal A}'_i  = \Omega {\cal A}_i \Omega^\dagger + i \ppp{\Omega}{\lambda^i}\Omega^\dagger\label{nonabgt}\ee
This is precisely the gauge transformation of a $U(N)$ connection in Yang-Mills theory. Similarly, we can also construct the {\it curvature} or {\it field strength} over the parameter space,
\be {\cal F}_{ij} = \ppp{{\cal A}_i}{\lambda^j} - \ppp{{\cal A}_j}{\lambda^i}-i[{\cal A}_i,{\cal A}_j]\nn\ee
This too lies in the $u(N)$ Lie algebra. In contrast to the Abelian case, the field strength is not gauge invariant. It transforms as
\be {\cal F}'_{ij} = \Omega {\cal F}_{ij} \Omega^\dagger\nn\ee
Gauge invariant combinations of the field strength can be formed by taking the trace over the matrix indices. For example, ${\rm tr}\,{\cal F}_{ij}$, which tells us only about the $U(1)\subset U(N)$ part of the Berry connection, or traces of higher powers such as ${\rm tr}\, {\cal F}_{ij}{\cal F}_{kl}$. However, the most important gauge invariant quantity is the unitary matrix $U$ determined by the differential equation \eqn{uab}.

\para
The solution to \eqn{uab} is somewhat more involved than in the Abelian case because of ordering ambiguities of the matrix ${\cal A}_i$ in the exponential: the matrix at one point of parameter space, ${\cal A}_i(\lambda)$, does not necessarily commute with the matrix at anther point ${\cal A}_i(\lambda')$. However, this is a problem that we've met in other areas of physics\footnote{See, for example, the discussion of Dyson's formula in Section 3.1 of the \href{http://www.damtp.cam.ac.uk/user/tong/qft.html}{\it Quantum Field Theory} notes, or the discussion of rotations in Sections 3.1 and 3.7 of the \href{http://www.damtp.cam.ac.uk/user/tong/dynamics.html}{\it Classical Dynamics} lecture notes}. The solution is 
\be
U= {\cal P}\exp\left(-i\oint {\cal A}_i\,d\lambda^i\right)\nn\ee
Here ${\cal A}_i\subset u(N)$ is an $N\times N$ matrix. The notation ${\cal P}$ stands for ``path ordering". It means that we Taylor expand the exponential and then order the resulting products so that matrices ${\cal A}_i(\lambda)$ which appear later in the path are placed to the right. The result is the unitary matrix  $U\subset U(N)$ which tells us how the states transform. This unitary matrix is called the {\it Berry holonomy}. The holonomy is gauge invariant: it does not change under a transformation of the form \eqn{nonabgt}. 

\para
The non-Abelian Berry holonomy does not play a role in the simplest quantum Hall systems. But it will be important in more subtle quantum Hall states which, for obvious reasons, are usually called {\it non-Abelian quantum Hall states}. These will be discussed in Section \ref{nonabsec}\footnote{There are also examples of non-Abelian Berry holonomies unrelated to quantum Hall physics. I have a soft spot for a simple quantum mechanics system whose Berry phase is the BPS 't Hooft-Polyakov monopole. This was described in J. Sonner and D. Tong, ``{\it Scheme for Building a 't Hooft-Polyakov Monopole}", Phys. Rev. Lett 102, 191801 (2009), \href{http://arxiv.org/abs/0809.3783}{arXiv:0809.3783}.}.

\newpage
\section{The Integer Quantum Hall Effect}\label{iqhesec}

In this section we discuss the integer quantum Hall effect. This phenomenon can be understood without taking into account the interactions between electrons. This means that we will assume that the quantum states for a single particle in a magnetic field that we described in Section \ref{llevelsec} will remain the quantum states when there are many particles present. The only way that one particle knows about the presence of others is through the Pauli exclusion principle: they take up space. In contrast, when we come to discuss the fractional quantum Hall effect in Section \ref{fqhesec}, the interactions between electrons will play a key role.

\subsection{Conductivity in Filled Landau Levels}\label{llcondsec}

Let's look at what we know. The experimental data for the Hall resistivity shows a number of plateaux labelled by an integer $\nu$. Meanwhile, the energy spectrum forms Landau levels, also labelled by an integer. Each Landau level can accommodate a large, but finite number of electrons.
\DOUBLEFIGURE{ihall.eps,width=140pt}{ll.eps,width=140pt}
{Integer quantum Hall effect}{Landau levels}

\noindent
It's tempting to think that these integers are the same: $\rho_{xy} =  2\pi \hbar/e^2\nu$ and  when precisely $\nu$ Landau levels are filled. And this is correct.

\para
Let's first check that this simple guess works. If know that on a plateau, the Hall resistivity takes the value
\be \rho_{xy} = \frac{2\pi\hbar}{e^2}\,\frac{1}{\nu}\nn\ee
with $\nu\in {\bf Z}$.  But, from our classical calculation in the Drude model, we have the expectation that the Hall conductivity should depend on the density of electrons, $n$
\be \rho_{xy} = \frac{B}{ne}\nn\ee
Comparing these two expressions, we see that the density needed to get the resistivity of the  $\nu^{\rm th}$ plateau is 
\be n = \frac{B}{\Phi_0}\nu\label{easyjet}\ee
with $\Phi_0 = 2\pi\hbar/e$. This is indeed the density of electrons required to fill $\nu$ Landau levels.

\para
Further, when $\nu$ Landau levels are filled, there is a gap in the energy spectrum: to occupy the next state  costs an  energy  $\hbar\omega_B$ where $\omega_B = e B/m$ is the cyclotron frequency. As long as we're at temperature $k_B T \ll \hbar \omega_B$, these states will remain empty. When we turn on a small electric field, there's nowhere for the electrons to move: they're stuck in place like in an insulator. This means that the scattering time $\tau\rightarrow \infty$ and we have $\rho_{xx}=0$ as expected.

\subsubsection*{Conductivity in Quantum Mechanics: a Baby Version}

The above calculation involved a curious mixture of quantum mechanics and the classical Drude mode. We can do better. Here we'll describe how to compute the conductivity for a single free particle. In section \ref{kubosec}, we'll derive a more general formula that holds for any many-body quantum system.

\para
We know that the velocity of the particle is given by
\be m\dot{\bf x} ={\bf p} + e {\bf A}\nn\ee
where $p_i$ is the canonical momentum. The current is ${\bf I} = -e \dot{\bf x}$, which means that, in the quantum mechanical picture, the total current is given by
\be {\bf I}  = -\frac{e}{m}\sum_{{\rm filled\ states}} \langle\psi| - i\hbar\nabla + e{\bf A}|\psi\rangle
\nn\ee
It's best to do these kind of calculations in Landau gauge, ${\bf A}= xB\hat{\bf y}$. We introduce an electric field $E$ in the $x$-direction so the Hamiltonian is given by \eqn{eham} and the states by \eqn{shiftedstate}. With the $\nu$ Landau levels filled, the current in the $x$-direction is
\be I_x &=& -\frac{e}{m}\sum_{n=1}^\nu \sum_{k} \langle\psi_{n,k}|-i\hbar \frac{\partial}{\partial x}|\psi_{n,k}\rangle =0 \nn\ee
This vanishes because it's computing the momentum expectation value of harmonic oscillator eigenstates. Meanwhile, the current in the $y$-direction is
\be
I_y &=& -\frac{e}{m}\sum_{n=1}^\nu\sum_k\langle\psi_{n,k}|-i\hbar\frac{\partial}{\partial y} + exB|\psi_{n,k}\rangle
=-\frac{e}{m}\sum_{n=1}^\nu\sum_k\langle\psi_{n,k}| \hbar k + eBx|\psi_{n,k}\rangle
\nn\ee
The second term above is computing the position expectation value $\langle x \rangle$ of the eigenstates. But we know from \eqn{landaupsi} and \eqn{shiftedstate} that these harmonic oscillator states are shifted from the origin, so that $\langle\psi_{n,k}|x|\psi_{n,k}\rangle = -\hbar k/eB + mE/eB^2$.  The first of these terms cancels the explicit $\hbar k$ term in the expression for $I_y$. We're left with 
\be I_y = -e\nu\sum_k \frac{E}{B} \label{allcontribute}\ee
The sum over $k$ just gives the number of electrons which we computed in \eqn{degeneracy} to be  $N = AB/\Phi_0$. We divide through by the area to get the current  density ${\bf J}$ instead of the current ${\bf I}$. The upshot of this is that 
\be {\bf E} = \left(\begin{array}{c} E \\ 0 \end{array}\right)\ \ \ \Rightarrow\ \ \ {\bf J} = \left(\begin{array}{c}0 \\ -e\nu E/{\Phi_0}\end{array}\right)\nn\ee
Comparing to the definition of the conductivity tensor \eqn{conductivity}, we have
\be  \sigma_{xx}=0\ \ \ {\rm and}\ \ \ \sigma_{xy} = \frac{e\nu}{\Phi_0} \ \ \ \Rightarrow\ \ \ \rho_{xx}=0\ \ \ {\rm and}\ \ \ \rho_{xy} = -\frac{\Phi_0}{e\nu} = - \frac{2\pi\hbar}{e^2\nu}
\label{hallquantised}\ee
This is exactly the conductivity seen on the quantum Hall plateaux. Although the way we've set up our computation we get a negative Hall resistivity rather than positive; for a magnetic field in the opposite direction, you get the other sign.

\subsubsection{Edge Modes}

\EPSFIGURE{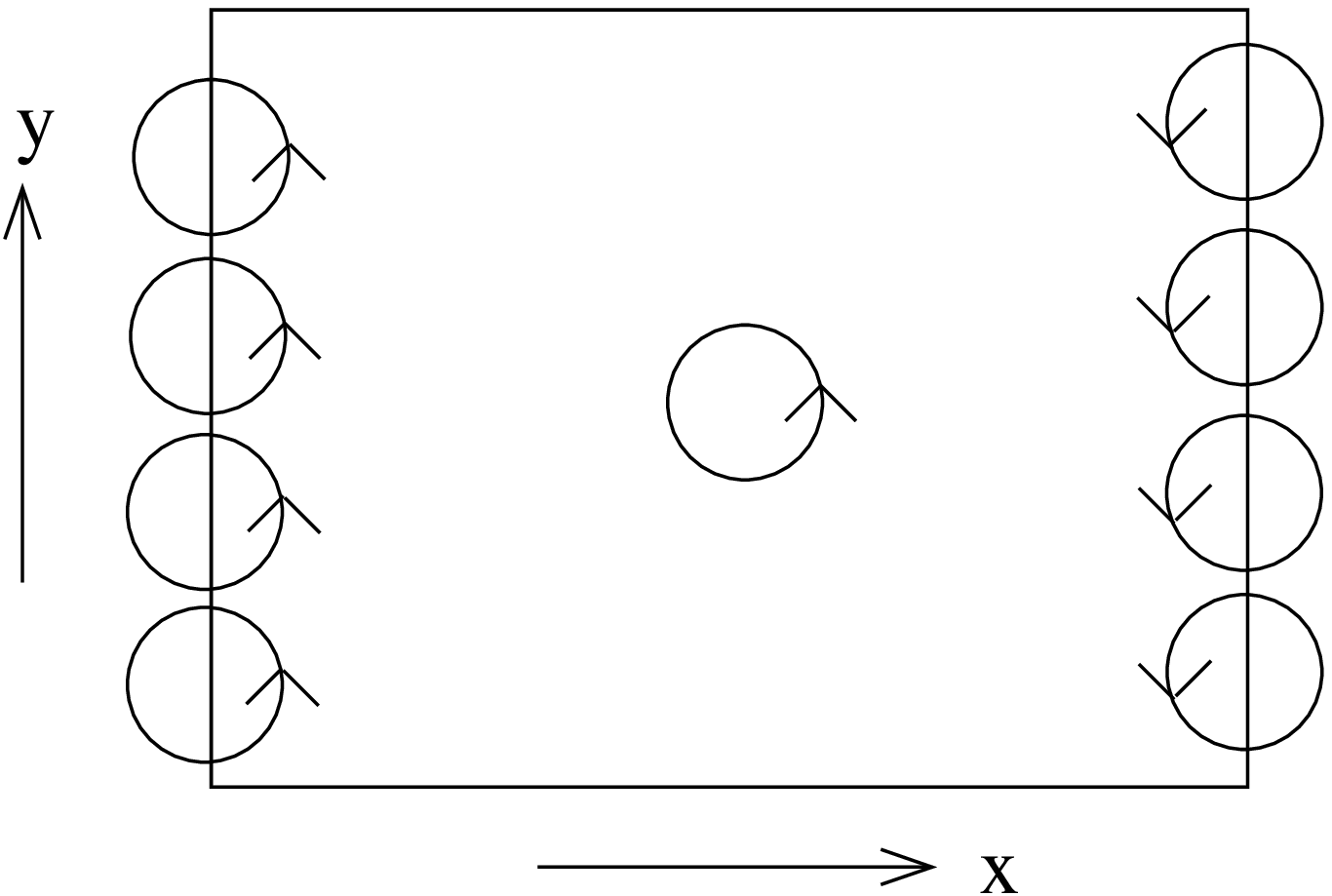,height=90pt}{}
There are a couple of aspects of the story which the simple description above does not capture. One of these is the role played by disorder; we describe this in Section \ref{disordersec}. The other is the special importance of modes at the edge of the system. Here we describe some basic facts about edge modes; we'll devote Section \ref{edgesec} to a more detailed discussion of edge modes in the fractional quantum Hall systems.

\para
The fact that something special happens along the edge of a quantum Hall system can be seen even classically. 
Consider particles moving  in circles in a magnetic field. For a fixed magnetic field, all particle motion is in one direction, say anti-clockwise. Near the edge of the sample, the orbits must collide with the boundary. As all motion is anti-clockwise, the only option open to these particles is to bounce back. The result is a skipping motion in which the particles along the one-dimensional boundary move only in a single direction, as shown in the figure. A particle restricted to  move in a single direction along a line is said to be {\it chiral}. 
Particles move  in one direction on one side of the sample, and in the other direction on the other side of the sample. We say that the particles have opposite {\it chirality} on the two sides. This ensures that the net current, in the absence of an electric field, vanishes.

\EPSFIGURE{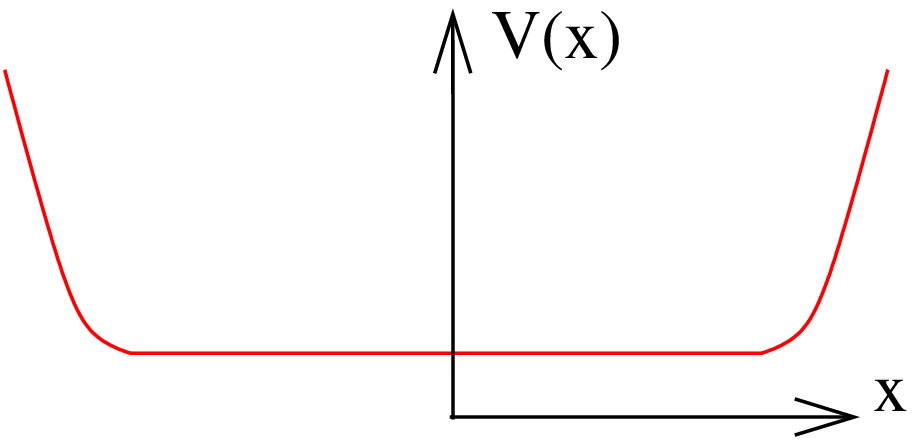,height=70pt}{}
\para
We can also see how the edge modes appear in the quantum theory. The edge of the sample is modelled by a potential which rises steeply as shown in the figure. We'll work in Landau gauge and consider a rectangular geometry which is finite only in the $x$-direction, which we model by $V(x)$. The Hamiltonian is
\be H = \frac{1}{2m}\left(p_x^2 + (p_y+eBx)^2\right) + V(x)\nn\ee
In the absence of the potential, we know that the wavefunctions are Gaussian of width $l_B$. If the potential is smooth over distance scales $l_B$, then, for each state,  we can Taylor expand the potential around its location  $X$. Each   wavefunction then experiences the potential $V(x) \approx V(X) + (\partial V/\partial x) (x-X)+\ldots$. We drop quadratic terms and, of course, the constant term can be neglected. We're left with a linear potential which  is exactly what we solved in Section \ref{turningesec} when we discussed Landau levels in a background electric field. The result is a drift velocity in the $y$-direction \eqn{drifting}, now given by
\be v_y = -\frac{1}{eB}\ppp{V}{x}\nn\ee
Each wavefunction, labelled by momentum $k$, sits at a different $x$ position, $x=-kl_B^2$ and has a different drift velocity. In particular,  the modes at each edge are both chiral, travelling in opposite directions: $v_y>0$ on the left, and $v_y<0$ on the right. This agrees with the classical result of skipping orbits. 
 
\para
Having a chiral mode is rather special. In fact, there's a theorem which says that you can't have charged chiral particles moving along a wire; there {\it has} to be particles which can move in the opposite direction as well. In the language of field theory, this follows from what's called the {\it chiral anomaly}. In the language of condensed matter physics, with particles moving on a lattice, it follows from the {\it Nielsen-Ninomiya theorem}. The reason that the simple example of a particle in a magnetic field avoids these theorems is because the chiral fermions live on the boundary of a two-dimensional system, rather than in a one-dimensional wire. This is part of a general story: there are physical phenomena which can only take place on the boundary of a system. This story plays a prominent role in the study of materials called {\it topological insulators}.

\para
Let's now look at what happens when we  fill the available states. We do this by introducing a chemical potential. The states are labelled by $y$-momentum $\hbar k$ but, as we've seen, this can equally well be thought of as the position of the state in the $x$-direction. This means that we're justified in drawing the filled states like this:
 \be   \raisebox{-6.5ex}{\epsfxsize=2.2in\epsfbox{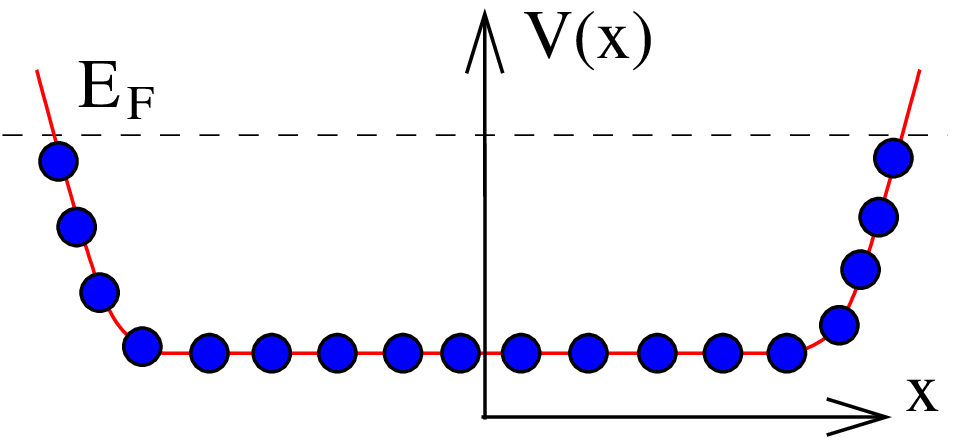}}\nn\ee
From our usual understanding of insulators and conductors, we would say that the bulk of the material is an insulator (because all the states in the band are filled) but the edge of the material is a metal. We can also think about currents in this language. We simply introduce a potential difference $\Delta\mu$ on the two sides of the sample. This means that we fill up more states on the right-hand edge than on the left-hand edge, like this:
 \be   \raisebox{-6.5ex}{\epsfxsize=2.4in\epsfbox{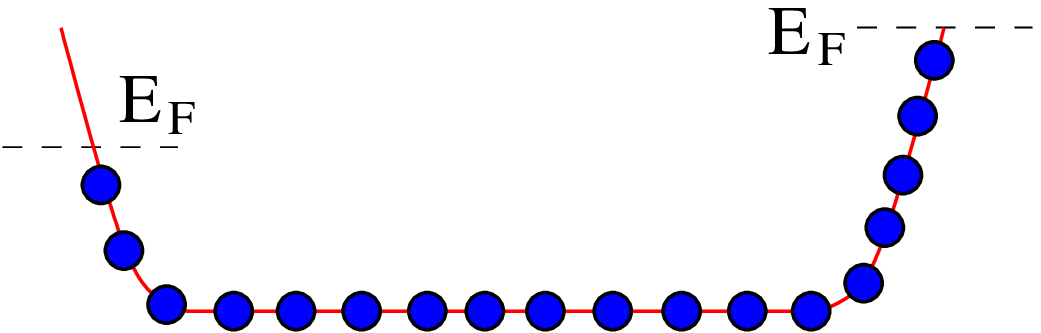}}\nn\ee
To compute the resulting current we simply need to sum over all filled states. But, at the level of our approximation, this is the same as integrating over $x$
\be I_y = -e\int \frac{dk}{2\pi} \, v_y(k) = \frac{e}{2\pi l_B^2} \int dx\, \frac{1}{eB}\ppp{V}{x} = \frac{e}{2\pi\hbar}\,\Delta \mu\label{vanything}\ee
The Hall voltage is $eV_H = \Delta\mu$, giving us the Hall conductivity 
\be \sigma_{xy} = \frac{I_y}{V_H} = \frac{e^2}{2\pi\hbar}\label{condyetagain}\ee
which is indeed the expected conductivity for a single Landau level.

\para
The picture above suggests that the current is carried entirely by the edge states, since the bulk Landau level is flat so these states carry no current. Indeed, you can sometimes read this argument in the literature. But it's  a little too fast:  in fact, it's even in  conflict with the computation that we did previously, where \eqn{allcontribute} shows that all states contribute equally to the current. That's because this calculation included the fact that the Landau levels are tilted by an electric field, so that the effective potential and the filled states looked  more like this:
 \be   \raisebox{1ex}{\epsfxsize=2.4in\epsfbox{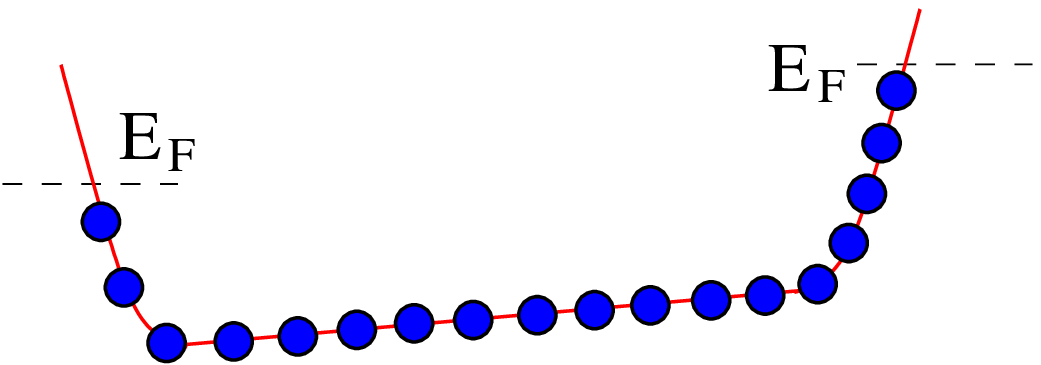}}\nn\ee
Now the current is shared among all of the states. However, the nice thing about the calculation \eqn{vanything} is that it doesn't matter what shape the potential $V$ takes. As long as it is smooth enough, the resulting Hall conductivity remains quantised as \eqn{condyetagain}. For example, you could consider the random potential like this
 \be   \raisebox{1ex}{\epsfxsize=2.4in\epsfbox{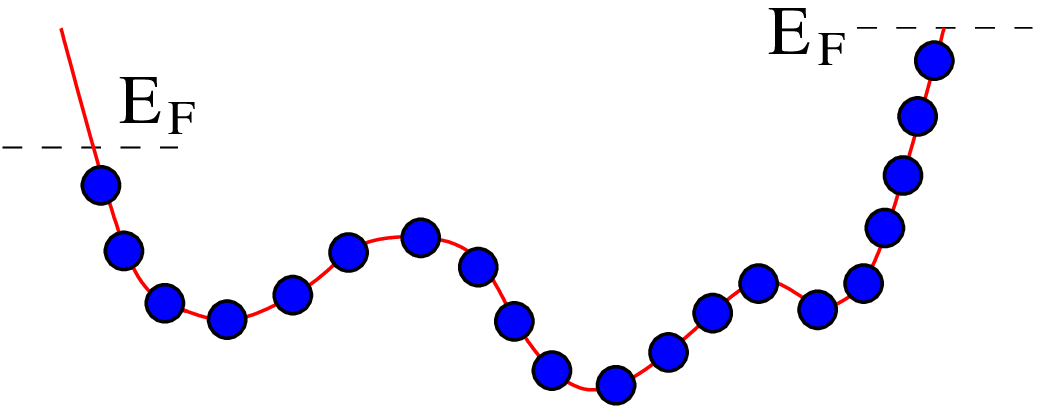}}\nn\ee
and you still get the quantised answer \eqn{vanything} as long as the random potential $V(x)$ doesn't extend above $E_F$. As we will describe in Section \ref{disordersec}, these kinds of random potentials introduce another ingredient that is crucial in understanding the quantised Hall plateaux.
\para
Everything we've described above holds for a single Landau level. It's easily generalised to multiple Landau levels. As long as the chemical potential $E_F$ lies between Landau levels, we have $n$ filled Landau levels, like this
 \be   \raisebox{0.5ex}{\epsfxsize=2.4in\epsfbox{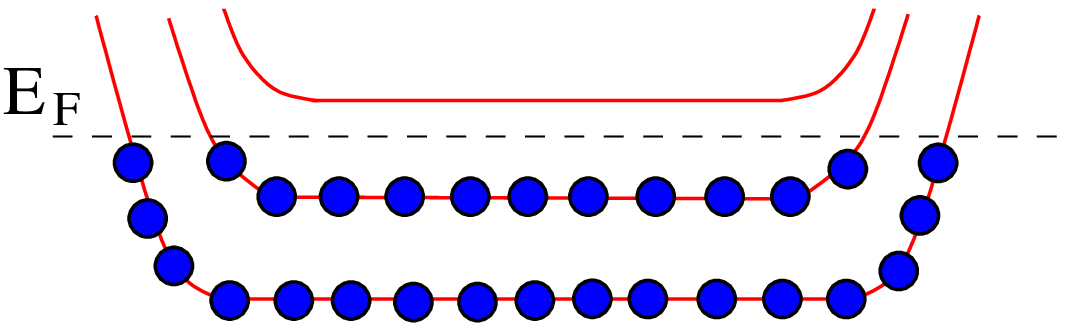}}\nn\ee
Correspondingly, there are $n$ types of chiral mode on each edge. 

\para
A second reason why chiral modes are special is that it's hard to disrupt them. If you add impurities to any system, they will scatter electrons. Typically such scattering makes the electrons bounce around in random directions and the net effect is often that the electrons don't get very far at all. But for chiral modes this isn't possible simply because all states move in the same direction. If you want to scatter a left-moving electron into a right-moving electron then it has to cross the entire sample. That's a long way for an electron and, correspondingly, such scattering is highly suppressed. It means that currents carried by chiral modes are immune to impurities. However, as we will now see, the impurities do play an important role in the emergence of the Hall plateaux.

\subsection{Robustness of the Hall State}

The calculations above show that {\it if} an integer number of Landau levels are filled, then the longitudinal and Hall resistivities are those observed on the plateaux. But it doesn't explain why these plateaux exist in the first place, nor why there are sharp jumps between different plateaux. 

\para
To see the problem, suppose that we fix the electron density $n$. Then we only completely  fill Landau levels when the magnetic field is exactly $B = n\Phi_0/\nu$ for some integer $\nu$. But what happens the rest of the time when  $B\neq n\Phi_0/\nu$? Now the final Landau level is only partially filled. Now when we apply a small electric field, there are accessible states for the electrons to scatter in to. The result is going to be some complicated, out-of-equilibrium distribution of electrons on this final Landau level. The longitudinal conductivity $\sigma_{xx}$ will surely be non-zero, while the Hall conductivity will differ from the quantised value \eqn{hallquantised}. 

\para
Yet the whole point of the quantum Hall effect is that the experiments reveal that the quantised values of the resistivity \eqn{hallquantised} persist over a range of magnetic field. How is this possible?

\subsubsection{The Role of Disorder}\label{disordersec}

It turns out that the plateaux owe their existence to one further bit of physics: disorder. This arises because experimental samples are inherently dirty. They contain impurities which can be modelled by adding a random potential $V({\bf x})$ to the Hamiltonian. As we now explain, this random potential is ultimately responsible for the plateaux observed in the quantum Hall effect. There's a wonderful irony in this: the glorious precision with which these integers $\nu$ are measured is due to the dirty, crappy physics of impurities. 

\DOUBLEFIGURE{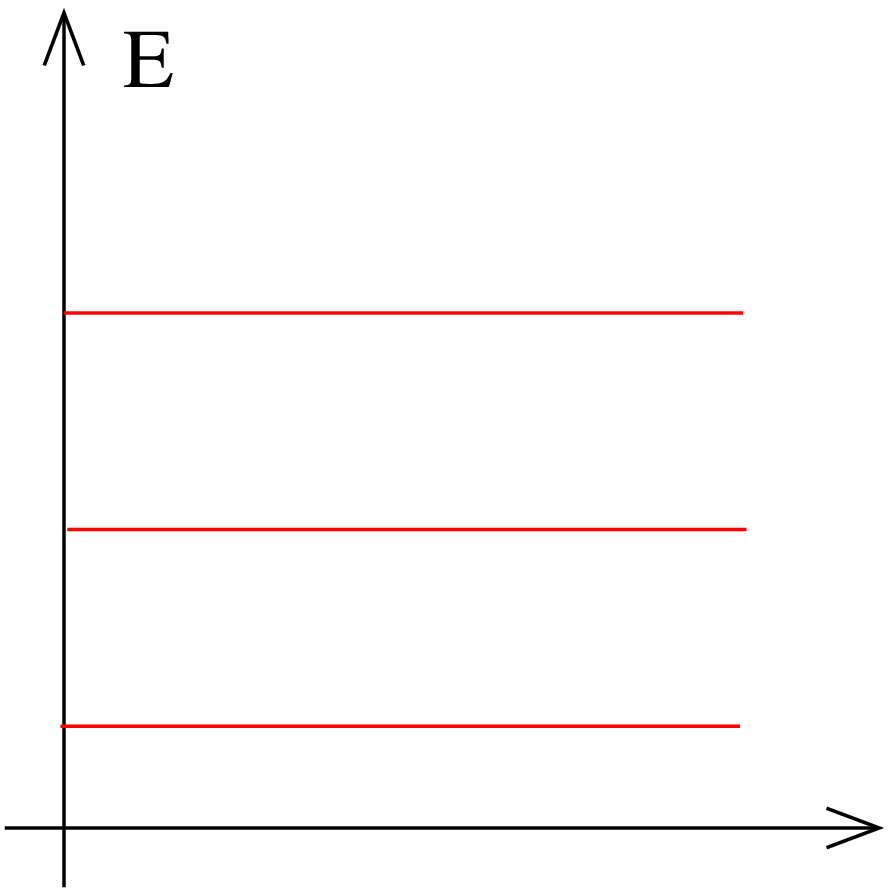,width=120pt}{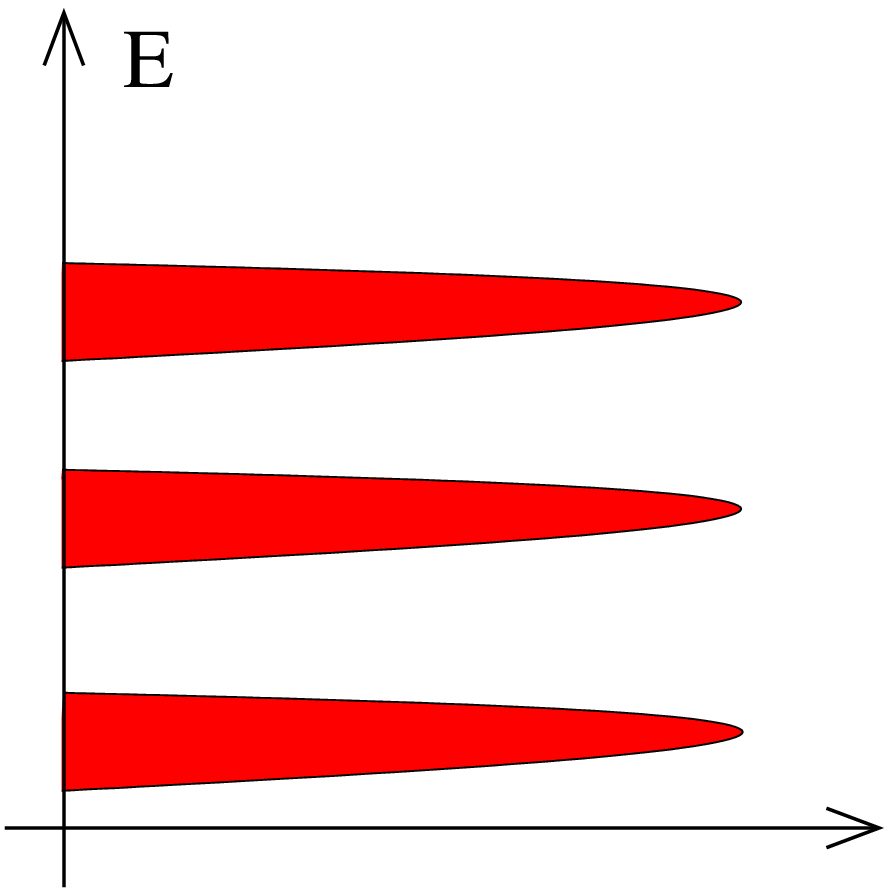,width=120pt}
{Density of states without disorder...}{...and  with disorder.}

\para
To see how this works, let's think about what disorder will likely do to the system. Our first expectation is that it will split the degenerate eigenstates that make up a Landau level. This follows on general grounds from quantum perturbation theory: any generic perturbation, which doesn't preserve a symmetry, will break degeneracies. We will further ask that the strength of disorder is small relative to the splitting of the Landau levels,
\be V \ll \hbar \omega_B\label{vsmallomega}\ee
In practice, this means that the samples which exhibit the quantum Hall effect actually have to be very clean. We need disorder, but not too much disorder! The energy spectrum in the presence of this weak disorder is the expected to change the quantised Landau levels from the familiar picture in the left-hand figure, to the more broad spectrum shown in the right-hand figure.

\para
There is a second effect of disorder: it turns many of the quantum states from {\it extended} to {\it localised}. Here, an extended state is spread throughout the whole system. In contrast, a localised state is restricted to lie in some region of space.  We can easily see the existence of these localised states in a semi-classical picture which holds if the potential, in addition to obeying \eqn{vsmallomega}, varies appreciably on distance scales much greater than the magnetic length $l_B$,
\be |\nabla V|\ll \frac{\hbar\omega_B}{l_B}\nn\ee
With this assumption, the cyclotron orbit of an electron takes place in a region of essentially constant potential. The centre of the orbit, ${\bf X}$ then drifts along equipotentials. 
To see this, recall that we can introduce quantum operators $(X,Y)$ describing the centre of the orbit \eqn{guiding},
\be X = x - \frac{\pi_y}{m\omega_B}\ \ \ {\rm and}\ \ \ Y = y + \frac{\pi_x}{m\omega_B}\nn\ee
\DOUBLEFIGURE{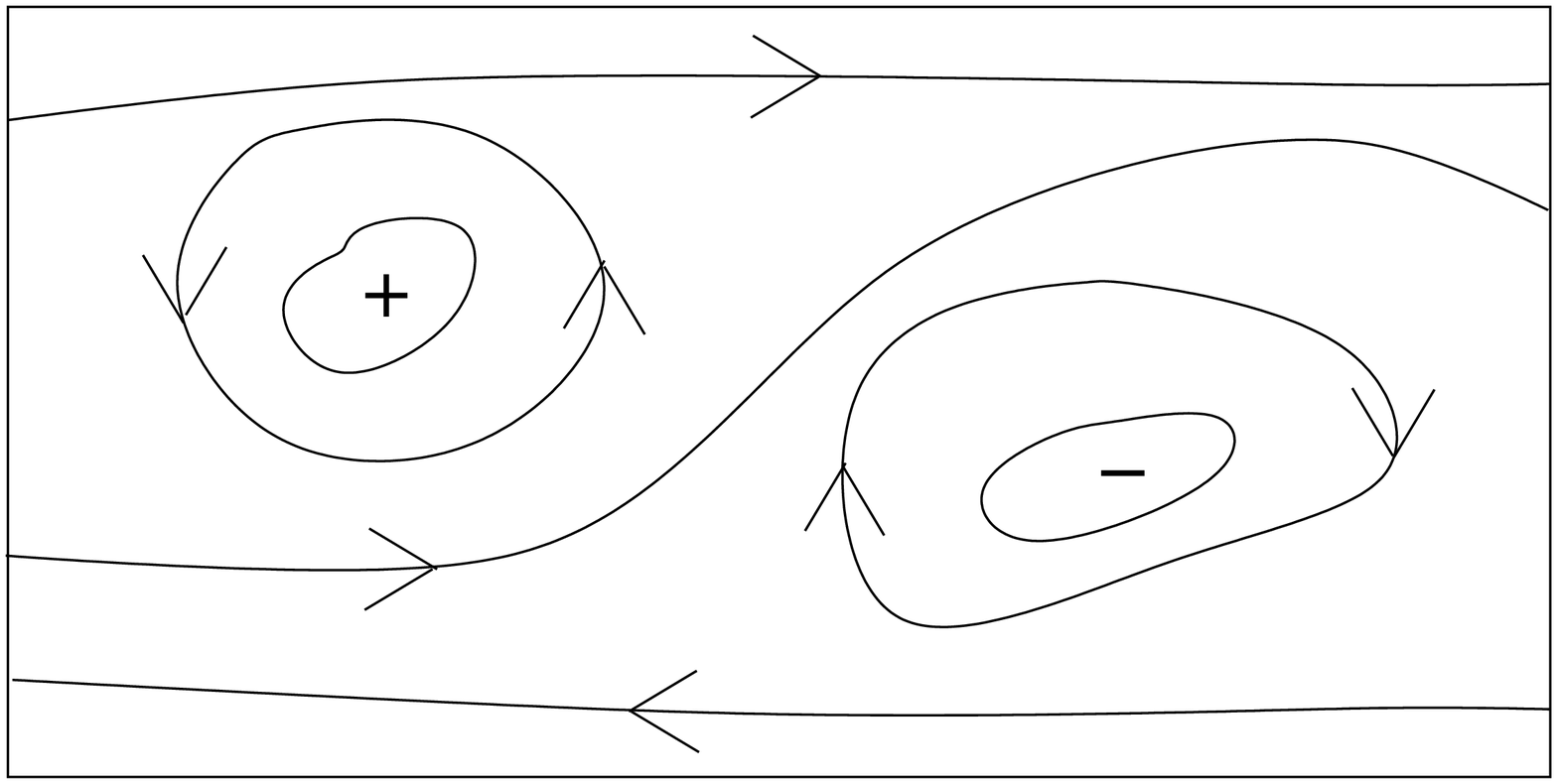,width=160pt}{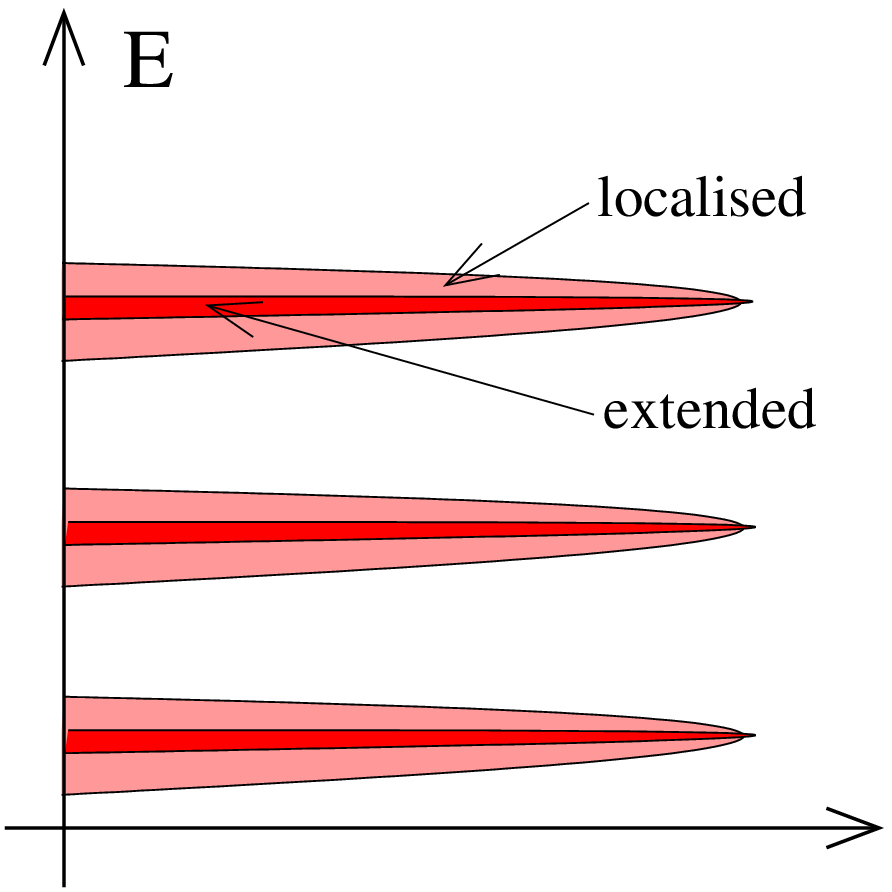,width=140pt}
{The localisation of states due to disorder.}{The resulting density of states.}

\noindent
with $\bpi$ the  mechanical momentum \eqn{mmom}. (Recall that, in contrast to the canonical momentum, $\bpi$ is gauge invariant).  The time evolution of these operators is given by
\be
i\hbar\dot{X} &=& [X,H+V] = [X,V] = [X,Y] \,\ppp{V}{Y}= il_B^2\ppp{V}{Y}  
\nn\\ i\hbar\dot{Y}&=& [Y,H+V] = [Y,V] = [Y,X] \ppp{V}{X} = -il_B^2 \ppp{V}{X}
\nn\ee
where we used the fact \eqn{guidingno} that, in the absence of a potential, $[X,H] = [Y,H] = 0$, together with the commutation relation $[X,Y] = il_B^2$ \eqn{xyl}.  This says that the centre of mass drifts in a direction $(\dot{X},\dot{Y})$ which is perpendicular to $\nabla V$; in other words, the motion is along equipotentials.

\para
Now consider what this means in a random potential with various peaks and troughs. We've drawn some contour lines of such a potential in the left-hand figure, with $+$ denoting the local maxima of the potential and $-$ denoting the local minima. The particles move anti-clockwise around the maxima and clockwise around the minima. In both cases, the particles are trapped close to the extrema. They can't move throughout the sample. In fact, equipotentials which stretch from one side of a sample to another are relatively rare. One place that they're guaranteed to exist is on the edge of the sample. 

\para
The upshot of this is that the states at the far edge of a band --- either of high or low energy --- are localised. Only the states close to the centre of the band will be extended. This means that the density of states looks schematically something like the right-hand figure.
%
%

\subsubsection*{Conductivity Revisited}

For conductivity, the distinction between localised and extended states is an important one. Only the extended states can transport charge from one side of the sample to the other. So only these states can contribute to the conductivity.

 \para
Let's now see what kind of behaviour we expect for the conductivity. Suppose that we've filled all the extended states in a given Landau level and consider what happens as we  decrease $B$ with fixed $n$. Each Landau level can accommodate fewer electrons, so the Fermi energy will increase. But rather than jumping up to the next Landau level, we now begin to populate the localised states. Since these states can't contribute to the current, the conductivity doesn't change. This leads to exactly the kind of plateaux that are observed, with constant conductivities over a range of magnetic field.

\para
So the presence of disorder explains the presence of plateaux. But now we have to revisit our original argument of why the resistivities take the specific quantised values \eqn{hallquantised}. These were computed assuming that all states in the Landau level contribute to the current. Now we know that many of these states are localised by impurities and don't transport charge. Surely we expect the value of the resistivity to be different. Right? Well, no. Remarkably, current carried by the extended states increases to compensate for the lack of current transported by localised states. This ensures that the resistivity remains quantised as \eqn{hallquantised} despite the presence of disorder. We now explain why.

\subsubsection{The Role of Gauge Invariance}\label{pumpingsec}

\EPSFIGURE{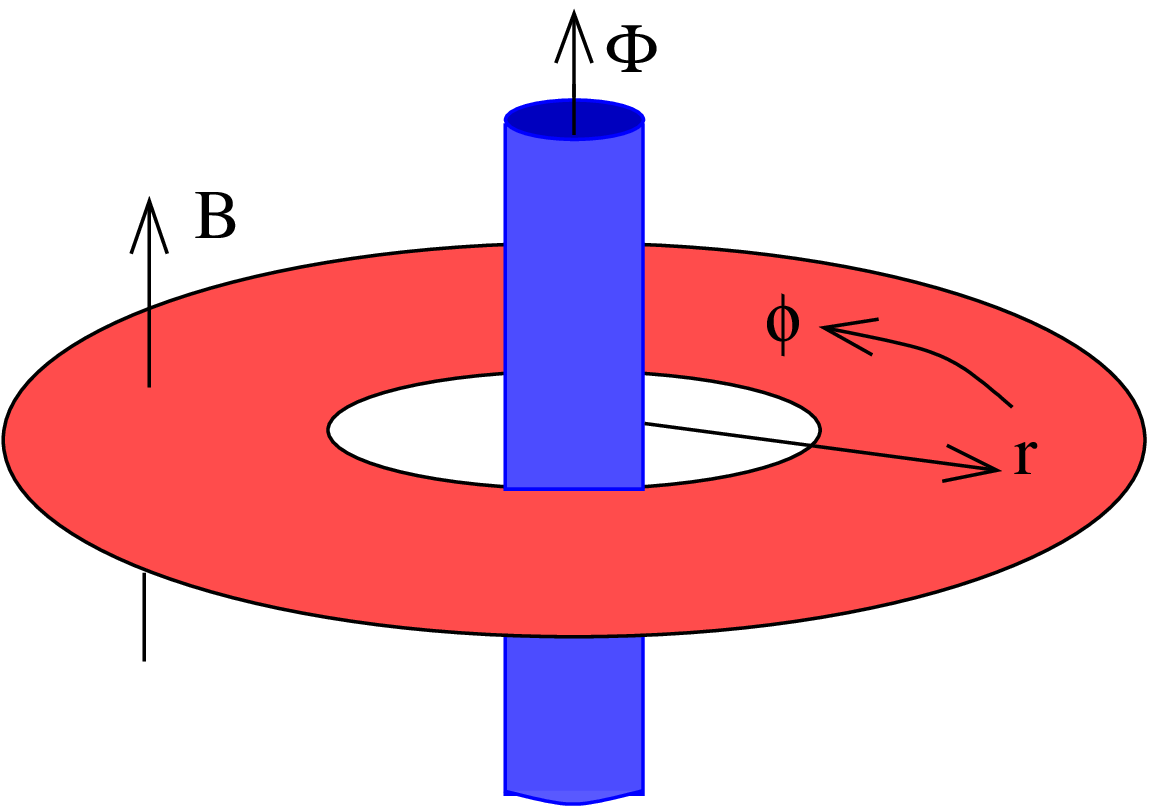,height=90pt}{}
\noindent
Instead of considering electrons moving in a rectangular sample, we'll instead consider electrons moving in the annulus shown in the figure. In this context, this is sometimes called a {\it Corbino ring}. We usually console ourselves by arguing that if the Hall conductivity is indeed quantised then it shouldn't depend on the geometry of the sample. (Of course, the flip side of this is that if we've really got the right argument, that shouldn't depend on the geometry of the sample either; unfortunately this argument does.)

\para
The nice thing about the ring geometry is that it provides us with an extra handle\footnote{This argument was first given by R. B. Laughlin in ``{\it Quantized Hall Conductivity in Two Dimensions}", \href{http://journals.aps.org/prb/abstract/10.1103/PhysRevB.23.5632}{Phys. Rev, {\bf B23} 5632 (1981)}. Elaborations on the role of edge states were given by  B.~I.~Halperin in ``{\it Quantized Hall conductance, current carrying edge states, and the existence of extended states in a two-dimensional disordered potential},''  \href{http://journals.aps.org/prb/abstract/10.1103/PhysRevB.25.2185}{Phys. Rev. {\bf B25} 2185 (1982)}.}. In addition to the background magnetic field $B$ which penetrates the sample, we can thread an additional flux $\Phi$ through the centre of the ring. Inside the ring, this $\Phi$ is locally pure gauge. Nonetheless, from our discussion in Section \ref{berrysec}, we known that $\Phi$ can affect the quantum states of the electrons. 

\para
Let's first see what $\Phi$ has to do with the Hall conductivity. Suppose that we slowly increase $\Phi$ from $0$ to $\Phi_0=2\pi\hbar/e$. Here ``slowly" means that we take a time $T \gg 1/\omega_B$. This induces an emf around the ring, ${\cal E} = -\partial \Phi/\partial t = -\Phi_0/T$.
Let's suppose that we can argue that $n$ electrons are transferred from the inner circle to the outer circle in this time. This would result in a radial current $I_r = -ne/T$. 
\be \rho_{xy} = \frac{{\cal E}}{I_r} = \frac{2\pi\hbar}{e^2}\frac{1}{n}\label{alexasleep}\ee
This is the result we want. Our task, therefore, is to argue that $n$ electrons are indeed transferred across the ring as the flux is increased to $\Phi_0$.

\subsubsection*{Spectral Flow in Landau Levels}

The key idea that we need is that of {\it spectral flow}, introduced in Section \ref{absec}. The spectrum of the Hamiltonian is the same whenever $\Phi$ is an integer multiple of $\Phi_0$. However, if we start with a particular energy eigenstate when $\Phi=0$, this will evolve into a different energy eigenstate with $\Phi=\Phi_0$. As the change is done suitably slowly, over a time $T\gg 1/\omega_B$, the adiabatic theorem ensures that the final energy eigenstate must lie in the same Landau level as the initial state.

\para
To illustrate this, let's first look at the situation with no disorder. 
For the ring geometry, it is sensible to use symmetric gauge and radial coordinates, $z=x-iy = re^{i\phi}$. The wavefunctions in the lowest Landau level are \eqn{lllwf},
\be \psi_m \sim z^m e^{-|z|^2/4l_B^2} = e^{im\phi} r^m e^{-r^2/4l_B^2 }\nn\ee
The $m^{\rm th}$ wavefunction is  strongly peaked at a radius $r\approx \sqrt{2ml_B^2}$ (where, of course, we must now chose $m\in {\bf Z}$ such that the wavefunction lies inside the annulus). From the discussion in Section \ref{absec}, we see that if we increase the flux from $\Phi=0$ to $\Phi=\Phi_0$, the wavefunctions shift from $m$ to $m+1$,
\be \psi_m(\Phi = 0) \ \longrightarrow \ \psi_m(\Phi=\Phi_0) = \psi_{m+1}(\Phi=0)\nn\ee
This means that each state moves outwards, from radius $r= \sqrt{2ml_B^2}$ to $r=\sqrt{2(m+1)l_B^2}$. The net result is that, if all states in the Landau level are filled, a single electron is transferred from the inner ring to the outer ring as the flux is increased from $\Phi=0$ to $\Phi=\Phi_0$. It is simple to check that the same result holds for higher Landau levels. If $n$ Landau levels are filled, then $n$ electrons are transferred from the inner to the outer ring and the Hall resistivity is given by \eqn{alexasleep} as required.

\subsubsection*{Spectral Flow in the Presence of Disorder}

The discussion above merely reproduces what we already know. Let's now see how it changes in the presence of disorder. In polar coordinates, the Hamiltonian takes the form
\be H_{\Phi=0} = \frac{1}{2m}\left[-\hbar^2\frac{1}{r}\ppp{}{r}\left(r\ppp{}{r}\right) + \left(-\frac{i\hbar}{r}\ppp{}{\phi} + \frac{eBr}{2}\right)^2 \right] + V(r,\phi)\nn\ee
where $V(r,\phi)$ is the random potential capturing the effects of disorder. Note that this depends on $\phi$, so angular momentum is no longer a good quantum number in this system. Adding the flux through the centre changes the Hamiltonian to 
\be H_{\Phi} = \frac{1}{2m}\left[-\hbar^2\frac{1}{r}\ppp{}{r}\left(r\ppp{}{r}\right) + \left(-i\frac{\hbar}{r}\ppp{}{\phi} + \frac{eBr}{2} + \frac{e\Phi}{2\pi r}\right)^2\right] + V(r,\phi)\nn\ee
Importantly, the flux $\Phi$ affects only the extended states. It does not change the localised states. To see this, we attempt to undo the flux  by a gauge transformation, 
\be
\psi(r,\phi) \rightarrow e^{-i e\Phi \phi/2\pi\hbar}\psi(r,\phi)\nn\ee
For the localised states, where $\psi$ is non-zero only in some finite region, there's no problem in doing this. However for the extended states, which wrap around the annulus, we also have the requirement that the wavefunction $\psi$ is single-valued as $\phi\rightarrow \phi + 2\pi$. We see that this is only true when $\Phi$ is an integer multiple of $\Phi_0= 2\pi\hbar/e$.

\para
The upshot of this argument is that the spectrum of the Hamiltonian is again left unchanged when $\Phi$ is an integer multiple of $\Phi_0$. But, this time, as the flux is increased from $0$ to $\Phi_0$, the localised states don't change. Only the extended states undergo spectral flow; these alone must map onto themselves.

\para
There are always at least two extended states: one near the inner ring and one near the outer ring. The spectral flow happens in 
the same heuristic manner as described above: an extended state localised at one radius is transformed into an extended state at the next available radius. The presence of disorder means that there are fewer extended states, but this doesn't change the overall conclusion: if all extended states in a given Landau level are filled, then the net effect of dialling the flux from $\Phi=0$ to $\Phi=\Phi_0$ is to transport one electron from the inner to the outer edge. If $n$ Landau levels are filled, we again get the result \eqn{alexasleep}.

\para
The arguments above involving gauge transformations start to give a hint of the topological nature of the quantum Hall effect. In fact, there are much deeper topological ideas underlying the quantisation of the Hall conductivity. We'll describe these in Section \ref{avronsec} and, in a slightly different context, in Section \ref{latticesec}. However, before we proceed we first need a basic result which gives an expression for the conductivity in any quantum mechanical system.

\subsubsection{An Aside: The Kubo Formula}\label{kubosec}

Before we get to anything related to topology, we first need to lay some groundwork. Our task in this section will be to derive a formula for the Hall conductivity $\sigma_{xy}$ in terms of quantum mechanical observables.  The expression that we're looking for is called the {\it Kubo formula}; it is part of  more general story that goes by the name of {\it linear response}\footnote{You can read about this story in the lecture notes on \href{http://www.damtp.cam.ac.uk/user/tong/kinetic.html}{\it Kinetic Theory} where a slightly more sophisticated discussion of the Kubo formula can be found in Sections 4.3 and 4.4. In particular, there is often an extra term proportional to ${\bf A}^2$ in $\Delta H$ which contributes to $\sigma_{xx}$ but not $\sigma_{xy}$ so is ignored in the present discussion.}.

\para
We'll derive the Kubo formula for a general, multi-particle Hamiltonian   $H_0$ where the subscript $0$  means that this is the unperturbed Hamiltonian before we turn on an electric field. At this point, $H_0$ could be that of many non-interacting particles each, for example, obeying the single-particle Hamiltonian  \eqn{ham} that we saw previously, or it could be something more complicated. Later,  we'll  apply the Kubo formula both to Hamiltonians which describe particles moving in the continuum and to Hamiltonians that describe particles moving on a lattice. 
We denote the energy eigenstates of $H_0$ as $|m\rangle$, with $H_0|m\rangle=E_m|m\rangle$.

\para
Now we add a background electric field. We work in the gauge with $A_t=0$ so that the electric field can be written as  ${\bf E} = -\partial_t {\bf A}$. The new Hamiltonian takes the form $H=H_0+\Delta H$ with 
\be \Delta H = -  {\bf J}\cdot{\bf A}\label{deltah}\ee
where ${\bf J}$ is the quantum operator associated to the electric current. For the simple Hamiltonians  that we considered in Section \ref{llevelsec},  ${\bf J}$ is equal (up to constants) to the mechanical momentum $\bpi ={\bf p} + e{\bf A} =  m\dot{\bf x}$ which we defined in equation \eqn{mmom}. However, we'll use more  general definitions of ${\bf J}$  in what follows.

\para
At this point, there's a couple of tricks that makes life simpler. First, we're ultimately interested in applying a constant, DC electric field. However, it turns out to be simpler to apply an AC electric field, ${\bf E}(t) = {\bf E}e^{-i\omega t}$ with frequency $\omega$, and to then take the limit $\omega\rightarrow 0$. Second, it's also somewhat simpler if we work with a complexified ${\bf A}$. There's nothing deep in this: it's just easier to write $e^{-i\omega t}$ than, say, $\cos(\omega t)$. Because all our calculations will be to linear order only, you can take the real part at any time. We then have
\be {\bf A} =   \frac{\bf E}{i\omega}e^{-i\omega t}\label{aise}\ee
Our goal is to compute the current $\langle {\bf J}\rangle$ that flows due to the perturbation $\Delta H$. We will assume that the electric field is small and proceed using standard perturbation theory. 

\para
We work in the interaction picture. This means that operators evolve as ${\cal O}(t) = V^{-1}{\cal O}V$ with $V=e^{-iH_0t/\hbar}$. In particular ${\bf J}$, and hence $\Delta H(t)$ itself, both vary in time in this way. Meanwhile states $|\psi(t)\rangle$, evolve by
\be |\psi(t)\rangle_I = U(t,t_0)|\psi(t_0)\rangle_I\nn\ee
where the unitary operator $U(t,t_0)$ is defined as
\be U(t,t_0) = T\exp\left(-\frac{i}{\hbar}\int^t_{t_0}\Delta H (t^\prime)\,dt^\prime\right)\label{unitary}\ee
Here $T$ stands for time ordering; it ensures that $U$ obeys the equation $i\hbar\,dU/dt =  \Delta H\,U$. 

\para
We're interested in systems with lots of particles. Later we'll only consider non-interacting particles but, importantly, the Kubo formula is more general than this. We prepare the system at time $t\rightarrow -\infty$ in  a specific many-body state  $|0\rangle$. This is usually taken to be the many-body ground state, although it needn't necessarily be. 
 Then, writing $U(t) = U(t,t_0\rightarrow-\infty)$, the expectation value of the current is given by
\be
\langle {\bf J}(t)\rangle &=&  \langle 0 (t)|\, {\bf J}(t)|0(t)\,\rangle \nn\\ &=& \langle 0 | \,U^{-1}(t) {\bf J}(t)U(t)\,|0 \rangle \nn\\  &\approx&  \langle 0| \left( {\bf J}(t) + \frac{i}{\hbar}\int_{-\infty}^t dt'\ [\Delta H(t'),{\bf J}(t)] \right) |0\rangle\nn\ee
where, in the final line, we've expanded the unitary operator \eqn{unitary}, keeping only the leading terms. The first term is the current in the absence of an electric field. We'll assume that  this term vanishes. Using the expressions \eqn{deltah} and \eqn{aise}, the current due to the electric field is then
\be \langle J_i(t)\rangle =  \frac{1}{\hbar \omega}\int_{-\infty}^t dt'\  \langle 0 | [J_j(t'),J_i(t)] |0 \rangle \, E_j\,e^{-i\omega t'} \nn\ee
Because the system is invariant under time translations, the correlation function above can only depend on $t''=t-t'$. We can then rewrite the expression above as
\be \langle J_i(t)\rangle =  \frac{1}{\hbar \omega}\left(\int_0^{\infty} dt''\ e^{i\omega t''}\,\langle 0| [J_j(0),J_i(t'')] |0 \rangle \right) E_je^{-i\omega t} \nn\ee
The only $t$ dependence in the formula above sits outside as $e^{-i\omega t}$. This is telling us that if you apply an electric field at frequency $\omega$, the current responds by oscillating at the same frequency $\omega$. This is the essence of {\it linear response}. The 
proportionality constant defines the frequency-dependent conductivity matrix $\sigma(\omega)$. The Hall conductivity is the off-diagonal part
\be \sigma_{xy}(\omega) = \frac{1}{\hbar \omega} \int_0^{\infty} dt\  e^{i\omega t}\,\langle 0 | [J_y(0),J_x(t)] | 0 \rangle \nn\ee
This is the {\it Kubo formula} for the Hall conductivity.

\para
We can massage the Kubo formula into a slightly more useful form. We use the fact that the current operator evolves as 
${\bf J}(t) = V^{-1}\,{\bf J}(0)\,V$ with $V=e^{-iH_0t/\hbar}$. We then evaluate $\sigma_{xy}(\omega)$ by inserting complete basis of energy eigenstates of $H_0$,
\be \sigma_{xy}(\omega) &=& \frac{1}{\hbar \omega}\int_0^\infty dt\ e^{i\omega t}\,\sum_{n}\left[\langle 0 |J_y|n\rangle\langle n|J_x| 0 \rangle e^{i(E_n-E_0)t/\hbar} - \langle 0| J_x|n\rangle\langle n|J_y|0\rangle e^{i(E_0-E_n)t/\hbar}\right]\nn\ee
We now perform the integral over $\int dt$. (There's a subtlety here: to ensure convergence, we should replace  $\omega \rightarrow \omega + i\epsilon$, with $\epsilon$ infinitesimal. There is a story related to causality and where poles can appear in the complex $\omega$ plane which you can learn more about  in the {\it Kinetic Theory} lecture notes.) Since the states with $|n\rangle=|0\rangle$ don't contribute to the sum, we get
\be \sigma_{xy}(\omega) &=&  -\frac{i}{\omega}\sum_{n \neq 0} \left[ \frac{\langle 0 |J_y|n\rangle\langle n|J_x| 0 \rangle}{\hbar\omega+E_n-E_0} -\frac{\langle 0| J_x|n\rangle\langle n|J_y|0\rangle }{\hbar\omega+E_0-E_n} \right] \label{gettingtheresigma}\ee
Now, finally, we can look at the DC $\omega\rightarrow 0$ limit that we're interested in. We expand the denominators as
\be \frac{1}{\hbar\omega + E_n-E_0} \approx \frac{1}{E_n-E_0} - \frac{\hbar\omega}{(E_n-E_0)^2}+{\cal O}(\omega^2)\ldots\nn\ee
and similar for the other term. The first term looks divergent. Indeed, such divergences do arise for longitudinal conductivities and tell us something physical, often that momentum is conserved due to translational invariance so there can be no DC resistivity. However, in the present case of the Hall conductivity, there is no divergence because this term vanishes. This can be shown on general grounds from gauge invariance or, equivalently, from the conservation of the current. Alternatively -- although somewhat weaker -- it can quickly seen by  rotational invariance which ensures that the expression should be invariant under $x\rightarrow y$ and $y\rightarrow -x$. We're then left only with a finite contribution in the limit $\omega\rightarrow 0$ given by
\be \sigma_{xy} = i\hbar \sum_{n \neq 0}  \frac{\langle 0 |J_y|n\rangle\langle n|J_x| 0 \rangle - \langle 0| J_x|n\rangle\langle n|J_y|0\rangle}{(E_n-E_0)^2} \label{kubo}\ee
This is the Kubo formula for Hall conductivity.

\para
Before we proceed, I should quickly apologise for being sloppy: the operator that we called ${\bf J}$ in \eqn{deltah} is actually the current rather than the current density. This means that the right-hand-side of \eqn{kubo} should, strictly speaking, be multiplied by the spatial area of the sample. It was simpler to omit this in the above derivation to avoid clutter.

\subsubsection{The Role of Topology}\label{avronsec}

In this section, we describe a set-up in which we can see the deep connections between topology and the Hall conductivity. The set-up is closely related to the gauge-invariance argument that we saw in Section \ref{pumpingsec}. However, we will consider the Hall system on a spatial torus ${\bf T}^2$. This can be viewed as a rectangle with opposite edges identified. We'll take the lengths of the sides to be  $L_x$ and $L_y$. 

\para
We thread a uniform magnetic field $B$ through the torus. The first result we need is that $B$ obeys the Dirac  quantisation condition,
\be B L_x L_y = \frac{2\pi \hbar}{e} \, n\ \ \ \ \ \ n\in {\bf Z}\label{diracq}\ee
This quantisation arises for the same reason that we saw in Section \ref{berryspinsec} when discussing the Berry phase. However, it's an important result so here we give an alternative derivation.

\para
We consider wavefunctions over the torus and ask: what periodicity requirements should we put on the wavefunction? The first guess is that we should insist that wavefunctions obey $\psi(x,y) = \psi(x+L_x,y) = \psi(x,y+L_y)$. But this turns out to be too restrictive when there is a magnetic flux through the torus. Instead, one has to work in patches; on the overlap between two different patches, wavefunctions must be related by a gauge transformation. 

\para
Operationally, there is a slightly simpler way to implement this. We introduce the magnetic translation operators, 
\be T({\bf d}) = e^{-i{\bf d}\cdot{\bf p}/\hbar}  = e^{-i{\bf d}\cdot(i\nabla + e{\bf A}/\hbar)}\nn\ee
These operators translate a state $\psi(x,y)$ by position vector ${\bf d}$. The appropriate boundary conditions will be that when a state is translated around a cycle of the torus, it comes back to itself. So $T_x\psi(x,y) = \psi(x,y)$ and $T_y\psi(x,y) = \psi(x,y)$ where  $T_x = T({\bf d} = (L_x,0))$ and $T_y = T({\bf d} = (0,L_y))$. 

\para
It is clear from the expression above that the  translation operators  are not gauge invariant: they depend on our choice of ${\bf A}$. We'll choose Landau gauge $A_x = 0$ and $A_y = Bx$. With this choice,  translations in the $x$ direction are the same as those in the absence of  a magnetic field, while translations in the $y$ direction pick up an extra phase. If we take a state $\psi(x,y)$, translated around a cycle of the torus, it becomes
\be T_x\psi(x,y) &=&   \psi(x + L_x,y) =\psi(x,y) \nn\\ 
 T_y\psi(x,y) &=& e^{-i eB L_y x/\hbar}\, \psi(x,y+L_y) = \psi(x,y)
\nn\ee
Notice that we can see explicitly in the last of these equations that the wavefunction is not periodic in the naive sense in the $y$ direction: $\psi(x,y+L_y)\neq \psi(x,y)$. Instead, the two wavefunctions agree only up to a gauge transformation. 

\para
However, these equations are not consistent for any choice of $B$. This follows by comparing  what happens if we translate around the $x$-cycle, followed by the $y$-cycle, or if we do these in the opposite order. We have
\be T_y T_x = e^{-ieB L_x L_y/\hbar} \, T_x T_y  \label{magcom}\ee
Since both are required to give us back the same state, we must have
\be \frac{eB L_x L_y}{\hbar} \in 2\pi{\bf Z}\nn\ee
This is the Dirac quantisation condition \eqn{diracq}. 

\para
There is an interesting story about solving for the wavefunctions of a free particle on a torus in the presence of a magnetic field. They are given by theta functions. We won't discuss these here.

\subsubsection*{Adding Flux}

\EPSFIGURE{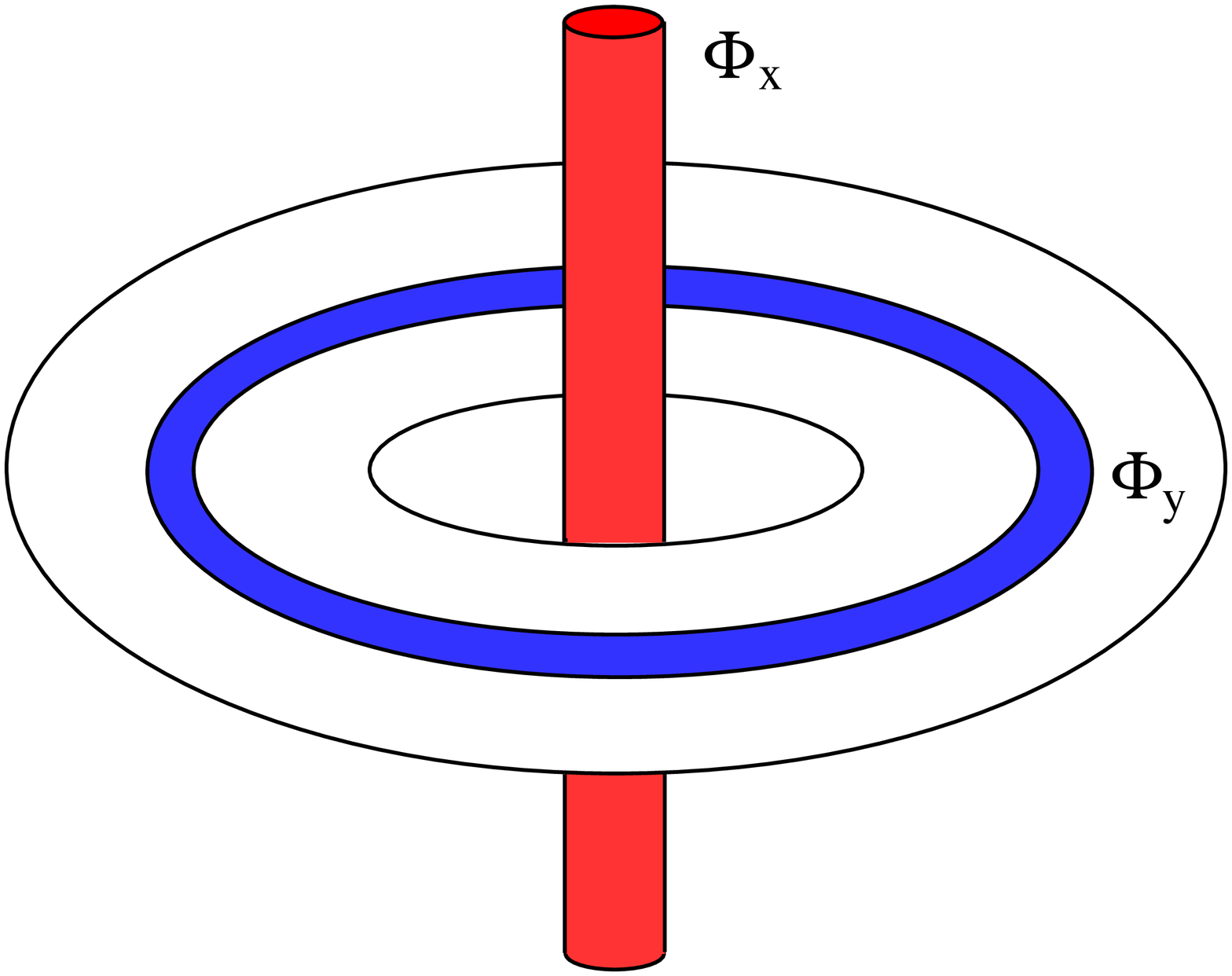,height=120pt}{}
\noindent
Now we're going to perturb the system. We do this by threading two fluxes, $\Phi_x$ and $\Phi_y$ through the $x$ and $y$-cycles of the torus respectively. This means that the gauge potential becomes
\be A_x = \frac{\Phi_x}{L_x}\ \ \ {\rm and}\ \ \ A_y = \frac{\Phi_y}{L_y} + Bx\nn\ee
This is the same kind of set-up that we discussed in Section \ref{pumpingsec}; the only difference is that now the geometry allows us to introduce two fluxes instead of one. Just as in our previous discussion, the states of the quantum system are only sensitive to the non-integer part of $\Phi_i/\Phi_0$ where $\Phi_0 = 2\pi \hbar/e$ is the quantum of flux. In particular, if we increase either $\Phi_i$ from zero to $\Phi_0$, then the spectrum of the quantum system must be invariant. However, just as in Section \ref{pumpingsec}, the system can undergo spectral flow.

\para
The addition of the fluxes adds an extra term to the Hamiltonian of the form \eqn{deltah},
\be \Delta H = -\sum_{i=x,y}  \frac{J_i\Phi_i}{L_i}\nn\ee
We want to see how this affects the ground state of the system which we will denote as $|\psi_0\rangle$. (We called this $|0\rangle$ when deriving the Kubo formula, but we'll want to differentiate it soon and the expression $\ppp{0}{\Phi}$ just looks too odd!). We'll assume that the ground state is non-degenerate and that there is a gap to the first excited state. Then, to first order in perturbation theory, the ground state becomes
\be |\psi_0\rangle' = |\psi_0\rangle + \sum_{n\neq \psi_0} \frac{\langle n|\Delta H |\psi_0\rangle}{E_n-E_0}\,|n\rangle\nn\ee
Considering infinitesimal changes of $\Phi_i$, we can write this as
\be |\ppp{\psi_0}{\Phi_i}\rangle =  - \frac{1}{L_i} \sum_{n\neq \psi_0} \frac{\langle n|J_i |\psi_0\rangle}{E_n-E_0}\,|n\rangle\nn\ee
But the right-hand-side is exactly the kind of expression that appeared in the Kubo formula \eqn{kubo}. This means that, including the correct factors of the spatial area, we can write the Hall conductivity as 
 \be \sigma_{xy} &=& i\hbar L_xL_y \sum_{n \neq \psi_0}  \frac{\langle \psi_0 |J_y|n\rangle\langle n|J_x| \psi_0 \rangle - \langle \psi_0| J_x|n\rangle\langle n|J_y|\psi_0\rangle}{(E_n-E_0)^2} 
 \nn\\ &=& i\hbar\left[ \langle \ppp{\psi_0}{\Phi_y}|\ppp{\psi_0}{\Phi_x}\rangle-  \langle \ppp{\psi_0}{\Phi_x}|\ppp{\psi_0}{\Phi_y}\rangle\right] \nn\\ &=& i\hbar\left[\ppp{}{\Phi_y} \langle {\psi_0}|\ppp{\psi_0}{\Phi_x}\rangle-  \ppp{}{\Phi_x}\langle \psi_0|\ppp{\psi_0}{\Phi_y}\rangle\right] 
 \nn\ee
As we now explain, this final way of writing the Hall conductivity provides a novel perspective on the integer quantum Hall effect.

\subsubsection*{Hall Conductivity and the Chern Number}

The fluxes $\Phi_i$ appear as parameters in the  perturbed Hamiltonian. As we discussed above, the spectrum of the Hamiltonian only depends on $\Phi_i$ mod $\Phi_0$, which means that these parameters should be thought of as periodic: the space of the flux parameters is itself a torus, ${\bf T}^2_{\Phi}$, where the subscript is there to distinguish it from the spatial torus that we started with. We'll introduce dimensionless angular variables, $\theta_i$ to parameterise this torus,
\be \theta_i = \frac{2\pi \Phi_i}{\Phi_0}\ \ \ \ \ \ \ {\rm with}\ \ \ \theta_i\in[0,2\pi)\nn\ee
As we discussed in Section \ref{berrysec}, given a parameter space it is natural to consider the Berry phase that arises as the parameters are varied. This is described by the Berry connection which, in this case, lives over ${\bf T}^2_\Phi$. It is
\be {\cal A}_i(\Phi) = -i\langle \psi_0|\ppp{}{\theta_i}|\psi_0\rangle\nn\ee
The field strength, or curvature, associated to the Berry connection is given by
\be {\cal F}_{xy} = \ppp{{\cal A}_x}{\theta_y} - \ppp{{\cal A}_y}{\theta_x} = -i \left[\ppp{}{\theta_y} \langle {\psi_0}|\ppp{\psi_0}{\theta_x}\rangle-  \ppp{}{\theta_x}\langle \psi_0|\ppp{\psi_0}{\theta_y}\rangle\right] 
\nn\ee
This is precisely our expression for the Hall conductivity! We learn that, for the torus with fluxes, we can write
\be \sigma_{xy} = -\frac{e^2}{\hbar}{\cal F}_{xy} \nn\ee
This is a nice formula. But, so far, it doesn't explain why $\sigma_{xy}$ is quantised. However, suppose that we average over all fluxes. In this case we integrate over the torus ${\bf T}^2_\Phi$ of parameters to get
\be \sigma_{xy} = -\frac{e^2}{\hbar}\int_{{\bf T}^2_\Phi} \frac{d^2\theta}{(2\pi)^2}\ {\cal F}_{xy} \nn\ee
The integral of the curvature over ${\bf T}^2_\Phi$, is  a number known as the {\it first Chern number} 
\be C = \frac{1}{2\pi} \int_{{\bf T}^2_\Phi} d^2\theta\ {\cal F}_{xy}\nn\ee
Importantly, this is always an integer: $C\in {\bf Z}$. This follows from the same kind of argument that we made in Section \ref{berrysec} (or, alternatively, the kind of argument that we made at the beginning of this section on Dirac quantisation). The net result is that if we average over the fluxes, the Hall conductivity is necessarily quantised as
\be
\sigma_{xy} = -\frac{e^2}{2\pi\hbar}\,C\label{tknn0}\ee
This, of course, is the integer quantum Hall effect. The relationship between the Hall conductivity and the Chern number is usually referred to as the {\it TKNN invariant} (after Thouless, Kohomoto, Nightingale and den Nijs) although, strictly speaking, this name should be reserved for a very similar expression that we'll discuss in the next section.

\subsection{Particles on a Lattice}\label{latticesec}

We saw in the previous section that there is a deep relationship between the Hall conductivity and a certain topological quantity called the Chern number that is related to the Berry phase. Here we'll continue to explore this relationship but in the slightly different context of  particles moving on  a lattice. The kind of  ideas that we will describe have had a resurgence in recent years when it was realised that they are the key to understanding the subject of {\it topological insulators}.

\para
The advantage of looking at the particle on a lattice is that its momentum lies on a torus ${\bf T}^2$, known as the {\it Brillouin zone}. It is this torus that will allow us to find interesting topological features. Indeed, it will play a very similar role to the parameter space ${\bf T}^2_\Phi$ that we met in the previous section. We'll learn that one can define a Berry connection over the Brillouin zone and that the associated Chern number determines the Hall conductivity.

\subsubsection{TKNN Invariants}\label{tknnsec}

We'll consider a particle moving on a rectangular lattice. The distance between lattice sites in the $x$-direction is $a$; the distance in the $y$-direction is $b$. Recall from earlier courses that the energy spectrum of this system form bands. Within each band,  states  are labelled by lattice momentum which takes values in the Brillouin zone, parameterised by
\be -\frac{\pi}{a} < k_x \leq \frac{\pi}{a}\ \ \ {\rm and}\ \ \ -\frac{\pi}{b}\ < k_y\leq \frac{\pi}{b}\label{bz}\ee
The states with momenta at the edges of the Brillouin zone are identified. This means that the Brillouin zone is a torus ${\bf T}^2$ as promised. The wavefunctions  in a given band can be written in Bloch form as
\be \psi_{\bf k}({\bf x}) = e^{i{\bf k}\cdot{\bf x}} \,u_{\bf k}({\bf x})\label{bloch}\ee
where $u_{\bf k}({\bf x})$ is usually periodic on a unit cell so that $u_{\bf k}({\bf x}+{\bf e}) = u_{\bf k}({\bf x})$ with either ${\bf e}=(a,0)$ or ${\bf e}=(0,b)$.

\para
We're now in a position to describe the topology underlying the quantum Hall effect. The results below are very general: they don't rely on any specific Hamiltonian, but rather apply to any system that satisfies a few simple criteria.

\begin{itemize}
\item
First, we will assume that the single particle spectrum decomposes into bands, with each band parameterised by a momentum label ${\bf k}$ which lives on a torus ${\bf T}^2$. This is obviously true for simple lattice models. As we explain in Section \ref{latticemagsec}, it is also true (under certain assumptions) for particles moving in a lattice in the presence of a magnetic field where the torus in question is slightly different concept called a  magnetic Brillouin zone. (In this case, the periodicity conditions on $u_{\bf k}$ are altered slightly but the formula we derive below still holds.) 
\item
Second, we'll assume that the electrons are non-interacting. This means that we get the multi-particle spectrum simply by filling up the single-particle spectrum, subject to the Pauli exclusion principle. 
\item
Finally, we'll assume that there is a gap between bands and that the Fermi energy  $E_F$ lies in one of these gaps. This means that all bands below $E_F$ are completely filled while all bands above $E_F$ are empty.  In band theory, such a situation describes an {\it insulator}. 
\end{itemize}

Whenever these three criteria are obeyed, one can assign an integer-valued topological invariant $C\in {\bf Z}$ to each band. The topology arises from the way the phase of the states winds as we move around the Brillouin zone ${\bf T}^2$. This is captured by a $U(1)$  Berry connection over ${\bf T}^2$, defined by
\be {\cal A}_i({\bf k}) =- i\langle u_{\bf k}|\ppp{}{k^i}| u_{\bf k}\rangle\nn\ee
There is one  slight conceptual difference from the type of  Berry connection we met previously. In Section \ref{berrysec}, the connections  lived on the space of parameters of the Hamiltonian; here the connection lives on the space of states itself. Nonetheless, it is simple to see that  many of the basic properties that we met in Section \ref{berrysec} still hold. In particular, a change of phase of the states $| u_{\bf k}\rangle$ corresponds to a change of gauge of the Berry connection.

\para
We can compute the field strength associated to ${\cal A}_i$. This is 
\be {\cal F}_{xy}  = \ppp{{\cal A}_x}{k^y} - \ppp{{\cal A}_y}{k^x} = -i \left< \ppp{u}{k^y}\right|\left.\ppp{u}{k^x}\right> + i \left< \ppp{u}{k^x}\right|\left.\ppp{u}{k^y}\right>\label{tknncurve}\ee
Once again, we can compute the first Chern number by integrating ${\cal F}$ over the Brillouin zone ${\bf T}^2$,
\be C = -\frac{1}{2\pi} \int_{{\bf T}^2} d^2k\ {\cal F}_{xy}\label{chern}\ee
 In the present context, it is usually referred to as the TKNN invariant\footnote{As we mentioned in the previous section, the initials stand for Thouless, Kohomoto, Nightingale and den Nijs. The original paper is ``{\it Quantized Hall Conductance in a Two-Dimensional Periodic Potential}", \href{http://journals.aps.org/prl/abstract/10.1103/PhysRevLett.49.405}{Phys. Rev. Lett. 49, 405 (1982)}.}. As we've seen before, the Chern number is always an integer: $C\in {\bf Z}$.  In this way, we can associate an integer $C_\alpha$ to each band $\alpha$.
 
\para
The Chern number once again has a beautiful physical manifestation: it is related to the Hall conductivity $\sigma_{xy}$ of a non-interacting band insulator by
\be \sigma_{xy} = \frac{e^2}{2\pi\hbar} \sum_\alpha C_\alpha\label{special}\ee
where the sum is over all filled bands $\alpha$ and $C_\alpha$ is the Chern class associated to that band. This is the famous TKNN formula. It is, of course, the same formula \eqn{tknn0} that we met previously, although the context here is rather different.

\para
Let's now prove the TKNN formula. Our starting point is the Kubo formula \eqn{kubo}. We previously wrote this in terms of multi-particle wavefunctions.  If we're dealing with non-interacting particles, then these can be written as tensor products of single particle wavefunctions, each of which is labelled by the band $\alpha$ and momentum ${\bf k} \in {\bf T^2}$. The expression for the Hall conductivity becomes
\be \sigma_{xy} = i\hbar\! \sum_{E_\alpha < E_F  < E_\beta} \int_{{\bf T}^2} \frac{d^2k}{(2\pi)^2} \
\frac{\langle u^\alpha_{\bf k} |J_y| u^\beta_{\bf k}\rangle\langle u^\beta_{\bf k}|J_x| u^\alpha_{\bf k} \rangle - \langle u^\alpha_{\bf k}| J_x| u^\beta_{\bf k}\rangle\langle u^\beta_{\bf k}|J_y|u^\alpha_{\bf k}\rangle}{(E_\beta({\bf k})-E_\alpha({\bf k}))^2} \nn\ee
where the index $\alpha$ runs over the filled bands and $\beta$ runs over the unfilled bands. We note that this notation is a little lazy; there are really separate momentum integrals for each band and no reason that the states in the expression have the same momentum ${\bf k}$. Our lazy notation will save us from adding yet more  annoying indices and not affect the result below.

\para
To make progress, we need to define what we mean by the current ${\bf J}$. For a single, free particle in the  continuum, the current carried by the particle was simply ${\bf J} = e\dot{\bf x}$ where the velocity operator is  $\dot{\bf x} = ({\bf p} + e {\bf A})/m$. Here we'll use a more general definition. We first look at the Schr\"odinger equation acting on single-particle wavefunctions of Bloch form \eqn{bloch},
\be H|\psi_{\bf k}\rangle = E_{\bf k}|\psi_{\bf k}\rangle \ \ \ &\Rightarrow&\ \ \ (e^{-i{\bf k}\cdot{\bf x}} H e^{i{\bf k}\cdot{\bf x}})|u_{\bf k}\rangle = E_{\bf k}|u_{\bf k}\rangle \nn\\ &\Rightarrow&\ \ \ \tilde{H}({\bf k})|u_{\bf k}\rangle = E_{\bf k}|u_{\bf k}\rangle \ \ \ \ {\rm with}  \ \ \ \tilde{H}({\bf k}) = e^{-i{\bf k}\cdot{\bf x}} H e^{i{\bf k}\cdot{\bf x}}\nn\ee
We then define the current in terms of the group velocity of the wavepackets,
\be {\bf J} = \frac{e}{\hbar}\ppp{\tilde{H}}{{\bf k}}\nn\ee
Before proceeding, it's worth checking that coincides with our previous definition. In the continuum, the Hamiltonian was simply $H=({\bf p}+ e{\bf A})^2/2m$, which gives $\tilde{H} = ({\bf p} + \hbar{\bf k} + e{\bf A})^2/2m$ and the current due to a single particle is ${\bf J} = e\dot{\bf x}$ as expected. 

\para
From now on it's merely a question of doing the algebra. The Kubo formula becomes
\be \sigma_{xy} = \frac{ie^2}{\hbar}\sum_{E_\alpha < E_F < E_\beta}\int_{{\bf T}^2}\frac{d^2k}{(2\pi)^2} \ \frac{\langle u^\alpha_{\bf k}  |\partial_{y}\tilde{H}| u^\beta_{\bf k}\rangle\langle u^\beta_{\bf k}| \partial_{x}\tilde{H}| u^\alpha_{\bf k} \rangle - \langle u^\alpha_{\bf k}|  \partial_{x}\tilde{H}|u^\beta_{\bf k}\rangle\langle u^\beta_{\bf k}| \partial_{y}\tilde{H}|u^\alpha_{\bf k}\rangle}{(E_\beta({\bf k})-E_\alpha({\bf k}))^2} \nn\ee
where $\partial_x$ and $\partial_y$ in this expression are derivatives with respect to momenta $k_x$ and $k_y$ respectively. 
We can then write
\be \langle u^\alpha_{\bf k}| \partial_i{\tilde{H}}|u^\beta_{\bf k}\rangle &=& \langle u^\alpha_{\bf k}| \partial_i\left(\tilde{H}|u^\beta_{\bf k}\rangle\right) - \langle u^\alpha_{\bf k}|\tilde{H}|\partial_i{u^\beta_{\bf k}}\rangle\nn\\
&=& (E_\beta({\bf k})-E_\alpha({\bf k}))\langle u^\alpha_{\bf k}|\partial_i{u^\beta_{\bf k}}\rangle\nn\\ &=& -(E_\beta({\bf k})-E_\alpha({\bf k}))\langle \partial_i u^\alpha_{\bf k}|{u^\beta_{\bf k}}\rangle \nn\ee
The  missing term, proportional to $\partial_i E_\beta$, doesn't  appear because $\alpha$ and $\beta$ are necessarily distinct bands. 
Substituting this into the Kubo formula gives
\be \sigma_{xy} = \frac{ie^2}{\hbar}\sum_{E_\alpha< E_F < E_\beta}\int_{{\bf T}^2}\frac{d^2k}{(2\pi)^2} \ \langle\partial_y u^\alpha_{\bf k}|u^\beta_{\bf k}\rangle\langle u^\beta_{\bf k}|\partial_x u^\alpha_{\bf k}\rangle - \langle \partial_x u^\alpha_{\bf k}|u^\beta_{\bf k}\rangle \langle u^\beta_{\bf k}|\partial_y u^\alpha_{\bf k}\rangle  \nn\ee
But now we can think of the sum over the unfilled bands as $\sum_\beta |u^\beta_{\bf k}\rangle \langle u^\beta_{\bf k}| = {\bf 1} - \sum_\alpha |u^\alpha_{\bf k}\rangle\langle u^\alpha_{\bf k}|$. The second term vanishes by symmetry, so we're left with 
\be  \sigma_{xy} = \frac{ie^2}{\hbar}\sum_{\alpha}\int_{{\bf T}^2}\frac{d^2k}{(2\pi)^2} \ \langle\partial_y u^\alpha_{\bf k}|\partial_x u^\alpha_{\bf k}\rangle - \langle \partial_x u^\alpha_{\bf k}|\partial_y u^\alpha_{\bf k}\rangle  \nn\ee
where now the sum is only over the filled bands $\alpha$. 
Comparing to \eqn{tknncurve}, we see that the Hall conductivity is indeed given by the sum of the Chern numbers of filled bands as promised,
\be \sigma_{xy} = -\frac{e^2}{2\pi\hbar} \sum_\alpha C_\alpha\nn\ee
The TKNN formula is the statement that the Hall conductivity is a topological invariant of the system. It's important because it goes some way to explaining the robustness of the integer quantum Hall effect. An integer, such as the Chern number $C$, can't change continuously. This means that if we deform  our system in some way then, as long as we retain the assumptions that went into the derivation above, the Hall conductivity can't change: it's  pinned at the integer value.

\para
The existence of the TKNN formula is somewhat surprising. The right-hand side is simple and pure. In contrast, conductivities are usually thought of as something complicated and messy, depending on all the intricate details of a system. The essence of the TKNN formula, and indeed the quantum Hall effect itself, is that this is not the case: the Hall conductivity is topological.

\subsubsection{The Chern Insulator}

Let's look at an example. Perhaps surprisingly, the simplest examples of lattice models with non-vanishing Chern numbers don't involve any magnetic fields at all. Such lattice models with filled bands are sometimes called {\it Chern insulators}, to highlight the fact that they do something interesting --- like give a Hall response --- even though they are insulating states. 

\para
The simplest class of Chern insulators involve just two bands. The single-particle Hamiltonian written, written in momentum space, takes the general form 
\be \tilde{H}({\bf k}) = \vec{E}({\bf k}) \cdot\vec{\sigma} + \epsilon(\bf k){\bf 1}\nn\ee
where ${\bf k}\in {\bf T}^2$ and $\vec{\sigma} = (\sigma_1,\sigma_2,\sigma_3)$ are the three Pauli matrices. The energies of the two states with momentum ${\bf k}$ are $\epsilon({\bf k}) \pm |\vec{E}({\bf k})|$. An insulator requires a gap between the upper and lower bands; we then fill the states of the lower band. An insulator can only occur when $\vec{E}({\bf k})\neq 0$ for all ${\bf k}$.

\para
For any such model, we can introduce a unit three-vector,
\be \vec{n}({\bf k}) = \frac{\vec{E}({\bf k})}{|\vec{E}({\bf k})|}\nn\ee
Clearly $\vec{n}$ describes a point on a two-dimensional sphere ${\bf S}^2$. This is the {\it Bloch sphere}. As we move in the Brillouin zone, $\vec{n}({\bf k})$ gives a map from ${\bf T}^2\rightarrow {\bf S}^2$ as shown in the figure. This  Chern number \eqn{chern} for this system can be written in terms of $\vec{n}$ as  
\be C = \frac{1}{4\pi} \int_{{\bf T}^2} d^2k\ \vec{n}\cdot\left(\ppp{\vec{n}}{k_x}\times \ppp{\vec{n}}{k_y}\right)\nn\ee
There is a particularly nice interpretation of this formula: it measures the area of the unit sphere (counted with sign) swept out as we vary ${\bf k}$ over ${\bf T}^2$. In other words, it counts how many times ${\bf T}^2$ wraps around ${\bf S}^2$.

\begin{figure}[htb]
\begin{center}
\epsfxsize=4in\leavevmode\epsfbox{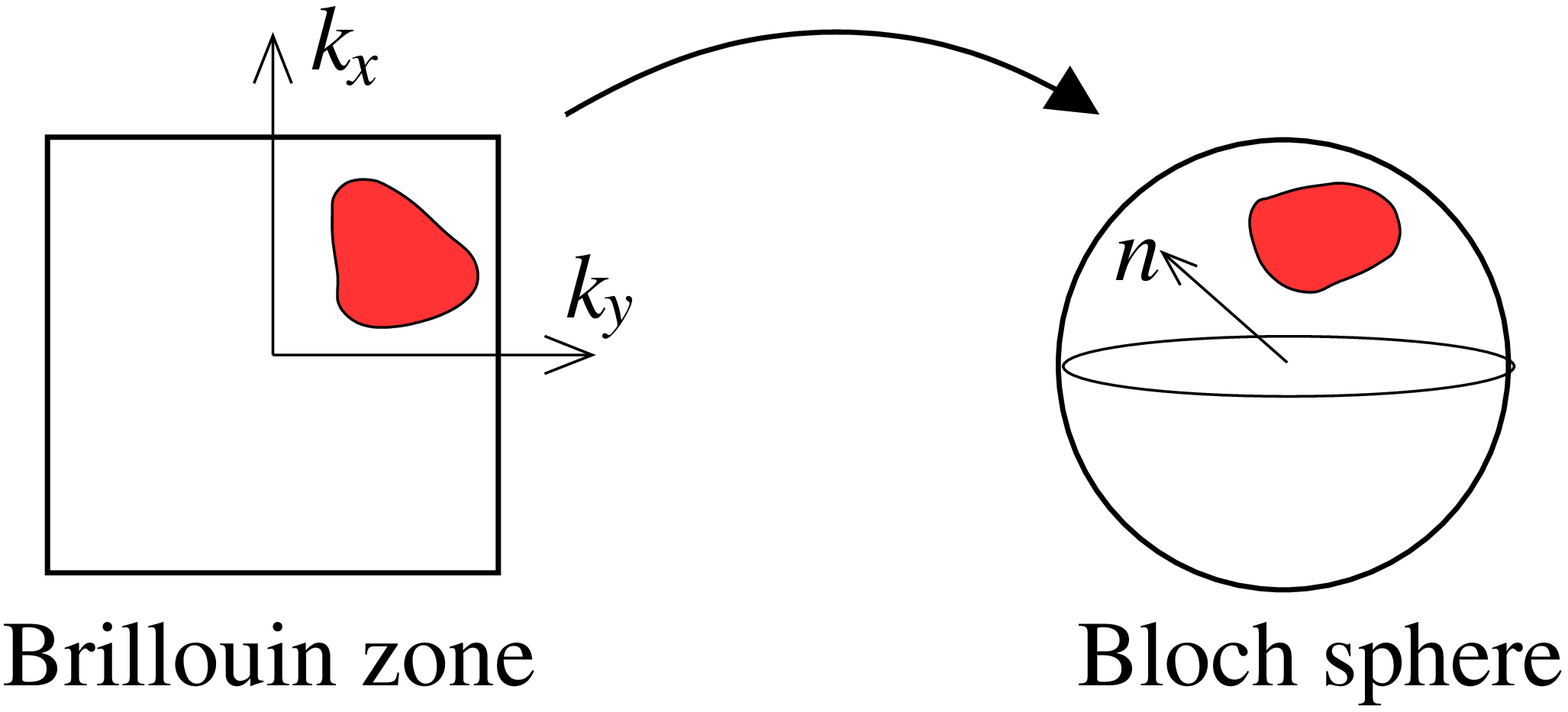}
\end{center}
\caption{The map from Brillouin zone to Bloch sphere}
\end{figure}

\para
Perhaps the simplest lattice model with a non-trivial Chern number arises on a square lattice, with the Hamiltonian in momentum space given by\footnote{This model was first constructed in Xiao-Liang Qi, Yong-Shi Wu and Shou-Cheng Zhang, ``{\it Topological quantization of the spin Hall effect in two-dimensional paramagnetic semiconductors}", \href{http://arxiv.org/abs/cond-mat/0505308}{cond-mat/0505308}. An earlier model of a quantum Hall effect without a magnetic field, involving a honeycomb lattice,  was described by Duncan Haldane, ``{\it Model for a Quantum Hall Effect without Landau Levels: Condensed Matter Realisation of the Parity Anomaly}",  \href{http://journals.aps.org/prl/abstract/10.1103/PhysRevLett.61.2015}{Phys. Rev. Lett. 61, 2015 (1988)}.}. 
\be \tilde{H}({\bf k}) = (\sin k_x)\sigma_1 + (\sin k_y)\sigma_2 + (m+\cos k_x + \cos k_y)\sigma_3\nn\ee
In the continuum limit, this becomes the Hamiltonian for a 2-component Dirac fermion in $d=2+1$ dimensions. For this reason, this model is sometimes referred to as a {\it Dirac-Chern} insulator. 

\para
For general values of $m$, the system is an insulator with a gap between the bands. There are three exceptions: the gap closes and the two bands touch at $m=0$ and at $m=\pm 2$. As $m$ varies, the Chern number --- and hence the Hall conductivity --- remains constant as long as the gap doesn't close. A direct computation gives
\be C = \left\{\begin{array}{ccc}  -1 & \ \ \ \ \ \ & -2 < m < 0 \\  1 && 0 <  m < 2 \\  0 &&|m|>2 \end{array}\right.\nn\ee

\subsubsection{Particles on a Lattice in a Magnetic Field}\label{latticemagsec}

So far, we've discussed the integer quantum Hall effect in lattice models but, perhaps surprisingly, we haven't explicitly introduced magnetic fields. In this section, we describe what happens when particles hop on a lattice in the presence of a magnetic field. As we will see, the physics is remarkably rich.

\para
To orient ourselves, first consider a particle hopping on two-dimensional square lattice in the absence of a magnetic field. We'll denote the distance between adjacent lattice sites as $a$. We'll work in the tight-binding approximation, which means that the position eigenstates $ |{\bf x}\rangle$ are restricted to the lattice sites ${\bf x} = a(m,n)$ with $m,n\in {\bf Z}$. The Hamiltonian is given by
\be H=  - t \sum_{{\bf x}}\sum_{j=1,2} |{\bf x}\rangle \langle {\bf x} + {\bf e}_j| + {\rm h.c.}\label{latticeham}\ee
where ${\bf e}_1 = (a,0)$ and ${\bf e}_2=(0,a)$ are the basis vectors of the lattice and $t$ is the hopping parameter. (Note: $t$ is the standard name for this parameter; it's not to be confused with time!) 
The lattice momenta ${\bf k}$ lie in the  Brillouin zone ${\bf T}^2$, parameterised by
\be -\frac{\pi}{a} < k_x \leq \frac{\pi}{a}\ \ \ {\rm and}\ \ \ -\frac{\pi}{a}\ < k_y\leq \frac{\pi}{a}\label{bz2}\ee
Suppose that we further make the lattice finite in spatial extent, with size $L_1\times L_2$. The momenta $k_i$ are now quantised in units of $1/2\pi L_i$. The total number of states in the Brillouin zone is then $(\frac{2\pi}{a} /  \frac{1}{2\pi L_1})\times (\frac{2\pi}{a} /  \frac{1}{2\pi L_1}) = L_1L_2/a^2$. This is the number of sites in the lattice which is indeed the expected number of states in the Hilbert space.

\para
Let's now add a background magnetic field to the story. The first thing we need to do is alter the Hamiltonian. The way to do this is to introduce a gauge field $A_j({\bf x})$ which lives on the links between the lattice sites. We take $A_1({\bf x})$ to be the gauge field on the link to the right of point ${\bf x}$, and $A_2({\bf x})$ to be the gauge field on the link above point ${\bf x}$, as shown in the figure. The Hamiltonian is then given by
\be H=  - t \sum_{{\bf x}}\sum_{j=1,2} |{\bf x}\rangle e^{-ie a A_j({\bf x})/\hbar} \langle {\bf x} + {\bf e}_j| + {\rm h.c.}\label{latticemagham}\ee
\EPSFIGURE{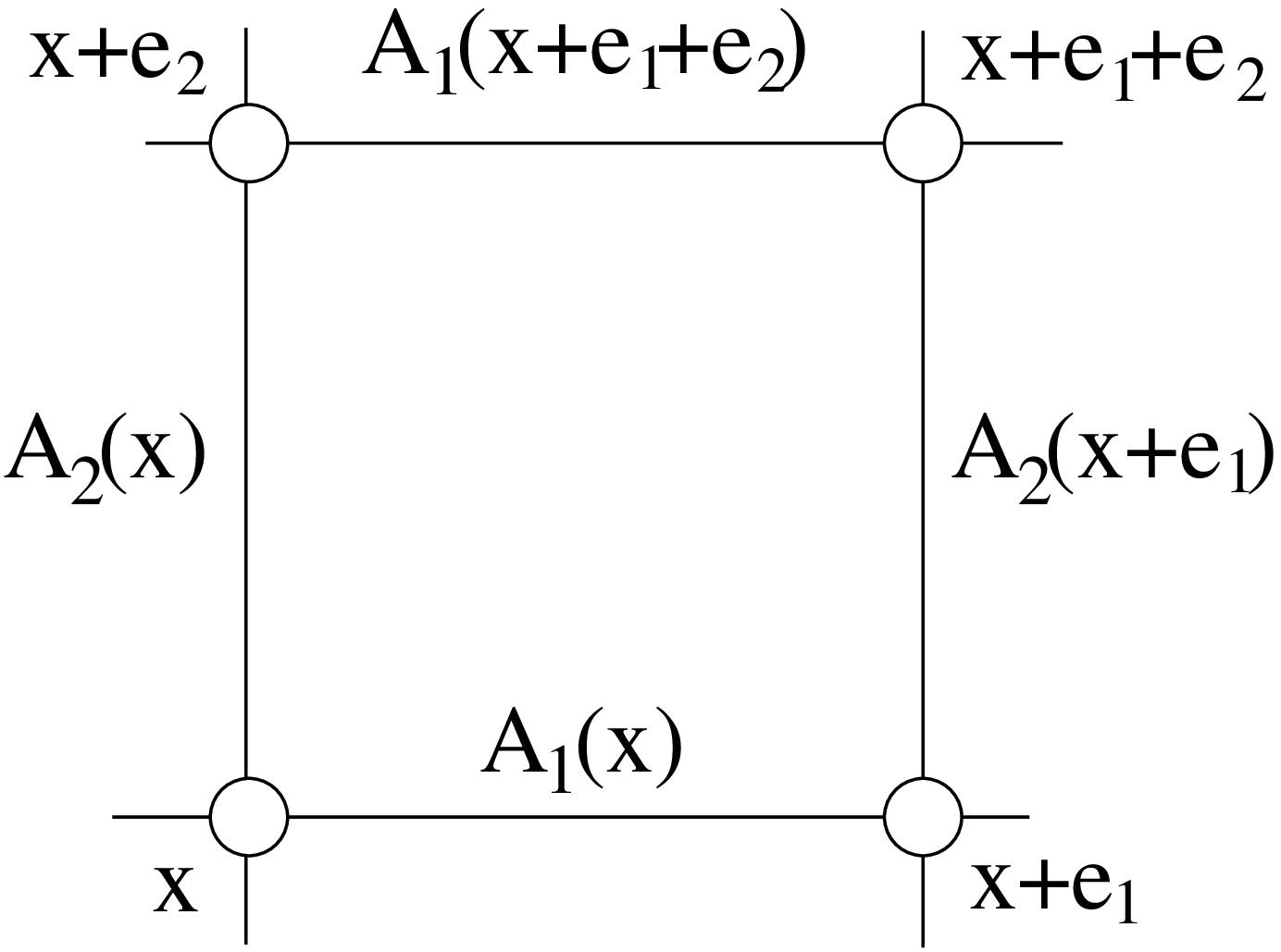,height=100pt}{}
\noindent
It might not be obvious that this is the correct way to incorporate a magnetic field. To gain some intuition, consider a particle which moves anti-clockwise around a plaquette. To leading order in $t$, it will pick up a phase $e^{-i\gamma}$, where
\be \gamma &=& \frac{ea}{\hbar} \left(A_1({\bf x}) + A_2({\bf x}+{\bf e}_1) - A_1({\bf x}+ {\bf e}_2) - A_2({\bf x})\right)
\nn\\ &\approx& \frac{ea^2}{\hbar}  \left(\ppp{A_2}{x^1}- \ppp{A_1}{x^2}\right)  = \frac{ea^2B}{\hbar}\nn\ee
where $B$ is the magnetic field which passes through the plaquette. This expression is the same as the Aharonov-Bohm phase \eqn{abphase1} for a particle moving around a flux $\Phi = Ba^2$. 

\para
Let's now restrict to a constant magnetic field. We  can again work in Landau gauge,
\be A_1 = 0\ \ \ {\rm and}\ \ \ A_2 = Bx_1\label{lgaugeagain}\ee
We want to understand the spectrum of the Hamiltonian \eqn{latticemagham} in this case and, in particular, what becomes of the Brillouin zone.

\subsubsection*{Magnetic Brillouin Zone}

We saw above that the key to finding topology in a lattice system was the presence of the Brillouin zone ${\bf T}^2$. 
 Yet it's not immediately obvious that the Brilliouin zone survives in the presence of the magnetic field. The existence of lattice momenta ${\bf k}$ are a consequence of the discrete translational invariance of the lattice.  But, as usual, the choice of gauge breaks the explicit translational invariance of the Hamiltonian, even though we expect the underlying physics to be translational invariant. 

\para
In fact, we'll see that the interplay between lattice effects and magnetic effects leads to some  rather surprising physics that is extraordinarily sensitive to the  flux $\Phi = Ba^2$ that threads each plaquette. In particular, we can define a magnetic version of the Brillouin zone whenever $\Phi$ is a rational multiple of $\Phi_0= 2\pi\hbar/e$, 
\be \Phi = \frac{p}{q}\Phi_0\label{rational}\ee
with $p$ and $q$ integers which share no common divisor. We will see that in this situation the spectrum splits up into $q$ different bands.
Meanwhile, if $\Phi/\Phi_0$ is irrational, there are no distinct bands in the spectrum: instead it takes the form of a Cantor set! Nonetheless, as we vary $\Phi/\Phi_0$, the spectrum changes continuously. 
Needless to say, all of this is rather odd!

\para
We start by defining the gauge invariant translation operators
\be T_j =  \sum_{{\bf x}}\ |{\bf x}\rangle e^{-ie a A_j({\bf x})/\hbar} \langle {\bf x} + {\bf e}_j| \nn\ee
This shifts each state by one lattice site; $T_1$ moves us to the left and $T_1^\dagger$ to the right, while $T_2$ moves us down and $T_2^\dagger$ up, each time picking up the appropriate phase from the gauge field. Clearly we can write the Hamiltonian as
\be H = -t\left(\sum_{j=1,2} \ T_j + T_j^\dagger\right)\nn\ee
These translation operators do not commute. Instead it's simple to check that they obey the nice algebra
\be T_2\, T_1 = e^{ie\Phi/\hbar\,} T_1\, T_2\label{talg}\ee
This is the discrete version of the magnetic translation algebra \eqn{magcom}. In the present context it means that $[T_i,H]\neq 0$ so, in the presence of a magnetic field, we don't get to label states by the naive lattice momenta which would be related to eigenvalues of $T_i$. This shouldn't be too surprising: the algebra \eqn{talg} is a reflection of the fact that the gauge invariant momenta don't commute in a magnetic field, as we saw in \eqn{picoms}.

\para
However, we can construct closely related operators that do commute with $T_j$ and, hence, with the Hamiltonian. These are defined by
\be \tilde{T}_j = \sum_{{\bf x}}\ |{\bf x}\rangle e^{-ie a \tilde{A}_j({\bf x})/\hbar} \langle {\bf x} + {\bf e}_j| \nn\ee
where the new gauge field $\tilde{A}_j$ is constructed to obey $\partial_k{\tilde{A}_j} = \partial_j {A_k}$. In Landau gauge, this means that we should take
\be \tilde{A}_1 = Bx_2\ \ \ {\rm and}\ \ \ \tilde{A_2} = 0\nn\ee
When this holds, we have
\be [T_j,\tilde{T}_k] = [T_j^\dagger,\tilde{T}_k]  = 0\ \ \ \Rightarrow\ \ \ [H,\tilde{T}_j]=0\nn\ee
These operators commute with the Hamiltonian, but do not themselves commute.  Instead, they too obey the algebra \eqn{talg}.
\be \tilde{T}_2\, \tilde{T}_1 = e^{ie\Phi/\hbar\,} \tilde{T}_1\, \tilde{T}_2\label{ttalg}\ee
This means that we could label states of the Hamiltonian by eigenvalues of, say, $\tilde{T}_2$ but not simultaneously by eigenvalues of $\tilde{T}_1$. This isn't enough to construct a Brillouin zone.

\para
At this point, we can see that something special happens when the flux is a rational multiple of $\Phi_0$, as in  \eqn{rational}. We can now build commuting operators by
\be [\tilde{T}_1^{n_1},\tilde{T}_2^{n_2} ] = 0\ \ \ {\rm whenever}\ \ \ \frac{p}{q}n_1n_2 \in{\bf Z}\nn\ee
This means in particular that we can label energy  eigenstates by their eigenvalue under $\tilde{T}_2$ and, simultaneously, their eigenvalue under $\tilde{T}_1^q$. We call these states $|{\bf k}\rangle$ with ${\bf k}=(k_1,k_2)$. They are Bloch-like eigenstates, satisfying
\be H|{\bf k}\rangle = E({\bf k})|{\bf k}\rangle\ \ \ {\rm with}\ \ \ T_1^q|{\bf k}\rangle = e^{iqk_1a}|{\bf k}\rangle\ \ \ {\rm and}\ \ \ T_2|{\bf k}\rangle = e^{ik_2a}|{\bf k}\rangle\nn\ee
Note that the momenta $k_i$ are again periodic, but now with the range
\be -\frac{\pi}{qa} < k_1  \leq \frac{\pi}{qa}\ \ \ {\rm and}\ \ \ -\frac{\pi}{a}<k_2\leq \frac{\pi}{a}\label{magbz}\ee
The momenta $k_i$ parameterise the {\it magnetic Brillouin zone}. It is again a torus ${\bf T}^2$, but $q$ times smaller than the original Brillouin zone \eqn{bz2}. Correspondingly, if the lattice has size $L_1\times L_2$, the number of states in each magnetic Brillouin zone is $L_1L_2/qa^2$. This suggests that the spectrum decomposes into $q$ bands, each with a different range of energies. For generic values of $p$ and $q$, this is correct. 

\para
The algebraic structure above also tells us that any energy eigenvalue in a given band  is $q$-fold degenerate. To see this, consider the state $\tilde{T}_1|{\bf k}\rangle$. Since $[H,\tilde{T}_1]=0$, we know that  this state has the same energy as $|{\bf k}\rangle$: $H\tilde{T}_1|{\bf k}\rangle = E({\bf k})\tilde{T}_1|{\bf k}\rangle$. But, using \eqn{ttalg}, the $k_y$ eigenvalue of this state is
\be \tilde{T}_2 (\tilde{T}_1|{\bf k}\rangle) = e^{ie\Phi/\hbar} \tilde{T}_1\tilde{T}_2|{\bf k}\rangle = e^{i(2\pi p/q + k_2a)}\tilde{T}_1|{\bf k}\rangle\nn\ee
We learn that $|{\bf k}\rangle$ has the same energy as $\tilde{T}_1|{\bf k}\rangle \sim |(k_1,k_2+2\pi p/qa)\rangle$.

\para
The existence of a Brillouin zone \eqn{magbz} is the main result we need to discuss Hall conductivities in this model.  
 However, given that we've come so far it seems silly not to carry on and describe what the spectrum of the Hamiltonian \eqn{latticemagham} looks like. Be warned, however, that the following subsection is a slight detour from our main goal. 

\subsubsection*{Hofstadter Butterfly}

To further understand the spectrum of the Hamiltonian \eqn{latticemagham}, we'll have to roll up our sleeves and work directly with the Schr\"odinger equation. Let's first look in position space. We can write the most general wavefunction as a linear combination of the position eigenstates $|{\bf x}\rangle$,
\be |\psi\rangle = \sum_{{\bf x}}\psi({\bf x})|{\bf x}\rangle\nn\ee
The Schr\"odinger equation $H|\psi\rangle = E|\psi\rangle$ then becomes an infinite system of coupled, discrete equations
\be \left[ \psi({\bf x}+{\bf e}_1) + \psi({\bf x}-{\bf e}_1) + e^{-i2\pi  px^1/qa} \psi({\bf x}+{\bf e}_2) + 
e^{+i2\pi px^1/qa} \psi({\bf x}-{\bf e}_2)\right] = -\frac{E}{t} \psi({\bf x})\nn\ee
We want to find the possible energy eigenvalues $E$.

\para
The way we usually solve these kinds of problems is by doing a Fourier transform of the wavefunction to work in momentum space, with
\be \tilde{\psi}({\bf k}) = \sum_{{\bf x}} \ e^{-i{\bf k}\cdot{\bf x}}\,\psi({\bf x})\label{firsttry}\ee
where, since ${\bf x}$ takes values on a discrete lattice, ${\bf k}$ takes values in the original Brillouin zone \eqn{bz2}.  In the absence of a magnetic field, modes with different momenta ${\bf k}$ decouple from each other. However, if you try the same thing in the presence of a magnetic field, you'll find that the modes with momentum ${\bf k} = (k_1,k_2)$ couple to modes with momentum $(k_1 + 2\pi p/qa,k_2)$. The reflects the fact that, as we have seen, the magnetic Brillouin zone \eqn{magbz} is $q$ times smaller.  For this reason, we instead split the wavefunction \eqn{firsttry} into $q$ different wavefunctions $\tilde{\psi}_r({\bf k})$, with $r=1,\ldots, q$ as
\be \tilde{\psi}_r({\bf k}) = \sum_{{\bf x}} \ e^{-i(k_1 + 2\pi pr/qa, k_2)\cdot{\bf x}}\,\psi({\bf x})\nn\ee
These contain the same information as \eqn{firsttry}, but now the argument ${\bf k}$ ranges over the magnetic Brillouin zone \eqn{magbz}. Given the wavefunctions $\tilde{\psi}_r$, we can always reconstruct $\psi({x})$ by the inverse Fourier transform,
\be \psi({\bf x}) = \sum_{r=1}^q \int_{-\pi/qa}^{+\pi/qa}\frac{dk_1}{2\pi} \int_{-\pi/a}^{+\pi/a} \frac{dk_2}{2\pi} \ e^{i{\bf k}\cdot{\bf x}}\,\tilde{\psi}_r({\bf k})\nn\ee
%
%
%
In this way, we see that we have a $q$-component vector of wavefunctions, $\tilde{\psi}_r({\bf k})$ living on the magnetic Brillouin zone.

\para
Taking the Fourier transform of the discrete Schr\"odinger equation in position space yields the following equation
\be 2\cos\left(k_1a + \frac{2\pi pr}{q}\right) \tilde{\psi}_r({\bf k}) +e^{ik_2a}\tilde{\psi}_{r+1}({\bf k}) + e^{-ik_2a}\tilde{\psi}_{r-1}({\bf k}) = -\frac{E({\bf k})}{t}\,\tilde{\psi}_r({\bf k})\nn\ee
This is known as the {\it Harper equation}.

\begin{figure}[htb]
\begin{center}
\epsfxsize=4in\leavevmode\epsfbox{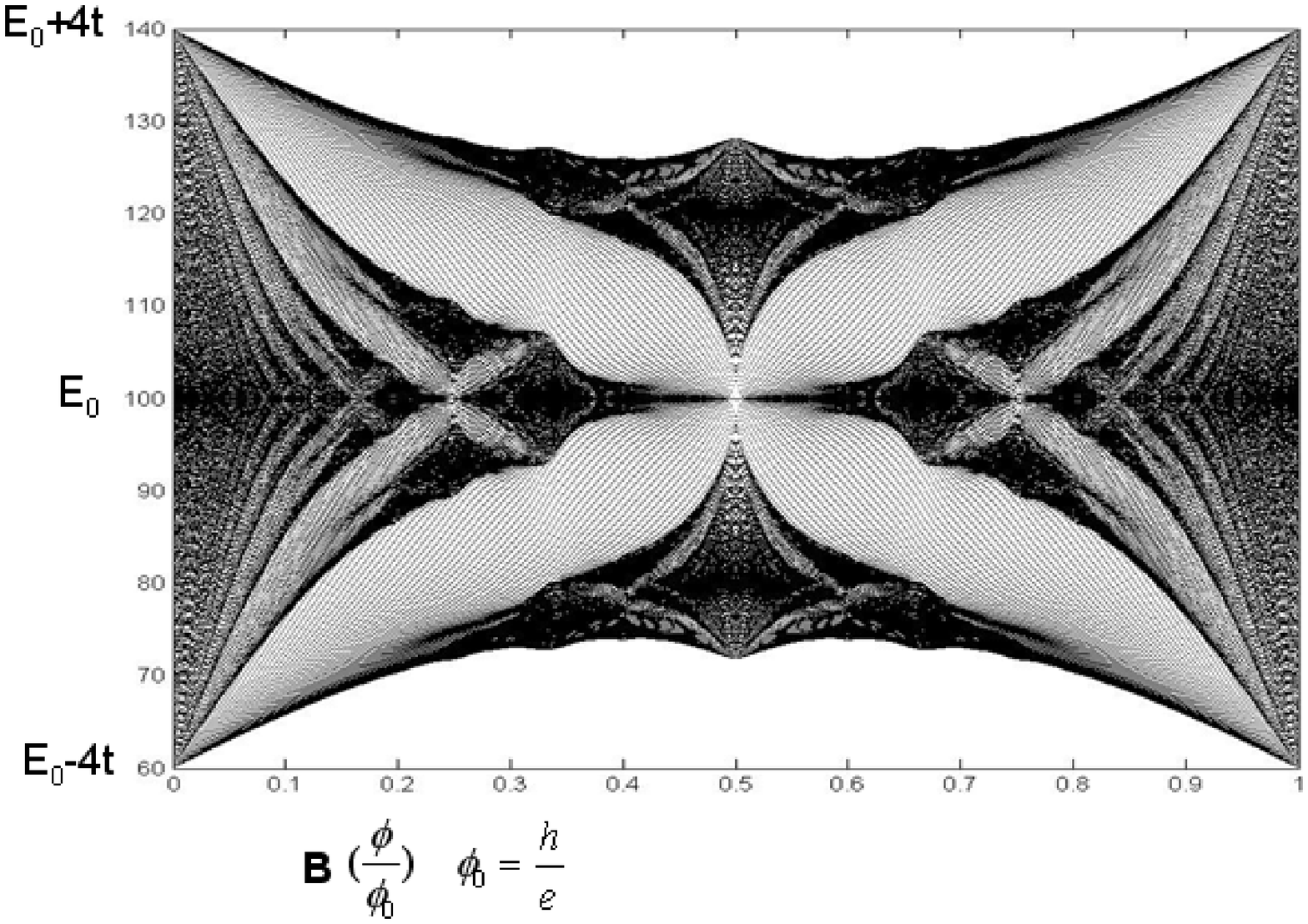}
\end{center}
\caption{The Hofstadter butterfly}
\label{butterflyfig}
\end{figure}

\para
The Harper equation can be solved numerically. The resulting spectrum is quite wonderful. For rational values,  $\Phi/\Phi_0 =  p/q$, the spectrum indeed decomposes into $q$ bands with gaps between them, as we anticipated above. Yet the spectrum also varies smoothly as we change $\Phi$. Obviously if we change   $\Phi/\Phi_0$ continuously it will pass through irrational values; when this happens the spectrum forms something like  a Cantor set. The result is a beautiful fractal structure called the {\it Hofstadter butterfly}\footnote{The spectrum was first solved numerically  by Douglas Hofstadter in  "{\it Energy levels and wave functions of Bloch electrons in rational and irrational magnetic fields}", \href{http://journals.aps.org/prb/abstract/10.1103/PhysRevB.14.2239}{Phys. Rev. B14, 2239 (1976)}. The picture of  the butterfly was taken from Or Cohen's webpage \href{http://phelafel.technion.ac.il/~orcohen/butterfly.html}{http://phelafel.technion.ac.il/$\sim$orcohen/butterfly.html} where you can find a nice description of the techniques used to generate it.}  shown in Figure \ref{butterflyfig}. Here, a point is drawn in black if there is a state with that energy. Otherwise it is white. To get a sense of the structure, you could look at the specific values $\Phi/\Phi_0 = 1/q$, above which you should see $q$ vertical bands of black.

\subsubsection*{TKNN Invariants for Particles on a Lattice in a Magnetic Field}

Finally we reach our main goal: to compute the Hall conductivity of the lattice model for a particle in a background magnetic field. We can only do this for rational fluxes $\Phi = p\Phi_0/q$ for which there exists a magnetic Brillouin zone. In this case, we can use the TKNN formula \eqn{special}, but with the Chern number, which used to be defined by integrating over the Brillouin zone, now arising by  integrating over the magnetic Brillouin zone. 

\para
The computation of the Chern numbers is not so straightforward. (You can find it in the original paper of TKNN or, in more detail, in the book by Fradkin). Here we just state the answer. Even this is not totally straightforward.

\para
First consider the $r^{\rm th}$  of the $q$ bands. Then, to compute the Chern number, you are invited to solve the linear Diophantine equation 
\be r = q s_r + p t_r\nn\ee
with $|t_r| \leq  q/2$. The Chern number of the $r^{\rm th}$ band is given by
\be C_r = t_r - t_{r-1}\nn\ee
where $t_0\equiv 0$. If the first $r$ bands are filled, so that $E_r < E_F < E_{r+1}$, then the Hall conductivity is given by
\be \sigma_{xy}= \frac{e^2}{2\pi\hbar}\, t_r\nn\ee
It's helpful to look at some examples. First, when $\Phi = p \Phi_0$, there is only a single band and the Hall conductivity vanishes. A more complicated, illustrative example is given by $p/q=11/7$. Here the solutions to the Diophantine equation are $(s_r,t_r) = (-3,2), (5,-3), (2,-1), (-1,1), (-4,3), (4,-2), (1,0)$. As we fill consecutive bands, the second number $t_r$ in these pairs determines the Hall conductivity. We see that the Hall conductivity varies in an interesting way, sometimes negative and sometimes positive.

\newpage
\section{The Fractional Quantum Hall Effect}\label{fqhesec}

We've come to a pretty good understanding of the integer quantum Hall effect and the reasons behind it's robustness. Indeed, some of the topological arguments in the previous chapter are so compelling that you might think the Hall resistivity of an insulator has to be an integer. But each of these arguments  has a subtle loophole and ultimately they hold only for non-interacting electrons. As we will now see, much more interesting things can happen when we include interactions. 

\para
As with the integer quantum Hall effect, these interesting things were first discovered by experimenters rather than theorists.  
Indeed, it came as a great surprise to the community when, in 1982, plateaux in the Hall resistivity were seen at non-integer filling fractions. These plateaux were first seen at filling fraction $\nu = \frac{1}{3} $ and $\frac{2}{3}$, and later at $\nu = \frac{1}{5}, \frac{2}{5},\frac{3}{7} , \frac{4}{9}, \frac{5}{9}, \frac{3}{5}, \ldots$ in the lowest Landau level  and  $\nu =  \frac{4}{3}, \frac{5}{3}, \frac{7}{5}, \frac{5}{2}$, $\frac{12}{5}$, $\frac{13}{5},\ldots$ in higher Landau levels, as well as many others. 
There are now around 80 quantum Hall plateaux that have been observed. A number of these are shown below\footnote{This data is from R. Willett, J. P. Eisenstein, H. L. Stormer,
D. C. Tsui, A. C. Gossard  and H. English ``{\it Observation of an Even-Denominator Quantum Number in the Fractional Quantum Hall Effect}", \href{http://journals.aps.org/prl/abstract/10.1103/PhysRevLett.59.1776}{Phys. Rev. Lett. 59, 15 (1987)}.}:
\be \raisebox{-1.1ex}{\epsfxsize=3.9in\epsfbox{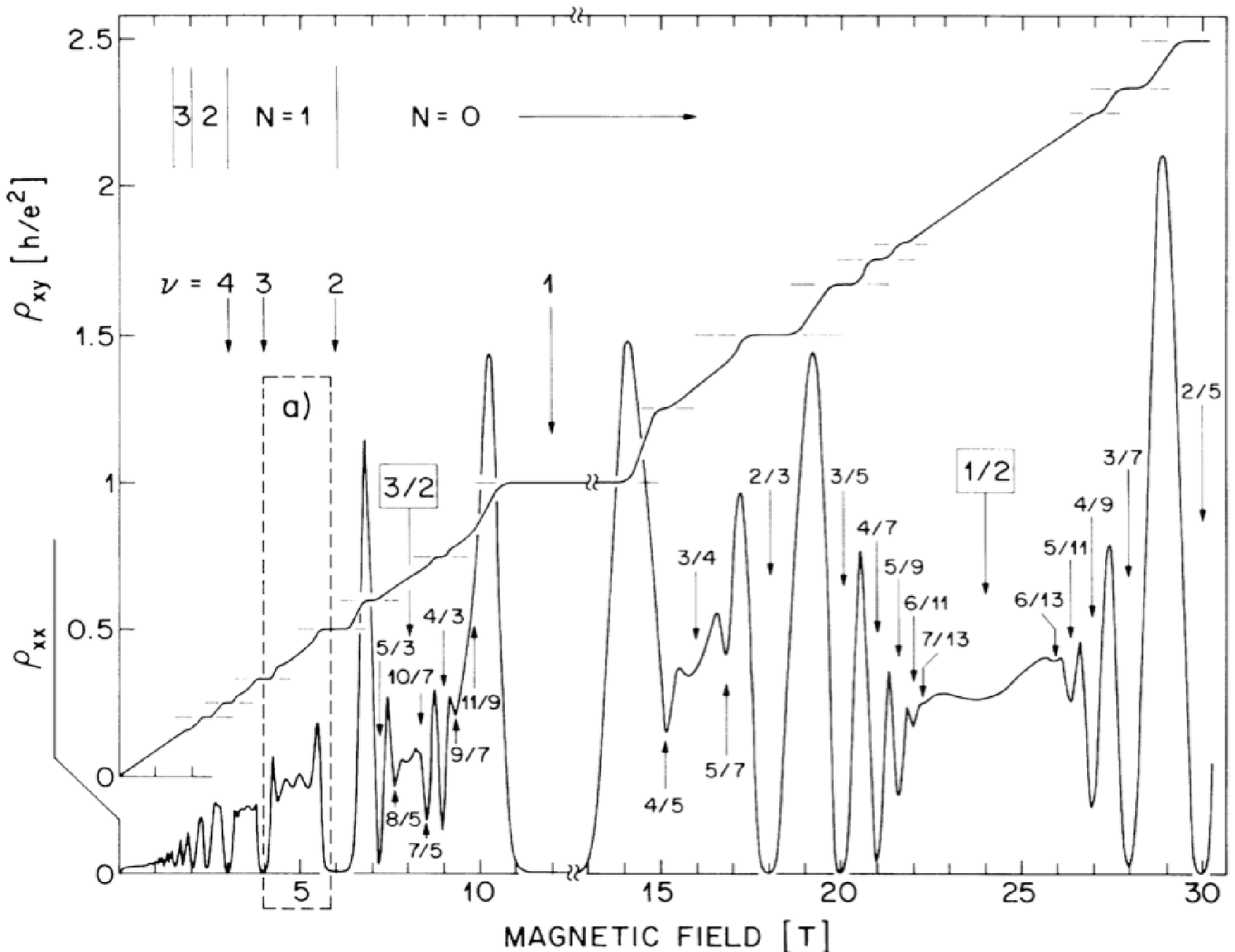}}\nn\ee
There's two things that we can say immediately. First, the interactions between electrons must be playing some role. And second, the answer to why these plateaux form is likely to be very hard. Let's see why.

\DOUBLEFIGURE{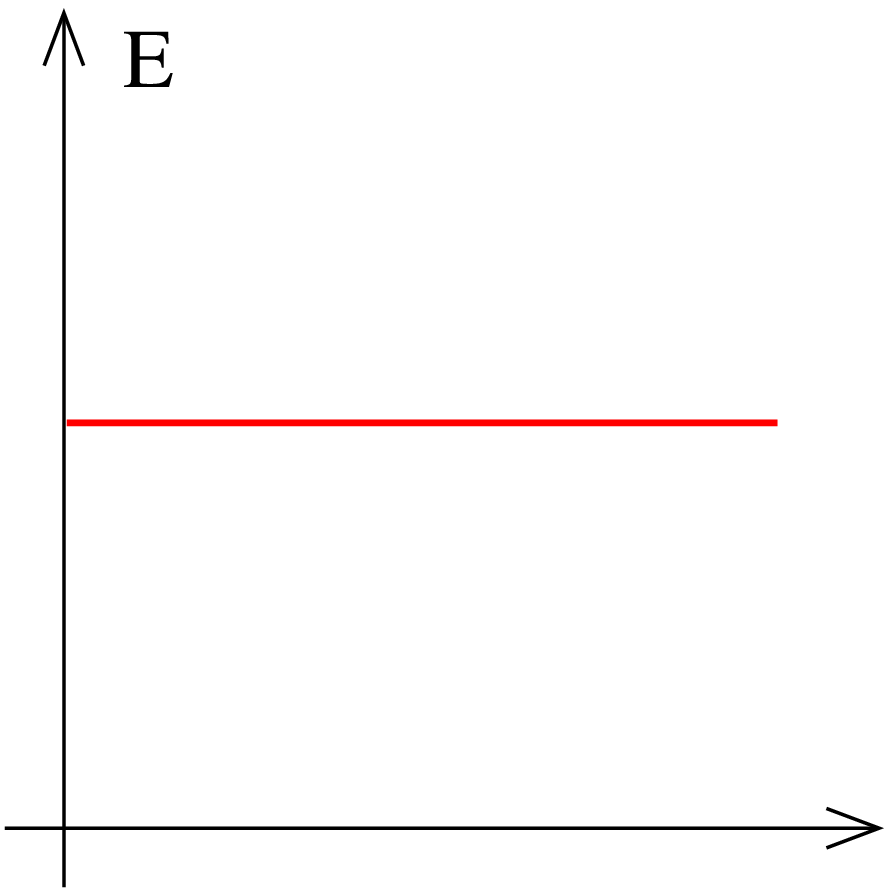,width=120pt}{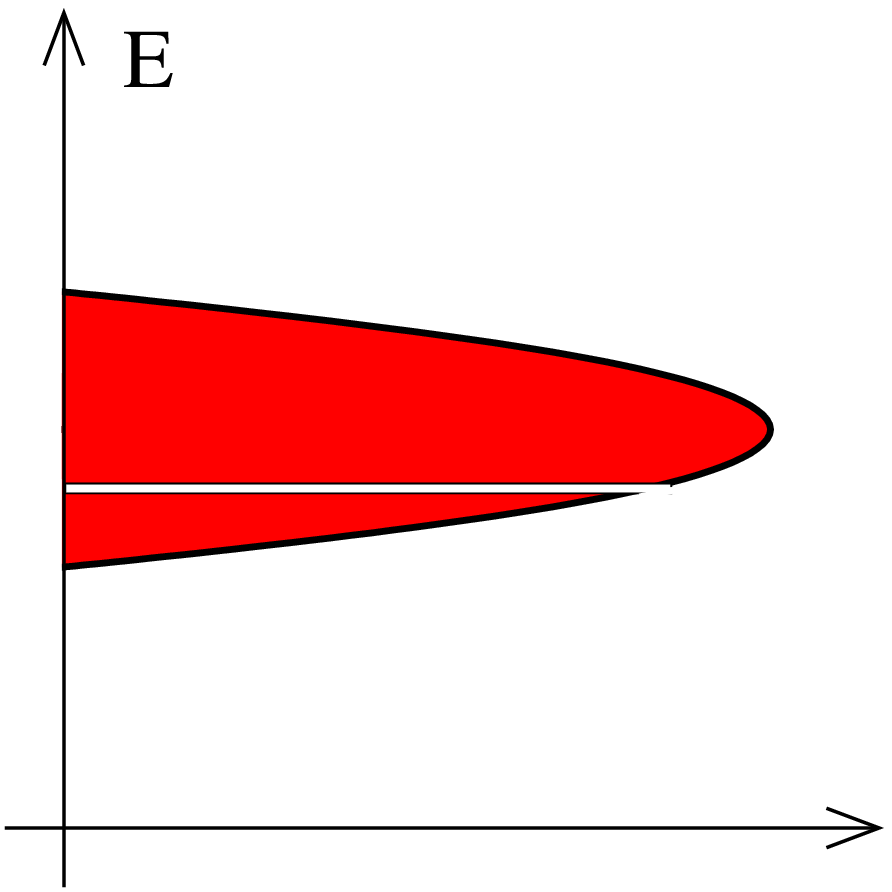,width=120pt}
{Density of states in the lowest Landau level without interactions}{...and  with interactions (with only a single gap at $\nu=1/3$ shown.}

\para
Suppose, for the sake of argument, that we have $\nu <1$ so that the lowest Landau level is partially filled. Each Landau level can house ${\cal N}= AB/\Phi_0$ (spin polarised) electrons, where $B$ is the magnetic field and $A$ is the area of the sample. This is a macroscopic number of electrons. The number of ways to fill $\nu {\cal N}$ of these states is ${{\cal N}}\choose{\nu {\cal N}}$ which, using Stirling's formula, is approximately $\left(\frac{1}{\nu}\right)^{\nu {\cal N}} \left(\frac{1}{1-\nu}\right)^{(1-\nu){\cal N}}$. This is a ridiculously large number: an exponential of an exponential. The ground state of any partially filled Landau level is wildly, macroscopically degenerate. 

\para
Now consider the effect of the Coulomb interaction between electrons,
\be V_{\rm Coulomb} = \frac{e^2}{4\pi\epsilon_0 |{\bf r}_i-{\bf r}_j|}\label{crepel}\ee
On general grounds, we would expect that such an interaction would lift the degeneracy of ground states. But how to pick the right one? The approach we're taught as undergraduates is to use perturbation theory. But, in this case, we're stuck with extraordinarily degenerate perturbation theory where we need to diagonalise a macroscopically large matrix. That's very very hard. Even numerically, no one can do this for more than a dozen or so particles. 

\para
We can, however, use the experiments to intuit what must be going on. As we mentioned above, we expect  the electron interactions to lift the degeneracy of the Landau level, resulting in a spectrum of states of width $\sim E_{\rm Coulomb}$. The data would be nicely explained if this spectrum had gaps at the filling fractions $\nu$ where Hall states are seen.  In the picture above, we've depicted just a single gap at $\nu=1/3$. Presumably though there are many gaps at different fractions: the more prominent the plateaux, the larger the gap. 

\para
Then we can just re-run the story we saw before: we include some disorder, which introduces localised states within the gap, which then gives rise both to the plateaux in $\rho_{xy}$ and the observed $\rho_{xx}=0$. The bigger the gap, the more prominent the  observed plateaux. This whole story requires the hierarchy of energy scales, 
\be \hbar \omega_B  \gg E_{\rm Coulomb} \gg V_{\rm disorder}\nn\ee
We will assume in what follows that this is the case. The question that we will focus on  instead is:  what is the physics of these fractional quantum Hall states? 

\para
In what follows, we will take advantage of the difficulty of a  direct theoretical attack on  the problem to give us  license to be more creative. As we'll see, the level of rigour in the thinking about the fractional quantum Hall effect is somewhat lower than that of the integer effect. Instead, we will paint a compelling picture, using a number of different approaches, to describe what's going on.

\subsection{Laughlin States}

The first approach to the fractional quantum Hall effect was due to Laughlin\footnote{The original paper is ``{\it Anomalous quantum Hall effect: An Incompressible quantum fluid with fractionally charged excitations}" , \href{http://journals.aps.org/prl/abstract/10.1103/PhysRevLett.50.1395}{Phys. Rev. Lett. 50 (1983) 1395}.}, who described the physics at filling fractions
\be \nu = \frac{1}{m} \nn\ee
with $m$ an odd integer. As we've explained above, it's too difficult to diagonalise the Hamiltonian exactly. Instead Laughlin did something very bold: he simply wrote down the answer. This was  motivated by a combination of physical insight and guesswork. As we will see, his guess isn't exactly right but, it's very close to being right. More importantly,  it captures all the relevant physics.

\subsubsection{The Laughlin Wavefunction}\label{lwavesec}

Laughlin's wavefunction didn't come out of nowhere. To motivate it, let's start by considering an illuminating toy model.

\subsubsection*{Two Particles}

Consider two particles interacting in the lowest Landau level. We take an arbitrary central potential between them,
\be V = V(|{\bf r}_1-{\bf r}_2|)\nn\ee
Recall that in our first courses on classical mechanics we solve problems like this by using the conservation of angular momentum. In quantum physics, this means that we work with eigenstates of angular momentum.  As we saw in Section \ref{llevelsec}, if we want to talk about angular momentum in Landau levels, we should work in symmetric gauge. The single particle wavefunctions in the lowest Landau level take the form \eqn{lllwf}
\be \psi_m  \sim z^m e^{-|z|^2/4l_B^2}\nn\ee
with $z=x-iy$. These states are localised on a ring of radius $r = \sqrt{2m} l_B$. The exponent $m$ of these wavefunctions labels the angular momentum. This can be seen by acting with the angular momentum operator \eqn{angmomop}, 
\be J = i\hbar\left(x\ppp{}{y} - y\ppp{}{x}\right)
= \hbar(z\partial -\bar{z}\bar{\partial}) \ \ \ \ \Rightarrow\ \ \ \ \  J\psi_m = \hbar m \,\psi_m\nn\ee
%
%
%
Rather remarkably, this information is enough to solve our two-particle problem for any potential $V$! As long as we neglect mixing between Landau levels (which is valid if $\hbar\omega_B \gg V$) then the two-particle eigenstates for any potential must take the form
\be \psi \sim (z_1+z_2)^M (z_1-z_2)^m e^{-(|z_1|^2 + |z_2|^2)/4l_B^2}\nn\ee
where $M, m$ are non-negative integers, with $M$ determining the angular momentum of the centre of mass, and $m$ the relative angular momentum. Note that here, and below, we've made no attempt to normalise the wavefunctions. 

\para
It's surprising that we can just write down eigenfunctions for a complicated potential $V(r)$ without having to solve the Schr\"odinger equation. It's even more surprising that all potentials $V(r)$ have the same energy eigenstates. It is our insistence that we lie in the lowest Landau level that allows us to do this.

\subsubsection*{Many-Particles}

Unfortunately, it's not possible to generalise arguments similar to those above to uniquely determine the eigenstates for $N>2$ particles. Nonetheless, on general grounds, any lowest Landau level wavefunction must take the form,
\be \psi(z_1,\ldots,z_n) = f(z_1,\ldots , z_N) e^{-\sum_{i=1}^N |z_i|^2/4l_B^2}\label{lllstate}\ee
for some analytic function $f(z)$. Moreover, this function must be anti-symmetric under exchange of any two particle $z_i\leftrightarrow z_j$, reflecting the fact that the underlying electrons are fermions.

\para
Laughlin's proposal for the ground state wavefunction at filling fraction $\nu = 1/m$ is:
\be \psi(z_i) = \prod_{i<j} (z_i-z_j)^me^{-\sum_{i=1}^n |z_i|^2/4l_B^2}\label{laughlin}\ee
Clearly this is anti-symmetric when $m$ is an odd integer. For $m$ an even integer, it can be thought of as a quantum Hall state for bosons. The pre-factor vanishes with a zero of order $m$ whenever two electrons come together. Meanwhile, the exponential factor  decreases quickly whenever the electrons get too far away from the origin. The wavefunction is peaked on configurations that balance these two effects.

\para
Let's first show that the wavefunction has the desired filling fraction. To do this, focus on what the wavefunction is telling us about a single particle, say $z_1$. The terms that depend on $z_1$ in the pre-factor of the Laughlin wavefunction are
\be \prod_{i<j} (z_i-z_j)^m \sim \prod_{i=2}^N (z_1-z_i)^m\nn\ee
which tells us that there are $m(N-1)$ powers of $z_1$. This, in turn, tells us that the maximum angular momentum of the first particle is $m(N-1)$ and so its maximum radius is $R\approx \sqrt{2mN}l_B$. Correspondingly, the area of the droplet is $A\approx 2\pi m Nl_B^2$ (where we've replaced $N-1$ with $N$). Recall that the number of states in the full Landau level  is $AB/\Phi_0 = A/2\pi l_B^2 \approx mN$. 
This argument gives us the filling fraction
\be \nu = \frac{1}{m}\label{nulaughlin}\ee
as promised.

\para
 It can be shown numerically that, at least for small numbers of particles, this wavefunction has greater than 99\% overlap with the true ground state arising arising from both the Coulomb repulsion \eqn{crepel}  as well as a number of other repulsive potentials $V$. Heuristically this occurs because the wavefunction has a zero of order $m$ whenever two electrons coincide. Of course, a single zero is guaranteed by Pauli exclusion, but the Laughlin wavefunction does more. It's as if each electron carves out a space around it which helps it minimise the energy for repulsive potentials. 
 
 \para
The high numerical overlap with the true ground state is often  put forward as strong evidence for the veracity of the Laughlin wavefunction. While it's certainly impressive, this isn't the reason that the Laughlin wavefunction is interesting. Finding the ground state numerically is difficult and can only be done for a couple of dozen particles. While this may provide 99.99\% overlap with the Laughlin wavefunction, by the time we get to $10^{11}$ particles or so, the overlap is likely to be essentially zero. Instead, we should think of the Laughlin wavefunctions as states which lie in the same ``universality class" as the true ground state. We will explain what this means in  Section \ref{quasisec} but, roughly speaking, it is the statement that the states have the same fractional excitations and the same topological order as the true ground states.

\subsubsection*{The Fully Filled Landau Level}

From the arguments above, the Laughlin state \eqn{laughlin} with $m=1$ should describe a completely filled Landau level. But this is something we can compute in the non-interacting picture and it provides a simple check on the Laughlin ansatz.

\para
Let us first review how to build the many-particle wavefunction for  non-interacting electrons. Suppose that $N$ electrons sit in states $\psi_i(x)$, with $1=1,\ldots,N$. Because the electrons are fermions, these states must be distinct. To build the many-particle wavefunction, we need to anti-symmetrise over all particles. This is achieved by the {\it Slater determinant},
\be \psi(x_i) = \left|\begin{array}{cccc} \psi_1(x_1) & \psi_1(x_2) & \ldots & \psi_1(x_N) \\  \psi_2(x_1) & \psi_2(x_2) & \ldots & \psi_2(x_N) \\ \vdots & & & \vdots \\ \psi_N(x_1) & \psi_N(x_2) & \ldots & \psi_N(x_N)  \end{array}\right|\label{slater}\ee
We can now apply this to the lowest Landau level, with the single-particle states built up with successive angular momentum quantum numbers
\be \psi_m(z) \sim z^{m-1} e^{-|z|^2/4l_B^2}\ \ \  \ \ \ \ m=1,\ldots,N\nn\ee
The resulting Slater determinant gives a state of the general form \eqn{lllstate}, with $f(z_i)$ given by a function known as the {\it Vandermonde determinant},
\be f(z_i) = \left|\begin{array}{cccc} z_1^0 & z_2^0 & \ldots & z_N^0 \\  z_1 & z_2 & \ldots & z_3 \\ \vdots & & & \vdots \\ z_1^{N-1} & z_2^{N-1} & \ldots & z_N^{N-1}  \end{array}\right|  = \prod_{i<j}(z_i-z_j)\nn\ee
To see that the determinant is indeed given by the product factor, note  that $\prod_{i<j} (z_i-z_j)$ is the lowest order, fully anti-symmetric polynomial (because any such polynomial must have a factor $(z_i-z_j)$ for each pair $i\neq j$). Meanwhile, the determinant is also completely anti-symmetric and has the same order as the product factor. This ensures that they must be equal up to an overall numerical factor which can be checked to be 1. We see that $m=1$ Laughlin state does indeed agree with the wavefunction for a completely filled lowest Landau level.

\subsubsection*{The Competing Phase: The Wigner Crystal}

The Laughlin state should be thought of as a liquid phase of electrons. In fact, strictly speaking, it should be thought of as an entirely new phase of matter, distinguished by a property called {\it topological order} which we'll discuss  in Section \ref{torussec}. But, if you're looking for a classical analogy, a liquid is the best. 

\para
There is a competing solid phase  in which the electrons form a two-dimensional triangular lattice, known as a {\it Wigner crystal}. Indeed, before the discovery of the quantum Hall effect, it was thought that this would be the preferred phase of electrons in high magnetic fields. It's now known that the Wigner crystal has lower energy than the Laughlin state only when the densities of electrons are low. It is observed for filling fractions $\nu \lesssim \frac{1}{7}$

\subsubsection{Plasma Analogy}\label{plasmasec}

The Laughlin wavefunctions \eqn{laughlin} are very easy to write down. But it's hard to actually compute with them. The reason is simple: they are wavefunctions for a macroscopic number of particles  which means that if we want to compute expectation values of operators, we're going to have to do a macroscopic number of integrals $\int d^2z_i$. And that's difficult.  

\para
For example, suppose that we want to figure out the average density of the quantum Hall droplet. We need to compute the expectation value of the density operator 
\be n(z) = \sum_{i=1}^N \delta(z-z_i)\nn\ee
This is given by
\be \langle \psi |n(z)|\psi\rangle = \frac{\int \prod_{i=1}^N d^2z_i\ n(z) P[z_i]}{\prod_{i=1}^N d^2z_i \ P[z_i]}\label{densexpt}\ee
 where we've introduced the un-normalised probability density associated to the Laughlin wavefunction
\be P[z_i] = \prod_{i<j} \frac{|z_i-z_j|^{2m}}{l_B^{2m}} e^{-\sum_i |z_i|^2/2l_B^2}\label{laughlinprob}\ee
The integrals in \eqn{densexpt} are hard. How to proceed?

\para
The key observation is that the expectation value \eqn{densexpt} has the same formal structure as the kind of things we compute in classical statistical mechanics, with the denominator interpreted as the partition function,
\be Z = \prod_{i=1}^N d^2 z_i\ P[z_i]\nn\ee
Indeed, we can make this analogy sharper by writing the  probability distribution \eqn{laughlinprob} so it looks like a Boltzmann distribution function,
\be P[z_i] = e^{-\beta U(z_i)}\nn\ee
with
\be \beta U(z_i)=  - 2m \sum_{i<j} \log\left(\frac{|z_i-z_j|}{l_B}\right) + \frac{1}{2l_B^2} \sum_{i=1}^N |z_i|^2 \nn\ee
Of course, this hasn't helped us do the integrals. But the hope is that perhaps we can interpret the potential $U(z_i)$ as something familiar from statistical physics which will at least provide us with some intuition for what to expect. And, indeed, this does turn out to be the case, but only if we pick $\beta$ --- which, in a statistical mechanics context is interpreted as inverse temperature --- to take the specific value 
\be \beta = \frac{2}{m}\label{beta}\ee
I stress that the quantum Hall state isn't placed at a finite temperature. This is an auxiliary, or fake, ``temperature". Indeed, you can tell it's not a real temperature because it's dimensionless! To compensate, the potential is also dimensionless, given by
\be U(z_i) =  -m^2  \sum_{i<j} \log\left(\frac{|z_i-z_j|}{l_B}\right)+ \frac{m}{4l_B^2} \sum_{i=1}^N |z_i|^2\label{plasmau}\ee
We'll now show  that this is the potential energy for a plasma of charged particles moving in two-dimensions, where each particle carried electric charge $q=-m$.

\para
The first term in \eqn{plasmau} is the Coulomb potential between two particles of charge $q$ when both the particle and the electric field lines are restricted to lie in a two-dimensional plane. To see this, note that Poisson equation in two dimensions tells us that the electrostatic potential generated by a point charge $q$ is 
\be -\nabla^2 \phi = 2\pi q \delta^2({\bf r})\ \ \ \Rightarrow\ \ \ \ \phi = -q \log\left(\frac{r}{l_B}\right)\nn\ee
The  potential energy between two particles of charge $q$ is then $U = q\phi$, which is indeed the first term in \eqn{plasmau}.

\para
The second term in \eqn{plasmau} describes a neutralising background of constant charge. A constant background of charge density $\rho_0$ would have electrostatic potential obeying $-\nabla^2\phi = 2\pi\rho_0$. Meanwhile, the second term in the potential obeys 
\be -\nabla^2 \left(\frac{|z|^2}{4l_B^2}\right) = -\frac{1}{l_B^2}\nn\ee
which tells us that each electron feels a  background charge density 
\be \rho_0 = -\frac{1}{2\pi l_B^2}\label{plasmaback}\ee
Note that this is equal (up to fundamental constants) to the background flux $B$ in the quantum Hall sample. 

\para
Now we can use our intuition about this plasma. To minimise the energy, the plasma will want to neutralise, on average, the background charge density. Each particle carries charge $q=-m$ which means that the compensating density of particles $n$ should be  $mn=\rho_0$, or
\be n = \frac{1}{2\pi l_B^2 m} \nn\ee
This is the expected density of a state at filling fraction $\nu=1/m$. This argument has also told us something new. Naively, the form of the Laughlin wavefunction makes it look as if the origin is special. But that's misleading. The plasma analogy tells us that the average density of particles is constant. 

\para
The plasma analogy can also help answer more detailed questions about the variation of the density \eqn{densexpt} on shorter distance scales. Intuitively, we might expect that at low temperatures (keeping the density fixed), the plasma forms a solid, crystal like structure, while at high temperatures it is a liquid. Alternatively, at low densities (keeping the temperature fixed) we would expect it to form a solid while, at high densities, it would be a liquid. To determine the structure of the Laughlin wavefunction, we should ask which phase the  plasma lies in at temperature $\beta = 2/m$ and density $n=1/2\pi l_B^2 m$.

\para
This is a question which can only be answered by numerical work. It turns out that the plasma is a solid when $m\gtrsim 70$. For the low $m$ of interest, in particular $m=3$ and $5$, the Laughlin wavefunction describes a liquid. (Note that this is not the same issue as whether the Wigner crystal wavefunction is preferred over the Laughlin wavefunction: it's a question of whether the Laughlin wavefunction itself describes a liquid or solid).

\subsubsection{Toy Hamiltonians}\label{toysec}

The Laughlin state \eqn{laughlin} is not the exact ground state of the Hamiltonian with Coulomb repulsion. However, it is possible to write down a toy Hamiltonian whose ground state is given by the Laughlin state. Here we explain how to do this, using some tools which will also provide us with a better understanding of the general problem.

\para
Let's go back to the problem of two particles interacting through a general central potential $V(|{\bf r}_1-{\bf r}_2|)$. As we saw in  Section \ref{lwavesec}, in the lowest Landau level the eigenstates for any potential are the same, characterised by two non-negative integers: the angular momentum of the centre of mass $M$ and the relative angular momentum $m$, 
\be |M,m\rangle \sim (z_1+z_2)^M(z_1-z_2)^me^{-(|z_1|^2 + |z_2|^2)/4l_B^2}\nn\ee
We should take $m$ odd if the particles are fermions, $m$ even if they are bosons.

\para 
The eigenvalues of the potential $V$ are given by
\be v_m  = \frac{\langle M,m|V|M,m\rangle}{\langle M,m|M,m\rangle}\label{pseudopot}\ee
These eigenvalues are sometimes referred to as {\it Haldane pseudopotentials}. For central potentials, they do not depend on the overall angular momentum $M$. 

\para
These eigenvalues capture a crude picture of the spatial profile of the potential. This is because, as we have seen, the wavefunctions $|M,m\rangle$ are peaked on a circle of radius $r \approx \sqrt{2m}l_B$. Correspondingly, the eigenvalues are roughly
\be v_m\approx V( r = \sqrt{2m}l_B)\label{vspace}\ee
This means that typically the $v_m$ are positive for a repulsive potential and negative for an attractive potential, in each case falling off as $V(r)$ as $m$ increases. 

\para
Importantly, however, the eigenvalues are discrete. This simple fact is telling us some interesting physics: it means that each of the states $|M,m\rangle$ can be thought of as a bound state of two particles, even if the potential is repulsive! This is in stark contrast to quantum mechanics in the absence of a magnetic field where there are no discrete-energy bound states for a repulsive potential, only scattering states with a continuous spectrum. But the magnetic field changes this behaviour.

\para
Given the eigenvalues $v_m$, we can always reconstruct the potential $V$. In this lowest Landau level, there is no kinetic energy and the potential is the only contribution to the Hamiltonian. It's useful to write it as
\be H = \sum_{m'} v_{m'} {\cal P}_{m'}\label{haldaneham}\ee
where ${\cal P}_m$ is the operator which projects onto states in which the two particles have relative angular momentum $m$. 

\para
Now we can just pick whatever $v_m$ we like to design our own Hamiltonians. Of course, they may not be very realistic when written in terms of $V(r)$ but we won't let that bother us too much. In this spirit, consider the choice 
\be v_{m'} = \left\{\begin{array}{lc} 1 \ \ \ \ \ & m' < m \\ 0 & m' \geq m\end{array}\right.\label{toyham}\ee
This Hamiltonian means that you pay an energy cost if the relative angular momentum of the particles dips below some fixed value $m$. But it costs you nothing to have a high angular momentum. In position space, the equation \eqn{vspace} tells us that there's a finite energy cost if the particles get too close. 

\subsubsection*{Toy Hamiltonians for Many Particles}

We can also use the pseudopotentials to construct Hamiltonians for $N$ particles. To do this, we introduce the operator ${\cal P}_m(ij)$. This projects the wavefunction onto the state in which the $i^{\rm th}$ and $j^{\rm th}$ particles have relative angular momentum $m$. We then construct the Hamiltonian as
\be H = \sum_{m'=1}^\infty \sum_{i<j} v_{m'} {\cal P}_{m'}(ij)\label{tinker}\ee
Note, however, that ${\cal P}_m(ij)$ and ${\cal P}_m(kj)$ do not commute with each other. This is what makes these many-particle Hamiltonians difficult to solve. 

\para
Now consider the many-particle Hamiltonian with $v_{m'}$ given by \eqn{toyham}. This time, you pay an energy cost whenever the relative angular momentum of any pair of particles is less than $m$. You can avoid this energy cost by including a factor of $(z_i-z_j)^m$ for each pair of particles, and writing down a  wavefunction of the form
\be \psi(z_i) = s(z_i)  \prod_{i<j} (z_i-z_j)^m \, e^{-\sum_i |z_i|^2/4 l_B^2}\label{toypsi}\ee
where  $s(z_i)$ can be any symmetric polynomial in the $z_i$ to preserve the statistics of the particles. All such wavefunctions have the vanishing energy.

\para
So far this doesn't pick out the Laughlin state, which has $s(z_i) =1$, as the ground state. But there is something special about this state: among all states  \eqn{toypsi}, it is the most compact. Indeed, we saw in Section \ref{lwavesec} that it takes up an area $A = 2\pi m N l_B^2$. Any state with $s(z_i)\neq 1$ necessarily spreads over a larger spatial area. This means that the Laughlin wavefunction will be the ground state if we also add a confining potential to the system.

\para
We can state this in a slightly different way in terms of angular momentum. We know that states with higher angular momentum sit at larger radius. This means that we can take the total  angular momentum operator $J$ as a proxy for the confining potential and consider the Hamiltonian
\be H = \sum_{m'=1}^{m-1} \sum_{i<j}  {\cal P}_{m'}(ij) +  \omega J \label{confiningtoy}\ee
The Laughlin wavefunction has the lowest energy: $E_0 = \frac{1}{2}\omega mN(N-1)$.  Any  wavefunction of the form \eqn{toypsi} with $s(z_i)\neq 1$ has spatial extent larger than the ground state, and hence higher angular momentum,  and so costs extra energy due to the second term; any wavefunction with spatial extent smaller than the Laughlin wavefunction necessarily has a pair of particles with relative angular momentum less than $m$ and so pays an energy cost due to the first term. 

\para
The fact that it costs a finite energy to squeeze the wavefunction is expected to hold for more realistic Hamiltonians as well. It is usually expressed by saying that the quantum Hall fluid is {\it incompressible}.  This is responsible for the gap in the bulk spectrum described in the introduction of this section. However, it turns out that the dynamics of states with $s(z_i)\neq 1$ contains some interesting information. We'll return to this in Section \ref{ledgesec}. 

\subsection{Quasi-Holes and Quasi-Particles}\label{quasisec}

So far, we've only discussed the ground state of the $\nu = 1/m$ quantum Hall systems. Now we turn to their excitations. There are two types of charged excitations, known as {\it quasi-holes} and {\it quasi-particles}.  We discuss them in turn.

\subsubsection*{Quasi-Holes}

The wavefunction describing a quasi-hole at position $\eta \in {\bf C}$ is 
\be \psi_{\rm hole}(z;\eta) = \prod_{i=1}^N(z_i-\eta)\prod_{k<l} (z_k-z_l)^m\,e^{-\sum_{i=1}^n |z_i|^2/4l_B^2}\label{hole}\ee
We see that the electron density now vanishes at the point $\eta$. In other words, we have created a ``hole" in the electron fluid. More generally,  we can introduce $M$ quasi-holes in the quantum Hall fluid at positions $\eta_j$ with $j=1,\ldots,M$, with wavefunction
\be \psi_{\rm M-hole}(z;\eta) =\prod_{j=1}^M \prod_{i=1}^N(z_i-\eta_j)\prod_{k<l} (z_k-z_l)^m\,e^{-\sum_{i=1}^n |z_i|^2/4l_B^2}\label{multihole}\ee
The quasi-hole has a remarkable property: it carries a fraction of the electric charge of the electron! In our convention, the electron has charge $-e$; the quasi-hole has charge $e^*=+e/m$.

\para
 A heuristic explanation of the fractional charge follows from noting that if we place $m$ quasi-holes at the same point $\eta$ then the wavefunction becomes
\be\psi_{\rm m-hole}(z;\eta) = \prod_{i=1}^N(z_i-\eta)^m\prod_{k<l} (z_k-z_l)^m\,e^{-\sum_{i=1}^n |z_i|^2/4l_B^2}\nn\ee
If $\eta$ was a dynamical variable, as opposed to a parameter, this is just the original wavefunction with an extra electron at position $\eta$. But because $\eta$ is not a dynamical variable, but instead a parameter, it's really a Laughlin wavefunction that describes a deficit of a single electron at position $\eta$. This means that  $m$ holes act like a deficit of a single electron, so a single quasi-hole is $1/m^{\rm th}$ of an electron. It should therefore carry charge $+e/m$.

\para
We can make exactly the same argument in the context of the plasma analogy for the quasi-hole wavefunction \eqn{hole}. 
The resulting plasma potential energy has an extra term compared to \eqn{plasmau}, 
\be
U(z_i) =  -m^2  \sum_{i<j} \log\left(\frac{|z_i-z_j|}{l_B}\right) - m\sum_i \log \left(\frac{|z_i-\eta|}{l_B}\right)+ \frac{m}{4l_B^2} \sum_{i=1}^N |z_i|^2\nn\ee
This extra term looks like an impurity in the plasma with charge 1. The particles in the plasma are expected to swarm around and screen this impurity. Each particle corresponds to a single electron, but has charge $q=-m$ in the plasma. The impurity carries $-1/m$ the charge of the electron. So the effective charge that's missing is $+1/m$; this is the charge of the quasi-hole.

\para
The existence of fractional charge is very striking. We'll discuss this phenomenon more in the following section, but we'll postpone a direct derivation of fractional charge until  Section \ref{fracstatsec} where we also discuss the related phenomenon of fractional statistics.

%
%
 %
%
%

\subsubsection*{Quasi-Particles}

There are also excitations of the quantum Hall fluid which carry charge $e^*=-e/m$, i.e. the same sign as the charge of an electron. These are {\it quasi-particles}.

\para
It seems to be somewhat harder to write down quasi-particle eigenstates compared to quasi-hole eigenstates. To see the problem, note that we want to increase the density of electrons inside the Hall fluid and, hence, decrease the relative angular momentum of some pair of electrons. In the case of the quasi-hole, it was simple enough to {\it increase} the angular momentum: for example, for a hole at the origin we simply need to multiply the Laughlin wavefunction by the factor $\prod_i z_i$.  But now that we want to decrease the angular momentum, we're not  allowed divide by  $\prod_i z_i$ as the resulting wavefunction is badly singular. Nor can we multiply by $\prod_i \bar{z}_i$ because, although this will decrease the angular momentum, the resulting wavefunction no longer sits in the lowest Landau level. Instead, a simple way to reduce the degree of a polynomial is to differentiate. This leads us to a candidate wavefunction for the quasi-particle,
\be \psi_{\rm particle}(z,\eta) =  \left[\prod_{i=1}^N\left(2\ppp{}{z_i}-\bar{\eta}\right)\prod_{k<l} (z_k-z_l)^m\right]\,e^{-\sum_{i=1}^n |z_i|^2/4l_B^2}\label{quasiparticle}\ee
Here the derivatives act only on the polynomial pre-factor; not on the exponential. The factor of $1/2$ in front of the position of the quasi-particle comes from a more careful analysis.

\para
The quasi-particle wavefunction \eqn{quasiparticle} is not quite as friendly as the quasi-hole wavefunction \eqn{hole}. For a start, the derivatives make it harder to work with and, for this reason, we will mostly derive results for quasi-holes in what follows. Further, the quasi-hole wavefunction \eqn{hole} is an eigenstate of the toy Hamiltonian \eqn{tinker} (we'll see why shortly) while \eqn{quasiparticle} is not. In fact, as far as I'm aware, the quasi-particle eigenstate of  the toy Hamiltonain is not known.

\DOUBLEFIGURE{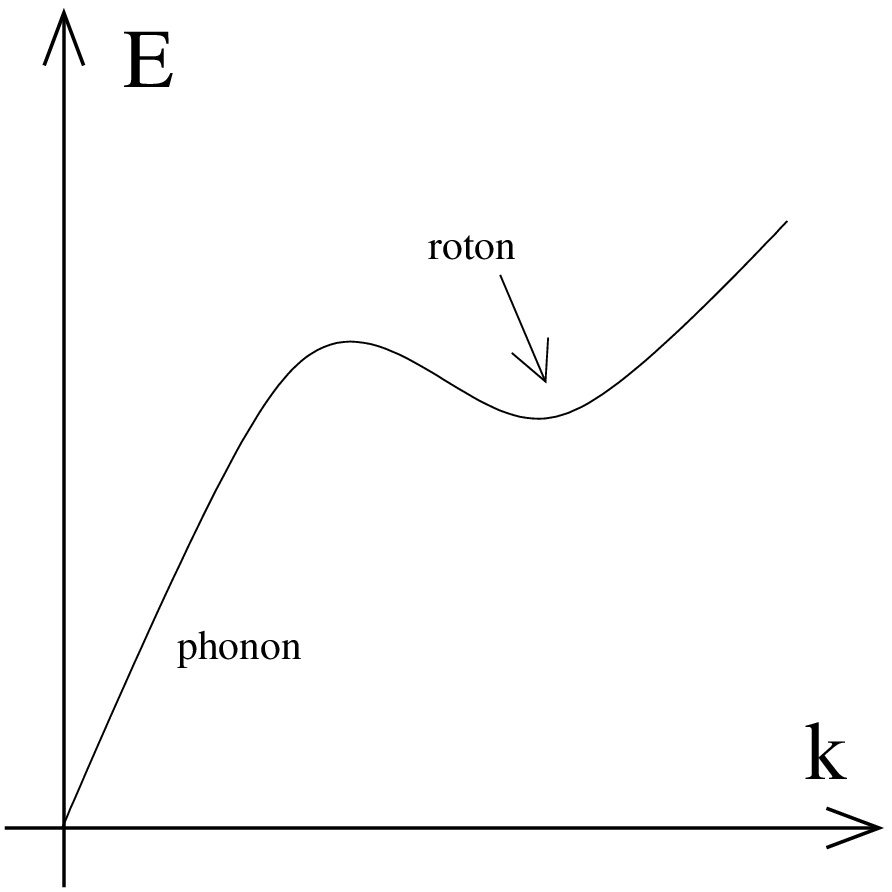,width=120pt}{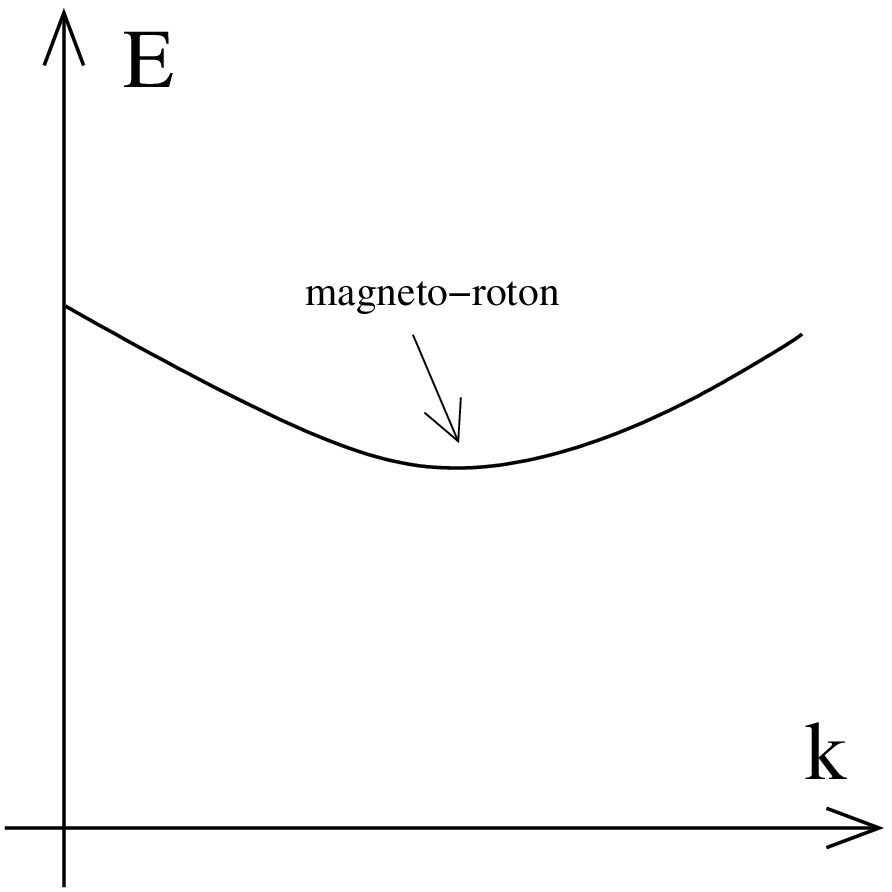,width=120pt}
{A cartoon of the dispersion relation in superfluids...}{...and  for neutral excitations in quantum Hall fluids.}

\subsubsection*{Neutral Excitations}

Before we proceed,  we mention in passing that there are also neutral, collective excitations of the quantum Hall fluid in which the density and charge ripples in wave-like behaviour over large distances. These are  similar to the phonon excitations in superfluids, except the energy cost does not vanish as the momentum $\hbar k\rightarrow 0$. The fact that these modes are gapped at $k=0$ is the statement that the quantum Hall liquid is incompressible. In both cases, the energy-momentum dispersion relation exhibits a minimum  at some finite wavevector $k$, referred to as a {\it roton} in superfluids and a {\it magneto-roton} in quantum Hall fluids. In both cases this is indicating the desire of the liquid to freeze to a solid -- which, for the quantum Hall fluid is a Wigner crystal. In both cases, this desire is ultimately thwarted by quantum fluctuations. 

\para
In the quantum Hall fluid, the minimum occurs at momentum $k\sim 1/l_B$. In recent years, experiment has shown there is a rich structure underlying this. In particular, at other filling fractions (which we will discuss in Section \ref{otherfillingsec}) more than one minima is observed. We will not discuss these neutral excitations in these lectures. 

\subsubsection{Fractional Charge}\label{fracchargesec}

The existence of an object which carries fractional electric charge is rather surprising. In this section, we'll explore some consequences.

\subsubsection*{Hall Conductivity Revisited}

The most basic question we should ask of the Laughlin state is: does it reproduce the right Hall conductivity? To see that it does, we can repeat  the Corbino disc argument of Section \ref{pumpingsec}. As before, we introduce a flux $\Phi(t)$ into the centre of the ring which 
we slowly increase from zero to $\Phi_0$.  This induces a spectral flow so that when we reach $\Phi=\Phi_0$ we sit in a new eigenstate of the Hamiltonian in which the angular momentum of each electron increased by one. This is achieved by multiplying the wavefunction by the factor $\prod_i z_i$. We could even do this procedure in the case where both the  inner circle and the inserted solenoid become vanishingly small. In this case, multiplying by $\prod_i z_i$ gives us precisely the quasi-hole wavefunction \eqn{hole} with $\eta =0$.

\para
As an aside, note that we can also make the above argument above tells us that the quasi-hole wavefunction with $\eta=0$ must be an eigenstate of the toy Hamiltonian \eqn{confiningtoy}, and indeed it is. (The wavefunction with $\eta\neq 0$ is also an eigenstate in the presence of the confining potential if we replace $\eta \rightarrow \eta e^{i\omega t}$, which tells us that the confining potential causes the quasi-hole to rotate).

\para
We learn that as we increase $\Phi$ from zero to $\Phi_0$, a  particle of charge $-e/m$ is transferred from the inner to the outer ring. This means that a whole electron is transferred only when the flux is increased by $m\Phi_0$ units. The resultant Hall conductivity is 
\be \sigma_{xy} = \frac{e^2}{2\pi\hbar} \frac{1}{m}\nn\ee
as expected.

\para
One can also ask how to reconcile the observed fractional Hall conductivity with the argument for integer quantisation based on Chern numbers when the Hall state is placed on a torus. This is slightly more subtle. It turns out that the ground state of the quantum Hall system on a torus is degenerate, hence violating one of the assumptions of the computation of the Chern number. We'll discuss this more in Section \ref{torussec}.

\subsubsection*{Measuring Fractional Electric Charge}

It's worth pausing to describe in what sense the quasi-particles of the quantum Hall fluid genuinely carry fractional charge. First, we should state the obvious: we haven't violated any fundamental laws of physics here. If you isolate the quantum Hall fluid and measure the total charge you will always find an integer multiple of the electron charge. 

\para
Nonetheless, if you inject an electron (or hole) into the quantum Hall fluid, it will happily split into $m$ seemingly independent quasi-particles (or quasi-holes). The states have a degeneracy labelled by the positions $\eta_i$ of the quasi-objects. Moreover, these positions will respond to outside influences, such a confining potentials or applied electric fields, in the sense that the you can build solutions to the Schr\"odginer equation by endowing the positions with suitable time dependence $\eta_i(t)$. All of this means that the fractionally charged objects truly act as independent particles. 

\para
The fractional charge can be seen experimentally in {\it shot noise} experiments. This is a randomly fluctuating current, where the fluctuations can be traced to the discrete nature of the underlying charge carriers. This allows a direct measurement\footnote{The experiment was first described in R. de-Picciotto, M. Reznikov, M. Heiblum, V. Umansky, G. Bunin, and D. Mahalu, ``{\it Direct observation of a fractional charge}", 
Nature 389, 162 (1997). \href{http://arxiv.org/abs/cond-mat/9707289}{cond-mat/9707289}.} 
 of the charge carriers which, for the $\nu=1/3$ state, were shown to indeed carry charge $e^\star = e/3$.

\subsubsection{Introducing Anyons}\label{anyonsec}

We're taught as undergrads that quantum particles fall into two categories: bosons and fermions. However, if particles are restricted to move in a two-dimensional plane then there is a loophole to the usual argument and, as we now explain, much more interesting things can happen\footnote{This possibility was first pointed out by Jon Magne Leinaas and Jan  Myrheim, ``{\it On the Theory of Identical Particles}", \href{http://www.ifi.unicamp.br/~cabrera/teaching/referencia.pdf}{Il Nuovo
Cimento  B37, 1-23 (1977)}. This was subsequently rediscovered by  Frank Wilczek in ``{\it Quantum Mechanics of Fractional-Spin Particles}",  \href{http://journals.aps.org/prl/abstract/10.1103/PhysRevLett.49.957}{Phys. Rev. Lett. 49 (14) 957 (1982)}.}.

\para
Let's first recall the usual argument that tells us we should restrict to boson and fermions. We take two identical particles described by the  wavefunction $\psi({\bf r}_1,{\bf r}_2)$. Since the particles are identical,  all probabilities must be the same if the particles are exchanged. This tells us that $|\psi({\bf r}_1,{\bf r}_2)|^2 = |\psi({\bf r}_2,{\bf r}_1)|^2$ so that, upon exchange, the wavefunctions differ by at most a phase
\be \psi({\bf r}_1,{\bf r}_2) = e^{i\pi \alpha} \psi({\bf r}_2,{\bf r}_1)\label{exchangephase}\ee
Now suppose that we exchange again. Performing two exchanges is equivalent to a rotation, so should take us back to where we started. This gives the condition
\be \psi({\bf r}_1,{\bf r}_2) = e^{2i\pi \alpha} \psi({\bf r}_1,{\bf r}_2)\ \ \ \ \Rightarrow\ \ \ \ \ e^{2\pi i\alpha} = 1\nn\ee
This gives the two familiar possibilities of bosons ($\alpha=0$) or fermions ($\alpha=1$). 

\para
So what's the loophole in the above argument? The weak point is the statement that when we rotate two particles by $360^\circ$ we should get back to where we came from. Why should this be true? The answer lies in thinking about the topology of the worldlines particles make in spacetime.


\para
In $d=3$ spatial dimensions (and, if you're into string theory, higher), the path that the pair of particles take in spacetime can always be continuously connected to the situation where the particles don't move at all. This is the reason the resulting state should be the same as the one before the exchange. But in $d=2$ spatial dimensions, this is not the case: the worldlines of particles now wind around each other. When particles are exchanged in an anti-clockwise direction, like this
\be   \raisebox{-6.5ex}{\epsfxsize=0.3in\epsfbox{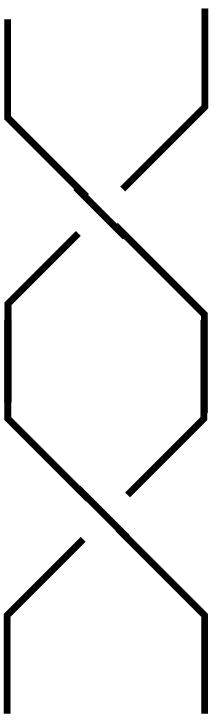}}\nn\ee
the worldlines get tangled. They can't be smoothly continued into the worldlines of particles which are exchanged clockwise, like this:
\be   \raisebox{-6.5ex}{\epsfxsize=0.3in\epsfbox{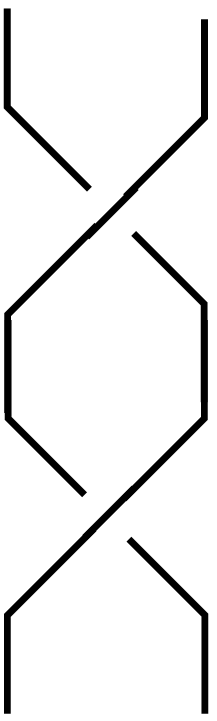}}\nn\ee
 Each winding defines a different topological sector. The essence of the loophole is that, after a rotation in the two-dimensions, the wavefunction may retain a memory of the path it took through the phase. This means that may have any phase $\alpha$ in \eqn{exchangephase}. In fact, we need to be more precise: we will say that after  an {\it anti-clockwise} exchange, the wavefunction is
\be \psi({\bf r}_1,{\bf r}_2) =e^{i\pi \alpha} \psi({\bf r}_2,{\bf r}_1)\label{anyon}\ee
After a clockwise exchange, the phase must be $e^{-i\pi\alpha}$. Particles with $\alpha \neq 0,1$ are referred to as {\it anyons}. 
 This whole subject usually goes by the name of {\it quantum statistics} or {\it fractional statistics}. But it has less to do with statistics and more to do with topology.  

 \subsubsection*{The Braid Group}

 Mathematically, what's going on is that in dimensions $d\geq 3$, the exchange of particles must be described by a representation of the permutation group. But, in $d=2$ dimensions,  exchanges are described a representation of the {\it braid group}.

\para
Suppose that we have $n$ identical particles sitting along a line. We'll order them $1,2,3,\ldots, n$. The game is that of a  street-magician: we shuffle the order of the particles. The image that their worldlines make in spacetime is called a braid. We'll only distinguish braids by their topological class, which means that two braids are considered the same if we can smoothly change one into the other without the worldlines crossing. All such braidings form an infinite group which we call $B_n$

\para
We can generate all elements of the braid group from a simple set of operations, $R_1,\ldots,R_{n-1}$ where $R_i$ exchanges the $i^{\rm th}$ and $(i+1)^{\rm th}$ particle in an anti-clockwise direction. The defining relations obeyed by these generators are
\be R_iR_j = R_jR_i\ \ \ \ |i-j|>2\nn\ee
together with the {\it Yang-Baxter relation},
\be R_iR_{i+1}R_i = R_{i+1}R_iR_{i+1} \ \ \ \ \ i=1,\ldots,n-1\nn\ee
This latter relation is most easily seen by drawing the two associated braids and noting that one can be smoothly deformed into the other.

\DOUBLEFIGURE{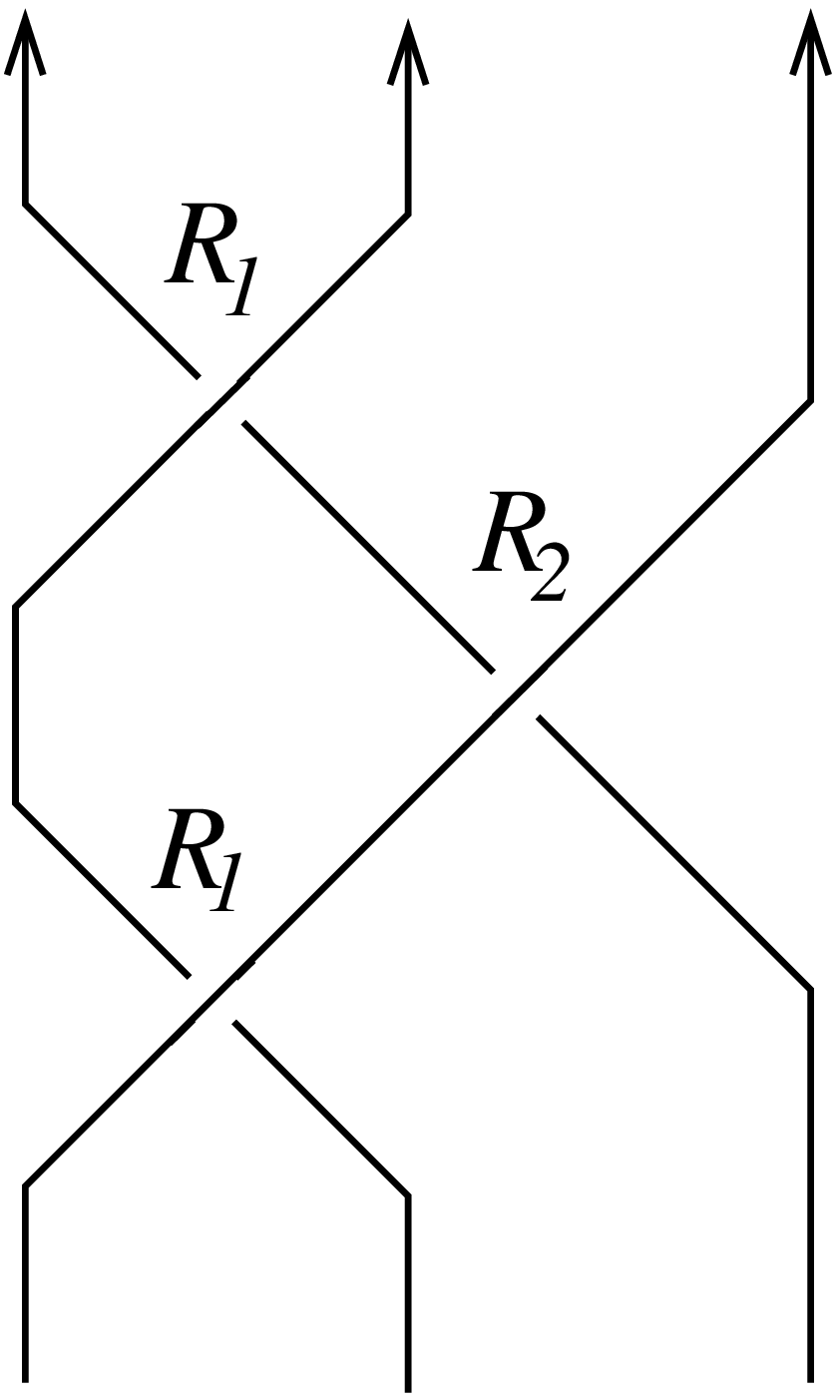,width=80pt}{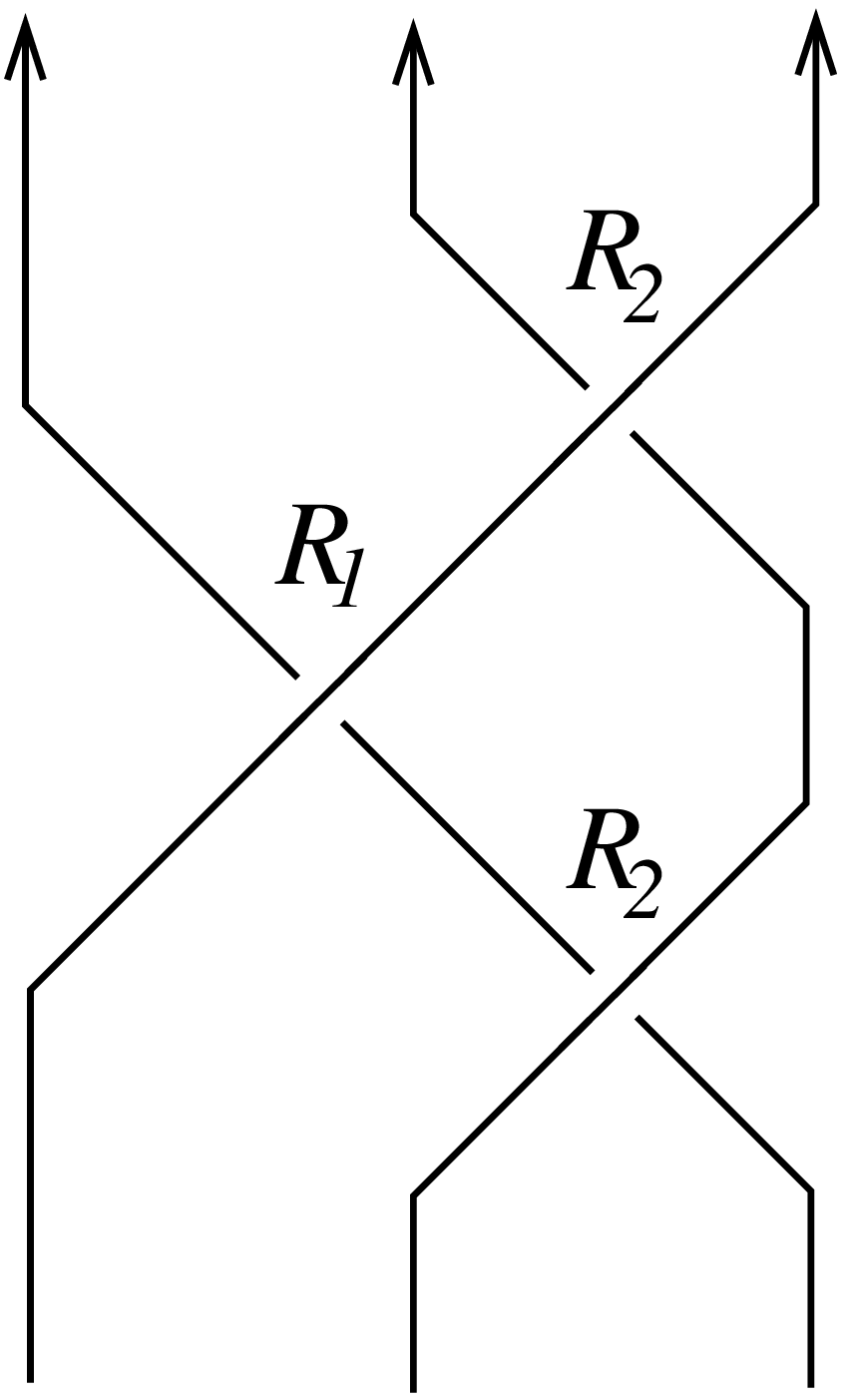,width=80pt}
{The left hand-side of the Yang-Baxter equation...}{...is topologically equivalent to the right-hand side.}

\para
In quantum mechanics, exchanges of particles act as  unitary operators on the Hilbert space. These will form representations of the braid group. The kind of anyons that we described above form a one-dimensional representation of the braid group in which each exchange just gives a phase: $R_i = e^{i\pi \alpha_i}$. The Yang-Baxter equation then  requires $e^{i\pi\alpha_i} = e^{i\pi\alpha_{i+1}}$ which simply tells us that all identical particles must have the same phase.

\para
One dimensional representation of the braid group are usually referred to as {\it Abelian anyons}. As we'll show below, these are the kind of anyons relevant for the Laughlin states. However, there are also more exotic, higher-dimensional representations of the braid group. These are called {\it non-Abelian anyons}. We will discuss some examples in  Section \ref{nonabsec}.

\subsubsection{Fractional Statistics}\label{fracstatsec}

We will now compute the quantum statistics of quasi-holes in the $\nu=1/m$ Laughlin state. In passing, we will also provide a more sophisticated argument for the fractional charge of the quasi-hole. Both computations involve the Berry phase that arises as quasi-holes move\footnote{The structure of this calculation was first described in Daniel Arovas, John Schrieffer and Frank Wilczek, ``{\it Fractional statistics and the quantum Hall effect}", \href{http://journals.aps.org/prl/abstract/10.1103/PhysRevLett.53.722}{Phys. Rev. Lett.  53, 772 (1984)}, although they missed the importance of working with normalised wavefunctions. This was  subsequently clarified by M. Stone. in the collection of reprints he edited called, simply, ``{\it The Quantum Hall Effect}".}.

\para
We consider a state of $M$ quasi-holes which we denote as $|\eta_1,\ldots,\eta_M\rangle$. The wavefunction is \eqn{multihole}
\be \langle z,\bar{z}|\eta_1,\ldots,\eta_M\rangle  =\prod_{j=1}^M \prod_{i=1}^N(z_i-\eta_j)\prod_{k<l} (z_k-z_l)^m\,e^{-\sum_{i=1}^n |z_i|^2/4l_B^2}\nn\ee
However, whenever we compute the Berry phase, we should work with the normalised states. We'll call this state $|\psi\rangle$, defined by
\be |\psi\rangle = \frac{1}{\sqrt{Z}} |\eta_1,\ldots,\eta_M\rangle\nn\ee
where the normalisation factor is defined as $Z= \langle \eta_1,\ldots,\eta_M|\eta_1,\ldots,\eta_M\rangle$, which reads
%
%
%
%
\be Z = \int \prod d^2z_i \ \exp\left({\sum_{i,j} \log|z_i-\eta_j|^2 +  m\sum_{k,l} \log|z_k-z_l|^2 - \frac{1}{2l_B^2}\sum_i |z_i|^2}\right)\ \ \ \ \ \label{zhole}\ee
This is the object which plays the role of the partition function in the plasma analogy, now in the presence of impurities localised at $\eta_i$.

\para
The holomorphic Berry connection is 
\be {\cal A}_\eta(\eta,\bar{\eta}) = -i\langle \psi| \ppp{}{\eta}|\psi\rangle = \frac{i}{2Z}\ppp{Z}{\eta} - \frac{i}{Z} \langle\eta|\ppp{}{\eta}|\eta\rangle\nn\ee
But because $|\eta\rangle$ is holomorphic, and correspondingly $\langle\eta|$ is anti-holomorphic, we have $\ppp{Z}{\eta} = \ppp{}{\eta}\langle\eta|\eta\rangle=\langle\eta|\ppp{}{\eta}|\eta\rangle$. So we can write
\be {\cal A}_\eta(\eta,\bar{\eta}) = -\frac{i}{2}\ppp{\log Z}{\eta}\nn\ee
Meanwhile, the anti-holomorphic Berry connection is
\be {\cal A}_{\bar{\eta}}(\eta,\bar{\eta}) = -i\langle \psi| \ppp{}{\bar{\eta}}|\psi\rangle = +\frac{i}{2}\ppp{\log Z}{\bar{\eta}}\nn\ee
So our task in both cases is to compute the derivative of the partition function \eqn{zhole}. This is difficult to do exactly. Instead, we will invoke our intuition for the behaviour of plasmas.

\para
Here's the basic idea. In the plasma analogy, the presence of the hole acts like a charged impurity. In the presence of such an impurity, the key physics is called {\it screening}\footnote{You can read about screening in the final section of the lecture notes on \href{http://www.damtp.cam.ac.uk/user/tong/em.html}{\it Electromagnetism}.}. This is the phenomenon in which the mobile charges -- with positions $z_i$ -- rearrange themselves to cluster around the impurity so that its effects cannot be noticed when you're suitably far away. More mathematically, the electric potential due to the impurity is modified by an exponential fall-off $e^{-r/\lambda}$ where $\lambda$ is called the {\it Debye screening length} and is proportional to $\sqrt{T}$. Note that, in order for us to use this argument, it's crucial that the artificial temperature \eqn{beta} is high enough that the plasma lies in the fluid phase and efficient screening can occur.

\para
Whenever such screening occurs, the impurities are effectively hidden at distances much greater than  $\lambda$. This means that the free energy of the plasma  is independent of the positions of the impurities, at least up to exponentially small corrections. This free energy is, of course, proportional to $\log Z$ which is the thing we want to differentiate. However, there are two ingredients missing:  the first is the energy cost  between the impurities and the constant background charge; the second is the Coulomb energy between the different impurities. The correct potential energy for the plasma with $M$ impurities should therefore be
\be
U(z_k;\eta_i) &=&  -m^2  \sum_{k<l} \log\left(\frac{|z_k-z_l|}{l_B}\right) - m\sum_{k,i} \log \left(\frac{|z_i-\eta_i|}{l_B}\right)-\sum_{i<j}\log \left(\frac{|\eta_i-\eta_j|}{l_B}\right) \nn\\ &&\ \ \ \ \ +\ \frac{m}{4l_B^2} \sum_{k=1}^N |z_k|^2 + \frac{1}{4l_B^2} \sum_{i=1}^M|\eta_i|^2\label{thegoodone}\ee
The corrected plasma partition function is then
\be \int \prod d^2z_i \ e^{-\beta U(z_i;\eta)}= \exp\left({-\frac{1}{m}\sum_{i<j} \log |\eta_i-\eta_j|^2  + \frac{1}{2ml_B^2}\sum_i|\eta_i|^2}\right) Z\nn\ee
As long as the distances between impurities $|\eta_i-\eta_j|$ are greater than the Debye  length $\lambda$, the screening argument tells us that this expression should be independent of the positions $\eta_i$ for high enough temperature.  In particular,  as we described previously, the temperature $\beta = 2/m$ relevant for the plasma analogy is  high enough for screening as long as $m\lesssim 70$ . This means that we must have
\be Z = C \exp\left({\frac{1}{m}\sum_{i<j}\log |\eta_i-\eta_j|^2-\frac{1}{2ml_B^2}\sum_i|\eta_i|^2}\right)\nn\ee
for some constant $C$ which does not depend on $\eta_i$. This gives some idea of the power of the plasma analogy. It looks nigh on impossible to perform the integrals in \eqn{zhole} directly; yet by invoking some intuition about screening, we are able to write down the answer, at least in some region of parameters.

\DOUBLEFIGURE{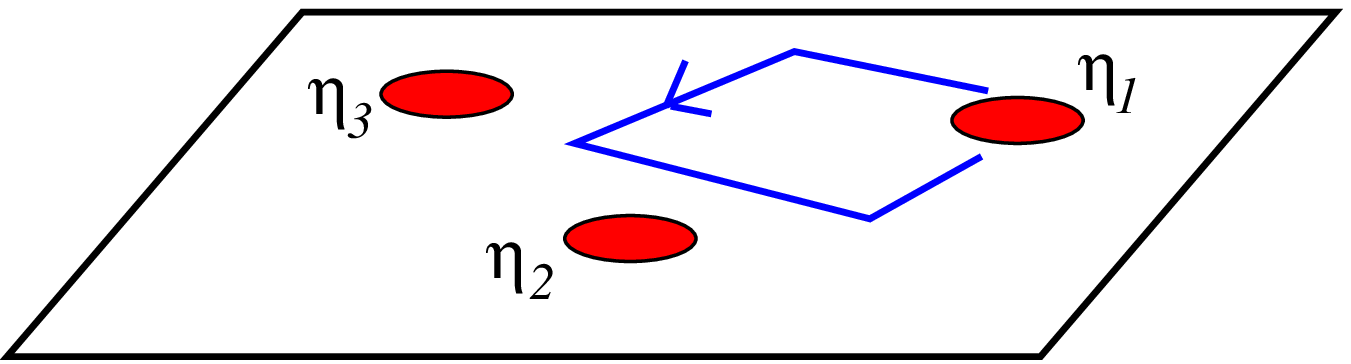,width=210pt}{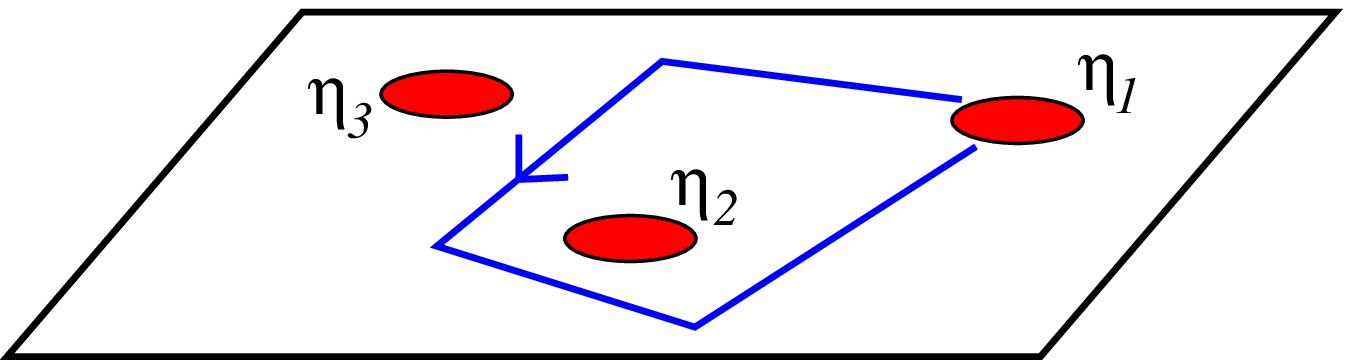,width=210pt}
{The path taken to compute the fractional charge of the quasi-hole...}{...and the path to compute the fractional statistics.}
\noindent

\para
The Berry connections over the configuration space of $M$ quasi-holes are then simple to calculate: they are 
\be {\cal A}_{\eta_i} =  -\frac{i}{2m}\sum_{j\neq i}\frac{1}{\eta_i-\eta_j} +\frac{i\bar{\eta_i}}{4ml_B^2}\label{aeta}\ee
and
\be {\cal A}_{\bar{\eta}_i} = +\frac{i}{2m}\sum_{j\neq i} \frac{1}{\bar{\eta}_i - \bar{\eta}_j}-\frac{i\eta_i}{4ml_B^2}\label{abareta}\ee
where we stress that these expressions only hold as long as the quasi-holes do not get too close to each other where the approximation of complete screening breaks down. We can now use these Berry connections to compute both the charge and statistics of the quasi-hole.

\subsubsection*{Fractional Charge}

Let's start by computing the charge of the anyon.  The basic idea is simple. We pick one of the quasi-holes --- say $\eta_1\equiv \eta$ ---  and move it on a closed path $C$. For now we choose a path which does not enclose any of the other anyons. This ensures that only the second term in the Berry phase contributes,
\be {\cal A}_\eta = \frac{i\bar{\eta}}{4ml_B^2}\ \ \ {\rm and}\ \ \ {\cal A}_{\bar{\eta}} = -\frac{i\eta}{4ml_B^2}\nn\ee
After traversing the path $C$, the quasi-hole will return with a phase shift of $e^{i\gamma}$, given by the Berry phase 
\be e^{i\gamma} = \exp\left(-i\oint_C {\cal A}_\eta d\eta + {\cal A}_{\bar{\eta}}d\bar{\eta}\right) \label{holoberry}\ee
This gives the Berry phase
\be \gamma = \frac{e\Phi}{m\hbar}\label{goesaround}\ee
where $\Phi$ is the total magnetic flux enclosed by the path $C$. But there's a nice interpretation of this result: it's simply the Aharonov-Bohm phase picked up by the particle. As  described in Section \ref{absec}, a particle of charge $e^\star$ will pick up  phase $\gamma =  {e^\star \Phi}/{\hbar}$. Comparing to \eqn{goesaround}, we learn that the charge of the particle is indeed
\be e^\star = \frac{e}{m}\nn\ee
as promised.

\subsubsection*{Fractional Statistics}

To compute the statistics, we again take a particular quasi-hole --- say $\eta_1$ --- on a journey, this time on  a path $C$ which encloses one other quasi-hole, which we'll take to be $\eta_2$. The phase is once again given by \eqn{holoberry} where, this time, both terms in the expressions \eqn{aeta} and \eqn{abareta} for ${\cal A}_\eta$ and ${\cal A}_{\bar{\eta}}$ contribute. The second term once again gives the Aharonov-Bohm phase; the first term tells us about the statistics. It is
\be e^{i\gamma} =\exp\left(-\frac{1}{2m} \oint_C\frac{d\eta_1}{\eta_1-\eta_2} + {\rm h.c.}\right) = e^{2\pi i/m} \nn\ee
This is the phase that arises from one quasi-hole encircling the other. But the quantum statistics comes from {\it exchanging} two objects, which can be thought of as a rotating by $180^\circ$ rather than $360^\circ$. This means that, in the notation of \eqn{anyon}, the phase above is 
\be e^{2\pi i\alpha} = e^{2\pi i/m} \ \ \ \Rightarrow \ \ \ \alpha =\frac{1}{m}\label{thisisanyon}\ee
Note that for a fully filled Landau level, with $m=1$, the quasi-holes are fermions. (They are, of course, actual holes). But for a fractional quantum Hall state, the quasi-holes are anyons.

\para
Suppose now that we put $n$ quasi-holes together and consider this as a single object. What are its statistics? If we exchange two such objects, then each quasi-hole in the first bunch gets exchanged with each quasi-hole in the second bunch. The net result is that the statistical parameter for $n$ quasi-holes is $\alpha=n^2/m$ (recall that the parameter $\alpha$ is defined mod 2). Note that $\alpha$ does not grow linearly with $n$. As a check, suppose that we put $m$ quasi-holes together to reform the original particle that underlies the Hall fluid. We get $\alpha = m^2/m  = m$ which is a boson for $m$ even and a fermion for $m$ odd.

\para
There's a particular case of this which is worth highlighting. The quasi-particles in the $m=2$ bosonic Hall state have statistical parameter $\alpha=1/2$. They are half-way between bosons and fermions and sometimes referred to as {\it semions}. Yet two semions do not make a fermion; they make a boson. 

\para
More generally, it's tempting to use this observation to argue that an electron can only ever split into an odd number of anyons. This argument runs as follow: if an electron were to split into an even number of constituents $n$,  each with statistical parameter $\alpha$, then putting these back together again would result in a particle with statistical parameter $n^2\alpha$. The argument sounds compelling. However, as we will see in Section \ref{nonabsec}, there is a loop hole!

\para
While the fractional charge of quasi-holes has been measured experimentally, a direct detection of their fractional statistics is more challenging. There have been a number of proposed (and performed) experiments using interferometry to demonstrate but their conclusions remain open to interpretation.

\subsubsection*{A Slightly Different Viewpoint}

There is a slightly different way of presenting the calculation. It will offer nothing new here, but often appears in the literature as it proves useful when discussing more complicated examples. The idea is that we consider a wavefunction that already has the interesting $\eta$ dependence built in. So, instead of \eqn{multihole}, we work with
\be \psi  = \prod_{a<b} (\eta_a-\eta_b)^{1/m}\prod_{a,i}^N(z_i-\eta_a)\prod_{k<l} (z_k-z_l)^m\,e^{-\sum_i |z_i|^2/4l_B^2 - \sum_a |\eta_a|^2/4 m l_B^2}\label{otherwf}\ee
This  wavefunction is cooked up so that the associated probability distribution is given precisely by the partition function with energy \eqn{thegoodone} and hence has no dependence on $\eta$ and $\bar{\eta}$. This means that the Berry connection for this wavefunction
has only the second terms in \eqn{aeta} and \eqn{abareta}, corresponding to the Aharonov-Bohm effect due to the background magnetic field. The term in the Berry connection that was responsible for fractional statistics is absent. But this doesn't mean that the physics has changed. Instead, this phase is manifest in the form of the wavefunction itself, which is no longer single-valued in $\eta_a$. Indeed, if $\eta_1$ encircles a neighbouring point $\eta_2$, the wavefunction pick up a phase $e^{2\pi i/m}$, so exchanging two quasi-holes gives the phase $e^{i\pi /m}$. 

\para
Of course, this approach doesn't alleviate the need to determine the Berry phase arising from the exchange. You still need to compute it to check that it is indeed zero.

\subsubsection{Ground State Degeneracy and Topological Order}\label{torussec}

In this section we describe a remarkable property of the fractional quantum Hall states which only becomes apparent when you place them on a compact manifold: the number of ground states depends on the topology of the manifold. As we now explain, this is intimately related to the existence of anyonic particles.

\para
Consider the following process on a torus. We create from the vacuum a quasi-particle -- quasi-hole pair. We then separate this pair, taking them around one of the two different cycles of the torus as shown in the figure, before letting them annihilate again. We'll call the operator that implements this process $T_1$ for the first cycle and $T_2$ for the second.

\DOUBLEFIGURE{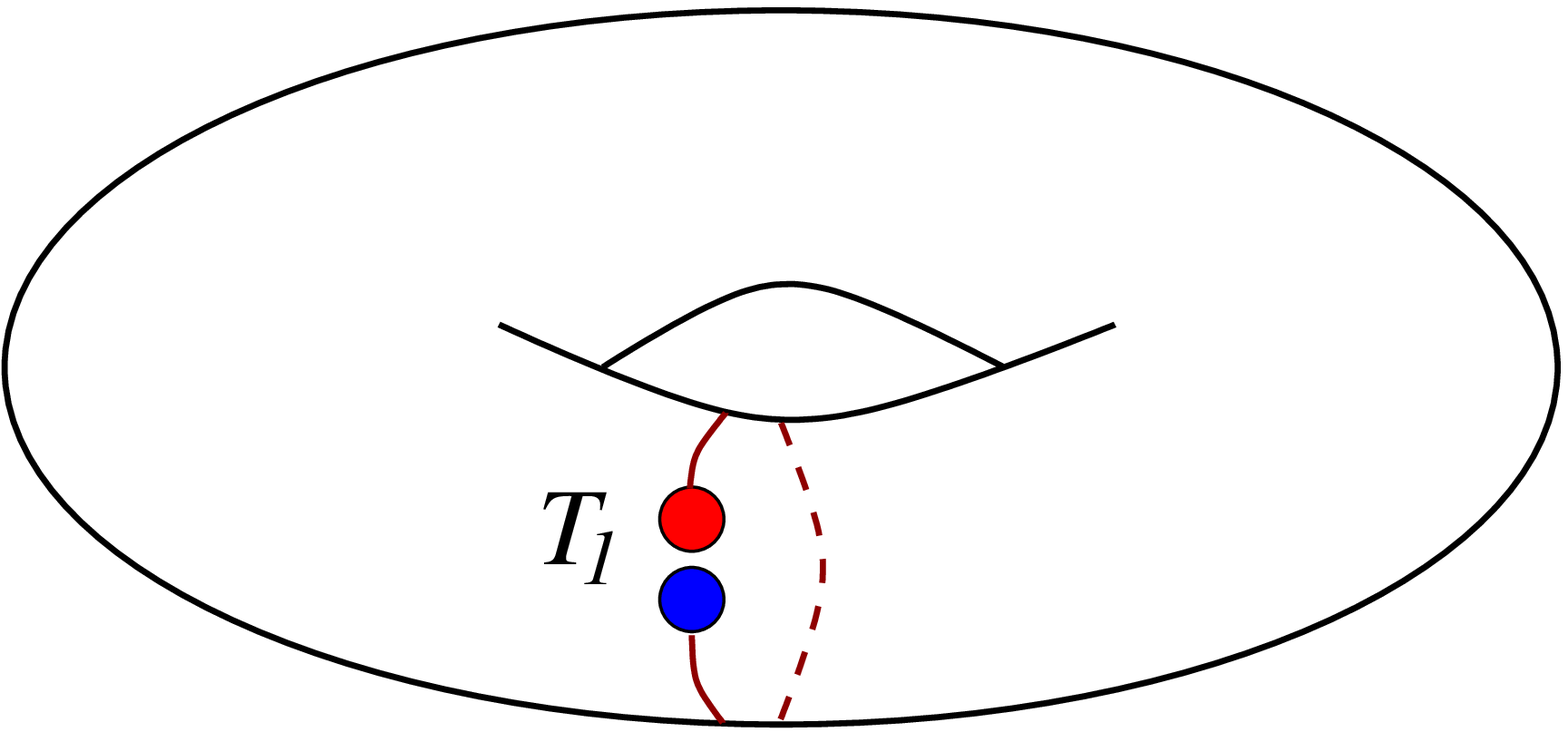,width=180pt}{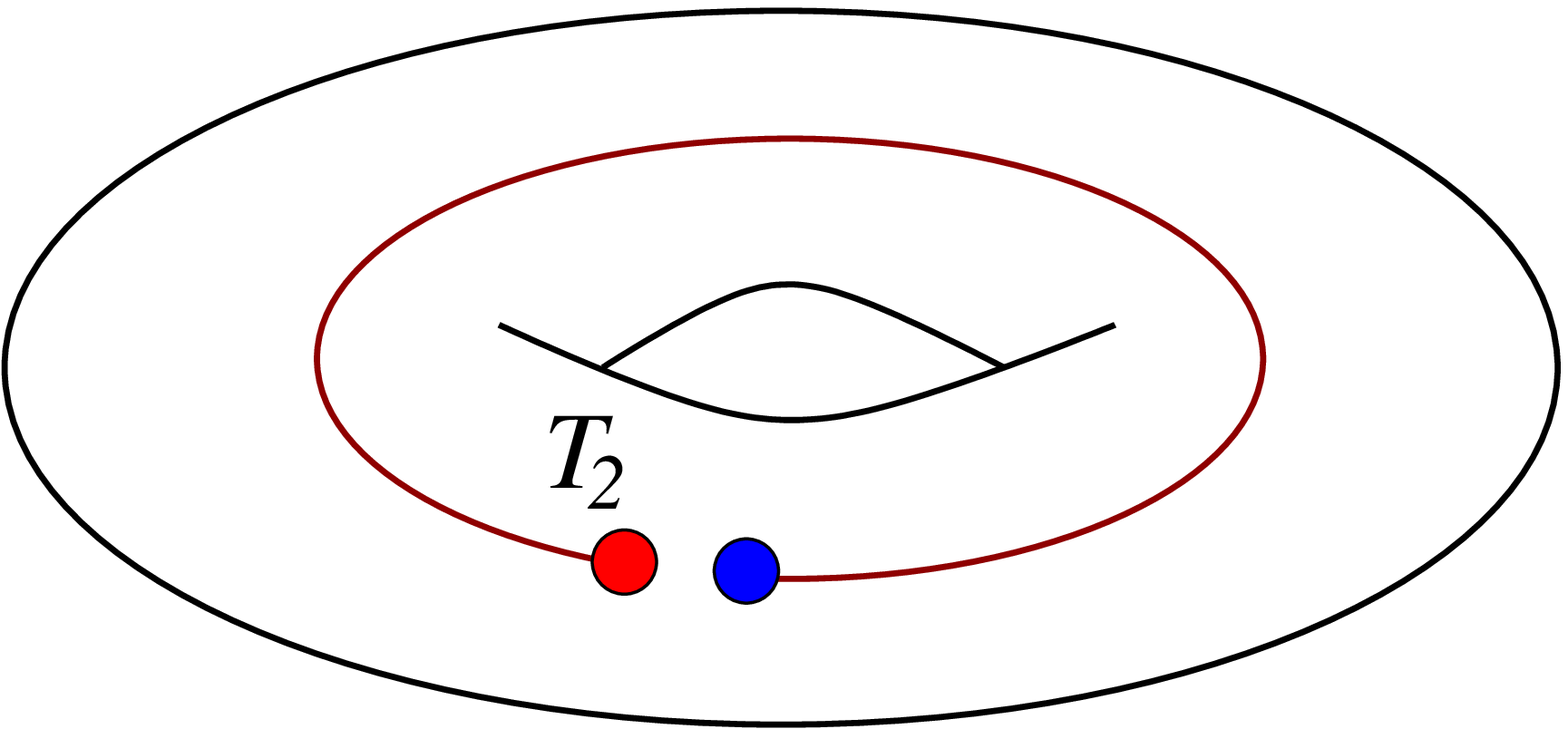,width=180pt}
{Taking a quasi-hole (red) and quasi-particle (blue) around one cycle of the torus}{...or around the other.}

\para
Now suppose we take the particles around one cycle and then around the other. Because the particles are anyons, the order in which we do this matters: there is a topological difference between the paths taken. Indeed, you can convince yourself that $T_1T_2T_1^{-1} T_2^{-1}$ is equivalent to taking one anyon around another: the worldlines have linking number one. This means that the $T_i$ must obey the algebra
\be T_1 T_2 = e^{2\pi i/m}\, T_2 T_1\label{torusalgebra}\ee
But such an algebra of operators can't be realised on a single vacuum state. This immediately tells us that the ground state must be degenerate. The smallest representation of \eqn{torusalgebra} has dimension $m$, with the action
\be  T_1|n\rangle  = e^{2\pi  n i/m}|n\rangle \nn\\ T_2 |n\rangle = |n+1\rangle\nn\ee
The generalisation of this  argument to a genus-$g$ Riemann surface tells us that the ground state must have degeneracy $m^g$. Notice that we don't have to say anything about the shape or sizes of these manifolds. The number of ground states depends only on the topology!

\para
It is also possible to explicitly construct the analog of the Laughlin states on a torus in terms of Jacobi theta functions and see that there are indeed $m$ such states.

\para
Before we proceed, we note that this resolves  a puzzle. In Section \ref{avronsec}, we described a topological approach to the integer quantum Hall effect which is valid when space is a torus. With a few, very mild, assumptions, we showed that the Hall conductivity is equal to a Chern number and must, therefore, be quantised. In particular, this calculation made no assumption that the electrons were non-interacting: it holds equally well for strongly interacting many-body systems. However, one of the seemingly mild assumptions was that the ground state was non-degenerate. As we've seen, this is not true for fractional quantum Hall states, a fact which explains how these states avoid having integer Hall conductivity. 

\subsubsection*{Topological Order}

We've seen in this section that the Laughlin states have a number of very special properties. One could ask: how can we characterise these states? This is an old and venerable question in condensed matter physics and, for most systems, has an answer provided by Landau. In Landau's framework, different states of matter are characterised by their symmetries, both those that are preserved by the ground state and those that are broken. This is described using order parameters of the kind that we met in the lectures on \href{http://www.damtp.cam.ac.uk/user/tong/statphys.html}{\it Statistical Physics} when discussing phase transitions.

\para
However, the quantum Hall fluids fall outside of this paradigm. There is no symmetry or local order parameter that distinguishes quantum Hall states. It turns out that there is a non-local order parameter, usually called ``off-diagonal long-range order" and this can be used to motivate a Ginzburg-Landau-like description. We  will describe this in Section \ref{csglsec} but, as we will see, it is not without its pitfalls. 

\para
Instead, Wen\footnote{The original paper is Xiao-Gang Wen, ``{\it Topological Orders in Rigid States}", Int. J. Mod. Phys. B4, 239 (1990), available at \href{http://dao.mit.edu/~wen/pub/topo.pdf}{Xiao-Gang's website}.} suggested that we should view quantum Hall fluids as a new type of matter, characterised by {\it topological order}. The essence of the proposal is that quantum states can be characterised their ground state degeneracy and the way in which these states transform among themselves under operations like \eqn{torusalgebra}.

\subsection{Other Filling Fractions}\label{otherfillingsec}

So far, we have only described the quantum Hall states at filling fraction $\nu = 1/m$. Clearly there are many more states that are not governed by the Laughlin wavefunction. As we now show, we can understand many of these by variants of the ideas above.

\subsubsection*{A Notational Convention}

Before we proceed, let's quickly introduce some new notation. All wavefunctions in the lowest Landau level come with a common exponential factor.  It gets tiresome writing it all the time, so define
\be \psi(z,\bar{z}) \sim \tilde{\psi}(z) e^{-\sum_{i=1}^n |z_i|^2/4l_B^2}\nn\ee
where $\tilde{\psi}(z)$ is a holomorphic function. In what follows we will often just write $\tilde{\psi}(z)$. Be warned that many texts drop the exponential factor in the wavefunctions but don't give the resulting object a different name.

\subsubsection{The Hierarchy}\label{hierarchysec}

We saw in Section \ref{fracchargesec} how one can induce quasi-hole (or quasi-particle) states by introducing a local excess (or deficit) of magnetic field  through a solenoid. We could also ask what happens if we change the magnetic field in a uniform manner so that the system as a whole moves away from $\nu=1/m$ filling. For definiteness, suppose that we increase $B$ so that  the filling fraction decreases. It seems plausible that for $B$ close to the initial Laughlin state, the new ground state of the system will contain some density of quasi-holes, arranged in some, perhaps complicated,  configuration. The key idea of this section is that these quasi-holes might  themselves form a  quantum Hall state. Let's see how this would work.

\para
We know that Laughlin states take the form
\be \tilde{\psi} \sim \prod_{i<j} (z_i-z_j)^m\nn\ee
where $m$ is odd for fermions and even for bosons.  What would a Laughlin state look like for anyons with positions $\eta_i$ and statistical parameter $\alpha$?
To have the right statistics, the wavefunctions must take the form
\be  \tilde{\psi}  \sim \prod_{i<j}^N (\eta_i-\eta_j)^{2p + \alpha}\nn\ee
with $p$ a positive integer. 
As we've seen, above the $\nu = 1/m$ state, quasi-holes have statistics $\alpha = 1/m$ while quasi-particles have statistics $\alpha = -1/m$. It's simple to  repeat our previous counting of the filling fraction, although now we need to be more careful about what we're counting.  The maximum angular momentum of a given quasi-excitation is $N(2p\pm \frac{1}{m})$ so the area of the droplet is  $A\approx 2\pi (2p\pm\frac{1}{m}) N (ml_B^2)$ where the usual magnetic length $l_B^2= \hbar/eB$ is now replaced by $ml_B^2$ because the charge of the quasi-excitations is $q=\pm e/m$. The number of {\it electron} states in a full Landau level is $AB/\Phi_0$ and each can be thought of as made of $m$ quasi-things. So the total number of quasi-thing states in a full Landau level is $mAB/\Phi_0 = (2p\pm \frac{1}{m}) m^2 N$. 

\para
The upshot of this is that the quasi-holes or quasi-particles give a contribution to the filling of electron states 
\be \nu_{\rm quasi}  = \mp \frac{1}{2pm^2 \pm m }\nn\ee
where the overall sign is negative for holes and positive for particles. Adding this to the filling fraction of the original  $\nu = 1/m$ state, we have
\be \nu = \frac{1}{m} \mp \frac{1}{2pm^2  \pm m} = \frac{1}{m \pm \frac{1}{2p}} \label{simplehier}\ee
Note that the filling fraction is decreased by quasi-holes and increased by quasi-particles.

\para
Let's look at some simple examples. We start with the $\nu=1/3$ state. The $p=1$ state for quasi-particles then gives $\nu =2/5$ which is one of the more prominent Hall plateaux. The $p=1$ state for quasi-hole gives $\nu = 2/7$ which has also been observed; while not particularly prominent, it's harder to see Hall states at these lower filling fractions. 

%
\para
Now we can go further. The quasi-objects in this new state can also form quantum Hall states. And so on. The resulting fillings are given by the continuous fractions
%
%
\be \nu = \cfrac{1}{m\pm \cfrac{1}{2p_1\pm \cfrac{1}{2p_2\pm\cdots}}}\label{contfrac}\ee
For example, building on the Hall state $\nu=1/3$, the set of continuous fractions for quasi-particles with $p_i=1$ leads to the sequence $\nu = 2/5$ (which is the  fraction \eqn{simplehier}), followed by $\nu = 3/7, \ 4/9, \ 5/11$ and $6/13$. This is precisely the sequence of Hall plateaux shown in the data presented at the beginning of this chapter.

\subsubsection{Composite Fermions}\label{compositesec}

We now look at an alternative way to think about the hierarchy known as {\it composite fermions}\footnote{This concept was first introduced by Jainendra Jain in the paper``{\it Composite-Fermion Approach to the Fractional Quantum Hall Effect}", \href{http://journals.aps.org/prl/abstract/10.1103/PhysRevLett.63.199}{Phys. Rev. Lett. 63 2 (1989)}. It is reviewed in some detail in his book called, appropriately, ``{\it Composite Fermions}". A clear discussion can also be found in the review ``{\it Theory of the Half Filled Landau Level}" by Nick Read, \href{http://arxiv.org/abs/cond-mat/9501090}{cond-mat/9501090}.}.   Although the starting point seems to be logically different from the ideas above, we will see the same filling fractions emerging. Moreover, this approach will allow us to go further ending, ultimately, in Section \ref{hfllsec} with a striking prediction for what happens at filling fraction $\nu=1/2$.

\para
First, some motivation for what follows.
It's often the case that when quantum systems become strongly coupled, the right degrees of freedom to describe the physics are not those that we started with. Instead new, weakly coupled degrees of freedom may emerge.  Indeed, we've already seen an example of this in the quantum Hall effect, where we start with electrons but end up with fractionally charged particles. 

\para
The idea of this section is to try to  find some new degrees of freedom --- these are the ``composite fermions". However, for the most part these won't be the degrees of freedom that are observed in the system. Instead, they play a role in the intermediate stages of the calculations. (There is an important exception to this statement which is the case of the half-filled Landau level, described in Section \ref{hfllsec}, where the observed excitations of the system are the composite fermions.) 
%
Usually it is difficult to identify the emergent degrees of freedom, and it's no different here. We won't be able to rigorously derive the composite fermion picture. Instead, we'll give some intuitive and, in parts, hand-waving arguments that lead us to a cartoon description of the physics. But the resulting cartoon is impressively accurate. It gives ansatze for wavefunctions which are in good agreement with the numerical studies and it  provides a useful and unified way to think about different classes of quantum Hall states.

\para
We start by introducing the idea of a {\it vortex}. Usually  a vortex is a winding in some complex order parameter. Here, instead, a vortex will mean a winding in the wavefunction itself. Ultimately we will be interested in vortices in the Laughlin wavefunction, but to understand the key physics it's simplest to revisit the quasi-hole whose wavefunction includes the factor
\be  \prod_i(z_i-\eta)\nn\ee
Clearly the wavefunction now has a zero at the position $\eta$. This does two things. First, it depletes the charge there. This, of course, is what gives the quasi-hole its fractional charge $e/m$. But because the lowest Landau level wavefunction is holomorphic, there is also  fixed angular dependence: the phase of the wavefunction winds once as the position of any particle moves around $\eta$. This is the vortex.

\para
The winding of the wavefunction is really responsible for the Berry phase calculations we did in Section \ref{fracstatsec} to determine the fractional charge and statistics of the quasi-hole. Here's  a quick and dirty explanation. The phase of the wavefunction changes by $2\pi$ as a particle moves around the quasi-hole. Which means that it should also change by $2\pi$ when the quasi-hole moves around the particle. So if we drag the quasi-hole around $N= \nu \Phi/\Phi_0$ particles, then the phase changes by $\gamma = 2\pi N = \nu e\Phi/\hbar$. This is precisely the result \eqn{goesaround} that we derived earlier. Meanwhile, if we drag one quasi-hole around a region in which there is another quasi-hole, the charge inside will be depleted by $e/m$, so the effective number of particles inside is now $N = \nu\Phi/\Phi_0 - 1/m$. This gives an extra contribution to the phase $\gamma = -2\pi/m$ which we associate the statistics of the quasi-holes: $\gamma = 2\pi \alpha = 2\pi/m$ so $\alpha = 1/m$, reproducing our earlier result \eqn{thisisanyon}. We stress that all of these results really needed only the vortex nature of the quasi-hole.


\para
Now let's turn to the Laughlin wavefunction itself
\be  \tilde{\psi}_m(z) \sim \prod_{i<j} (z_i-z_j)^m\nn\ee
For now we focus on $m$ odd so that the wavefunction is anti-symmetric and we're dealing with a Hall state of fermions. One striking feature is that the wavefunction has a zero of order $m$ as two electrons approach. This means that each particle can be thought of as $m$ vortices. Of course, one of these zeros was needed by the Pauli exclusion principle. Moreover, we needed $m$ zeros per particle to get the filling fraction right. But nothing forced us to have the other $m-1$ zeros sitting at  exactly the same place. This is something special about the Laughlin wavefunction.


\para
Motivated by this observation, we define a {\it composite fermion} to be an electron (which gives rise to one vortex due to anti-symmetry) bound to $m-1$ further vortices. The whole thing is a fermion when $m$ is odd.
You'll sometimes hear composite fermions described as electrons attached to flux. We'll describe this picture in the language of Chern-Simons theory in Section \ref{cssec} but it's not overly useful at the moment.  In particular,  it's important to note that the composite fermions don't carry real magnetic flux with them. This remains uniform. Instead, as we will see later, they carry a different, emergent flux. 

\para
Let's try to treat this as an object in its own right and see what behaviour we find. Consider placing some density $n= \nu B/\Phi_0$ of electrons in a magnetic field and subsequently attaching these vortices to make composite fermions. We will first show that these composite fermions experience both a different magnetic field $B^\star$ and different filling fraction $\nu^\star$ than the electrons. To see this, we repeat our Berry phase argument where we move the composite fermion along a path encircling an area $A$. The resulting Berry phase has two contributions,
\be \gamma = 2\pi\left(\frac{AB}{\Phi_0} - (m-1)n A\right)\label{compositeflux}\ee
with $n$ the density of electrons. The first term is the usual Aharonov-Bohm phase due to the total flux inside the electron path. The second term is the contribution from the electron encircling the vortices: there are $m-1$ such vortices attached to each of the $\rho A$ electrons.

\begin{figure}[htb]
\begin{center}
\epsfxsize=5.7in\leavevmode\epsfbox{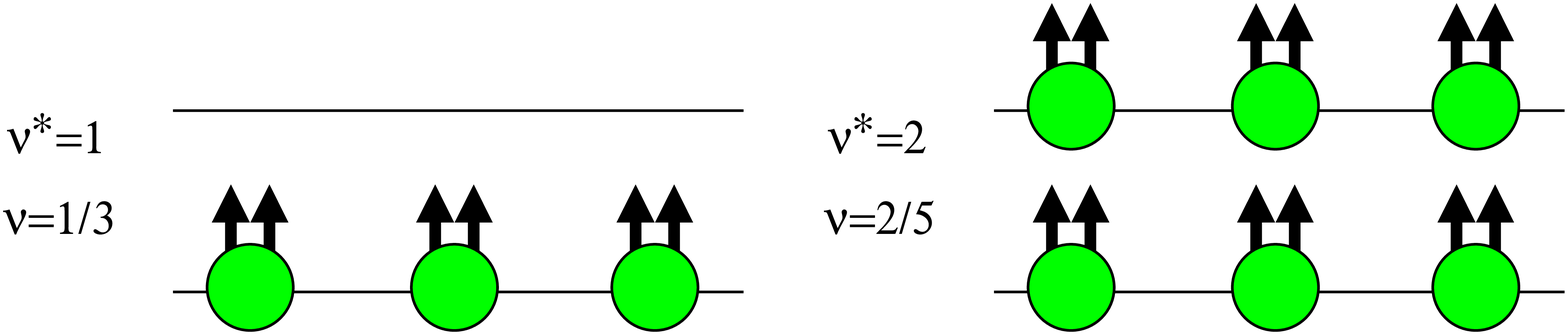}
\end{center}
\caption{The composite fermion picture describes a hierarchy of  plateaux around, starting with $\nu=1/3$, in terms of the integer quantum Hall effect 
for electrons bound to two vortices.}
\end{figure}
\noindent

\para
When we discussed quasi-holes, we also found a different Aharonov-Bohm phase. In that context, we interpreted this as a different charge of quasi-particles. In the present context, one usually interprets the result  \eqn{compositeflux} in a different (although ultimately equivalent) way: we say that the composite fermions experience a different magnetic field which we call $B^\star$. The Aharonov-Bohm phase should then be
\be \gamma = \frac{2\pi AB^\star}{\Phi_0}\ \ \ \Rightarrow\ \ \ B^\star = B - (m-1) n \Phi_0\label{jainb}\ee
Because there is one electron per composite fermion, the  density is the same. But because the magnetic fields experienced by electrons and composite fermions differ, the filling fractions must also differ: we must have $n = \nu^\star B^\star/\Phi_0 = \nu B/\Phi_0$. This gives
\be \nu = \frac{\nu^\star}{1 + (m-1)\nu^\star}\label{jainfilling}\ee
This is an interesting equation! Suppose that we take the composite fermions to completely fill their lowest Landau level, so that $\nu^\star = 1$. Then we have
\be \nu^\star = 1\ \ \ \ \Rightarrow\ \ \ \ \nu= \frac{1}{m}\nn\ee
In other words, the fractional quantum Hall effect can be thought of as an integer quantum Hall effect for composite fermions. That's very cute! Indeed, we can even see some hint of this in the Laughlin wavefunction itself which we can trivially rewrite as
\be  \tilde{\psi}_m(z) \sim \prod_{i<j} (z_i-z_j)^{m-1}\,\prod_{k<l} (z_k-z_l)\label{sillydecomp}\ee
The second term in this decomposition is simply the wavefunction for the fully-filled lowest Landau level. We're going to think of the first term  as attaching $m-1$  vortices to each position $z_i$ to form the composite fermion.

\EPSFIGURE{lotsplat.eps,height=250pt}{The fractional Hall plateaux....again}
\noindent

\para
So far we've said a lot of words, but we haven't actually derived anything new from this perspective. But we can extract much more from \eqn{jainfilling}. Suppose that we fill the first $\nu^\star$ Landau levels  to get an integer quantum Hall effect  for composite fermions with $\nu^\star>1$. (The Landau levels for composite fermions are sometimes referred to $\Lambda$ levels.) Then we find filling fractions that are different from the Laughlin states. For example, if we pick $m=3$, then the sequence of states arising from \eqn{jainfilling} is $\nu = 1/3, \ 2/5, \ 3/7, \ 4/9,\ldots$. These is the same sequence that we saw in the hierarchy construction and is clearly visible in the data shown in the figure. Inspired by the form of  \eqn{sillydecomp}, we will write down a guess for the wavefunction, usually referred to as {\it Jain states},
\be \tilde{\psi}_\nu (z) = {\cal P}_{LLL} \left[\prod_{i<j} (z_i-z_j)^{m-1}\,\Psi_{\nu^\star}(z,\bar{z})\right]\label{jain}\ee
Here $\Psi_{\nu^\star}$ is the wavefunction for $\nu^\star\in {\bf Z}$ fully-filled Landau levels while the $\prod (z_i-z_j)^{m-1}$ factor attaches the $(m-1)$ vortices to each electron.  The wavefunction $\Psi_{\nu^\star}$ can be easily constructed by a Slater determinant of the form \eqn{slater} except that, this time, we run into a problem. The electrons have filling fraction $\nu<1$ and so are supposed to lie in the lowest Landau level. Meanwhile, the  integer quantum Hall states $\Psi_{\nu^\star}$ are obviously not lowest Landau level wavefunctions: they depend on $\bar{z}_i$ as well as $z_i$. This is what the mysterious symbol ${\cal P}_{LLL}$ is doing in the equation \eqn{jain}: it means ``project to the lowest Landau level".

\para
Operationally, ${\cal P}_{LLL}$ is defined by moving all factors of $\bar{z}_i$ in $[\ldots]$ to the left. We then make the substitution 
\be \bar{z}_i \rightarrow 2l_B^2\ppp{}{z_i}\label{zbarsub}\ee
Note that this is the same kind of substitution we made in constructing the quasi-particle wavefunction  \eqn{quasiparticle}. For a small number of particles ($N\approx 20$ or so) one can compute numerically the exact wavefunctions in different filling fractions: the wavefunctions \eqn{jain} built using the procedure described above have an overlap of around 99\% or so.

\para
Note that it's also possible to have $B^\star<0$. In this case, we have $\rho = -\nu^\star B^\star/\Phi_0$ and the relationship \eqn{jainfilling} becomes 
\be \nu = \frac{\nu^\star}{(m-1)\nu^\star - 1}\nn\ee
Then filling successive Landau levels $\nu^\star \in {\bf Z}$ gives the sequence $\nu = 1,\ 2/3,\ 3/5,\ 4/7,\ 5/9,\ldots$ which we again see as the prominent sequence of fractions sitting to the left of $\nu=1/2$ in the data. 

\para
We can also use the projection trick \eqn{jain} to construct excited quasi-hole and quasi-particle states in these new filling fractions. For each, we can determine the charge and statistics.  We won't do this here, but we will later revisit this question in Section \ref{cshiersec} from the perspective of Chern-Simons theory.

\subsubsection{The Half-Filled Landau Level}\label{hfllsec}

The composite fermion construction does a good job of explaining the observed plateaux. But arguably its greatest success lies in a region where no quantum Hall state is observed: $\nu=1/2$. (Note that the Laughlin state for $m=2$ describes bosons at half filling; here we are interested in the state of fermions at half filling). Looking at the data, there's no sign of a plateaux in the Hall conductivity at $\nu=1/2$. In fact, there seems to be a distinct absence of Hall plateaux in this whole region.    What's going on?!

\para
The composite fermion picture gives a wonderful and surprising answer to this. Consider a composite fermion consisting of an electron bound to two vortices. If $\nu=1/2$, so that the electrons have density $n = B/2\Phi_0$ then the effective magnetic field experienced by the composite fermions is \eqn{jainb}
\be B^\star = B - 2n\Phi_0 = 0\label{redmag}\ee
According to this, the composite fermions shouldn't feel a magnetic field. That seems kind of miraculous. Looking at the data, we see that the $\nu=1/2$ quantum Hall state occurs at a whopping $B\approx 25\ T$ or so. And yet this cartoon picture we've built up of composite fermions suggests that the electrons dress themselves with vortices so that they don't see any magnetic field at all.

\para
So what happens to these fermions? Well, if they're on experiencing a magnetic field, then they must pile up and form a Fermi sea. The resulting state is simply the compressible state of a two-dimensional metal. The wavefunction describing a Fermi sea of non-interacting fermions is  well known. If we have $N$ particles, with position ${\bf r}_i$, and the $N$ lowest momentum modes are ${\bf k}_i$, then we place particles in successive plane-wave states $e^{i{\bf k}_i\cdot{\bf r}_i}$ and subsequently anti-symmetrise over particles. The resulting slater determinant wavefunction is
\be {\psi}_{\rm Fermi\ Sea} = \det\left(e^{i{\bf k}_i\cdot{\bf r}_j}\right)\label{hlr}\ee
The Fermi momentum is defined to be the highest momentum i.e. $k_F\equiv |{\bf k}_N|$. 
Once again, this isn't a lowest Landau level wavefunction since, in complex coordinates,  ${\bf k}\cdot {\bf r} = \frac{1}{2}(k\bar{z} + \bar{k}z)$. This is cured, as before, by the projection operator giving us  the ground state wavefunction  at $\nu=1/2$,
\be \tilde{\psi}_{\nu=\frac{1}{2}} = {\cal P}_{LLL}\left[ \prod_{i<j} (z_i-z_j)^2\,\det\left(e^{i{\bf k}_m\cdot{\bf r}_l}\right)\right]\label{cfsea}\ee
where, as before, the $(z_i-z_j)^2$ factor captures the fact that each composite fermion contains two vortices.
This state, which describes an interacting Fermi sea, is sometimes called the {\it Rezayi-Read} wavefunction. (Be warned:  we will also describe a different class of wavefunctions in Section \ref{rrsec} which are called Read-Rezayi states!).
 There is a standard theory, due to Landau, about what happens when you add interactions to a Fermi sea known as Fermi liquid theory. The various properties of the state at $\nu=1/2$ and its excitations were studied in this context   by Halperin, Lee and Read, and is usually referred to as the HLR theory\footnote{The paper is ``{\it Theory of the half-filled Landau level}", 
\href{http://journals.aps.org/prb/abstract/10.1103/PhysRevB.47.7312}{Phys. Rev. B 47, 7312 (1993)}.}.

\para
There is overwhelming experimental evidence that the $\nu=1/2$ state is indeed a Fermi liquid. The simplest way to see this comes when we change the magnetic field slightly away from $\nu=1/2$. Then the composite fermions will experience a very small magnetic field $B^\star$ as opposed to the original $B$.  We can then see the Fermi surface and measure $k_F$ through standard techniques such as de Haas-van Alphen oscillations. Perhaps the cleanest demonstration is then to look at excitations above the Fermi surface.  Using  simple classical physics, we expect that the particles will move in the usual cyclotron circles, with $x+iy =Re^{i\omega t}$ where $\omega = eB^\star/m^\star$. The slight problem here is that we don't know $m^\star$. But if we differentiate, we can relate the radius of the circle to the momentum of the particle which, in the present case, we can take to be $\hbar k_F$. We then get the simple prediction
\be R = \frac{\hbar k_F}{e B^\star}\nn\ee
which has been confirmed experimentally.

\subsubsection*{The Dipole Interpretation}

Usually when we build  a Fermi sea by filling successive momentum states, it's obvious where the momentum comes from. But not so here. The problem is that the electrons are sitting in the lowest Landau level where  all kinetic energy is quenched. The entire Hamiltonian is governed only by the interactions between electrons, 
\be H = V_{\rm int}(|{\bf r}_i-{\bf r}_j|)\nn\ee
%
Typically we take this to be the Coulomb repulsion \eqn{crepel} or some toy Hamiltonian of the kind described in Section \ref{toysec}. How can we get something resembling momentum out of such a set-up?

\para
A potential answer comes from looking at the wavefunction \eqn{hlr} in more detail. The plane wave state is $e^{\frac{i}{2}(\bar{k}z+ k\bar{z}})$. Upon making the substitution \eqn{zbarsub}, this includes the term
\be \exp\left(i k l_B^2\ppp{}{z}\right)\nn\ee
But this is simply a translation operator. It acts by shifting $z\rightarrow z + ikl_B^2$. It means that in this case we can rewrite the wavefunction \eqn{cfsea} explicitly in holomorphic form, 
\be {\psi}_{\nu=\frac{1}{2}} = {\cal A} \left[\prod_i e^{i\bar{k}_iz_i - |z_i|^2/4l_B^2} \right]  \prod_{i<j} \Big((z_i+ i k_i l_B^2) -(z_j+ i k_jl_B^2)\Big)^2\nn\ee
\DOUBLEFIGURE{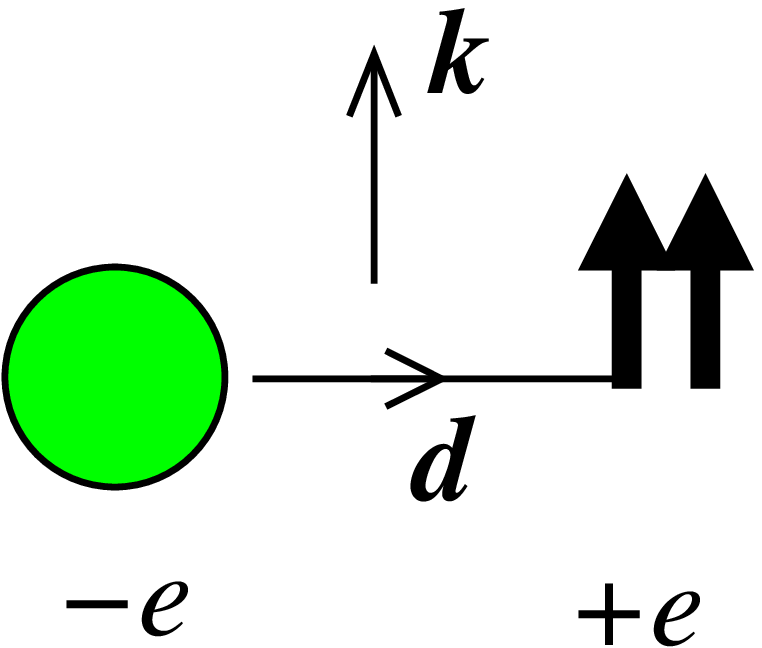,width=100pt}{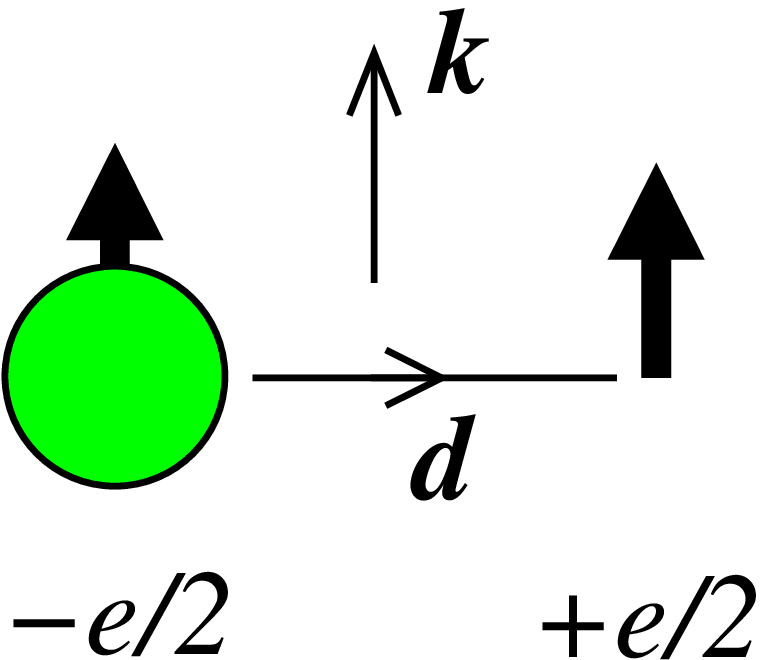,width=100pt}
{The composite fermion is a dipole like this.}{Or perhaps like this.}
where ${\cal A}$ is what's left of the determinant, and means that we should anti-symmetrise over all different ways of pairing up $k_i$ and $z_i$. Note that, for once, we've written the wavefunction including the exponential factor. The net result is that the zeros of the wavefunction --- which are the vortices --- are displaced by  a distance $|kl_B^2|$ from the electron, in the direction perpendicular to $\vec{k}$.

\para
As with much of the discussion on composite fermions, the ideas above are no more than suggestive. But they have turned out to be useful.   Now that we have an extended object, thinking in terms of a reduced magnetic field is perhaps not so useful since the  two ends can experience different magnetic fields. Instead, we can return to our original quasi-hole interpretation in which the vortices carry charge. One end then has  two vortices, each  with  charge   $+e/2$.  The other end consists of an electron with charge $-e$. The net result is the symmetric,  dipole configuration shown in the figure with a dipole moment $\vec{d}$, with magnitude proportional to $\vec{k}$, such that $\vec{d}\cdot\vec{k}=0$ and $|\vec{d}| = ekl_B^2$.

\para
The energy needed to produce such a dipole separation now comes entirely from the Coulomb interaction $V(|{\bf d}|)$ which we now interpret as $V(|\vec{k}|)$.  On rotational grounds, the expansion of the potential energy  should start with a term $\sim |\vec{d}|^2$ for small $\vec{d}$. This is the origin of the kinetic energy. The electron will drift along equipotentials of $V|{\vec{k}}|)$, while the vortices experience it as a magnetic field. The net effect is that both ends of the dipole move in the same direction, $\vec{k}$ with velocity $\partial V/\partial \vec{k}$ as expected.

\para
We note that, more recently it's been suggested that it's better to  think of the displacement as acting on just one of the two vortices bound to the electron\footnote{This was proposed by Chong Wang and Senthil in ``{\it Half-filled Landau level, topological insulator surfaces, and three dimensional quantum spin liquids}", \href{http://arxiv.org/abs/1507.08290}{arXiv:1507.08290}.}.  This can be justified on the grounds that each electron always accompanies a single zero because of Pauli exclusion. The end with a single vortex has charge $+e/2$, while the end that consists of an electron bound to a single vortex has charge $-e+e/2= -e/2$. We get the same qualitative physics as before, but with 
$|\vec{d}| = ekl_B^2/2$ as shown in the figure. The only difference between these two possibilities lies in the Berry phase that the dipole acquires as it moves around the Fermi surface. This helps resolve an issue about particle-hole symmetry at half-filling which we will discuss briefly in Section \ref{hlrsec}.

\subsubsection{Wavefunctions for Particles with Spin}\label{multiwfsec}

Until now we've neglected the role of spin in the quantum Hall states, arguing that the Zeeman effect is sufficient to polarise the spin of the electron. Here we describe a simple generalisation of the Laughlin wavefunction for particles that carry spin\footnote{These wavefunctions were first introduced by Bert Halperin in ``{\it Theory of the quantized Hall conductance}", Helv. Phys. Acta, 56 (1983).}.

\para
We split our particles into two sets. The first set has spin-up, with positions $z_1,\ldots z_{N^\uparrow}$. The second set has spin-down, with positions $w_1,\ldots,w_{N^\downarrow}$. Note that each electron has a fixed spin which is an eigenvalue of $S_z$: we don't allow the spin to fluctuate, nor do we allow the spin to be misaligned from the $z$-axis. We'll relax this condition shortly.

\para
If the two sets of particles didn't talk to each other, we can trivially take the product of two Laughlin wavefunctions,
\be \psi(z,w) = \prod_{i<j}^{N^\uparrow}(z_i-z_j)^{m_1}\prod_{k<l}^{N^\downarrow}(w_k-w_l)^{m_2}\,e^{-\sum |z_i|^2/4l_B^2-\sum |\omega_i|^2/4l_B^2}\nn\ee
Such a state would have filling fraction $\nu^\uparrow = 1/m_1$ and $\nu^\downarrow = 1/m_2$, giving total filling fraction $\nu = \nu^\uparrow + \nu^\downarrow$.

\para
Clearly there's nothing new in these wavefunctions. What's more, they miss the interesting physics. As we saw above, the Coulomb interactions are what drives the state to the Laughlin wavefunction. But these Coulomb interactions are blind to spin. They must also give correlations between the two sets of electrons. Halperin proposed to capture this with the simple wavefunction
\be \tilde{\psi}(z,w) = \prod_{i<j}^{N^\uparrow}(z_i-z_j)^{m_1}\prod_{k<l}^{N^\downarrow}(w_k-w_l)^{m_2}\prod_{i,k} (z_i-w_k)^n\label{halperin}\ee
where  now $\tilde{\psi}$ means that we're dropping the exponential factors for both variables. This set of wavefunctions are characterised by the three integers and  usually referred to as the $(m_1,m_2,n)$  states, or sometimes as {\it Halperin states}.

\para
These wavefunctions have very similar properties to the Laughlin states. In particular, the relative angular momentum  is never less than $m_1$ between two spin-up particles, never less than $m_2$ for two down-spin particles and never less than $n$ for particles of opposite spin. This kind of intuition allows us to build toy Hamiltonians, similar to those of Section \ref{toysec}, which have these wavefunctions as ground states.

\para
Let's now compute the filling fractions of these wavefunctions. Following our calculation in Section \ref{lwavesec}, we'll look at the highest power of a given spin-up electron, say $z_1$. We see that this has maximum angular momentum $m_1N^\uparrow + n N^\downarrow$ and hence fills out an area 
\be A^\uparrow = 2\pi(m_1N^\uparrow + n N^\downarrow) l_B^2\nn\ee
Meanwhile, the same computation for the spin-down particles gives us the area
\be A^\downarrow=  2\pi(m_2N^\downarrow + n N^\uparrow) l_B^2\nn\ee
If we want to focus on the places where both spin-up and spin-down particles intermingle, we should take $A^\uparrow = A^\downarrow$. Clearly for a given state $(m_1,m_2,n)$ this puts a constraint on, say, $N_\downarrow$ given $N_\uparrow$.  The filling fractions are then
\be \nu^\uparrow  &=& \frac{N^\uparrow}{m_1N^\uparrow + nN^\downarrow} = \frac{m_2-n}{m_1m_2-n^2}\nn\\ \nu^\downarrow &=& \frac{N^\downarrow}{m_2N^\downarrow + n N^\uparrow} = \frac{m_1-n}{m_1m_2-n^2}\nn\ee
%
where, in the second equality, we have used the constraint that follows from choosing $A^\uparrow = A^\downarrow$. The total filling fraction is then
\be \nu = \nu^\uparrow +\nu^\downarrow = \frac{m_1+m_2 -2n}{m_1m_2-n^2}\label{halpfill}\ee
The most prominent states of this kind have the form $(m,m,n)$. These have filling fractions $\nu^\uparrow = \nu^\downarrow = \nu/2$ with
\be \nu = \frac{2}{m+n}\ee
%
%
%
\noindent
Interesting examples include
\begin{itemize} 
\item 
$(3,3,1)$ with $\nu=1/2$. Note that this is a genuine quantum Hall state at $\nu=1/2$, as opposed to the Fermi liquid state described in Section \ref{hfllsec}. It has been seen in bi-layer samples, in which the $z$ and $w$ coordinate refer to the positions of particles in the two different layers\footnote{See  Y. Suen et. al, ``{\it Observation of a $\nu=1/2$ Fractional Quantum Hall State in a Double-Layer Electron System}", \href{http://journals.aps.org/prl/abstract/10.1103/PhysRevLett.68.1379}{Phys. Rev. Lett 68 9 (1992)}.}.
\item  $(3,3,2)$ with $\nu =2/5$. This state competes with the spin-polarised Jain state that occurs at the same filling. 
\end{itemize}
\para
Given these states, we could now start to construct quasi-hole and quasi-particle states for these multi-component wavefunctions. The quasi-holes in the $(m,m,n)$ state turn out to have charge $e/(m+n)$. We'll postpone this discussion to Section \ref{cssec}, where we'll see that we can describe both the $(m_1,m_2,n)$ states and the Jain states of Section \ref{compositesec} in a unified framework.

\subsubsection*{Putting Spin Back In}

So far, we've been calling the different sets of particles ``spin-up" and ``spin-down", but the wavefunctions \eqn{halperin} don't really carry the spin information. For example, there's no way to measure the spin of the particle in along the $x$-axis, as opposed to the $z$-axis. However, there's a simple way to remedy this. We just add the spin information, $\sigma=\uparrow$ or $\downarrow$ for each particle and subsequently anti-symmetrise (for fermions) over all $N=N_\uparrow+N_\downarrow$ particles. For $(m,m,n)$ states,  with $m>n$ and $N^\uparrow = N^\downarrow = N/2$, this is written as
%
%
%
\be \tilde{\psi}(z,\sigma) = {\cal A}\left[\prod_{i<j}^{N}(z_i-z_j)^{n}  \!\!\! \!\!\!\prod_{1<i<j<N/2} \!\!\!\!\!\! (z_i-z_j)^{m-n} \!\!\!\!\!\! \prod_{N/2+1<k<l<N} \!\!\!\!\!\! (z_k-z_l)^{m-n}\,|\uparrow\ldots\uparrow\,\downarrow\ldots \downarrow\rangle\right]\nn\ee
%
%
%
where ${\cal A}$ stands for anti-symmetrise over all particles,  exchanging both positions and spins.  Since the spin state above is symmetric in the first $N/2$ spins and the second $N/2$ spins, we must have  $m$  odd. (For bosons we could symmetrise over all particles providing $m$ is even).

\para
A particularly interesting class of wavefunction are spin singlets. Given a bunch of $N$ spins, one simple way to form a spin singlet state is to choose a  pairing of particles --- say $(12)$ and $(34)$ and so on --- and, for each pair, forming the spin singlet
\be |12\rangle  = \frac{1}{\sqrt{2}}\Big(|\uparrow_1\downarrow_2\rangle - |\downarrow_1\uparrow_2\rangle\Big)\nn\ee
Then the spin state $|\Psi\rangle = |12\rangle|34\rangle \ldots |N-1,N\rangle$ is a spin singlet. 

\para
An Aside: Of course, the spin singlet constructed above  is not unique. 
The number of spin singlet states is given by the {\it Catalan number}, $N!/(N^\uparrow +1)! N^\uparrow !$ where $N=2N^\uparrow$.

\para
We now want to write a spin singlet quantum Hall wavefunction. (Note that this is the opposite limit to the Laughlin wavefunctions which were fully spin polarised). Since the spin singlet state is itself anti-symmetric, we now require, in addition to having $m$ odd, that $n$ is even. 
It is then straightforward to construct a spin singlet version of the $(n+1,n+1,n)$ Halperin state by writing 
\be \tilde{\psi}(z,w,\sigma) = {\cal A}\left[\prod_{i<j}^{N}(z_i-z_j)^{n}  \prod_{i<j\,{\rm odd}} (z_i-z_j)  \prod_{k<l\,{\rm even}} (z_k-z_l)\ |12\rangle|34\rangle \ldots |N-1,N\rangle\right]\nn\ee
 It can be seen to be a spin singlet because the last two factors are just Slater determinants for spin up and spin down respectively, which is guaranteed to form a spin singlet. Meanwhile, the first factor is a symmetric polynomial and doesn't change the spin. 
A stronger statement, which would require somewhat more group theory to prove, is that the $(n+1,n+1,n)$ Halperin states are the only spin singlets.


\para
There is much more interesting physics in these quantum Hall states with spin. In particular, for the case $m=n$,   the Halperin states become degenerate with others in which the spins do not lie in along the $z$-direction and the spin picks up its own dynamics. The resulting physics is much studied and  associated to the phenomenon of quantum Hall ferromagnetism

\newpage
\section{Non-Abelian Quantum Hall States}\label{nonabsec}

\EPSFIGURE{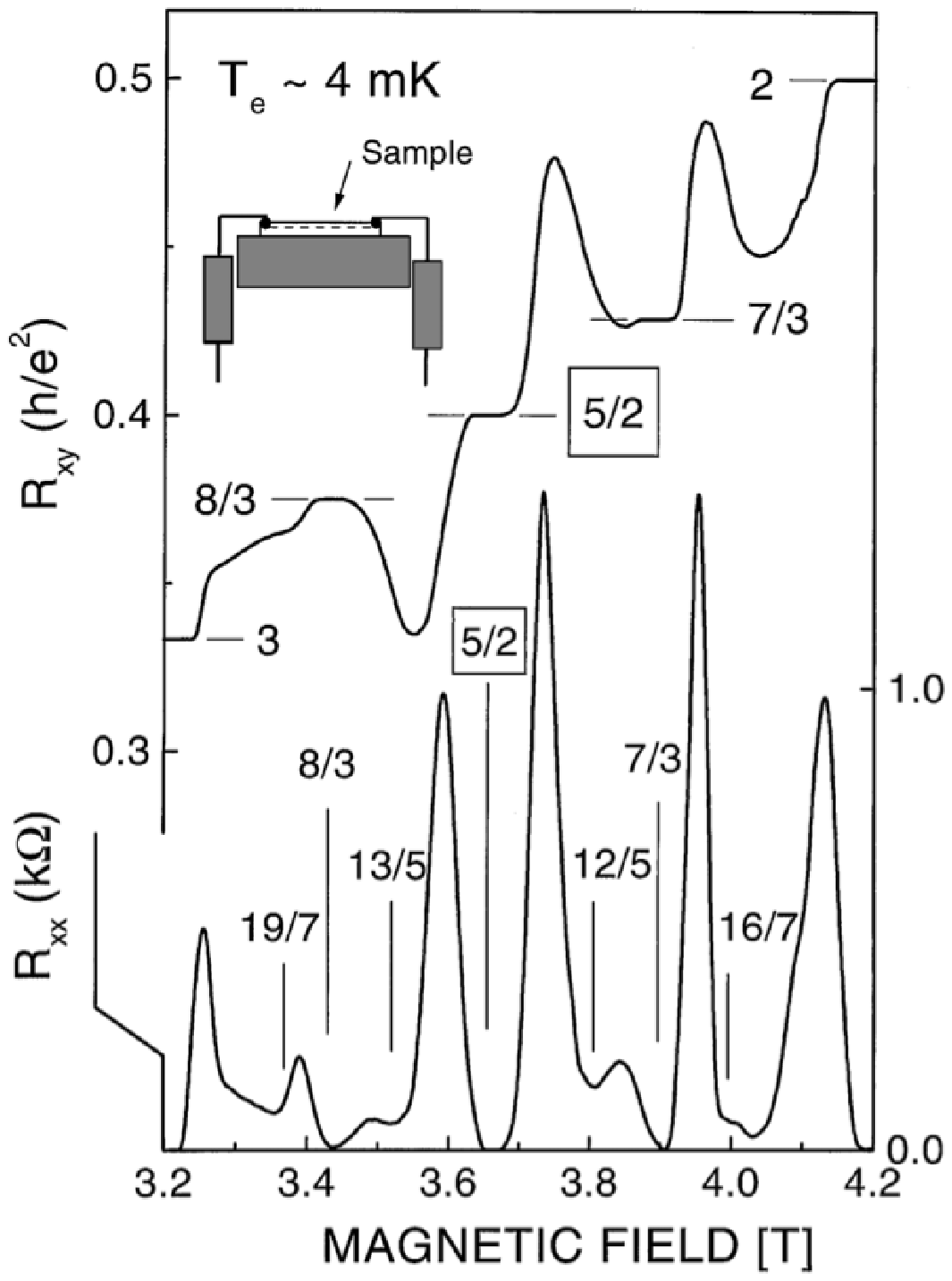,height=175pt}{}
\noindent
The vast majority of the observed quantum Hall plateaux sit at fractions with odd denominator. As we've seen above, this can be simply understood from the fermonic nature of electrons and the corresponding need for anti-symmetric wavefunctions. But there are some exceptions. Most prominent among them is the very clear quantum Hall state observed at $\nu =5/2$, shown in the figure\footnote{This state was first obseved by R. Willett, J. P. Eisenstein, H. L. Stormer,
D. C. Tsui, A. C. Gossard  and H. English ``{\it Observation of an Even-Denominator Quantum Number in the Fractional Quantum Hall Effect}", \href{http://journals.aps.org/prl/abstract/10.1103/PhysRevLett.59.1776}{Phys Rev Lett 59, 15 (1987)}. The data shown is from W. Pan et. al. 
Phys. Rev. Lett. 83, 17 (1999), \href{http://arxiv.org/abs/cond-mat/9907356}{cond-mat/9907356}.}. A similar quantum Hall state is also seen at $\nu=7/2$. 

\para
The $\nu=5/2$ state is thought to consist of fully filled lowest Landau levels for both spin up and spin down electrons, followed by a spin-polarised Landau level at half filling. The best candidate for this state turns out to have a number of extraordinary properties that opens up a whole new world of interesting physics involving {\it non-Abelian anyons}. The purpose of this section is to describe this physics.

\subsection{Life in Higher Landau Levels}

Until now, we've only looked at states in the lowest Landau level. These are characterised by holomorphic polynomials and, indeed, the holomorphic structure has been an important tool for us to understand the physics. Now that we're talking about quantum Hall states with $\nu>1$, one might think that we lose this advantage. Fortunately, this is not the case. As we now show,  if we can neglect the coupling between different Landau level then there's a way to map the physics back down to the lowest Landau level. 

\para
The first point to make is that there is a one-to-one map between Landau levels. We saw this already in Section \ref{llevelsec} where we introduced the creation and annihilation operators $a^\dagger$ and $a$ which take us from one Landau level to another. Hence, given a one-particle state in the lowest Landau level, 
\be |m\rangle \sim z^m e^{-|z|^2/4l_B^2}\nn\ee
we can construct a corresponding state $a^{\dagger\, n} |m\rangle$ in the $n^{\rm th}$ Landau level. (Note that the counting is like the British way of numbering floors rather than the American: if you go up one flight of stairs you're on the first floor or, in this case, the first Landau level). Similarly, a state of two particles in the lowest Landau level decomposes into a centre of mass part and a relative part, written as 
\be |M,m\rangle \sim (z_i+z_j)^M(z_i-z_j)^me^{-(|z_i|^2 + |z_j|^2)/4l_B^2}\nn\ee
 We can also again construct the corresponding state $a_1^{\dagger\,n}a_2^{\dagger\, n}|M,m\rangle$  in which each particle now sits in the $n^{\rm th}$ Landau level.

\para
We've already seen in Section \ref{toysec} that, if we focus attention to the lowest Landau level, then the interactions between particles can be characterised by pseudopotentials, defined by \eqn{pseudopot}
\be v_m  = \frac{\langle M,m|V(|{\bf r}_i-{\bf r_j}|)|M,m\rangle}{\langle M,m|M,m\rangle}\nn\ee
For a potential of the form  $V(|{\bf r}_1-{\bf r}_2|)$ which is both translationally and rotationally invariant, these pseudopotentials depend only on a single integer $m$. 

\para
However, this same argument also holds for higher Landau levels. Once again we can define pseudopotentials, now given by
\be v^{(n)}_m  = \frac{\langle M,m|a_i^na_j^nV(|{\bf r}_i-{\bf r_j}|)a_i^{\dagger\, n}a_j^{\dagger\, n}|M,m\rangle}{\langle M,m|M,m\rangle}\label{pseudonew}\ee
Of course, these differ from the $v_n$, but otherwise the resulting problem is the same. The upshot of this is that we can think of particles in the $n^{\rm th}$ Landau level, interacting through a potential $V$ as equivalent to particles in the lowest Landau level interacting with a potential given by \eqn{pseudonew}. Typically one finds that the values of $v^{(n)}_m$ are smaller than the values of $v_m$ for low $m$. This means that there's less of a penalty paid for particles coming close.

\para
Practically speaking, all of this provides us with a handy excuse to continue to work with holomorphic wavefunctions, even though we're dealing with higher Landau levels. Indeed, you may have noticed that we've not exactly been careful about what potential we're working with! 
Solving the Schr\"odinger equation for any realistic potential is way beyond our ability. Instead, we're just at the stage of making up reasonable looking wavefunctions. Given this, the fact that we have  to deal with a different potential is not going to be much of a burden.

\para
Moreover, these ideas also explain how  $\nu=5/2$ can be an incompressible quantum Hall state while, as we've seen,  $\nu=1/2$ is a compressible Fermi liquid state. Both states must be possible at half filling, but which is chosen depends on the detailed interactions that the electrons experience. Our first task, then, is to write down the quantum Hall state for electrons at half filling. In fact, we've already seen an example of this: the $(3,3,1)$ state described in Section \ref{multiwfsec}. But, as we now explain, there is also another, much more interesting candidate.

\subsection{The Moore-Read State}\label{mrsec}

\newcommand{\Pf}{{\rm Pf}}

The {\it Moore-Read}, or {\it Pfaffian} state describes an even number of particles, $N$, with filling fraction $\nu = 1/m$. It is given by\footnote{This state was proposed by Greg Moore and Nick Read in ``{\it NonAbelions in the Fractional Quantum Hall Effect}", Nucl. Phys {\bf B360} 362  (1991) which can be \href{http://www.physics.rutgers.edu/~gmoore/MooreReadNonabelions.pdf}{found here}. This important paper also introduces the relationship between wavefunctions and conformal field theory described later in these lectures.}
\be \tilde{\psi}_{MR}(z) = {\rm Pf}\left(\frac{1}{z_i-z_j}\right)\prod_{i<j} (z_i-z_j)^m\label{mr}\ee
In contrast to the Laughlin state, the wavefunction is anti-symmetric, and hence describes fermions, for $m$ even. It is symmetric for $m$ odd. To see this, we first need to answer the question:

\subsubsection*{What's a Pfaffian?}

Consider a n $N\times N$ anti-symmetric matrix, $M_{ij}$. The determinant of such a matrix vanishes when $N$ is odd, but when $N$ is even the determinant can be written the square of an object known as the {\it Pfaffian},
\be \det(M) = {\rm Pf}(M)^2\nn\ee
The Pfaffian is itself a polynomial of degree $N/2$ in the elements of the matrix, with integer coefficients. 

\para
There are a number of alternative expressions for the Pfaffian. Perhaps the simplest is to partition $N$ into $N/2$ pairs of numbers. For, example the simplest such partition is $(12)$, $(34)$, \ldots, $(N-1,N)$. The Pfaffian then takes the form
\be {\rm Pf}(M) = {\cal A}\left[M_{12} M_{34}\ldots M_{N-1,N}\right]
\label{pfafa}\ee
where all the details are hidden in the notation ${\cal A}$ which means {\it anti-symmetrise} on the indices, i.e. sum over all $\frac{(N)!}{2^{N/2} (N/2)!}$ partitions  with $\pm$ signs. Equivalently, can be written as
\be {\rm Pf}(M) = \frac{1}{2^{N/2} (N/2)!} \sum_\sigma\,{\rm sign}(\sigma)\,\prod_{k=1}^{N/2} M_{\sigma(2k-1),\sigma(2k)}\nn\ee
where the sum is over all $\sigma \in S_{N}$, the symmetric group, and ${\rm sign}(\sigma)$ is the signature of $\sigma$.

\para
For example, if we have four particles then
\be \Pf\left(\frac{1}{z_i-z_j}\right) = \frac{1}{z_1-z_2}\frac{1}{z_3-z_4} + \frac{1}{z_1-z_3}\frac{1}{z_4-z_2} + \frac{1}{z_1-z_4}\frac{1}{z_2-z_3}\nn\ee
Of course, the expressions rapidly get longer as $N$ increases. For 6 particles, there are 12 terms; for 8 particles there are 105. 

\subsubsection*{What's the Physics?}

The Pfaffian removes factors of $z_i-z_j$ compared to the Laughlin wavefunction, but in a clever way so that $\tilde{\psi}$ is never singular: whenever two particles approach, the Pfaffian diverges but is compensated by the $\prod(z_i-z_j)^m$ factor.

\para
 In particular, for the bosonic $m=1$ state, the wavefunction doesn't vanish when a pair of particles coincides, but it does vanish when the positions of three particles become coincident. This means that the $m=1$ state is a zero-energy ground state of the 3-body toy Hamiltonian,
 \be H = A\sum_{i<j<k} \delta^2(z_i-z_j)\delta^2(z_i-z_k)\label{2clusterham}\ee
Similar toy Hamiltonians can be constructed that have the general-$m$ Moore-Read state as their ground state. 
 
\para
The presence of the Pfaffian means that the Moore-Read state has fewer zeros than the Laughlin state, suggesting that the particles are more densely packed. However, the difference is irrelevant in the thermodynamic $N\rightarrow \infty$ limit. To see this, we compute the  filling fraction. There are $m(N-1)$ powers of $z_1$ in the Laughlin-like factor and a single $1/z_1$ factor from the Pfaffian. This tells us that the area of the droplet in the large $N$ limit is the same as the area of the Laughlin droplet with $N$ particles. We again have
\be \nu = \frac{1}{m}\nn\ee
as promised.

\para
The case of $m=1$ describes a fully filled Landau level of bosons and may be realisable using cold atoms in a rotating trap. The case of $m=2$ describes a half-filled Landau level of fermions. This will be our primary focus here.

\subsubsection*{The View from the Composite Fermion}

The Moore-Read wavefunction is crying out to be interpreted in terms of composite fermions.  In this language, the $\prod (z_i-z_j)^m$ factor attaches $m$ vortices to each electron. If $m$ is even, then the underlying electron was a fermion. Attaching an even number of vortices leaves it as a fermion. In contrast, if $m$ was odd then the underlying ``electron" was a boson. Attaching an odd number of vortices now turns it into a fermion. Either, way, the combined object of electron + $m$ vortices is a fermion. 

\para
We saw in  Section \ref{hfllsec} that for $m=2$, attaching the vortices  results in a composite fermion in an effectively vanishing magnetic field. The question is: how should we interpret the Pfaffian in this language?
In fact, there is a very natural interpretation: the Moore-Read state describes composite fermions which pile up to form a Fermi liquid and subsequently suffer a BCS pairing instability to superconductivity

\para
More meat can be put on this proposal. Here we  skip the meat and offer only some pertinent facts\footnote{This idea was proposed by Martin Greiter, Xiao-Gang Wen and Frank Wilzcek in ``{\it On Paired Hall States}", with all the details provided in the paper by Nick Read and Dmitry Green, ``{\it Paired states of fermions in two dimensions with breaking of parity and time-reversal symmetries, and the fractional quantum Hall effect}", \href{http://arxiv.org/abs/cond-mat/9906453}{cond-mat/9906453}.}. 
In a conventional superconductor, the spins of the electrons form a spin singlet. This provides the necessary anti-symmetry of the wavefunction so that the angular momentum part is symmetric. The simplest choice is that the electron pair condense in the s-wave. However, our composite fermions all have the same spin so the anti-symmetry must now come from the angular momentum. The simplest choice is now that the composite fermion pair condenses in the $p$-wave. In fact, the relevant choice of spherical harmonics gives what's known as a $p_x+ip_y$ superconductor. The appropriate BCS wavefunction for such a superconductor, in the weak pairing limit, indeed takes the form of the Pfaffian factor in \eqn{mr}.

\subsubsection{Quasi-Holes}

We can now look at excitations of the Moore-Read state. We will focus on quasi-holes. One obvious thing to try is to simply repeat what we did for the Laughlin quasi-hole \eqn{hole} and propose the wavefunction,
\be
\tilde{\psi}(z) = \prod_k (z_k-\eta)\,{\rm Pf}\left(\frac{1}{z_i-z_j}\right)\prod_{i<j} (z_i-z_j)^m\label{holeno}\ee
and, indeed, there's nothing wrong with this. By the same arguments we used before, the resulting object has charge $e/m$ and can be thought of as the addition of a single flux quantum or, in the language of \eqn{compositesec}, a single vortex.

\para
However, in the Moore-Read state (much) more interesting things can happen. The Laughlin quasi-hole, described by \eqn{holeno}, can itself split into two! We describe this by building  the positions of the new objects into the Pfaffian part of the wavefunction like so:
\be \tilde{\psi}(z) = \Pf\left(\frac{(z_i-\eta_1)(z_j-\eta_2) + (z_j-\eta_1)(z_i-\eta_2)}{z_i-z_j}\right)\prod_{i<j}(z_i-z_j)^m\label{mrhole}\ee
Note that the argument of the Pfaffian remains anti-symmetric, as it must. 
Multiplying out the Pfaffian, we see that this state contains the same number of $(z-\eta)$ factors as \eqn{holeno}, but clearly encodes the positions $\eta_1$ and $\eta_2$ of two independent objects. We will refer to these smaller objects as the quasi-holes. When these two quasi-holes coincide, so $\eta_1=\eta_2$, we get back the state \eqn{holeno}. 

\para
This means that the individual quasi-holes in \eqn{mrhole} can each be thought of as a half-vortex. They have charge%
\be e^\star = \frac{e}{2m}\nn\ee
In particular, for the $m=2$ state at half-filling, the quasi-holes should have charge $e/4$. There are claims that this prediction has been confirmed in  the $\nu=5/2$ state by shot-noise experiments\footnote{M. Dolev, M. Heiblum, V. Umansky, A. Stern and  D. Mahalu, ``{\it Observation of a Quarter of an Electron Charge at the $\nu=5/2$ Quantum Hall State}",  \href{http://www.nature.com/nature/journal/v452/n7189/full/nature06855.html}{Nature 452, 829-834 (2008)}.}, although the results remain somewhat controversial and are certainly less clean than the analogous experiments in the Abelian quantum Hall states.

\subsubsection*{How Many States with 4 Quasi-Holes?}

What about multiple quasi-holes? This is where things start to get interesting. Suppose that we want to write down a wavefunction for $4$ quasi-holes. Clearly we need to include the positions $\eta_\alpha$, $\alpha=1,2,3,4$ into the elements of the Pfaffian. One simple guess is the following expression
\be  \tilde{\psi}_{(12),(34)}(z) = \Pf_{(12),(34)}(z)\,\prod_{i<j}(z_i-z_j)^m\label{4qh}\ee
%
%
where we've defined
%
%
\be 
\Pf_{(12),(34)}(z) = \Pf\left(\frac{(z_i-\eta_1)(z_i-\eta_2)(z_j-\eta_3)(z_j-\eta_4) + (i\leftrightarrow j)}{z_i-z_j}\right)\nn\ee
Indeed, \eqn{4qh} is a fine quasi-hole state. But it's not unique: there was an arbitrariness in the way split the four quasi-particles into the two groups $(12)$ and $(34)$.  This makes it look as if there are two further states that we can write down,
\be \tilde{\psi}_{(13),(24)}(z)\ \ \ {\rm and}\ \ \ \tilde{\psi}_{(14),(23)}(z)\nn\ee
So it looks as if there are three states describing 4 quasi-holes. But this isn't right. It turns out that these states are not all linearly independent.

\para
It's a little fiddly to derive the linear dependence of quasi-hole states, but it's important. Here we'll derive the result for the simplest case of 4 quasi-holes  and then just state the result for the general case of $2n$ quasi-holes\footnote{The proof was first given by Chetan Nayak and Frank Wilczek in ``{\it $2n$ Quasihole States Realize $2^{n-1}$-Dimensional Spinor Braiding Statistics in Paired Quantum Hall States}, \href{http://arxiv.org/abs/cond-mat/9605145}{cond-mat/9605145}. The derivation above for 4 particles also follows this paper.}. The first step is to note the relation
\be (z_1-\eta_1)(z_1-\eta_2)(z_2-\eta_3)(z_2-\eta_4) &-& (z_1-\eta_1)(z_1-\eta_3)(z_2-\eta_4)(z_2-\eta_2) + (1\leftrightarrow 2) \nn\\  &=& (z_1-z_2)^2(\eta_1-\eta_4)(\eta_2-\eta_3)\label{cwrelation}\ee
which is simplest to see by noting that the left-hand side indeed vanishes on the roots. To save space, we introduce some new notation. Define $\eta_{\alpha\beta} = \eta_\alpha - \eta_\beta$ and
\be (12,34) \equiv (z_1-\eta_1)(z_1-\eta_2)(z_2-\eta_3)(z_2-\eta_4)  + (1\leftrightarrow 2)\nn\ee
So that \eqn{cwrelation} reads
\be (12,34) - (13,24) = (z_1-z_2)^2\eta_{14}\eta_{23}\nn\ee
%
%
Then, using the definition of the Pfaffian \eqn{pfafa}, we have
\be \Pf_{(13),(24)}(z) &=& {\cal A}\left(\frac{(13,24)}{z_1-z_2}\frac{(13,24)}{z_3-z_4}\ldots\right) 
\nn\\ &=& {\cal A}\left(\frac{(12,34) - (z_1-z_2)^2\eta_{14}\eta_{23}}{z_1-z_2}\,\frac{(12,34) - (z_3-z_4)^2\eta_{14}\eta_{23}}{z_3-z_4}\ldots\right) \nn\\ &=& {\cal A}\left(\frac{(12,34)}{z_1-z_2}\frac{(12,34)}{z_3-z_4}\ldots\right) - {\cal A}\left((z_1-z_2)\eta_{14}\eta_{23}\,\frac{(12,34) \eta_{14}\eta_{23}}{z_3-z_4}\ldots\right) \nn\\ && +\ {\cal A}\left((z_1-z_2)\eta_{14}\eta_{23}(z_3-z_4)\eta_{14}\eta_{23}\, 
\frac{(12,34) \eta_{14}\eta_{23}}{z_5-z_6}\ldots\right) + \ldots
\nn\ee
where the terms that we didn't write down have factors like $(z_1-z_2)(z_3-z_4)(z_5-z_6)$ and so on. However, in the last term, the anti-symmetrisation acts on the $(z_1-z_2)(z_3-z_4)$ factor which vanishes. Indeed, for all the remaining terms we have to anti-symmetrise a polynomial which is linear in each factor and this too vanishes.  We're left with
\be \Pf_{(13),(24)}(z) = \Pf_{(12),(34)}(z) -
{\cal A}\left((z_1-z_2)\eta_{14}\eta_{23}\,\frac{(12,34) \eta_{14}\eta_{23}}{z_3-z_4}\ldots\right) \nn\ee
The same kind of calculation also gives
\be \Pf_{(14),(23)}(z) = \Pf_{(12),(34)}(z) -
{\cal A}\left((z_1-z_2)\eta_{13}\eta_{24}\,\frac{(12,34) \eta_{14}\eta_{23}}{z_3-z_4}\ldots\right) \nn\ee
But this gives the result that we want: it says that there is a linear relation between the three different Pfaffian wavefunctions.
\be  \Pf_{(12),(34)}(z) - \Pf_{(13),(24)}(z) = \frac{\eta_{14}\eta_{23}}{\eta_{13}\eta_{24}}\Big(\Pf_{(12),(34)}(z)-\Pf_{(14),(23)}(z)\Big)\nn\ee
There are two lessons to take from this. The first is that if we fix the positions $\eta_\alpha$ of the four quasi-holes, then there is not a unique state that describes them. Instead, the state is degenerate. But it's not as degenerate as we might have thought. There are only 2 states describing four quasi-holes, rather than the 3 that a naive counting gives.

\subsubsection*{How Many States with Multiple Quasi-holes?}

We can now repeat this for the general situation of $2n$ quasi-hole.  To build a suitable wavefunction, we first decompose these quasi-particles into two groups of $n$. For example let's pick $(1\ldots n)$ and $(n+1\ldots 2n)$ as a particularly obvious choice. Then the wavefunction takes the form \eqn{4qh}, but with the Pfaffian component replaced by
\be \Pf\left(\frac{(z_i-\eta_1)(z_i-\eta_2)\ldots (z_i-\eta_n)(z_j-\eta_{n+1})(z_j-\eta_{n+2})\ldots(z_j-\eta_n) + (i\leftrightarrow j)}{z_i-z_j}\right)\ \ \ \ \ \label{manymrhole}\ee
Clearly this again depends on the choice of grouping. The number of ways of placing $2n$ elements into two groups is
\be \frac{1}{2}\frac{(2n)!}{n!n!}\nn\ee
but, as our previous discussion shows, these states are unlikely to be linearly independent. The question is: how many linearly independent states are there? It turns out that the answer is:
\be \mbox{dimension of Hilbert space} = 2^{n-1}\label{mrcounting}\ee
Obviously this agrees with our answer of 2 when we have four quasi-holes.

\para
A moments thought shows that the counting \eqn{mrcounting} is very peculiar. We're quite used to the Hilbert space for a  group of particles having a degeneracy when each particle has an internal degree of freedom. For example, if we have $N$ particles each of spin-1/2 then the total Hilbert space has dimension $2^{N}$. But that can't be what's going on with our quasi-holes. We have $2n$ quasi-holes but an internal Hilbert space of dimension $2^{n-1}$. Even ignoring the factor of $2^{-1}$ for now, we have many fewer states than could be accounted for by each particle having it's own  internal degree of freedom.

\para
This simple observation is really the key bit of magic captured by the Moore-Read excitations. The ``internal" degrees of freedom described by the Hilbert space of dimension $2^{n-1}$ are not associated to any individual quasi-hole and they can't be seen by looking at any local part of the wavefunction.  Instead they are a property of the entire collection of particles. It is information stored non-locally in the wavefunction.

\subsubsection*{Quasi-Holes are Non-Abelian Anyons}

Let's now think about what happens when the quasi-holes are exchanged. As we have seen, if we have $2n$ quasi-holes then there are $2^{n-1}$ possible ground states. When we take any closed path in the configuration space of quasi-holes, the state of the system can come back to itself up to a unitary $U(2^{n-1})$ rotation. This is an example of the non-Abelian Berry holonomy discussed in Section \ref{nonabberrysec}. The quasi-holes are referred to as {\it non-Abelian anyons}. (The original suggested name was ``{\it non-Abelions}", but it doesn't seem to have caught on.)

\para
Our task is to figure out the unitary matrices associated to the exchange of particles. 
Conceptually, this task is straightforward. We just need to construct an orthonormal set of $2^{n-1}$ wavefunctions and compute the non-Abelian Berry connection \eqn{nonabconnection}. In practice, that's easier said than done.  Recall that in the computation of the Berry connection for Laughlin quasi-holes we relied heavily on the plasma analogy. This suggests that to make progress we would need to develop a similar, but more involved, plasma analogy for the Moore-Read state. The resulting calculations are quite long\footnote{The results were conjectured in the '96 paper by Nayak and Wilczek, but a full proof had to wait until the work of   Parsa Bonderson, Victor Guarie and Chetan Nayak, ``{\it Plasma Analogy and Non-Abelian Statistics for Ising-type Quantum Hall States}", \href{http://arxiv.org/abs/1008.5194}{arXiv:1008.5194}. }.

\para
The good news is that although the calculation is somewhat involved, the end result is quite simple However, this also suggests that there might be a more physical way to get to this result. And, indeed there is: it involves returning to the composite fermion picture.

\subsubsection{Majorana Zero Modes}\label{majoranasec}

Recall that, at $\nu=1/2$, composite fermions are immune to the background magnetic field and instead form a Fermi sea. The Moore-Read state arises when these composite fermions pair up and condense, forming a p-wave superconductor.

\para
This viewpoint provides a very simple way to understand the non-Abelian statistics. Moreover, the results are general and apply to any other $(p_x+ip_y)$ superconductor. The unconventional superconductor  $Sr_2RuO_4$ is thought to fall into this class, and it may be posisible to construct these states in cold atom systems. (A warning: this last statement is usually wheeled out for almost anything that people don't really know how to build.)

\para
To proceed, we will need a couple of facts about the p-wave superconducting state that I won't prove. The first is that, in common with all superconductors, there are vortices, in which the phase of the condensate winds around the core. Because the composite electrons condense in pairs, the simplest vortex can carry $\Phi_0/2e$ flux as opposed to $\Phi_0/e$. For this reason, it's sometimes called a half-vortex, although we'll continue to refer to it simply as the vortex. This will be our quasi-hole.

\para
The second fact that we'll need is the crucial one, and is special to $p_x+ip_y$ superconductors. The vortices have {\it zero modes}. These are  solutions to the equation for the fermion field in the background of a vortex. They can be thought of as a fermion bound to the vortex. Importantly, for these p-wave superconductors, this zero mode is {\it Majorana}\footnote{A very simple explanation of Majorana fermions in different contexts can be found in Frank Wilczek's nice review ``{\it Majorana Returns}",  \href{http://www.nature.com/nphys/journal/v5/n9/full/nphys1380.html}{Nature Physics {\bf 5} 614 (2009)}. }. 

\subsubsection*{A Hilbert Space from Majorana Zero Modes}

To explain what a Majorana mode means, we'll have to work in the language of creation and annihilation operators for particles which is more familiar in the context of quantum field theory. We start by reviewing these operators for standard fermions. We define $c_i^\dagger$ to be the operator that creates an electron (or, more generally a fermion). Here the index $i$ labels any other quantum numbers of the electron, such as momentum or spin. Meanwhile, the conjugate operator $c_i$ annihilates an electron or, equivalently, creates a hole. (In high-energy physics, we'd call this an anti-particle.) These fermionic creation and annihilation operators obey 
\be \{ c_i,c_j^\dagger\} = \delta_{ij}\ \ \ {\rm and}\ \ \ \{c_i,c_j\}= \{c_i^\dagger,c_j^\dagger\} =0\label{calg}\ee
which can be thought of as the manifestation of the Pauli exclusion principle.

\para
A {\it Majorana} particle is a fermion which is its own anti-particle. It can be formally created by the operator
\be \gamma_i = c_i+c_i^\dagger\label{gammacc}\ee
which clearly satisfies the condition $\gamma_i=\gamma_i^\dagger$. From \eqn{calg}, we see that these Majorana operators satisfy
\be \{\gamma_i,\gamma_j\} = 2\delta_{ij}\label{clifford}\ee
This is known as the {\it Clifford algebra}.

\para
While it's simple to write down the equation \eqn{gammacc}, it's much harder to cook up a physical system in which these excitations exist as eigenstates of the Hamiltonian. For example, if we're talking about real electrons then $c^\dagger$  creates a particle of charge $-e$ while $c$ creates a hole of charge $+e$. This means that $\gamma$ creates a particle which is in a superposition of different charges. Usually, this isn't allowed. However, the environment in a superconductor makes it possible. Electrons have paired up into Cooper pairs to form a boson which subsequently condenses. The ground state then contains a large reservoir of particles which can effectively absorb any $\pm 2e$ charge. This means that in a superconductor, charge is conserved only mod 2. The electron and hole then have effectively the same charge.

\para
Suppose now that we have $2n$ well-separated vortices, each with their Majorana zero mode $\gamma_i$. (We'll see shortly why we restrict to an even number of vortices.) We fix the positions of the vortices. What is the corresponding Hilbert space? To build the Hilbert space, we need to take two Majorana modes and, from them, reconstruct a complex fermion zero mode. To do this, we make an arbitrary choice to pair the Majorana mode associated to one vortex with the Majorana mode associated to a different vortex. There's no canonical way to pair vortices like this but any choice we make will work  fine. For now, let's pair $(\gamma_1,\gamma_2)$ and $(\gamma_3,\gamma_4)$ and so on. 
 We then define the complex zero modes
\be \Psi_k = \frac{1}{2}\left(\gamma_{2k-1} + i\gamma_{2k}\right)\ \ \ \ k=1,\ldots,n\label{arbpair}\ee
These obey the original fermionic commutation relations
\be \{\Psi_k,\Psi_l^\dagger\} = \delta_{kl}\ \ \ {\rm and}\ \ \ \{\Psi_k,\Psi_l\}=\{\Psi_k^\dagger,\Psi_l^\dagger\} = 0\nn\ee
The Hilbert space is then constructed in a way which will be very familiar if you've taken a first course on quantum field theory. We first introduce a ``vacuum", or reference state $|0\rangle$  which obeys $\Psi_k|0\rangle=0$ for all $k$. We then construct the full Hilbert space by acting with successive creation operators, $\Psi_k^\dagger$ to get
\be |0\rangle \nn\\ \Psi_k^\dagger |0\rangle \nn\\ \Psi_k^\dagger\Psi_l^\dagger|0\rangle\label{majhilbert}\\ \vdots\ \ \ \ \nn\\ \Psi_1^\dagger\ldots\Psi_n^\dagger |0\rangle\nn\ee
The sector with $p$ excitations has ${p}\choose{n}$ possible states. The dimension of the full Hilbert space is
\be \mbox{dimension of Hilbert space} = 2^n\nn\ee
Note, firstly, that the same comments we made for quasi-hole wavefunctions also apply here. There's no way to think of this Hilbert space as arising from local degrees of freedom carried by each of the $2n$ vortices. Indeed, one advantage of this approach is that it demonstrates very clearly the non-local nature of the Hilbert space. 
 Each individual vortex carries only a Majorana zero mode. But a single Majorana zero mode doesn't buy you anything: you need two of them to form a two-dimensional Hilbert space.

\para
The dimension of the Hilbert space we've found here is twice as big as the dimension \eqn{mrcounting} that comes from counting linearly independent wavefunctions. But it turns out that there's a natural way to split this Hilbert space into two. As we'll see shortly, the braiding of vortices mixes states with an even number of $\Psi^\dagger$ excitations among themselves. Similarly, states with an odd number of $\Psi^\dagger$ excitations also mix among themselves. Each of these Hilbert spaces has dimension $2^{n-1}$. The linearly independent quasi-hole excitations \eqn{manymrhole} can be thought of as spanning one of these smaller Hilbert spaces. 

\subsubsection*{Braiding of Majorana Zero Modes}

The Majorana zero modes give us a simple way to construct the Hilbert space for our non-Abelian anyons. They also give us a simple way to see the braiding\footnote{This calculation was first done by Dimitry Ivanov in ``{\it Non-abelian statistics of half-quantum vortices in p-wave superconductors}",  \href{http://arxiv.org/abs/cond-mat/0005069}{cond-mat/0005069}.}.

\para
Recall from Section \ref{anyonsec} that the  braid group is generated by $R_i$,  with $i=1,\ldots, 2n-1$, which exchanges the $i^{\rm th}$ vortex with the $(i+1)^{\rm th}$ vortex in an anti-clockwise direction. The action of this braiding on the Majorana zero modes is
\be R_i:\left\{\begin{array}{cc} \gamma_i\rightarrow \gamma_{i+1} & \\ \gamma_{i+1}\rightarrow -\gamma_i \\  \gamma_j\rightarrow\gamma_j &\ \ \ \ \ j\neq i,i+1\end{array}\right.\nn\ee
where the single minus sign corresponds to the fact that the phase of a Majorana fermion changes by $2\pi$ as it encircles a vortex.

\para
We want to represent this action by a unitary operator --- which, with a slight abuse of notation we will also call $R_i$ --- such that the effect of a braid can be written as  $R_i\gamma_j R_i^\dagger$. It's simple to write down such an operator,
\be R_i =   \exp\left(\frac{\pi}{4}\gamma_{i+1}\gamma_i\right) e^{i\pi\alpha}= \frac{1}{\sqrt{2}} ( 1 + \gamma_{i+1}\gamma_i) e^{i\pi\alpha}\nn\ee
To see that these two expressions are equal, you need to use the fact that $(\gamma_{i+1}\gamma_i)^2=-1$, together with $\sin(\pi/4) = \cos(\pi/4)=1/\sqrt{2}$. The phase factor $e^{i\pi\alpha}$ captures the Abelian statistics which is not fixed by the Majorana approach. For the Moore-Read states at filling fraction $\nu=1/m$, it turns out that this statistical phase is given by
\be \alpha = \frac{1}{4m}\label{mralpha}\ee
Here, our interest lies more in the non-Abelian part of the statistics. For any state in the Hilbert space, the action of the braiding is 
\be |\Psi\rangle \rightarrow R_i|\Psi\rangle\nn\ee
Let's look at how this acts in some simple examples.

\subsubsection*{Two Quasi-holes}

Two quasi-holes give rise to two states, $|0\rangle$ and $\Psi^\dagger|0\rangle$. Written in terms of the complex fermions, the exchange operator becomes
\be R = \frac{1}{\sqrt{2}}(1+i-2i\Psi^\dagger \Psi)e^{i\pi\alpha}\nn\ee
from which we can easily compute the action of exchange on the two states
\be R\,|0\rangle = e^{i\pi/4}e^{i\pi\alpha} |0\rangle\ \ \ {\rm and}\ \ \ R\,\Psi^\dagger|0\rangle = e^{-i\pi/4}e^{i\pi\alpha}\Psi^\dagger|0\rangle\label{ri}\ee
Alternatively, written as a $2\times 2$ matrix, we have  $R = e^{i\pi\sigma^3/4} e^{i\pi\alpha}$ with $\sigma^3$ the third Pauli matrix. We see that each state simply picks up a phase factor as if they were Abelian anyons.

\subsubsection*{Four Quasi-holes}

For four vortices, we have four states: $|0\rangle$, $\Psi_k|0\rangle$ for $k=1,2$, and $\Psi^\dagger_1\Psi_2^\dagger|0\rangle$. Meanwhile, there three generators of the braid group. For the exchanges $1\leftrightarrow 2$ and $3\leftrightarrow 4$, the corresponding
operators involve only a single complex fermion, 
\be R_{1} = \frac{1}{\sqrt{2}} ( 1 + \gamma_{2}\gamma_1) e^{i\pi\alpha} =  \frac{1}{\sqrt{2}}(1+i-2i\Psi_1^\dagger \Psi_1)e^{i\pi\alpha}\nn\ee 
and 
\be R_{3} = \frac{1}{\sqrt{2}} ( 1 + \gamma_{4}\gamma_3) e^{i\pi\alpha} =  \frac{1}{\sqrt{2}}(1+i-2i\Psi_2^\dagger \Psi_2)e^{i\pi\alpha}\nn\ee 
This is because each of these exchanges vortices that were paired in our arbitrary choice \eqn{arbpair}. This means that, in our chosen basis of states, these operators give rise to only Abelian phases, acting as
\be R_1 = \left(\begin{array}{cccc}  e^{i\pi/4}  &&& \\ & e^{-i\pi/4} & & \\ &&e^{i\pi/4} & \\ & &&e^{-i\pi/4}    \end{array} \right)e^{i\pi\alpha}
\ \ \ {\rm and}\ \ \  R_3 = \left(\begin{array}{cccc}  e^{-i\pi/4}  &&& \\ & e^{-i\pi/4} & & \\ &&e^{i\pi/4} & \\ & &&e^{i\pi/4}    \end{array} \right)e^{i\pi\alpha}\nn\ee 
Meanwhile, the generator $R_2$ swaps $2\leftrightarrow 3$. This is more interesting because these two vortices sat in different pairs in our construction of the basis states using \eqn{arbpair}. This means that the operator  involves both $\Psi_1$ and $\Psi_2$,
\be R_2 = \frac{1}{\sqrt{2}}( 1 + \gamma_3\gamma_2) = \frac{1}{\sqrt{2}}\left(1-i(\Psi_2+\Psi_2^\dagger)(\Psi_1-\Psi_1^\dagger)\right)\nn\ee
and, correspondingly, is not diagonal in our chosen basis. Instead, it is written as 
\be R_2 = \frac{1}{\sqrt{2}} \left(\begin{array}{cccc}  1 \ & 0\ & 0\ & -i\  \\ 0\ & 1\ & -i\ & 0\ \\ 0\ & -i\ & 1\ & 0\ \\ i\ &0\ & 0\ & 1\ \end{array}\right)\label{nonabrot}\ee
Here we see  the non-Abelian nature of exchange. Note that, as promised, the states $\Psi_k|0\rangle$ with an odd number of $\Psi$ excitations transform into each other, while the states $|0\rangle$ and $\Psi_1^\dagger\Psi_2^\dagger|0\rangle$ transform into each other. This property persists with an arbitrary number of anyons because the generators $R_i$ defined in \eqn{ri} always contain one creation operator $\Psi^\dagger$ and one annihilation operator $\Psi$. It means that we are really describing two classes of non-Abelian anyons, each with Hilbert space of dimension $2^{n-1}$. 

\para
The non-Abelian anyons that we have described above are called {\it Ising anyons}.  The name is strange as it's not at all clear at this stage what these anyons have to do with the Ising model. We will briefly explain the connection in Section \ref{fermisec}. 

\subsubsection*{Relationship to $SO(2n)$ Spinor Representations}

The discussion above has a nice interpretation in terms of the spinor representation of the rotation group $SO(2n)$. This doesn't add anything new to the physics, but it's simple enough to be worth explaining. 

\para
As we already mentioned, the algebra obeyed by the Majorana zero modes \eqn{clifford} is called the Clifford algebra. It is well known to have a unique irreducible representation of dimension $2^n$. This can be built from $2\times 2$ Pauli matrices, $\sigma^1, \sigma^2$ and $\sigma^3$ by
\be \gamma^1 &=& \sigma^1\otimes\sigma^3\otimes\ldots\otimes\sigma^3\nn\\
\gamma^2 &=& \sigma^2\otimes \sigma^3\otimes\ldots\sigma^3\nn\\ &\vdots&\nn\ee\be
\gamma^{2k-1} &=& {\bf 1}\otimes \ldots\otimes {\bf 1}\otimes \sigma^1 \otimes \sigma^3 \otimes\ldots\otimes\sigma^3\nn\\
\gamma^{2k} &=& {\bf 1}\otimes \ldots\otimes {\bf 1}\otimes \sigma^2 \otimes \sigma^3 \otimes\ldots\otimes\sigma^3\nn\\
&\vdots&\nn\\
\gamma^{2n-1} &=& {\bf 1}\otimes\ldots\otimes{\bf 1}\otimes \sigma^1\nn\\
\gamma^{2n} &=& {\bf 1}\otimes\ldots\otimes{\bf 1}\otimes \sigma^2\nn\ee
The Pauli matrices themselves obey $\{\sigma^a,\sigma^b\} = 2\delta^{ab}$ which ensures that the gamma-matrices defined above obey the Clifford algebra. 

\para
From the Clifford algebra, we can build generators of the Lie algebra $so(2n)$. The rotation in the $(x^i,x^j)$ plane is generated by the anti-symmetric matrix
\be {T}_{ij} = \frac{i}{4}[\gamma^i,\gamma^j]\label{diracbraid}\ee
This is called the {\it (Dirac) spinor} representation of $SO(2n)$. The  exchange of the $i^{\rm th}$ and $j^{\rm th}$ particle is represented on the Hilbert space by a $\pi/2$ rotation in the
 $(x^i,x^{i+1})$ plane,
\be R_{ij} = \exp\left(- \frac{i\pi}{2} T_{ij}\right)\nn\ee
For the generators $R_i = R_{i,i+1}$, this coincides with our previous result \eqn{ri}.

\para
The spinor representation \eqn{diracbraid} is not irreducible. To see this, note that there is one extra gamma matrix,
\be \gamma^{2n+1} &=& \sigma^3\otimes\sigma^3\otimes\ldots\otimes\sigma^3\nn\ee
which anti-commutes with all the others, $\{\gamma^{2n+1},\gamma^i\}=0$ and hence commutes with the Lie algebra elements $[\gamma^{2n+1},{T}_{ij}] = 0$. Further, we have $(\gamma^{2n+1})^2={\bf 1}_{2n}$, so $\gamma^{2n+1}$ has eigenvalues $\pm 1$.  By symmetry, there are $n$ eigenvalues $+1$ and $n$ eigenvalues $-1$. We can then construct two irreducible {\it chiral spinor} representations of $so(2n)$ by projecting onto these eigenvalues. These are the representation of non-Abelian anyons that act on the Hilbert space of dimension $2^{n-1}$.

\para
This, then, is the structure of Ising anyons, which are excitations of the Moore-Read wavefunction. The Hilbert space of $2n$ anyons has dimension $2^{n-1}$. The act of braiding two anyons acts on this Hilbert space in the chiral spinor representation of $SO(2n)$, rotating by an angle $\pi/2$ in the appropriate plane.

\subsubsection{Read-Rezayi States}\label{rrsec}

In this section, we describe an extension of the Moore-Read states. Let's first give the basic idea. We've seen that the $m=1$ Moore-Read state has the property that it vanishes only when three or more particles come together. It can be thought of as a zero-energy ground state of the simple toy Hamiltonian,
\be H = A\sum_{i<j<k}\delta^2(z_i-z_j)\delta^2(z_j-z_k)\nn\ee
This suggests an obvious generalisation to wavefunctions which only vanish when some group of $p$ particles come together. These would be the ground states of the toy Hamiltonian
\be H = A\sum_{i_1<i_2<\ldots<i_p} \delta^2(z_{i_1}-z_{i_2})\delta^2(z_{i_2}-z_{i_3})\ldots\delta^2(z_{i_{p-1}}-z_{i_p})\nn\ee
The resulting wavefunctions are called {\it Read-Rezayi} states. 

\para
To describe these states, let us first re-write the Moore-Read wavefunction in a way which allows a simple generalisation. We take $N$ particles and arbitrarily divide them up into two groups. We'll label the positions of the particles in the first group by $v_1,\ldots,v_{N/2}$ and the position of particles in the second group by $w_1,\ldots,w_{N/2}$. Then we can form the wavefunction
\be \tilde{\psi}_{CGT}(z) = {\cal S}\left[\prod_{i<j}(v_i-v_j)^2(w_i-w_j)^2\right]
\nn\ee
where ${\cal S}$ means that we symmetrise over all ways of diving the electrons into two groups, ensuring that we end up with a bosonic wavefunction. The claim is that
\be \psi_{MR}(z) = \tilde{\psi}_{CGT}(z)\prod_{i<j}(z_i-z_j)^{m-1}\nn\ee
We won't prove this claim here\footnote{The proof isn't hard but it is a little fiddly. You can find it in the paper by Cappelli, Georgiev and Todorov, ``{\it Parafermion Hall states from coset projections of abelian conformal theories}", \href{http://arxiv.org/abs/hep-th/0009229}{hep-th/0009229}.}. But let's just do a few sanity checks. At $m=1$, the Moore-Read wavefunction is a polynomial in $z$ of degree $N(N/2-1)$, while any given coordinate -- say $z_1$ -- has at most power $N-2$. Both of these properties are easily seen to hold for $\tilde{\psi}_{CGT}$. Finally, and most importantly, $\tilde{\psi}_{CGT}(z)$ vanishes only if three particles all come together since two of these particles must sit in the same group.

\para
It's now simple to generalise this construction. Consider $N=pd$ particles. We'll separate these into $p$ groups of $d$ particles whose positions we label as $w_1^{(a)},\ldots,w_d^{(a)}$ where $a=1,\ldots,p$ labels the group. We then form the Read-Rezayi wavefunction\footnote{The original paper  ``{\it Beyond paired quantum Hall states: parafermions and incompressible states in the first excited Landau level},  \href{http://arxiv.org/abs/cond-mat/9809384}{cond-mat/9809384}, presents the wavefunction is a slightly different, but equivalent form.}
\be \tilde{\psi}_{RR}(z) = {\cal S}\left[ \prod_{i<j} (w^{(1)}_i-w^{(1)}_j)^2\ldots  \prod_{i<j}(w^{(p)}_i-w^{(p)}_j)^2\right] \prod_{k<l}(z_k-z_l)^{m-1}\nn\ee
where, again, we symmetrise over all possible clustering of particles into the $p$ groups. This now has the property that the $m=1$ wavefunction vanishes only if the positions of $p+1$ particles coincide. For this reason, these are sometimes referred to as $p$-clustered states, while the original Moore-Read wavefunction is called a paired state.

\para
Like the Moore-Read state, the Read-Rezayi state describes fermions for $m$ even and bosons for $m$ odd. The filling fraction can be computed in the usual manner by looking at the highest power of some given position. We find
\be \nu = \frac{p}{p(m-1)+2}\nn\ee
The fermionic $p=3$-cluster state at $m=2$  has filling fraction $\nu=3/5$ and is a promising candidate for the observed Hall plateaux at $\nu=13/5$. One can also consider the particle-hole conjugate of this state which would have filling fraction $\nu=1-3/5=2/5$. There is some hope that this describes the observed plateaux at $\nu=12/5$.

\subsubsection*{Quasi-Holes}

One can write down quasi-hole excitations above the Read-Rezayi state. Perhaps unsurprisingly, such quasi-holes necessarily come in groups of $p$. The simplest such state is
\be \tilde{\psi}(z) = {\cal S}\left[   \prod_{a=1}^p\prod_{i=1}^{N/p} (w_i^{(a)}-\eta_a)\ \prod_{a=1}^p \prod_{i<j} (w^{(a)}_i-w^{(a)}_j)^2 \right] \prod_{k<l}(z_k-z_l)^{m-1}\nn\ee
As with the Moore-Read state, when the positions of all $p$ quasi-holes coincide, we get a Laughlin quasi-hole factor $\prod(z_i-\eta)$.
This combined object should have charge $\nu e$, so the individual quasi-holes of the Read-Rezayi state have charge
\be e^\star = \frac{\nu}{p} = \frac{1}{p(m-1)+2}\nn\ee
What about for more quasi-holes? We can easily write down some candidate wavefunctions simply by including more of the $\prod (w-\eta)$ type factors in the wavefunction. But we still have the hard work of figuring out how many of these are linearly independent. To my knowledge, this has never been shown from a direct analysis of the wavefunctions. However, the result is known through more sophisticated techniques involving conformal field theory that we will briefly describe  in Section \ref{edgesec}. Perhaps the most interesting is the case $p=3$. Here, the number of linearly independent states of $3n$ quasi-holes can be shown to be $d_{3n-2}$, where $d_i$ are the Fibonacci numbers: $d_1=1$, $d_2=2$ and $d_{n+1} = d_n + d_{n-1}$. For this reason, the anyons in the $p=3$ Read-Rezayi state are referred to as {\it Fibonacci anyons}.

\para
Like their Moore-Read counterparts, the Fibonacci anyons are also non-Abelian. In fact, it turns out that they are the simplest possible non-Abelian anyons. Rather than describe their properties here, we instead take a small diversion and describe the general abstract theory behind non-Abelian anyons. We'll use the Fibonacci and Ising anyons throughout as examples to illustrate the main points. We will postpone to Section \ref{edgesec} any further explanation of how we know that these are the right anyons to describe the quasi-holes in quantum Hall states.

%
%
%
%
%
%
%

\subsection{The Theory of Non-Abelian Anyons}\label{nonabanyonsec}

This section is somewhat tangential to the main theme of these lectures. Its purpose is to review a general, somewhat formal, theory  that underlies non-Abelian anyons\footnote{More details can be found in Chapter 9 of the beautiful set of lectures on Quantum Computation by John Preskill: 
\href{http://www.theory.caltech.edu/people/preskill/ph229/}{http://www.theory.caltech.edu/people/preskill/ph229/}}. We'll see that there is an intricate structure imposed on any model arising  from the consistency of exchanging different groups of anyons. As we go along, we'll try to make contact with the non-Abelian anyons that we've seen arising in quantum Hall systems.

\para
The starting point of this abstract theory is simply a list of the different types of anyons that we have in our model. We'll call them $a$, $b$, $c$, etc. We include in this list a special state which has no particles. This is called the vacuum and is denoted as $1$.

\subsubsection{Fusion}

The first important property we need is the idea of {\it fusion}. When we bring two anyons together, the object that we're left with must, when viewed from afar, also be one of the anyons on our list. The subtlety is that we need not be  left with a unique type of anyon when we do this. We denote the possible types of anyon that can arise as $a$ and $b$ are brought together --- of fused --- as
\be a \star b = \sum_c N_{ab}^c \ c\label{fusion}\ee
where $N_{ab}^c$ is an integer that tells us how many different ways there are to get the anyon of type $c$.  It doesn't matter which order we fuse anyons, so $a\star b =b\star a$ or, equivalently, $N_{ab}^c = N_{ba}^c$. We can also interpret the equation the other way round: if a specific anyon $c$ appears on the right of this equation, then there is a way for it to split into anyons of type $a$ and $b$.

\para
The vacuum $1$ is the trivial state in the sense that 
\be a\star 1 = a\nn\ee
for all $a$.

\para
The idea that we can get different states when we bring two particles together is a familiar concept from the theory of angular momentum.
For example, when we put two spin-1/2 particles together we can either get a particle of spin 1 or a particle of spin 0. However, there's an important difference between this example and the non-Abelian anyons. Each spin 1/2 particle had a Hilbert space of dimension 2. When we tensor two of these together, we get a Hilbert space of dimension 4 which we decompose as 
\be {\bf 2}\times {\bf 2} = {\bf 3} + {1}\nn\ee
Such a simple interpretation is not available for non-Abelian anyons. Typically, we don't think of a single anyon as having any internal degrees of freedom and, correspondingly, it has no associated Hilbert space beyond its position degree of freedom. Yet a pair of anyons do carry extra information. Indeed, \eqn{fusion} tells us that the Hilbert space ${\cal H}_{ab}$ describing the ``internal" state of a  pair of anyons has dimension
\be {\rm dim}({\cal H}_{ab}) = \sum_c N_{ab}^c\nn\ee
The anyons are called {\it non-Abelian} whenever $N_{ab}^c \geq 2$ for some $a$, $b$ and $c$. The information contained in this Hilbert space is not carried by any local degree of freedom. Indeed, when the two anyons $a$ and $b$ are well separated, the wavefunctions describing different states in ${\cal H}_{ab}$ will typically look more or less identical in any local region.  The information is carried by more global properties of the wavefunction. For this reason, the Hilbert space ${\cal H}_{ab}$ is sometimes called the {\it topological Hilbert space}.

\para
All of this is very reminiscent of the situation that we met when discussing the quasi-holes for the Moore-Read state, although there we only found an internal Hilbert space when we introduced 4 or more quasi-holes. We'll see the relationship shortly.

\para
Suppose now that we  bring three or more anyons together. We will insist that the Hilbert space of final states is independent of the order in which we bring  them together. Mathematically, this means that fusion is associative,
\be  (a\star b)\star c = a\star (b\star c) \nn\ee
With this information, we can extrapolate to bringing any number of   $n$ anyons, $a_1, a_2,\ldots , a_n$ together. The resulting states can be figured out by iterating the rules above: each $c$ that can be formed from $a_1\times a_2$ can now fuse with $a_3$ and each of their products can fuse with $a_4$ and so on. The dimension of the resulting Hilbert space ${\cal H}_{a_1\ldots a_n}$ is 
\be {\rm dim}({\cal H}_{a_1\ldots a_n}) = \sum_{b_1,\ldots,b_{n-2}} N_{a_1a_2}^{b_1}N_{b_1 a_3}^{b_2}\ldots N_{b_{n-2}a_n}^{b_{n-1}}\label{dimh}\ee
In particular, we can bring $n$ anyons of the same type $a$ together. The asymptotic dimension of the resulting Hilbert space ${\cal H}_a^{(n)}$ is written as
\be {\rm dim}({\cal H}_a^{(n)})  \rightarrow (d_a)^n\ \ \ \ \ \ \ {\rm as}\ n\rightarrow \infty\nn\ee
Here $d_a$ is called the quantum dimension of the anyon. They obey $d_a\geq 1$. The vacuum anyon $1$ always has $d_1=1$. Very roughy speaking, the quantum dimension should be thought of as the number of degrees of freedom carried by in a single anyon. However, as we'll see, these numbers are typically non-integer reflecting the fact that, as we've stressed above, you can't really think of the information as being stored on an individual anyon.

\para
There's a nice relationship obeyed by the quantum dimensions. From \eqn{dimh}, and using the fact that $N_{ab}^c = N_{ba}^c$, we can write the dimension of ${\cal H}_a^{(n)}$ as
\be {\rm dim}({\cal H}_a^{(n)}) = \sum_{b_1,\ldots,b_{n-2}} N_{aa}^{b_1}N_{ab_1 }^{b_2}\ldots N_{ab_{n-2}}^{b_{n-1}} = \sum_b [N_a]^n_{ab}\nn\ee
where $N_a$ is the matrix with components $N_{ab}^c$ and in the expression above it is raised to the $n^{\rm th}$ power.  But, in the $n\rightarrow\infty$, such a product is dominated by the largest eigenvalue of the matrix $N_a$. This eigenvalue is the quantum dimension $d_a$. There is therefore an eigenvector ${\bf e}=(e_1,\ldots,e_n)$ satisfying
\be N_a {\bf e} = d_a {\bf e}\ \ \ \ \Rightarrow\ \ \ \ N_{ab}^c e_c = d_a e_b\nn\ee
For what it's worth, the Perron-Frobenius theorem in mathematics deals with eigenvalue equations of this type. Among other things, it states that all the components of $e_a$ are strictly positive. In fact, in the present case the symmetry of $N_{ab}^c = N_{ba}^c$ tells us what they must be. For the right-hand-side to be symmetric we must have $e_a=d_a$. This means that the quantum dimensions obey
\be d_ad_b = \sum_c N_{ab}^c d_c\nn\ee
Before we proceed any further with the formalism, it's worth looking at two examples of non-Abelian anyons.

\subsubsection*{An Example: Fibonacci Anyons}

\para
Fibonacci anyons are perhaps the simplest\footnote{A simple introduction to these anyons can be found in the paper by S. Trebst, M. Troyer, Z. Wang and A. Ludwig in ``{\it A Short Introduction to Fibonacci Anyon Model}",  \href{http://arxiv.org/abs/0902.3275}{arXiv:0902.3275}.}. They have, in addition to the vacuum $1$, just a single type of anyon which we denote as $\tau$. The fusion rules consist of the obvious $\tau\star 1  = 1\star \tau = \tau$ together with
\be \tau\star \tau  = 1\oplus\tau\label{fibnfuse}\ee
So we have ${\rm dim}({\cal H}_\tau^{(2)}) = 2$. Now if we add a third anyon, it can fuse with the single $\tau$ to give 
\be  \tau\star \tau \star \tau  = 1\oplus\tau \oplus \tau\nn\ee
with ${\rm dim}({\cal H}_\tau^{(3)}) = 3$. For four anyons we have ${\rm dim}({\cal H}_\tau^{(4)}) = 5$. In general, one can show that ${\rm dim}({\cal H}_\tau^{(n+1)}) = {\rm dim}({\cal H}_\tau^{(n)}) + {\rm dim}({\cal H}_\tau^{(n-1)})$. This is the Fibonacci sequence and is what gives the anyons their name.

\para
The  matrix  $N_\tau$, with components $N_{\tau b}^c$ can be read off from the fusion rules
\be N_\tau = \left(\begin{array}{cc} 0 & 1 \\ 1 & 1 \end{array}\right)\nn\ee
The quantum dimension is the positive eigenvalue of this matrix which turns out to be the golden ratio. 
\be d_\tau = \frac{1}{2}(1+\sqrt{5})\label{dtau}\ee
This, of course, is well known to be the limiting value of ${\rm dim}({\cal H}_\tau^{(n+1)})/{\rm dim}({\cal H}_\tau^{(n)})$.

\subsubsection*{Another Example: Ising Anyons}

\para
Ising anyons contain, in addition to the vacuum, two types which we  denote as $\sigma$ and $\psi$. The fusion rules are
\be \sigma \star \sigma = 1 \oplus \psi\ \ \ ,\ \ \ \sigma\star \psi = \sigma\ \ \ ,\ \ \ \psi\star \psi = 1\label{ising}\ee
The $\psi$ are somewhat boring; they have ${\rm dim}({\cal H}_\tau^{(n)}) =1$ for all $n$. The dimension of the Hilbert space of multiple $\sigma$ anyons is more interesting; it depends on whether there are an even or odd number of them. It's simple to check that
\be {\rm dim}({\cal H}_\sigma^{(2n)}) = {\rm dim}({\cal H}_\sigma^{(2n+1)}) = 2^n\label{isingdim}\ee
so we have
\be d_\psi = 1 \ \ \ \ {\rm and}\ \ \ \ d_\sigma = \sqrt{2}\nn\ee
Of course, we've seen this result before. This is the dimension of the Hilbert space of anyons constructed from Majorana zero modes described in Section \ref{majoranasec}. In this language, we saw that a pair of vortices share a single complex zero mode, leading to the states $|0\rangle$ and $\Psi^\dagger|0\rangle$. These are identified with the vacuum 1 and the fermion $\psi$ respectively.   The fusion rule $\psi\star\psi=1$ then reflects the fact that pairs of composite fermions have condensed in the ground state.

\subsubsection{The Fusion Matrix}

Let's now return to the general theory. The fusion rules \eqn{fusion} aren't all we need to specify a particular theory of non-Abelian anyons. 
There are two further ingredients. The first arises by considering the order in which we fuse particles together.

\para
Suppose that we have three anyons, $a$, $b$ and $c$. We first fuse $a$ and $b$ together and, of all the possibilities allowed by the fusion rules, we get some specific anyon $i$. We subsequently fuse $i$ with $c$ and end up with a specific anyon $d$. All of this is captured by a {\it fusion tree} which looks like this:
\be   \raisebox{-5ex}{\epsfxsize=1.0in\epsfbox{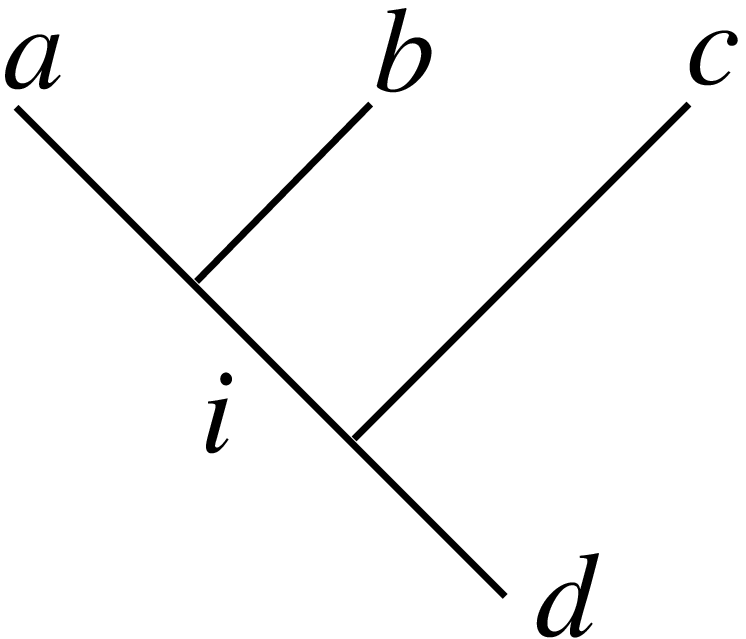}}\label{fusion1}\ee
We list the anyons that we start with at the top and then read the tree by working downwards to see which anyons fuse to which. Alternatively, you could read the tree by starting at the bottom and thinking of anyons as splittng. Importantly, there can be several different anyons $i$ that appear in the intermediate channel.

\para
Now suppose that we do the fusing in a different order: we first fuse $b$ with $c$ and subsequently fuse the product with $a$. We ask that the end product  will again be the anyon $d$. But what will the intermediate state be? There could be several different possibilities $j$.
%
\be   \raisebox{0.5ex}{\epsfxsize=1.0in\epsfbox{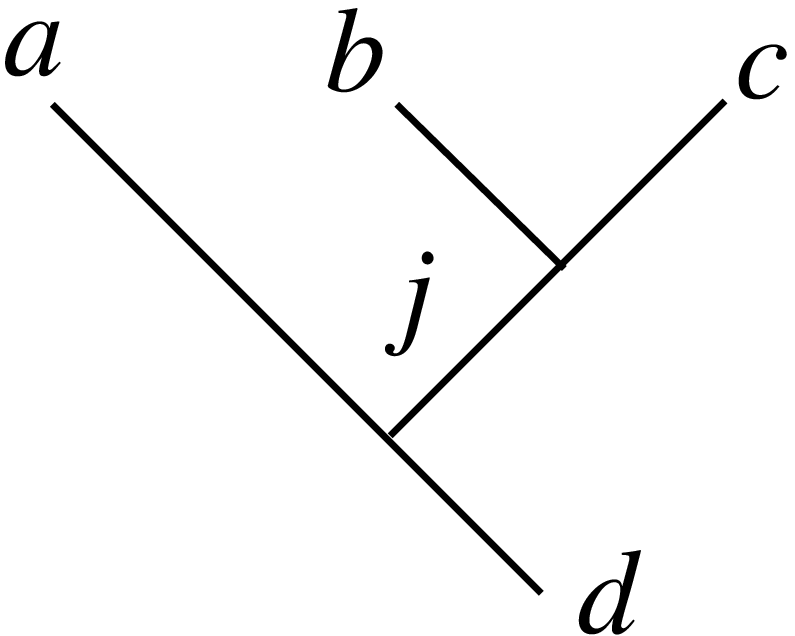}}\nn\ee
The question we want to ask is: if we definitely got state $i$ in the first route, which of the states $j$ appear in the second route. In general, there won't be a specific state $j$, but rather a linear combination of them. This is described graphically by the equation
\be \raisebox{-5ex}{\epsfxsize=1.0in\epsfbox{ftree1.eps}}\ \ \  = \sum_j\ (F_{abc}^d)_{ij} \ \ \ \raisebox{-5ex}{\epsfxsize=1.0in\epsfbox{ftree2.eps}}\label{fusionmatrix}\ee
where the coefficients $(F_{abc}^d)_{ij}$ are  thought of as the coefficients of a unitary matrix, $F_{abc}^d$, specified by the four anyons $a$, $b$, $c$ and $d$. This is called the {\it fusion matrix}.

\para
A comment: in our attempt to keep the notation concise, we've actually missed an important aspect here. If there are more than one ways in which the anyons $j$ can appear in intermediate states then we should sum over all of them and, correspondingly, the fusion matrix should have more indices. More crucially, sometimes there will be  multiple ways in which the final state $d$ can appear. This will happen whenever $N_{aj}^d\geq 2$ for some $j$. In this case, the the process on the left will typically give a linear combination of the different $d$ states on the right. The fusion matrix should also include indices which sum over these possibilities. 

\para
The fusion matrices are extra data needed to specify the structure of non-Abelian anyons. However, they can't be chosen arbitrarily: there are consistency relations which they must satisfy. For some simple theories, this is sufficient to determine the fusion matrix completely given the fusion rules. 

\para
The consistency condition comes from considering four anyons fusing to an end product. To avoid burgeoning alphabetical notation, we'll call the initial anyons $1,2,3$ and $4$ and the final anyon $5$. (The notation is not ideal because  the anyon 1 does not mean the vacuum here!) We start with some fusion process in which the anyons are fused in order,
with fixed intermediate states $i$ and $j$, like this
\be   \raisebox{0.5ex}{\epsfxsize=1.3in\epsfbox{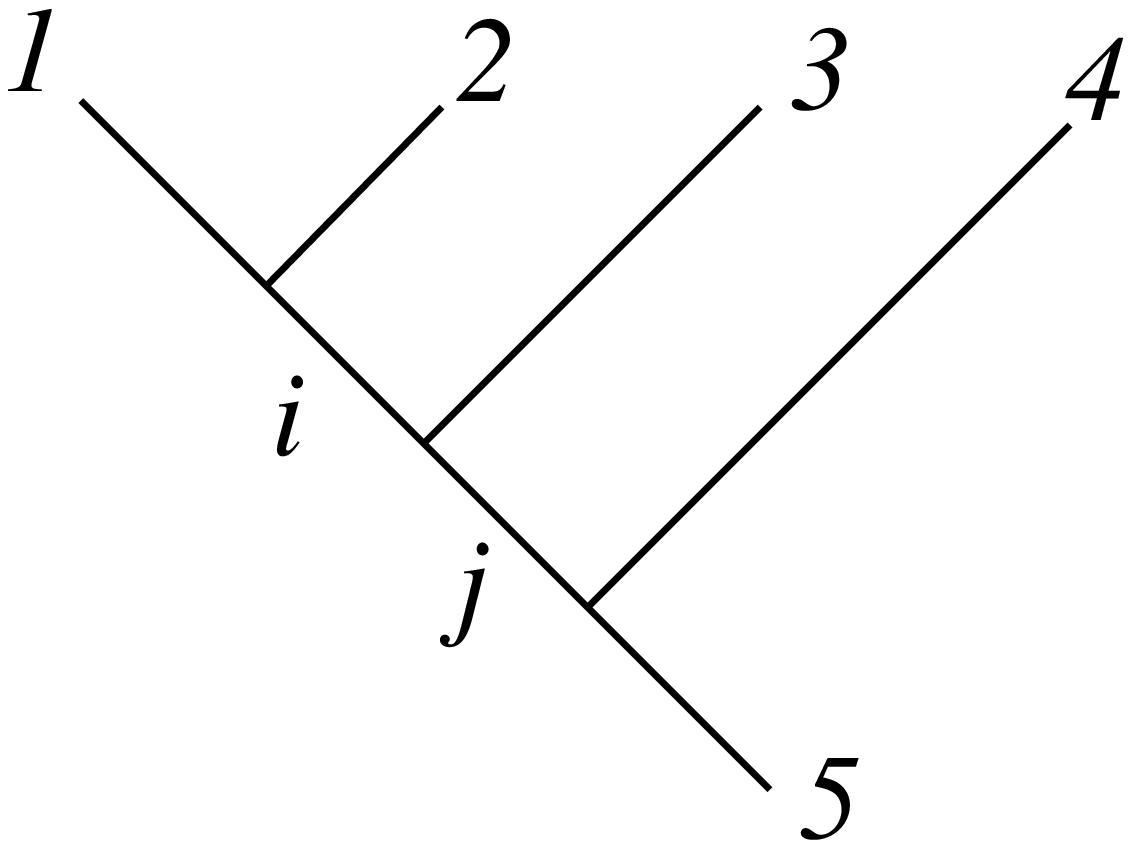}}\nn\ee
Now we consider reversing the order of fusion. We can do this in two different paths which is simplest to depict in a graphical notation, known as the {\it pentagon diagram},
\be   \raisebox{0.5ex}{\epsfxsize=6.0in\epsfbox{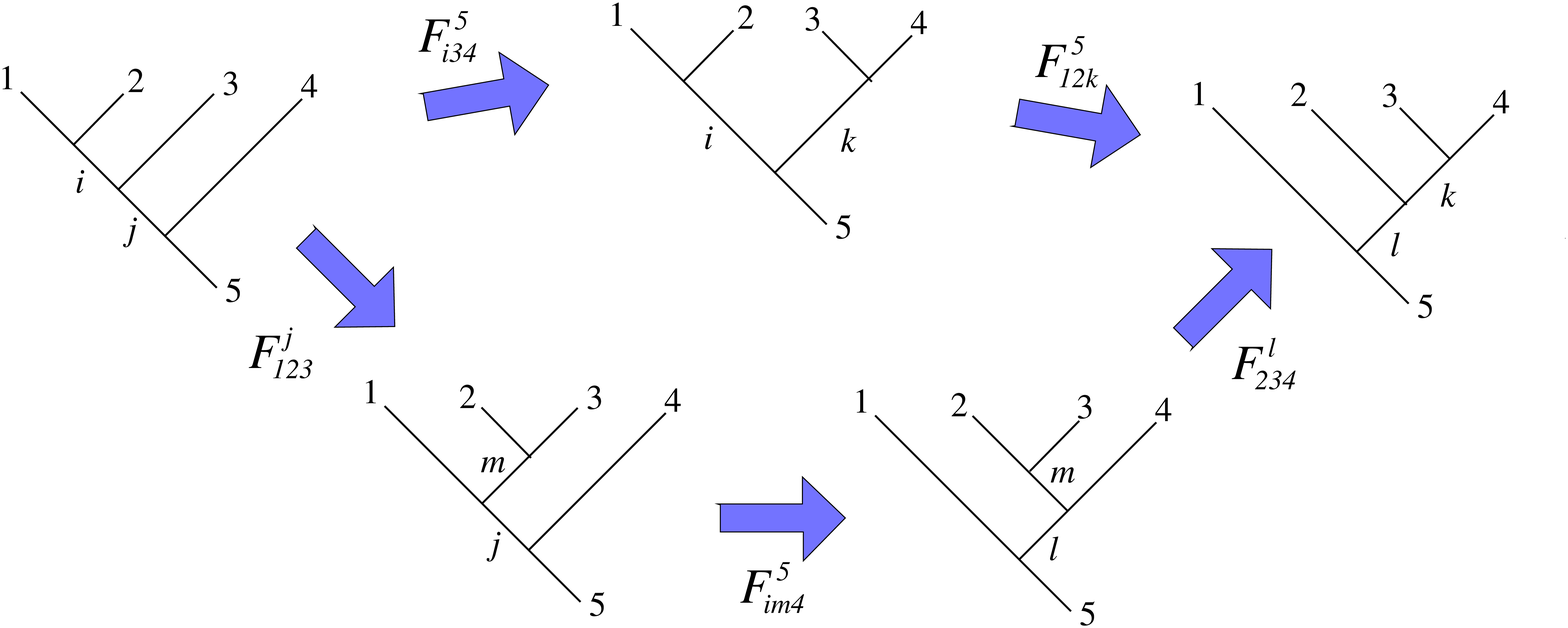}}\nn\ee
The fact that the upper and lower paths in the diagram give the same result means that the fusion matrix must obey
\be (F_{12k}^5)_{il}(F_{i34}^5)_{jk} = \sum_m(F_{234}^l)_{mk}(F_{1m4}^5)_{jl}(F_{123}^j)_{im}\label{pentagon}\ee
These are simply sets of polynomial relations for the coefficients of the fusion matrix. One might think that fusing more anyons together gives further consistency rules. It turns out that these all reduce to the pentagon condition above. Let's look at what this means for our two favourite examples.

\subsubsection*{The Fusion Matrix for Fibonacci Anyons}

For Fibonacci anyons, the interesting constraint comes from when all external particles are $\tau$. The pentagon equation \eqn{pentagon} then reads
\be (F_{\tau\tau k}^\tau)_{il}(F_{i\tau\tau}^\tau)_{jk} = \sum_m(F_{\tau\tau\tau}^l)_{mk}(F_{\tau m\tau}^\tau)_{jl}(F_{\tau\tau\tau}^j)_{im}\nn\ee
Things simplify further by noting that all fusion matrices $F_{abc}^d$ are simply given by the identity whenever $a$, $b$, $c$ or $d$ are equal to the vacuum state. (This is always true when $a$, $b$ or $c$ is equal to the vacuum state and, for Fibonacci anyons, holds also when $d$ is the vacuum state). The only non-trivial matrix is $F_{\tau\tau\tau}^\tau$. If we set $j,k=\tau$ and $i,l=1$ in the above equation, we get
\be (F_{\tau\tau\tau}^\tau)_{11} = (F_{\tau\tau\tau}^\tau)_{\tau 1} (F_{\tau\tau\tau}^\tau)_{1\tau}\nn\ee
Combined with the fact that $F_{\tau\tau\tau}^\tau$ is unitary, this constraint is sufficient to determine the fusion matrix completely. It is
\be F_{\tau\tau\tau}^\tau =  \left(\begin{array}{cc} d_\tau^{-1} & d_\tau^{-1/2} \\ d^{-1/2}_\tau & -d_\tau^{-1}\end{array}\right)\label{fmatrixfib}\ee
where we previously calculated \eqn{dtau} that the quantum dimension $d_\tau = (1+\sqrt{5})/2$, the golden ratio.

\subsubsection*{The Fusion Matrix for Ising Anyons}

The pentagon constraint can also be studied for Ising anyons. It's a little more complicated\footnote{Details can be found in Alexei Kitaev's ``{\it Anyons in an exactly solved model and beyond}", \href{http://arxiv.org/abs/cond-mat/0506438}{cond-mat/0506438}.}. You can check that a  solution to the pentagon equation \eqn{pentagon} is given by fusion matrices  $F_{\sigma\psi\sigma}^\sigma = F_{\psi\sigma\psi}^\sigma = -1$ and
\be (F_{\sigma\sigma\sigma}^\sigma)_{ij} = \frac{1}{\sqrt{2}}\left(\begin{array}{cc} 1& 1 \\ 1 & -1 \end{array}\right)\label{isingfmatrix}\ee
where the $i,j$ indices run over the vacuum state $1$ and the fermion $\psi$. 

\para
We'd now like to make contact with what we learned in Section \ref{mrsec}. How do we think about this fusing matrix in the context of, say, Majorana zero modes? In fact, there seems to be mismatch from the off, because the fusion matrix starts with three anyons fusing to one, while the Majorana zero modes naturally came in pairs, meaning that we should start with an even number of vortices. 

\para
We can, however, interpret the original fusion diagram \eqn{fusion1} in a slightly different way. We fuse $a$ and $b$ to get anyon $i$, but (tilting out heads), the diagram also says that fusing $c$ and $d$ should give the same type of anyon $i$.  What does this mean in terms of our basis of states \eqn{majhilbert}? The  obvious interpretation is that state $\oket$ is where both have fused to $1$; the state $\Psi_1^\dagger \oket$ is where the first and second anyon have fused to give $\psi$ while the third and fourth have fused to give $1$; the state $\Psi_1^\dagger \oket$ is the opposite; and the state $\Psi_1^\dagger\Psi_2^\dagger|0\rangle$ is where both have fused to give $\psi$ anyons. All of this means that the diagram \eqn{fusion1} with $i=1$ is capturing the state $\oket$ of four anyons, while the diagram with $i=\psi$ is capturing the state $\Psi_1^\dagger\Psi_2^\dagger|0\rangle$.

\para
Now let's think about the right-hand side of equation \eqn{fusionmatrix}. This time anyons $a$ and $d$ fuse together to give a specific anyon $j$, while $b$ and $c$ fuse together to give the same anyon $j$. In terms of Majorana zero modes, we should now rebuild our Hilbert space, not using the original pairing \eqn{arbpair}, but instead using 
\be \tilde{\Psi}_1 = \frac{1}{2}(\gamma_1+i\gamma_4)\ \ \ {\rm and}\ \ \ \tilde{\Psi}_2 = \frac{1}{2} ( \gamma_3 - i\gamma_2)\nn\ee
and we now construct a Hilbert space built starting from $|\tilde{0}\rangle$ satisfying $\tilde{\Psi}_k |\tilde{0}\rangle=0$. The diagram with $j=1$ corresponds to $|\tilde{0}\rangle$ while the diagram with $j=\psi$ corresponds to $\tilde{\Psi}_1^\dagger\tilde{\Psi}_2^\dagger |\tilde{0}\rangle$. We want to find the relationship between these basis. It's simple to check that the unitary map is indeed given by the fusion matrix \eqn{isingfmatrix}. 
%
%

\subsubsection{Braiding}

The second important process is a braiding of two anyons. We can do this in two different ways:
\be 
\mbox{clockwise}\ \ \ \raisebox{-5.5ex}{\epsfxsize=0.4in\epsfbox{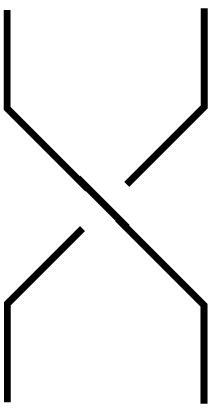}} \ \ \ \mbox{or anti-clockwise}\ \ \ \raisebox{-5.5ex}{\epsfxsize=0.4in\epsfbox{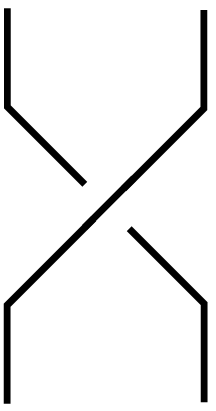}}\nn\ee
Suppose that we fuse two anyons $a$ and $b$ together to get $c$. We then do this again, but this time braiding the two anyons in an anti-clockwise direction before fusing. The resulting states are related by the {\it R-matrix}, defined by
\be  \raisebox{-7.5ex}{\epsfxsize=0.4in\epsfbox{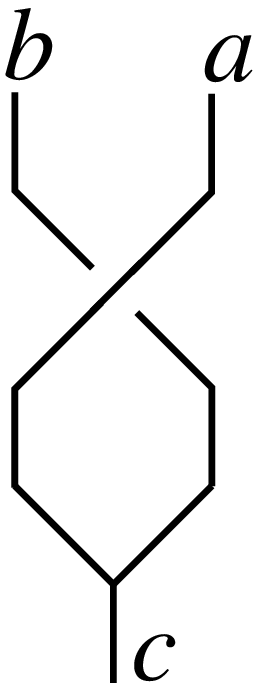}} \ \ = R_{ab}^c \ \ \raisebox{-5.5ex}{\epsfxsize=0.4in\epsfbox{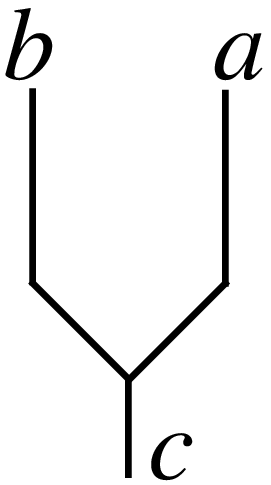}} \nn\ee
If $N_{ab}^c=1$, so that there only a single option for the final anyon, then $R_{ab}^c$ is simply a complex phase. However, if $N_{ab}^c\geq 2$, so that there are several different ways of getting the final anyon $c$, then there's no reason we should get the same state after the exchange. In this case, the R-matrix is a genuine matrix of size $N_{ab}^c\times N_{ab}^c$ and we should be summing over all possible final states on the right-hand side.
 
\para
There are further consistency relations that come from reversing the operations of fusion and braiding. Again, these are best described graphically although the resulting pictures tend to have lots of swirling lines unless we first introduce some new notation. We'll write the left-hand side of the R-matrix equation above as
\be  \raisebox{-7.5ex}{\epsfxsize=0.4in\epsfbox{fwindfuse.eps}} \ \  \equiv\  \ \ \raisebox{-5.5ex}{\epsfxsize=0.55in\epsfbox{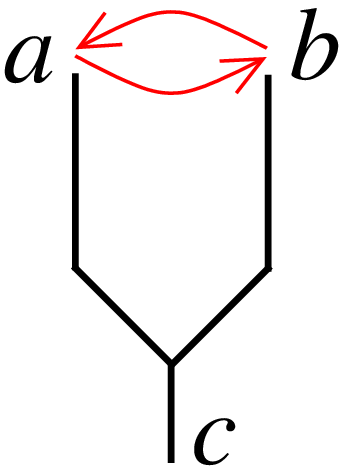}}\ \  = R_{ab}^c \ \ \raisebox{-5.5ex}{\epsfxsize=0.4in\epsfbox{ffuse.eps}} \nn\ee
Now the consistency relation between fusion matrices and R-matrices arise from the following {\it hexagon diagram}
\be   \raisebox{0.5ex}{\epsfxsize=5.9in\epsfbox{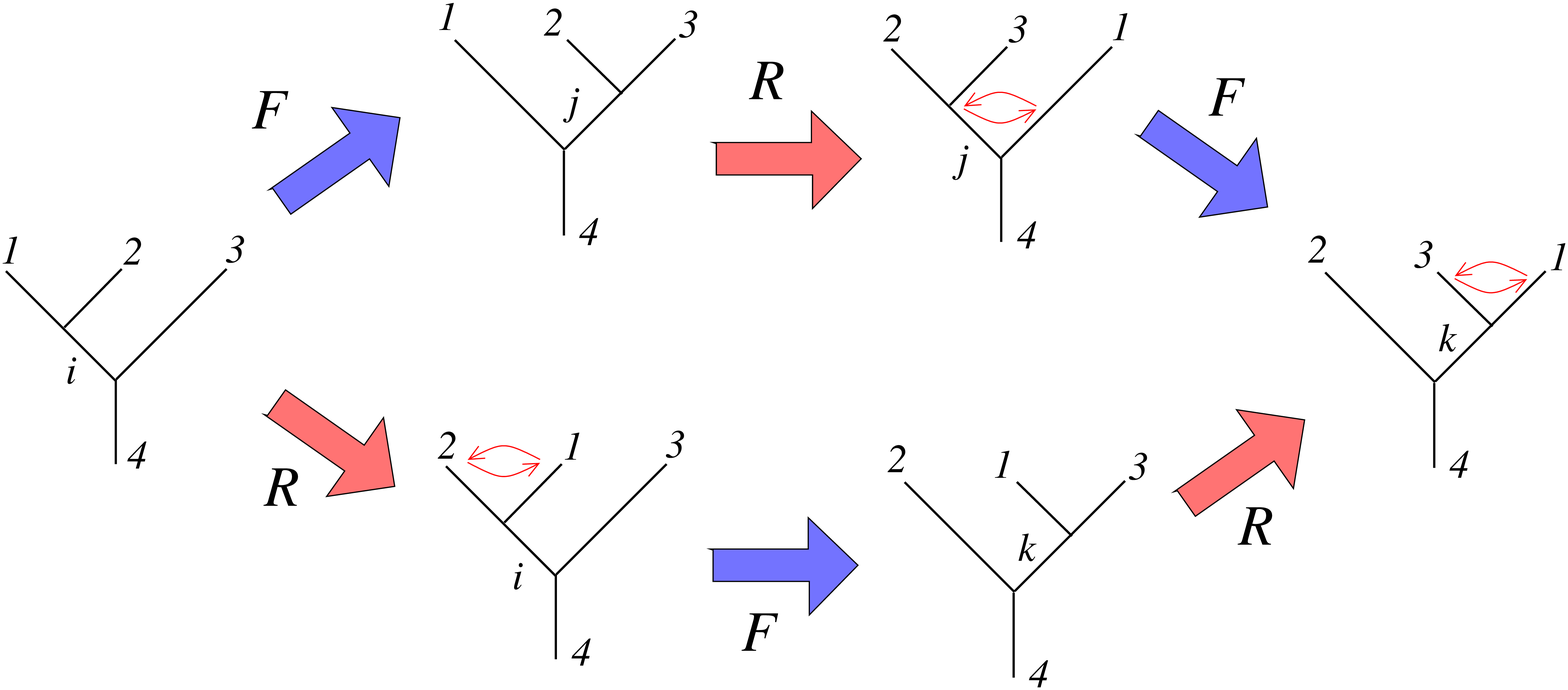}}\nn\ee
In equations, this reads

\be R_{13}^k (F_{213}^4)_{ki} R_{12}^i =\sum_j (F_{231}^4)_{kj} R^4_{j1} (F^4_{123})_{ji}\label{hexagon}\ee
It turns out that the pentagon  \eqn{pentagon} and hexagon \eqn{hexagon} equations are the only constraints that we need to impose on the system. If, for a given set of fusion rules \eqn{fusion}, we can find solutions to these sets of polynomial equations then we have a consistent theory of non-Abelian anyons.

\subsubsection*{The R-Matrix for Fibonacci Anyons}

Let's see how this works for Fibonacci anyons. We want to compute two phases: $R_{\tau\tau}^1$ and $R_{\tau\tau}^\tau$. (When either of the lower indices on $R$ is the vacuum state, it is equal to 1.) We computed the fusion matrix $F=F^\tau_{\tau\tau\tau}$ in  \eqn{fmatrixfib}. The left-hand side of the equation is then
\be R_{\tau\tau}^k F_{ki} R^i_{\tau\tau} = F_{k1}F_{1i} + F_{k\tau}F_{\tau i}  R_{\tau\tau}^\tau
\nn\ee
Note also our choice of notation has become annoying: in the equation above 1 means the vacuum, while in \eqn{hexagon} it refers to whatever external state we chose to put there. (Sorry!) The equation above must hold  for each $k$ and $i$; we don't sum over these indices. This means that it is three equations for two unknowns and there's no guarantee that there's a solution. This is the non-trivial part of the consistency relations. For Fibonacci anyons, it is simple to check that there is a solution. The phases arising from braiding are:
\be R^1_{\tau\tau} = e^{4\pi i/5} \ \ \ {\rm and}\ \ \ R^\tau_{\tau\tau} = -e^{2\pi i/5}\nn\ee

\subsubsection*{The R-Matrix for Ising Anyons}

For Ising anyons, the consistency relations give
\be R_{\sigma\sigma}^1 =  e^{-i\pi/8}\ \ \ {\rm and}\ \ \ \ R_{\sigma\sigma}^\sigma = e^{-3\pi i/8}\nn\ee
Note firstly that these are just Abelian phases; the non-Abelian part of exchange that was described by \eqn{nonabrot} for Majorana zero modes is really captured by the fusion matrix in this more formal notation. 

\para
Note also that this doesn't agree with the result for anyons computed in Section \ref{majoranasec} since these results depended on the additional Abelian statistical parameter $\alpha$. (In fact, the results do agree if we take $\alpha  = \pm 1/8$ or, equivalently filling factor $\nu = 1/2$.) For general filling factor, the non-Abelian anyons in the Moore-Read state should be thought of as attached to further Abelian anyons which shifts this phase.

\subsubsection{There is a Subject Called Topological Quantum Computing}

There has been a huge surge of interest in non-Abelian anyons over the past 15 years, much of it driven by the possibility of using these objects to build a quantum computer. The idea is that the Hilbert space of non-Abelian anyons should be thought of as the collection of qubits,  while the  braiding and fusion operations that we've described above are the unitary operations that act as quantum gates. The advantage of using non-Abelian anyons is that, as we've seen, the information is not stored locally. This means that it is immune to decoherence and other errors which mess up calculations since this noise, like all other physics, arises from local interactions\footnote{This proposal was first made by A. Kitaev in ``{\it Fault tolerant quantum computation by anyons }", \href{http://arxiv.org/abs/quant-ph/9707021}{quant-ph/9707021}.}. This subject goes by the name of {\it topological quantum computing}. I'll make no attempt to explain this vast subject here. A wonderfully clear introduction can be found in the lecture notes by John Preskill.

\newpage
\section{Chern-Simons Theories}\label{cssec}

So far we've approached the quantum Hall states from a microscopic perspective, looking at the wavefunctions which describe individual electrons. In this section, we take a step back and describe the quantum Hall effect on a more coarse-grained level. Our goal is to construct effective field theories which capture the response  of the quantum Hall ground state to low-energy perturbations. These effective theories are known as   {\it Chern-Simons theories}\footnote{Two  reviews on the Chern-Simons approach to the quantum Hall effect are  Tony Zee, ``{\it Quantum Hall Fluids}", \href{http://arxiv.org/abs/cond-mat/9501022}{cond-mat/9501022}, Xiao-Gang Wen ``{\it Topological Orders and Edge Excitations in FQH States}", \href{http://arxiv.org/abs/cond-mat/9506066}{cond-mat/9506066}. The discussion here follows the spirit of the beautiful lectures by  Edward Witten, ``{\it Three Lectures on Topological Phases of Matter}", \href{http://arxiv.org/abs/1510.07698}{arXiv:1510.07698}.}. They have many interesting properties and, in addition to their role in the quantum Hall effect, play a starring role in several other stories.

\para
Throughout this section, we'll make very general statements about the kind of low-energy effective behaviour that is possible, with very little input about the microscopic properties of the model. As we will see, we will be able to reconstruct many of the phenomena that we've met in the previous chapters.

\para
We will treat the gauge potential $A_\mu$ of electromagnetism as a background gauge field. This means that $A_\mu$ is not dynamical; it is only a parameter of the theory which tells us which electric and magnetic fields we've turned on. Further, we will not include in $A_\mu$ the original background magnetic field which gave rise the Hall effect to begin with. Instead, $A_\mu$ will describe only perturbations around a given Hall state, either by turning on an electric field, or by perturbing the applied magnetic field but keeping the kind of state (i.e. the filling fraction) fixed.

\para
In the field theory context, $A_\mu$ always couples to the dynamical degrees of freedom through the appropriate current $J_\mu$, so that the action includes the term
\be S_{A} = \int d^3x\ J^\mu A_\mu\label{sa}\ee
This is the field theoretic version of \eqn{deltah}.
Note that the measure $\int d^3x$ means that we've assumed that the current lives in a $d=2+1$ dimensional slice of spacetime; it couples to the gauge field $A_\mu$ evaluated on that slice. The action $S_A$ is invariant under gauge transformations $A_\mu \rightarrow A_\mu + \partial_\mu\omega$ on account of the conservation of the current
\be \partial_\mu J^\mu =0\nn\ee
These two simple equations will be our starting point for writing down effective field theories that tell us how the system responds when we perturb it by turning on a background electric or magnetic field.

\subsection{The Integer Quantum Hall Effect}\label{csiqhesec}

We start by looking at the integer quantum Hall effect. We will say nothing about electrons or Landau levels or anything microscopic. Instead, in our attempt to talk with some generality, we will make just one, seemingly mild, assumption: at low-energies, there are no degrees of freedom that can affect the physics when the system is perturbed.

\para
Let's think about what this assumption means. The first, and most obvious, requirement is that there is a gap to the first excited state.  In other words, our system is an insulator rather than a  conductor. We're then interested in the physics at energies below this gap.

\para
Naively, you might think that this is enough to ensure that there are no relevant low-energy degrees of freedom. However, there's also a more subtle requirement  hiding in our assumption. This is  related to the existence of so-called ``topological degrees of freedom". We will ignore this subtlety for now, but return to it in Section \ref{fraccssec} when we discuss the fractional quantum Hall effect.

\para
As usual in quantum field theory, we want to compute the partition function. This is not a function of the dynamical degrees of freedom  since these are what we integrate over. Instead, it's a function of the sources which, for us, is the electromagnetic potential $A_\mu$. We write the partition function schematically as 
\be Z[A_\mu] = \int D({\rm fields}) \ e^{iS[{\rm fields};A]/\hbar}\label{zamu}\ee
where ``fields" refer to all dynamical degrees of freedom. The action  $S$ could be anything at all, as long as it satisfies our assumption above and includes the coupling to $A_\mu$ through the current \eqn{sa}. We now want to integrate out all these degrees of freedom, to leave ourselves with a theory of the ground state which we write as
\be Z[A_\mu] = e^{iS_{\rm eff}[A_\mu]/\hbar}\label{csz}\ee
Our goal is to compute $S_{\rm eff}[A_\mu]$, which is usually referred to as the {\it effective action}. Note, however, that it's not the kind of action you meet in classical mechanics. It depends on the parameters of the problem rather than dynamical fields. We don't use it to compute Euler-Lagrange equations since there's no dynamics in $A_\mu$. Nonetheless, it does contain important information since, from the coupling \eqn{sa}, we have
\be \frac{\delta S_{\rm eff}[A]}{\delta A_\mu(x)} = \langle J^\mu(x)\rangle\label{j}\ee
This is telling us that the effective action encodes the response of the current to electric and magnetic fields.

\para
Since we don't know what the microscopic Lagrangian is, we can't explicitly do the path integral in \eqn{zamu}. Instead, our strategy is just to write down all possible terms that can arise and then focus on the most important ones. Thankfully, there are many restrictions on what the answer can be which means that there are just a handful of terms we need to consider.  The first restrictions is that the effective action $S_{\rm eff}[A]$ must be gauge invariant. One simple way to achieve this is to construct it out of electric and magnetic fields, 
\be {\bf E} = -\frac{1}{c}\nabla A_0- \ppp{{\bf A}}{t}\ \  \ {\rm and}\ \ \ {\bf B}=\nabla\times {\bf A}\nn\ee
The kinds of terms that we can write down are then further restricted by other symmetries that our system may (or may not) have, such as rotational invariance and translational invariance.

\para
Finally,  if we care only about long distances,  the effective action should be a local functional, meaning that we can write is as $S_{\rm eff}[A] = \int d^dx\ldots\,$. This property is extremely  restrictive. It holds because we're working with a theory with a gap $\Delta E$ in the spectrum. The non-locality will only arise at distances comparable to $\sim v\hbar/\Delta E$ with $v$ a characteristic velocity. (This is perhaps most familiar for relativistic theories where the appropriate scale is the Compton wavelength $\hbar/mc$). To ensure that  the gap isn't breached, we should also restrict to suitably small electric and magnetic fields.

\para
Now we just have to write down all terms in the effective action that satisfy the above requirements. There's still an infinite number of them but there's a simple organising principle. Because we're interested in small electric and magnetic fields, which vary only over long distances, the most important terms will be those with the fewest powers of $A$ and the fewest derivatives. Our goal is simply to write them down.

\para
Let's first see what all of this means in the context of  $d=3+1$ dimensions. If we have rotational invariance then we can't write down any terms linear in ${\bf E}$ or ${\bf B}$. The first terms that we can write down are instead
\be S_{\rm eff}[A] = \int d^4x\ \epsilon {\bf E}\cdot{\bf E} - \frac{1}{\mu} {\bf B}\cdot{\bf B}\label{4daction}\ee
There is also the possibility of adding a ${\bf E}\cdot{\bf B}$ term although, when written in terms of $A_i$ this is a total derivative and so doesn't contribute to the response. (This argument is a little bit glib; famously the ${\bf E}\cdot{\bf B}$ term plays an important role in the subject of 3d topological insulators but this is beyond the scope of these lectures.) The response \eqn{j} that follows from this effective action is essentially that of free currents. Indeed, it only differs from the familiar Lorentz invariant Maxwell action by the susceptibilities $\epsilon$ and $\mu$ which are the free parameters characterising the response of the system.  (Note that the response captured by \eqn{4daction}   isn't quite the same as Ohm's law that we met in Section \ref{basicsec} as there's no dissipation in our current framework).

\para
The action \eqn{4daction} has no Hall conductivity because this is ruled out in $d=3+1$ dimensions on rotational grounds. But, as we have seen in great detail, a Hall conductivity is certainly possible in $d=2+1$ dimensions. This means that there must be another kind of term that we can write in the effective action. And indeed there is....

\subsubsection{The Chern-Simons Term}

The thing that's special in $d=2+1$ dimension is the existence of the epsilon symbol $\epsilon_{\mu\nu\rho}$ with $\mu,\nu,\rho=0,1,2$. We can then write down a new term, consistent with rotational invariance. The resulting effective action is $S_{\rm eff}[A] = S_{CS}[A]$ where
\be  S_{CS}[A]= \frac{k}{4\pi} \int d^3x\ \epsilon^{\mu\nu\rho}A_\mu \partial_\nu A_\rho\label{csA}\ee
This is the famous Chern-Simons term. The coefficient $k$ is sometimes called the {\it level} of the Chern-Simons term.

\para
At first glance, it's not obvious that the Chern-Simons term is gauge invariant since it depends explicitly on $A_\mu$. However, under a gauge transformation, $A_\mu \rightarrow A_\mu + \partial_\mu \omega$, we have
\be S_{CS}[A] \rightarrow S_{CS}[A]+ \frac{k}{4\pi}\int d^3x\ \partial_\mu \left(\omega\epsilon^{\mu\nu\rho} \partial_\nu A_\rho\right)\nn\ee
The change is a total derivative. In many situations we can simply throw this total derivative away and the Chern-Simons term is gauge invariant. However, there are some situations where the total derivative does not vanish. Here we will have to think a  little harder about what additional restrictions are necessary to ensure that $S_{CS}[A]$ is gauge invariant. We see that the Chern-Simons term is flirting with danger. It's very close to failing the demands of gauge invariance and so being disallowed. The interesting and subtle  ways on which it succeeds in retaining gauge invariance will lead to much of the interesting physics.

\para
The Chern-Simons term \eqn{csA} respects rotational invariance, but breaks both parity and time reversal. Here we focus on parity which, in $d=2+1$ dimensions, is defined as 
\be  x^0\rightarrow x^0\ \ \ ,\ \ \ x^1\rightarrow -x^1\ \ \ ,\ \ \ x^2\rightarrow x^2\nn\ee
and, correspondingly, $A_0\rightarrow A_0$, $A_1\rightarrow -A_1$ and $A_2\rightarrow A_2$. The measure $\int d^3 x$ is invariant under parity (recall that although $x_1\rightarrow -x_1$, the limits of the integral also change). However,  the integrand is not invariant: $\epsilon^{\mu\nu\rho}A_\mu\partial_\nu A_\rho \rightarrow - \epsilon^{\mu\nu\rho}A_\mu\partial_\nu A_\rho$. This means that the Chern-Simons effective action with $k\neq 0$ can only arise in systems that break parity. Looking back at the  kinds of systems we met in Section \ref{iqhesec} which exhibit a Hall conductivity, we see that they all break parity, typically because of  a background magnetic field.

\para
Let's look at the physics captured by the Chern-Simons term using \eqn{j}. First, we can compute the current that arises from Chern-Simons term. It is
\be J_i = \frac{\delta S_{CS}[A]}{\delta A_i} = -\frac{k}{2\pi}\epsilon_{ij} E_i\nn\ee
In other words, the Chern-Simons action describes a Hall conductivity with
\be \sigma_{xy} = \frac{k}{2\pi}\label{hsxy}\ee
This coincides with the Hall conductivity of $\nu$ filled Landau levels if we identify the Chern-Simons level with $k =e^2\nu/\hbar$.

\para
We can also compute the charge density $J^0$. This is given by
\be J_0 = \frac{\delta S_{CS}[A]}{\delta A_i} = \frac{k}{2\pi} B\label{justsxy}\ee
Recall that we should think of $A_\mu$ as the additional gauge field over and above the original magnetic field. Correspondingly, we should think of $J^0$ here as the change in the charge density over and above that already present in the ground state. Once again, if we identify  $k = e^2\nu/\hbar$ then this is precisely the result we get had we kept $\nu$ Landau levels filled while varying $B({\bf x})$.

\para
We see that the Chern-Simons term captures the basic physics of the integer quantum Hall effect, but only if we identify the level $k=e^2\nu/\hbar$. But this is very restrictive because $\nu$ describes the number of filled Landau levels and so can only take integer values. Why should $k$ be quantised in this way?

\para
Rather remarkably, we don't have to assume that $k$ is quantised in this manner; instead, it is {\it obliged} to take values that are integer multiples of $e^2/\hbar$. This follows from the  ``almost" part of the almost-gauge invariance of the Chern-Simons term. The  quantisation in the Abelian Chern-Simons term \eqn{csA} turns out to be somewhat subtle. (In contrast, it's much more direct to see the corresponding quantisation for the non-Abelian Chern-Simons theories that we introduce in Section \ref{nonabcssec}). 
To see how it arises, it's perhaps simplest to place the theory at finite temperature and compute the corresponding partition function, again with $A_\mu$ a source. To explain this, we first need a small aside about how should think about the equilibrium properties of  field theories at finite temperature.

\subsubsection{An Aside: Periodic Time Makes Things Hot}

In this small aside we will look at the connection between the thermal partition function that we work with in statistical mechanics and the quantum partition function that we work with in quantum field theory. To explain this, we're going to go right back to basics. This means the dynamics of a single particle.

\para
Consider a quantum particle of mass $m$ moving in one direction with coordinate $q$. Suppose it moves in a potential $V(q)$. The statistical mechanics partition function is 
\be Z[\beta] = {\rm Tr}\,e^{-\beta H}\label{thermalz}\ee
where $H$ is, of course, the Hamiltonian operator 
and $\beta=1/T$ is the inverse temperature (using conventions with $k_B=1$). We would like to write down a path integral expression for this thermal partition function.

\para
We're more used to thinking of path integrals for time evolution in quantum mechanics.  Suppose the particle sits at some point $q_i$ at time $t=0$. The Feynman path integral provides an expression  for the amplitude for the particle to evolve  to position $q=q_f$ at a time $t$ later,
\be \langle q_f|e^{-iHt}|q_i\rangle = \int_{q(0)=q_i}^{q(t)=q_f}{\cal D}q\,e^{iS}\label{path101}\ee
where $S$ is the classical action, given by
\be S = \int_0^t\,dt'\ \left[\frac{m}{2}\left(\frac{dq}{dt'}\right)^2 - V(q)\right]\nn\ee
Comparing \eqn{thermalz} and \eqn{path101}, we see that they look tantalisingly similar. 
Our task is to use \eqn{path101} to derive an expression for the thermal partition function \eqn{thermalz}. We do this in three steps. We start by getting rid of the factor of $i$ in the quantum mechanics path integral. This is accomplished  by Wick rotating, which just means working with the Euclidean time variable
\be \tau = it\nn\ee
With this substitution, the action becomes
\be iS = \int_0^{-i\tau}d\tau'\ \left[-\frac{m}{2}\left(\frac{dq}{d\tau}\right)^2 - V(q)\right] \equiv -S_E\nn\ee
where $S_E$ is the {\it Euclidean action}. 

\para
The second step is to introduce the temperature. We do this by requiring the particle propagates for a (Euclidean) time $\tau =\beta$, so that the quantum amplitude becomes,
\be \langle q_f|e^{- H\beta}|q_i\rangle = \int_{q(0)=q_i}^{q(\beta)=q_f}{\cal D}q\,e^{-S_E}\nn\ee
Now we're almost there. All that's left is to implement the trace. This simply means a sum over a suitable basis of states. For example, if we choose to  sum over the initial position, we have
\be \Tr\ \cdot = \int dq_i\ \langle q_i|\cdot|q_i\rangle\nn\ee
We see that taking the trace means we should insist that $q_i=q_f$ in the path integral, before integrating over all $q_i$. We can finally write 
\be \Tr\,e^{-\beta H} &=& \int dq_i\ \langle q_i|e^{-H\beta}|q_i\rangle \nn\\ 
&= &\int dq_i \int_{q(0)=q_i}^{q(\beta)=q_i}{\cal D}q\,e^{-S_E}\nn\\
&=& \int_{q(0)=q(\beta)}{\cal D} q \, e^{-S_E}\nn\ee
The upshot is that we have to integrate over all trajectories with the sole requirement $q(0)=q(\beta)$, with no constraint on what this starting point is. All we have to impose is that the particle comes back to where it started after Euclidean time $\tau =\beta$. This is usually summarised by simply saying that the Euclidean time direction is compact: $\tau$  should be thought of as parameterising a circle, with periodicity
\be \tau \equiv \tau + \beta\label{tauperiod}\ee
Although we've walked through this simple example of a quantum particle, the general lesson that we've seen here holds for all field theories. If you take a quantum field theory that lives on Minkowski space ${\bf R}^{d-1,1}$  
and want to compute the thermal partition function, then all you have to do is consider the Euclidean path integral, but with the theory now formulated on the Euclidean space ${\bf R}^{d-1}\times {\bf S}^1$, 
where the circle is parameterised by $\tau\in [0,\beta)$. There is one extra caveat that you need to know. While all bosonic field are periodic in the time direction (just like $q(\tau)$ in our example above), fermionic fields should be made anti-periodic: they pick up a minus sign as you go around the circle. 

\para
All of this applies directly to the thermal partition function for our quantum Hall theory, resulting in an effective action $S_{\rm eff}[A]$ which itself lives on ${\bf R}^2\times {\bf S}^1$. However, there's one small difference for Chern-Simons terms. The presence of the $\epsilon_{\mu\nu\rho}$ symbol in \eqn{csA} means that the action in Euclidean space picks up an extra factor of $i$. The upshot is that, in both Lorentzian and Euclidean signature, the term in the path integral takes the form $e^{iS_{CS}/\hbar}$. This will be important in what follows.

\subsubsection{Quantisation of the Chern-Simons level}\label{csquantsec}

We're now in a position to understand the quantisation of the Chern-Simons level $k$ in \eqn{csA}.  As advertised earlier, we look at the partition function at finite temperature by taking time to be Euclidean ${\bf S}^1$, parameterised by $\tau$ with periodicity \eqn{tauperiod}.

\para
Having a periodic ${\bf S}^1$ factor in the geometry allows us to do something novel with gauge transformations, $A_\mu \rightarrow A_\mu + \partial_\mu \omega$. Usually, we work with functions $\omega(t,{\bf x})$ which are single valued. But that's actually too restrictive: we should ask only that  the physical fields are single valued. The electron wavefunction (in the language of quantum mechanics) or field (in the language of, well, fields) transforms as $e^{ie\omega/\hbar}$. So the real requirement is not that $\omega$ is single valued, but rather that $e^{ie\omega/\hbar}$ is single valued. And, when the background geometry has a ${\bf S}^1$ factor, that allows us to do something novel where the gauge transformations ``winds" around the circle, with
\be \omega = \frac{2\pi \hbar  \tau}{e\beta}\label{window}\ee
which leaves the  exponential $e^{ie\omega/\hbar}$ single valued as required. These are sometimes called {\it large} gauge transformations; the name is supposed to signify that they cannot be continuously connected to the identity. Under such a large gauge transformation, the temporal component of the gauge field is simply shifted by a constant  
\be A_0\rightarrow A_0 + \frac{2\pi \hbar}{e\beta}\label{aoshift}\ee
Gauge fields that are related by gauge transformations should be considered physically equivalent. This means that we can think of $A_0$ (strictly speaking, its zero mode) as being a periodic variable, with periodicity $2\pi \hbar/e\beta$, inversely proportional to the radius $\beta$ of the ${\bf S}^1$. Our interest is in how the Chern-Simons term fares under gauge transformations of the type \eqn{window}.

\para
To get something interesting, we'll also need to add one extra ingredient. We think about the spatial directions as forming a sphere ${\bf S}^2$, rather than a plane ${\bf R}^2$. (This is reminiscent of the kind of set-ups we used in Section \ref{iqhesec}, where all the general arguments we gave for quantisation involved some change of the background geometry, whether an annulus or torus or lattice). We take advantage of this new geometry by  threading a background magnetic flux through the spatial ${\bf S}^2$,  given by
\be \frac{1}{2\pi}\int_{{\bf S}^2} F_{12} = \frac{\hbar}{e}\label{ffflux}\ee
where $F_{\mu\nu} = \partial_\mu A_\nu - \partial_\nu A_\mu$.This is tantamount to placing a Dirac magnetic monopole inside the ${\bf S}^2$. The flux above is the minimum amount allowed by the Dirac quantisation condition. Clearly this experiment is hard to do in practice. It involves building a quantum Hall state on a sphere which sounds tricky. More importantly, it also requires the discovery of a magnetic monopole! However, there should be nothing wrong with doing this in principle. And we will only need the {\it possibility} of doing this to derive constraints on our quantum Hall system.

\para
We now evaluate the Chern-Simons term \eqn{csA} on a configuration with constant $A_0= a$ and spatial field strength \eqn{ffflux}. Expanding  \eqn{csA}, we find
\be S_{CS} = \frac{k}{4\pi}\int d^3x\ A_0 F_{12} + A_1 F_{20} + A_2 F_{01} \nn\ee
Now it's tempting to throw away the last two terms when evaluating this on our background. But we should be careful as it's topologically non-trivial configuration. We can safely set all terms with $\partial_0$ to zero, but integrating by parts on the spatial derivatives we get an extra factor of 2, 
\be S_{CS} = \frac{k}{2\pi}\int d^3x\ A_0 F_{12}  \nn\ee
Evaluated on the flux \eqn{ffflux} and constant $A_0=a$, this gives
\be S_{CS} = \beta a \frac{\hbar k}{e}\label{harlow}\ee
The above calculation was a little tricky: how do we know that we needed to integrate by parts before evaluating? The reason we got different answers is that we're dealing with a topologically non-trivial gauge field. To do a proper job, we should think about the gauge field as being defined locally on different patches and glued together in an appropriate fashion. (Alternatively, there's a way to think of the Chern-Simons action as living on the boundary of a four dimensional space.) We won't do this proper job here. But the answer \eqn{harlow} is the correct one.

%
%
%
%

\para
Now  that we've evaluated the Chern-Simons action on this particular configuration, let's see how it fares under gauge transformations \eqn{aoshift} which shift  $A_0$. We learn that the Chern-Simons term is not quite gauge invariant after all. Instead, it transforms as
\be S_{CS} \rightarrow S_{CS} + \frac{2\pi\hbar^2 k}{e^2}\nn\ee
This looks bad. However, all is not lost. Looking back, we see that the Chern-Simons term should really be interpreted as a quantum effective action,
\be Z[A_\mu] = e^{iS_{\rm eff}[A_\mu]/\hbar}\nn\ee
It's ok if the Chern-Simons term itself is not gauge invariant, as long as the partition function 
$e^{iS_{CS}/\hbar}$ is. We see that we're safe provided
\be \frac{\hbar k}{e^2} \in {\bf Z}\nn\ee
This is exactly the result that we wanted. We now write, $k = e^2\nu/\hbar$ with $\nu\in {\bf Z}$. Then the Hall conductivity \eqn{hsxy} is
\be \sigma_{xy} = \frac{e^2}{2\pi\hbar}\,\nu\nn\ee
which is precisely the conductivity seen in the integer quantum Hall effect. Similarly, the charge density \eqn{justsxy} also agrees with that of the integer quantum Hall effect.

\para
This is a lovely result. We've reproduced the observed quantisation of the integer quantum Hall effect without ever getting our hands dirty. We never needed to discuss what underlying theory we were dealing with. There was no mention of Landau levels, no mention of whether the charge carriers were fermions or bosons, or whether they were free or strongly interacting. Instead, on very general grounds we showed that the Hall conductivity {\it has} to be quantised.  This nicely complements  the kinds of microscopic arguments we met in Section \ref{iqhesec} for the quantisation of $\sigma_{xy}$

\subsubsection*{Compact vs. Non-Compact}

Looking back at the derivation, it seems to rely on two results. The first is the periodic nature of gauge transformations, $e^{ie\omega/\hbar}$, which means that the topologically non-trivial gauge transformations \eqn{window} are allowed. Because the charge appears in the exponent, an implicit assumption here is that all fields transform with the same charge. We can, in fact, soften this slightly and one can repeat the argument whenever charges are rational multiples of each other.  Abelian gauge symmetries with this property are sometimes referred to as {\it compact}. It is an experimental fact, which we've all known since high school, that the gauge symmetry of Electromagnetism is compact (because the charge of the electron is minus the charge of the proton). 

\para
Second, the derivation required there to be a minimum flux quantum \eqn{ffflux}, set by the Dirac quantisation condition. Yet a close inspection of the Dirac condition shows that this too hinges on the compactness of the gauge group. In other words, the compact nature of Electromagnetism is all that's needed to ensure the quantisation of the Hall conductivity. 

\para
In contrast, Abelian gauge symmetries which are non-compact --- for example, because they have charges which are irrational multiples of each other --- cannot have magnetic monopoles, or fluxes of the form \eqn{ffflux}. We sometimes denote their gauge group as ${\bf R}$ instead of $U(1)$ to highlight this non-compactness. For such putative non-compact gauge fields, there is no topological restriction on the Hall conductivity.

\subsection{The Fractional Quantum Hall Effect}\label{fraccssec}

In the last section, we saw very compelling arguments for why the Hall conductivity must be quantised. Yet now that leaves us in a bit of a bind, because we somehow have to explain the fractional quantum Hall effect where this quantisation is not obeyed. Suddenly, the great power and generality of our previous arguments seems quite daunting!

\para
If we want to avoid the conclusion that the Hall conductivity takes integer values, our only hope is to violate one of the assumptions that went into our previous arguments. Yet the only thing we assumed is that there are no dynamical degrees which can affect the low-energy energy physics when the system is perturbed. And, at first glance, this looks rather innocuous: we might naively expect that this is true for any system which has a gap in its spectrum, as long as the energy of the perturbation is smaller than that gap. Moreover, the fractional quantum Hall liquids certainly have a gap. So what are we missing?

\para
What we're missing is a  subtle and beautiful piece of physics that has many far reaching consequences. It turns out that there can be degrees of freedom which are gapped, but nonetheless affect the physics at arbitrarily low-energy scales. These degrees of freedom are sometimes called  ``topological". Our goal in this section is to describe the topological degrees of freedom relevant for the fractional quantum Hall effect.

\para
Let's think about what this means. We want to compute the partition function
\be Z[A_\mu] = \int D({\rm fields}) \ e^{iS[{\rm fields};A]/\hbar}\nn\ee
where $A_\mu$ again couples to the fields through the current \eqn{sa}. However, this time, we should not integrate out all the fields if we want to be left with a local effective action. Instead, we should retain the topological degrees of freedom. The tricky part is that these topological degrees of freedom can be complicated combinations of the original fields and it's usually  very difficult to identify in advance what kind of emergent fields will arise in a given system.   So, rather than work from first principles, we will first think about what kinds of topological degrees of freedom may arise. Then we'll figure out the consequences. 

\para
In the rest of this section, we describe the low-energy effective theory relevant to Laughlin states with $\nu=1/m$. In subsequent sections, we'll generalise this to other filling fractions.

\subsubsection{A First Look at Chern-Simons Dynamics}

In $d=2+1$ dimensions, the simplest kind of topological field theory involves a $U(1)$ dynamical gauge field $a_\mu$.  We stress that this is not the gauge field of electromagnetism, which  we'll continue to denote as $A_\mu$. Instead $a_\mu$ is an {\it emergent} gauge field, arising from the collective behaviour of many underlying electrons. You should think of this as something analogous to the way phonons arise as the collective motion of many underlying atoms.  We will see the direct relationship between $a_\mu$ and  the electron degrees of freedom later.

\para
We're used to thinking of gauge fields as describing massless degrees of freedom (at least classically). Indeed, their dynamics is usually described by the Maxwell action, 
\be S_{\rm Maxwell}[a] = -\frac{1}{4g^2}\int d^3x\ f_{\mu\nu}f^{\mu\nu}\label{maxf}\ee
where $f_{\mu\nu} = \partial_\mu a_\nu - \partial_\nu a_\mu$ and $g^2$ is a coupling constant. The resulting equations of motion are $\partial_\mu f^{\mu\nu}=0$. They admit wave solutions, pretty much identical to those we met in the \href{http://www.damtp.cam.ac.uk/user/tong/em.html}{\it Electromagnetism} course except that in $d=2+1$ dimensions there is only a single allowed polarisation. In other words, $U(1)$ Maxwell theory in $d=2+1$ dimension describes a single massless degree of freedom. 

\para
However, as we've already seen, there is a new kind of action that we can write down for gauge fields in $d=2+1$ dimensions. This is the Chern-Simons action
\be S_{CS}[a] = \frac{k}{4\pi} \int d^3x\ \epsilon^{\mu\nu\rho}a_\mu \partial_{\nu}a_{\rho}\label{csa}\ee
 The arguments of the previous section mean that $k$ must be integer (in units of $e^2/\hbar$) if the emergent $U(1)$ symmetry is compact.

\para
Let's see how the Chern-Simons term  changes the classical and quantum dynamics\footnote{An introduction to Chern-Simons theory can be found in G. Dunne, ``{\it Aspects of Chern-Simons Theory}", \href{http://arxiv.org/abs/hep-th/9902115}{hep-th/9902115}.}. Suppose that we take as our action the sum of the two terms
\be S = S_{\rm Maxwell} + S_{CS}\nn\ee
The equation of motion for $a_\mu$ now becomes
\be \partial_\mu f^{\mu\nu} + \frac{kg^2}{2\pi}\epsilon^{\nu\rho\sigma}f_{\rho\sigma}=0\nn\ee
Now this no longer describes a massless photon. Instead, any excitation decays exponentially.  Solving the equations is not hard and one finds that the presence of the Chern-Simons term gives the photon mass $M$. Equivalently, the spectrum has an energy gap  $E_{\rm gap} = Mc^2$. A short calculation shows that it is given by
\be E_{\rm gap} =\frac{kg^2}{\pi}\nn\ee
 (Note: you need to divide by $\hbar$ on the right-hand side to get something of the right dimension).

\para
 In the limit $g^2\rightarrow \infty$, the photon becomes infinitely massive and we're left with no physical excitations at all.  This is the situation described by the Chern-Simons theory \eqn{csa} alone. One might wonder what the Chern-Simons theory can possibly describe given that there are no propagating degrees of freedom. The purpose of this section is to answer this!

\subsubsection*{Chern-Simons Terms are Topological}

Before we go on, let us point out one further interesting  and important property of \eqn{csa}: it doesn't depend on the metric of the background spacetime manifold. It depends only on the topology of the manifold. To see this, let's first look at the Maxwell action \eqn{maxf}. If we are to couple this to a background metric $g_{\mu\nu}$, the action becomes
\be S_{\rm Maxwell} = -\frac{1}{4g^2}\int d^3x\ \sqrt{-g}\,g^{\mu\rho}g^{\nu\sigma} f_{\mu\nu} f_{\rho\sigma}\nn\ee
We see that the metric plays two roles: first, it is needed to raise the indices when contracting $f_{\mu\nu}f^{\mu\nu}$; second it provides a measure $\sqrt{-g}$ (the volume form) which allows us to integrate in a diffeomorphism invariant way.

\para
In contrast, neither of these are required when generalising \eqn{csa} to curved spacetime. This is best stated in the language of differential geometry: $a\wedge da$ is a 3-form, and we can quite happily integrate this over any three-dimensional manifold
\be S_{CS} = \frac{k}{4\pi}\int a\wedge da\nn\ee
The action is manifestly independent of the metric. In particular, recall from our \href{http://www.damtp.cam.ac.uk/user/tong/qft.html}{\it Quantum Field Theory} lectures, that we can compute the stress-energy tensor of any theory by differentiating with respect to the metric,
\be T^{\mu\nu} = \frac{2}{\sqrt{-g}}\frac{\partial {\cal L}}{\partial g_{\mu\nu}}\nn\ee
For Chern-Simons theory, the stress-energy tensor vanishes. This means that the Hamiltonian vanishes. It is an unusual kind of theory.

\para
However, will see in Section \ref{cstorussec} that the topology of the underlying manifold does play an important role in Chern-Simons theory. This will be related to the ideas of topological order that we introduced in Section \ref{torussec}.
Ultimately, it is this topological nature of the Chern-Simons interaction which means that we can't neglect it in low-energy effective actions. 

\subsubsection{The Effective Theory for the Laughlin States}\label{laughlinfoffsec}

Now we're in a position to describe the effective theory for the $\nu=1/m$ Laughlin states. These Hall states have a single emergent, compact $U(1)$ gauge field $a_\mu$. This is a dynamical field, but we should keep it in our effective action. The partition function can then be written as
\be Z[A_\mu] = \int {\cal D}a_\mu\ e^{iS_{\rm eff}[a;A]/\hbar}\nn\ee
where ${\cal D}a_\mu$ is short-hand for all the usual issues involving gauge-fixing that go into defining a path integral for a gauge field. 

\para
Our goal now is to write down $S_{\rm eff}[a;A]$. However, to get something interesting we're going to need a coupling between $A_\mu$ and $a_\mu$. Yet we know that $A_\mu$ has to couple to the electron current $J^\mu$. So if this is going to work at all, we're going to have to find a relationship between $a_\mu$ and $J^\mu$.

\para
Thankfully, conserved currents are hard to come by and there's essentially only one thing that we can write down. The current is given by
\be J^\mu = \frac{e^2}{2\pi \hbar}\epsilon^{\mu\nu\rho}\,\partial_{\nu}a_\rho\label{jisa}\ee
The conservation of the current, $\partial_\mu J^\mu=0$, is simply an identity when written like this.  This relation means that the magnetic flux of $a_\mu$ is interpreted as the electric charge that couples to $A_\mu$. The normalisation follows directly if we take the  emergent $U(1)$ gauge symmetry to be compact, coupling to particles with charge $e$. In this case, the minimum allowed flux is given by the Dirac quantisation condition
\be \frac{1}{2\pi} \int_{{\bf S}^2} f_{12} = \frac{\hbar}{e}\label{fdirac}\ee
The relationship \eqn{jisa} then ensures that the minimum charge is $\int J^0 = e$ as it should be. (Picking different signs of the flux $f_{12}$ corresponds to electrons and holes in the system).

\para
We then postulate the following  effective action,
\be S_{\rm eff}[a;A] = \frac{e^2}{\hbar}\int d^3x\ \,\frac{1}{2\pi} \,\ep A_\mu\partial_\nu a_\rho - \frac{m}{4\pi}\,\ep a_\mu \partial_\nu a_\rho +\ldots\label{csaa}\ee
The first term is a ``mixed" Chern-Simons term which comes from the $A_\mu J^\mu$ coupling; the second term is the simplest new term that we can write down. By the same arguments that we used before, the level must be integer: $m\in {\bf Z}$. As we will see shortly, it is no coincidence that we've called this integer $m$.  The $\ldots$ above include more irrelevant terms, including the Maxwell term \eqn{maxf}. At large distances, none of them will play any role and we will ignore them in what follows. We could also add a Chern-Simons $\epsilon^{\mu\nu\rho} A_\mu \partial_\nu A_\rho$ for $A$ itself but we've already seen what this does: it simply gives an integer contribution to the Hall conductivity. Setting the coefficient of this term to zero will be equivalent to working in the lowest Landau level.

\para
Let's start by computing the Hall conductivity. The obvious way to do this is to reduce the effective action to something which involves only $A$ by explicitly integrating out the dynamical field $a$. Because the action is quadratic in $a$, this looks as if it's going to be easy to do. We would naively just replace such a field with its equation of motion which, in this case, is
\be f_{\mu\nu} = \frac{1}{m}F_{\mu\nu}\label{stupidfish}\ee
The solution to this equation is $a_\mu  = A_\mu/m$ (up to a gauge transformation). Substituting this back into the action \eqn{csaa} gives
\be  S_{\rm eff}[A] = \frac{e^2}{2\pi}\int d^3x\ \,\frac{1}{4\pi m } \,\ep A_\mu\partial_\nu A_\rho \label{fishy}\ee
This is now the same kind of action \eqn{csA} that we worked with before and we can immediately see that the Hall conductivity is 
\be \sigma_{xy} = \frac{e^2}{2\pi\hbar}\,\frac{1}{m}\label{swimming}\ee
as expected for the Laughlin state. 

\para
Although we got the right answer for the Hall conductivity, there's something very fishy about our derivation. The kind of action \eqn{fishy} that we ended up lies in the class that we previously argued wasn't allowed by gauge invariance if our theory is defined on a sphere!  Our mistake was that we were too quick in the integrating out procedure.  The gauge field $a_\mu$ is constrained by the Dirac quantisation condition \eqn{fdirac}. But this is clearly incompatible with the equation of motion \eqn{stupidfish} whenever $F$ also has a single unit of flux \eqn{ffflux}. In fact, it had to be this way. If it was possible to integrate out $a_\mu$, then it couldn't have been playing any role in the first place!

\para
Nonetheless, the final answer \eqn{swimming} for the Hall conductivity is correct. To see this, just consider the theory on the plane with $F_{12}=0$ where there are no subtleties with \eqn{stupidfish} and the calculation above goes through without a hitch.   However, whenever we want to compute something where monopoles are important, we can't integrate out $a_\mu$. Instead, we're obliged to work with the full action \eqn{csaa}.

\subsubsection*{Quasi-Holes and Quasi-Particles}

The action \eqn{csaa} describes the quantum Hall state at filling $\nu=1/m$. Let's now add something new to this. We will couple the emergent gauge field $a_\mu$ to its own current, which we call $j^\mu$, through the additional term
\be \Delta S = \int d^3x\ a_\mu j^\mu\nn\ee
To ensure gauge invariance, $j^\mu$ must be conserved: $\partial_\mu j^\mu=0$. We will now show that the current $j^\mu$ describes the quasi-holes and quasi-particles in the system.

\para
First, we'll set the background gauge field $A_\mu$ to zero. (It is, after all, a background parameter at our disposal in this framework).  The equation of motion for $a_\mu$ is then
\be \frac{e^2}{2\pi \hbar} f_{\mu\nu} = \frac{1}{m}\epsilon_{\mu\nu\rho}j^{\rho}\label{attachthis}\ee
The simplest kind of current we can look at is a static charge which we place at the origin. This is described by  $j^1=j^2=0$ and $j^0=e\delta^2(x)$. Note that the fact these particles have charge $e$ under the gauge field $a_\mu$ is related to our choice of Dirac quantisation \eqn{fdirac}.  The equation of motion above then becomes
\be \frac{1}{2\pi} f_{12} = \frac{\hbar}{e m} \, \delta^2(x)\label{fluxattach}\ee
This is an important equation. We see that the effect of the Chern-Simons term is to attach flux $\hbar/ em$ to each particle of charge $e$. 
From this we'll see that the particle has both the fractional charge and fractional statistics appropriate for the Laughlin state. The fractional charge follows immediately by looking at the electron current $J^\mu$ in \eqn{jisa} which, in this background, is
\be J^0= \frac{e^2}{2\pi\hbar} f_{12} = \frac{e}{m}\delta^2(x)\nn\ee
This, of course, is the current appropriate for a stationary particle of electric charge $e/m$. 

\para
Note: the flux attachment \eqn{fluxattach} doesn't seem compatible with the Dirac quantisation condition \eqn{fdirac}. Indeed, if we were on a spatial sphere ${\bf S}^2$ we would be obliged to add $m$ quasi-particles, each of charge $e/m$. However, these particles can still roam around the sphere independently of each other so they should still be considered as individual object. On the plane ${\bf R}^2$, we need not be so fussy: if we don't have a multiple of $m$ quasi-holes, we can always think of the others as being somewhere off at infinity. 

\para
To see how the fractional statistics emerge, we just need the basic Aharonov-Bohm physics that we reviewed in Section \ref{absec}. Recall that a particle of charge $q$ moving around a flux $\Phi$ picks up a phase $e^{iq\Phi/\hbar}$. But because of flux attachment \eqn{fluxattach}, our quasi-particles necessarily carry both charge $q=e$ and flux $\Phi= 2\pi\hbar/em$.  If we move one particle all the way around another, we will get a phase $e^{iq\Phi/\hbar}$. But the statistical phase is defined by exchanging particles, which consists of only half an orbit (followed by a translation which contributes no phase). So the expected statistical phase is $e^{i\pi\alpha} = e^{iq\Phi/2\hbar}$. For our quasi-holes, with $q=e$ and $\Phi = 2\pi \hbar /em$, we get
\be \alpha = \frac{1}{m}\nn\ee
which is indeed the expected statistics of quasi-holes in the Laughlin state.

\para
The attachment of the flux to the quasi-hole is reminiscent of the composite fermion ideas that we met in Section \ref{compositesec}, in which we attached vortices (which were zeros of the wavefunction) to quasi-holes.

\subsubsection*{Fractional Statistics Done Better}

The above calculation is nice and quick and gives the right result. But there's a famously annoying factor of 2 that we've swept under the rug. Here's the issue. As the charge $q$ in the first particle moved around the flux $\Phi$ in the second, we picked up a phase $e^{iq\Phi/\hbar}$. But you might think that the  flux $\Phi$ of the first particle also moved around the charge $q$ of the second. So surely this should give another factor of $e^{iq\Phi/\hbar}$. Right? Well, no. To see why, it's best to just do the calculation.

\para
For generality, let's take $N$ particles sitting at positions ${\bf x}_a(t)$ which, as the notation shows, we allow to change with time. The charge density and currents are
\be j^0({\bf x},t) = e\sum_{a=1}^N \delta^2({\bf x}-{\bf x}_a(t))\ \ \ {\rm and}\ \ \ {\bf j}(x,t) = e\sum_{a=1}^N\dot{{\bf x}}_a\,
\delta^2({\bf x}-{\bf x}_a(t))\nn\ee
The equation of motion \eqn{attachthis} can be easily solved even in this general case. If we work in the Coulomb gauge $a_0=0$ with 
$\partial_i a_i =0$ (summing over spatial indices only), the solution is given by
\be a_i ({\bf x},t) = \frac{\hbar}{em}\sum_{a=1}^N \epsilon^{ij} \,\frac{x^j-x^j_a(t)}{|{\bf x}- {\bf x}_a(t)|^2}\label{littlea}\ee
This follows from the standard methods that we know from our {\it Electromagnetism} course, but this time using the  Green's function for the Laplacian in two dimensions: $\nabla^2 \,\log|{\bf x}-{\bf y}| = 2\pi \delta^2({\bf x}-{\bf y})$. This solution is again the statement that each particle carries flux $\hbar/em$. However, we can also use this solution directly to compute the phase change when one particle -- say, the first one -- is transported along a curve $C$. It is simply
\be \exp\left(ie\oint_C{\bf a}\cdot d{\bf x}_1\right)\nn\ee
If the curve $C$ encloses one other particle, the resulting phase change can be computed to be $e^{2\pi i/m}$. As before, if we exchange two particles, we get half this phase, or $e^{\i\pi\alpha} = e^{i\pi/m}$. This, of course, is the same result we got above.

\para
It's worth pointing out that this Chern-Simons computation ended up looking exactly the same as the original Berry phase calculation for the Laughlin wavefunctions that we saw in Section \ref{fracstatsec}. For example, the connection \eqn{littlea} is identical to the relevant part of the Berry connections \eqn{aeta} and \eqn{abareta}. (The seeming difference in the factor of 2 can be traced to our previous normalisation for complex connections).

\subsubsection*{Breathing Life into the Quasi-Holes}

In the calculations above,  we've taken $j^\mu$ to be some fixed, background current describing the quasi-particles. But the framework of effective field theory also allows us to make the quasi-particles dynamical. We simply need to introduce a new bosonic field $\phi$ and include it in the effective action, coupled minimally to $a_\mu$. We then endow $\phi$ with its own dynamics. Exactly what dynamics we choose is up to us at this point. For example, if we wanted the quasi-holes to have a relativistic dispersion relation, we would introduce the action
\be S_{\rm eff}[a,\phi] = \int d^3x\ \frac{e^2 m}{4\pi\hbar} \ep a_\mu \partial_\nu a_\rho +  |{\cal D}_\mu\phi|^2 - V(\phi)\nn\ee
where the relativistic form of the action also implies  that $\phi$ will describe both particle and anti-particle (i.e. hole) excitations. Here $V(\phi)$ is a potential that governs the mass and self-interactions of  interactions of the quasi-particles. Most important, the covariant derivative ${\cal D}_\mu = \partial_\mu - iea_\mu$ includes the coupling to the Chern-Simons field. By the calculations above, this ensures that the excitations of $\phi$ will have the correct anyonic statistics to describe quasi-particles, even though the field $\phi$ itself is bosonic.

\para
We'll see a different way to make the current $j^\mu$ dynamical in Section \ref{cshiersec} when we discuss other filling fractions.

\subsubsection{Chern-Simons Theory on a Torus}\label{cstorussec}

In Section \ref{torussec}, we argued that if we place a fractional quantum Hall state on a compact manifold, then the number of ground states depends on the topology of that manifold. In particular, we showed that the existence of anyons alone was enough to ensure $m$ ground states on a torus and $m^g$ ground states on a genus-$g$ surface. This is the essence of what's known as {\it topological order}. 

\DOUBLEFIGURE{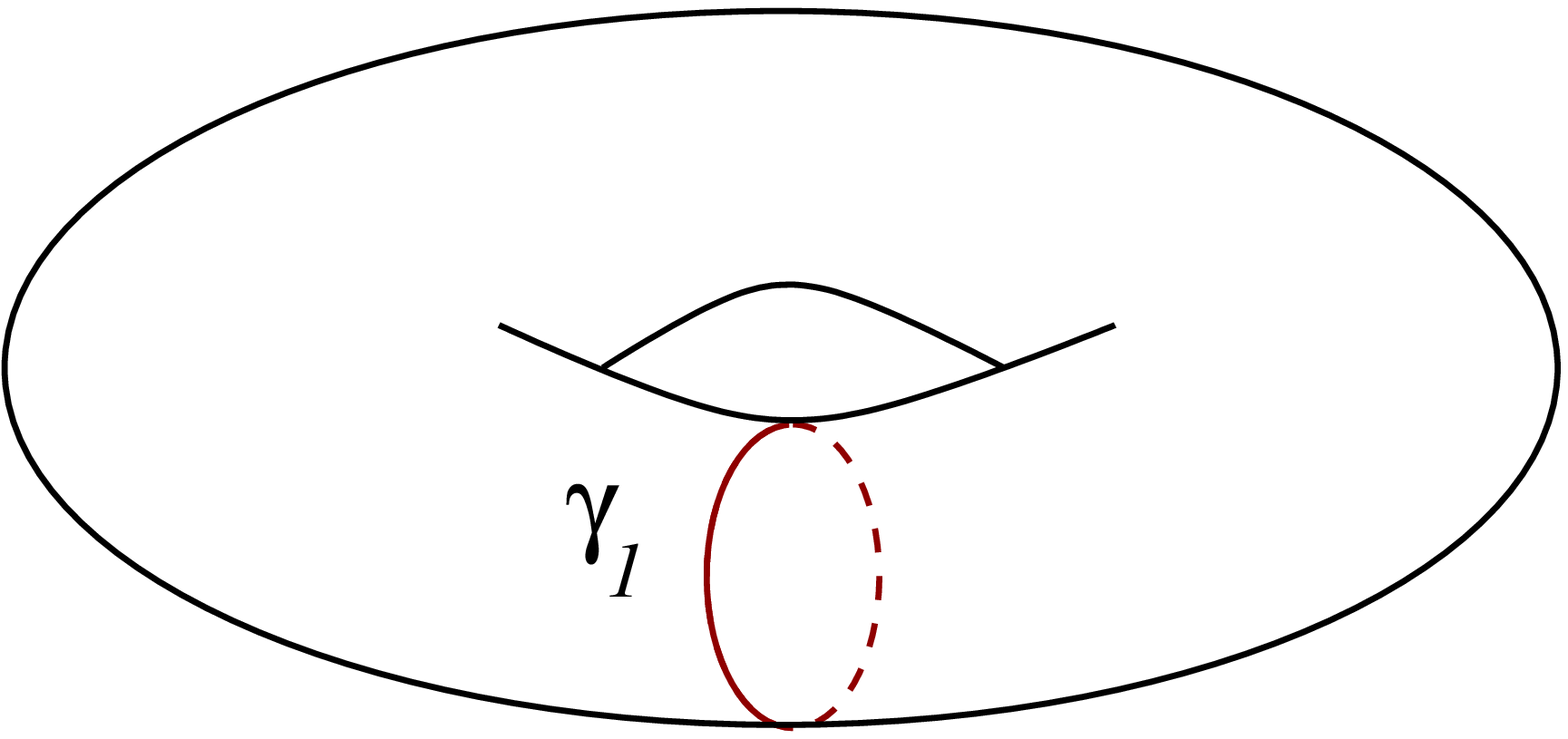,width=180pt}{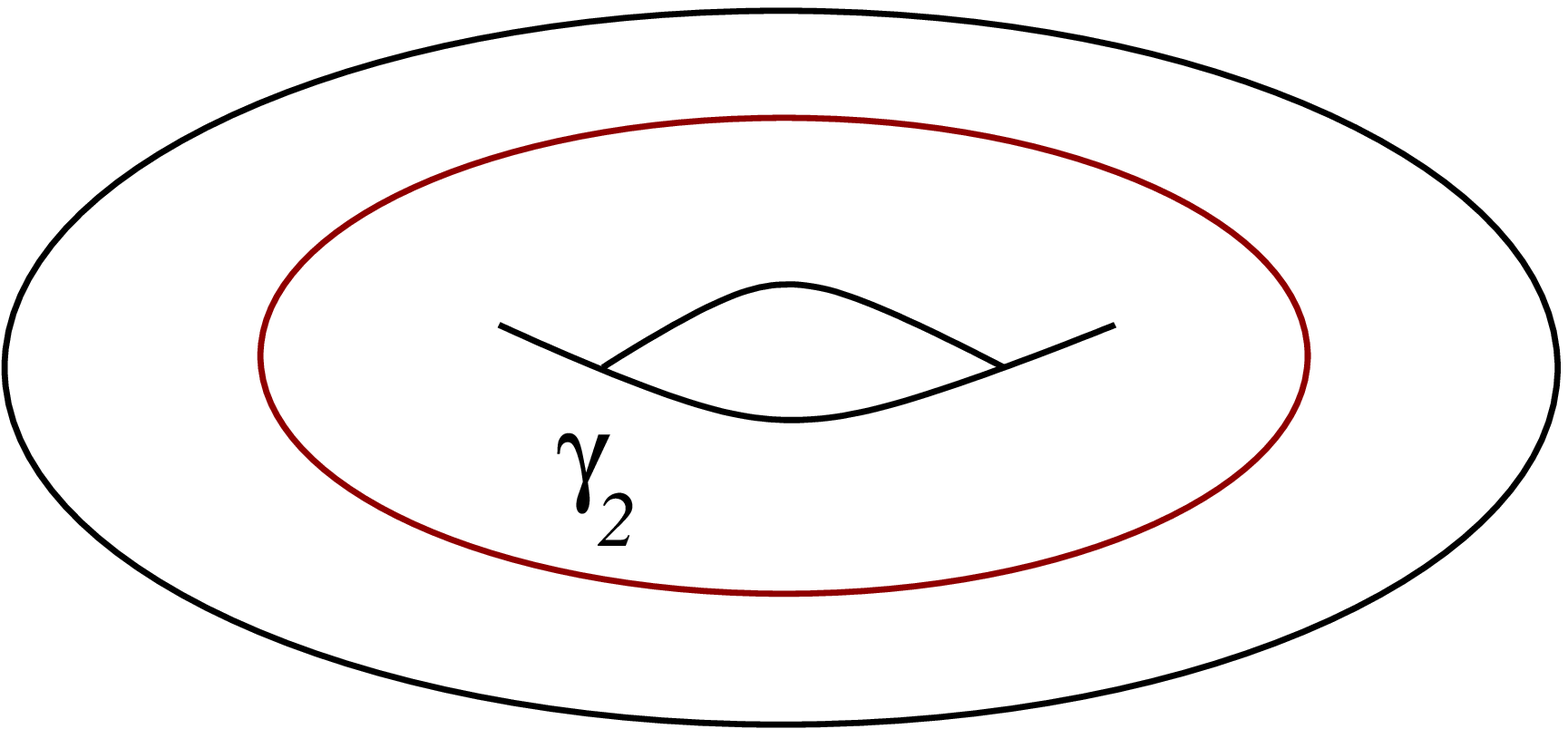,width=180pt}
{}{}
\noindent

\para
Here we show how this is reproduced by the Chern-Simons theory. If we live on the plane ${\bf R}^2$ or the sphere ${\bf S}^2$, then Chern-Simons theory has just a single state. But if we change the background manifold to be anything more complicated, like a torus, then there is a degeneracy of ground states. 

\para
To see this effect, we can turn off the background sources and focus only on the dynamical part of the effective theory,
\be S_{CS} = \frac{e^2}{\hbar}\int d^3x\ \frac{m}{4\pi}\ep \, a_\mu\partial_\nu a_\rho\label{justcs}\ee
The equation of motion for $a_0$, known, in analogy with electromagnetism, as Gauss' law, is
\be f_{12} =0\nn\ee
Although this equation is very simple, it can still have interesting solutions if the background has some non-trivial topology. These are called, for obvious reason, {\it flat connections}. It's simple to see that such solutions exist  on the torus ${\bf T}^2$, where one example is to simply set each $a_i$ to be constant. Our first task is to find a gauge-invariant way to parameterise this space of solutions.

\para
We'll denote the radii of the two circles of the  torus ${\bf T}^2  = {\bf S}^1\times {\bf S}^1$ as $R_1$ and $R_2$. We'll denote two corresponding non-contractible curves shown in the figure as $\gamma_1$ and $\gamma_2$. The simplest way to build a gauge invariant object from a gauge connection is to integrate
\be w_i = \oint_{\gamma_i} dx^j\,a_j\nn\ee
This is invariant under most gauge transformations, but not those that wind around the circle. By the same kind of arguments that led us to \eqn{aoshift}, we can always construct gauge transformations which shift $a_j \rightarrow a_j + \hbar/eR_j$, and hence $w_i\rightarrow w_i + 2\pi \hbar/e$. The correct gauge invariant objects to parameterise the solutions are therefore the Wilson loops
\be W_i = \exp\left(i\frac{e}{\hbar}\oint_{\gamma_i} a_j dx^j\right)= e^{iew_i/\hbar}\nn\ee
Because the Chern-Simons theory is first order in time derivatives, these Wilson loops are really parameterising the phase space of solutions, rather than the configuration space. Moreover, because the Wilson lines are complex numbers of unit modulus, the phase space is compact. 
On general grounds, we would expect that when we quantise a compact phase space, we get a finite-dimensional Hilbert space. Our next task is to understand how to do this. 

\para
The canonical commutation relations can be read off from the Chern-Simons action \eqn{justcs}
\be [a_1(x),a_2(x')] =  \frac{2\pi i}{m}\frac{\hbar^2 }{e^2}\,\delta^2(x-x')\ \ \ \Rightarrow\ \ \ [w_1,w_2]= \frac{2\pi i}{m}\frac{\hbar^2 }{e^2}\nn\ee
The algebraic relation obeyed by the Wilson loops then follows from the usual Baker-Campbell-Hausdorff formula,
\be e^{iew_1/\hbar}e^{iew_2/\hbar} = e^{e^2[w_1,w_2]/2\hbar^2} e^{ie(w_1+w_2)/\hbar}\nn\ee
Or, in other words, 
\be W_1 W_2 = e^{2\pi i/m}\, W_2 W_1\label{wtorus}\ee
But this is exactly the same as the algebra \eqn{torusalgebra} that we met when considering anyons on a torus! This is not surprising: one interpretation of the Wilson loop is for a particle charged under $e$ to propagate around the cycle of the torus. And that's exactly how we introduced the operators $T_i$ that appear in \eqn{torusalgebra}.

\para
From Section \ref{torussec}, we know that the smallest representation of the algebra \eqn{wtorus} has dimension $m$. This is the number of ground states of the Chern-Simons theory on a torus.   The generalisation of the above calculation to a genus-$g$ surface gives a ground state degeneracy of $m^g$.

\subsubsection{Other Filling Fractions and $K$-Matrices}\label{cshiersec}

It's straightforward to generalise the effective field theory approach to other filling fractions. We'll start by seeing how the hierarchy of states naturally emerges. To simplify the equations in what follows, we're going to use units in which $e=\hbar=1$. (Nearly all other texts resort to such units long before now!)

\subsubsection*{The Hierarchy}

The effective field theory for the Laughlin states that we saw above can be summarised as follows: we write the electron current as
\be J^\mu = \frac{1}{2\pi }\epsilon^{\mu\nu\rho}\,\partial_{\nu}a_\rho\label{melatime}\ee
where $a_\mu$ is an emergent field. We then endow $a_\mu$ with a Chern-Simons term.

\para
Now we'd like to repeat this to implement the hierarchy construction described in Section \ref{hierarchysec} in which the quasi-particles themselves form a new quantum Hall state. But that's very straightforward. We simply write the quasi-particle current $j^\mu$ as
\be
j^\mu = \frac{1}{2\pi }\epsilon^{\mu\nu\rho}\,\partial_{\nu}\tilde{a}_\rho\label{newbell}\ee
where $\tilde{a}_\mu$ is a second emergent gauge field whose dynamics are governed by a second Chern-Simons term. The final action is
\be S_{\rm eff}[a,\tilde{a};A] = \int d^3x\ \,\frac{1}{2\pi} \,\ep A_\mu\partial_\nu a_\rho - \frac{m}{4\pi}\,\ep a_\mu \partial_\nu a_\rho + \frac{1}{2\pi} \ep a_\mu \partial_\nu \tilde{a}_\rho -  \frac{\tilde{m}}{4\pi}\,\ep \tilde{a}_\mu \partial_\nu \tilde{a}_\rho \nn\ee
To compute the Hall conductivity, we can first integrate out $\tilde{a}$ and then integrate out $a$. We find that this theory describes a Hall state with filling fraction
\be \nu = \frac{1}{m - \frac{1}{\tilde{m}}}\nn\ee
When $\tilde{m}$ is an even integer, this coincides with our expectation  \eqn{simplehier}  for the first level of the hierarchy. 

\para
We can also use this approach to quickly compute the charge and statistics of quasi-particles in this state.  There are two such quasi-holes, whose currents couple to $a$ and $\tilde{a}$ respectively. For a static quasi-hole which couples to $a$, the equations of motion read
\be 
m f_{12} - \tilde{f}_{12} = 2\pi\delta^2(x) \ \ \ {\rm and}\ \ \ \tilde{m}\tilde{f}_{12} - f_{12} = 0 \ \ \ \Rightarrow\ \ \ f_{12} = \frac{2\pi}{m-1/\tilde{m}} \delta^2(x)\nn\ee
while, if the quasi-hole couples to $\tilde{a}$, the equations of motion are
\be 
m f_{12} -  \tilde{f}_{12} =0 \ \ \ {\rm and}\ \ \ \tilde{m}\tilde{f}_{12} - f_{12} = 2\pi\delta^2(x)\ \ \ \Rightarrow\ \ \ f_{12} = \frac{2\pi}{m\tilde{m}-1} \delta^2(x)\nn\ee
The coefficients of the right-hand side of the final equations tell us the electric charge. 

\para
For example, the $\nu=2/5$ state has $m=3$ and $\tilde{m}=2$. The resulting charges of the quasi-holes are $e^\star=2/5$ and  $e^\star  = 1/5$. This has been confirmed experimentally. Using the results from either Section \ref{torussec} or Section \ref{cstorussec}, we learn that the the $\nu=2/5$ state has a 5-fold degeneracy on the torus.

\para
Now it's obvious how to proceed: the quasi-particles of the new state are described by a current $j_2(x)$ which couples to $\tilde{a}_\mu$. We write this in the form \eqn{newbell} and introduce the new, third, emergent gauge field with a Chern-Simons term. And so on and so on. The resulting states have filling fraction
\be \nu = \cfrac{1}{m\pm \cfrac{1}{\tilde{m}_1\pm \cfrac{1}{\tilde{m}_2\pm\cdots}}}\nn\ee
which is the result that we previously stated \eqn{contfrac} without proof. 

\subsubsection*{$K$-Matrices}

Using these ideas, we can now write down the effective theory for the most general Abelian quantum Hall state. We introduce $N$ emergent gauge fields $a^i_\mu$, with $i=1,\ldots,N$. The most general theory is 
\be S_{K}[a^i,A] = \int d^3x\ \frac{1}{4\pi}K_{ij}\ep a^i_\mu \partial_\nu a_\rho^j + \frac{1}{2\pi} t_i \ep A_\mu \partial_\nu a_\nu^i\label{kmatrix}\ee
It depends on the {\it $K$-matrix}, $K_{ij}$, which specifies the various Chern-Simons couplings, and the charge vector $t_i$ which specifies the linear combination of currents that is to be viewed as the electron current. We could also couple different quasi-holes currents to other linear combinations of the $a^i$

\para
The $K$-matrix and $t$-vector encode much of the physical information that we care about. The Hall conductance is computed 
 by integrating out the gauge fields and is given by
\be \sigma_{xy} = (K^{-1})^{ij} t_it_j\nn\ee
the charge of the quasi-hole which couples to the gauge field $a^i$ is 
\be (e^\star)^i = (K^{-1})^{ij}t_j\nn\ee
and the statistics between quasi-holes that couple to $a^i$ and those that couple to $a^j$ is 
\be \alpha^{ij} = (K^{-1})^{ij}\nn\ee
One can also show, by repeating the kinds of arguments we gave in Section \ref{cstorussec}, that the ground state degeneracy on a genus-$g$ surface is $|{\rm det}K|^g$.

\para
We've already met the $K$-matrix associated to the hierarchy of states. It is
\be K = \left(\begin{array}{cccc} m\ & -1\ & 0\ & \ldots\\  -1 &\ \ \tilde{m}_1\  & \ -1\ & \\ 0\  &\ -1\ &\ \  \tilde{m}_2\ & \\ \vdots & &  &\ddots \end{array}\right)\ \ \ {\rm and}\ \ \ t = (1,0,0\ldots)\nn\ee
But we can also use the $K$-matrix approach to describe other Hall states. For example, the $(m_1,m_2,n)$ Halperin states that we met in Section \ref{multiwfsec} have $K$-matrices given by
\be K =\left(\begin{array}{cc} m_1\ & n \\ n\ & m_2 \end{array}\right) \ \ \ {\rm and}\ \ \ t = (1,1)\nn\ee
Using our formula above, we find that the filling fraction is 
\be \nu = (K^{-1})^{ij} t_it_j = \frac{m_1+m_2-2n}{m_1m_2-n^2}\nn\ee
in agreement with our earlier result \eqn{halpfill}. The ground state degeneracy on a torus is $|m_1m_2-n^2|$.

\para
Restricting now to the $(m,m,n)$ states, we can compute the charges and statistics of the two quasi-holes. From the formulae above, we can read off straightaway that the two quasi-holes have charges $e^\star = 1/(m+n)$ and $\alpha = m/(m^2-n^2)$. We can also take appropriate bound states of these quasi-holes that couple to other linear combinations of $a^1$ and $a^2$


\subsubsection*{Relating Different $K$-Matrices}

Not all theories \eqn{kmatrix} with different $K$-matrices and $t$-vectors describe different physics. We could always rewrite the theory in terms of different linear combinations of the gauge fields. After this change of basis, 
\be K \rightarrow S K S^T\ \ \ {\rm and}\ \ \ t\rightarrow St\label{ktransform}\ee
However, there's an extra subtlety. The gauge fields in \eqn{kmatrix} are all defined such that their fluxes on a sphere are integer valued: $\frac{1}{2\pi}\int_{{\bf S}^2} f^i_{12} \in {\bf Z}$, just as in \eqn{fdirac}. This should be maintained under the change of basis. This holds as long as the matrix $S$ above lies in $SL(N,{\bf Z})$.

\para
The pair $(K,t)$, subject to the equivalence \eqn{ktransform}, are almost enough to classify the possible Abelian quantum states. It turns out, however, that there's one thing missing. This is known as the {\it shift}. It is related to the degeneracy when the Hall fluid is placed on manifolds of different topology; you can read about this in the reviews by Wen or Zee. More recently, it's been realised that the shift is also related to the so-called Hall viscosity of the fluid.

\subsection{Particle-Vortex Duality}

The effective field theories that we've described above were not the first attempt to use Chern-Simons theory as a description of the quantum Hall effect. Instead, the original attempts tried to write down local order parameters for the quantum Hall states and build a low-energy effective theory modelled on the usual Ginzburg-Landau approach that we met in the {\it Statistical Physics} lectures.

\para
It's now appreciated that the more subtle topological aspects of the quantum Hall states that we've described above are not captured by a Ginzburg-Landau theory. Nonetheless, this approach provides a framework in which many detailed properties of the quantum Hall states can be computed. We won't provide all these details here and this section will be less comprehensive than others. Its main purpose is to explain how to construct these alternative theories and provide some pointers to the literature. Moreover, we also take this opportunity to advertise a beautiful property of quantum field theories in $d=2+1$ dimensions known as particle-vortex duality.

\subsubsection{The $XY$-Model and the Abelian-Higgs Model}\label{xysec}

In $d=2+1$ dimensional field theories, there are two kinds of particle excitations that can appear. The first kind is the familiar excitation that we get when we quantise a local field. This is  that kind that we learned about in our {\it Quantum Field Theory} course. The second kind of particle is a {\it vortex}, defined by the winding of some local order parameter. These arise as solitons of the theory. 

\para
Often in $d=2+1$ dimensions, it's possible to write down two very different-looking theories which describe the same physics. This is possible because the particles of one theory are related to the vortices of the other, and vice versa. We start by  explaining how this works in the simplest example, first proposed in the 70's by Peskin and early  '80's by Dasgupta and Halperin. 

\subsubsection*{\underline{Theory A: The $XY$-Model}}

Our first theory consists only of a complex scalar field $\phi$ with action
\be S_A = \int d^3x\ |\partial_\mu \phi|^2 - a |\phi|^2 - b|\phi|^4 + \ldots
\label{theorya}\ee
The theory has a global $U(1)$ symmetry which acts by rotations of the form $\phi\rightarrow e^{i\theta}\phi$. The different phases  of this theory, and the corresponding physical excitations, can be characterised by symmetry breaking of this $U(1)$. There are three different possibilities which we'll characterise by the sign of $a$ (assuming that $b>0$),
\begin{itemize}
\item $a>0$: In this phase, the $U(1)$ symmetry is unbroken and the  $\phi$ excitations are massive. 
\item $a<0$: In this phase, $\phi$ gets a vacuum expectation value and the $U(1)$ global symmetry is broken. We can write $\phi = \rho e^{i\sigma}$. The fluctuations of $\rho$ are massive, while the $\sigma$ field is massless: it is the Goldstone mode for the broken $U(1)$. This phase is sometimes called the ``{\it XY model}\," (as it also arises from lattice models of spins which can rotate freely in the $(x,y)$-plane).

\para
In this phase, the theory also has vortex excitations. These arise from the phase of $\phi$ winding asymptotically. The winding is measured by
\be \oint dx^i\ \partial_i\sigma = 2\pi n\nn\ee
with $n\in {\bf Z}$ countrs the number of vortices (or anti-vortices for $n<0$). Note that $n$ is quantised for topological reasons.
These vortices are gapped. Indeed, if you compute their energy from the action \eqn{theorya}, you'll find that it is logarithmically divergent. Said another way, there is a logarithmically increasing attractive force between a vortex and an anti-vortex. The vortices are sometimes said to be ``logarithmically confined". 
\item $a=0$: Lying between the two phases above is a critical point. We are being a little careless in describing this as $a=0$; strictly, you should tune both $a$ and the other parameters to sit at this point. Here, the low-energy dynamics is described by a conformal field theory. 
 \end{itemize}

We now compare this to the physics that arises in a very different theory:

\subsubsection*{\underline{Theory B: The Abelian-Higgs Model}}

 Our second theory again consists of a complex scalar field, which we now call $\tilde{\phi}$. This time the scalar is coupled to a dynamical gauge field $\alpha_\mu$. The action is
\be S _B= \int d^3x\  -\frac{1}{4g^2}\tilde{f}_{\mu\nu} \tilde{f}^{\mu\nu}+|{\cal D}_\mu\tilde{\phi}|^2 - a'|\tilde{\phi}|^2 - b'|\tilde{\phi}|^4 + \ldots\label{theoryb}\ee
with $\tilde{f}_{\mu\nu} = \partial_\mu\alpha_\nu - \partial_\nu\alpha_\mu$. 
At first glance, Theory A and Theory B look very different. Nonetheless, as we now explain, they describe the same physics. Let's start by matching the symmetries. 

\para
Theory B clearly has a $U(1)$ gauge symmetry. This has no counterpart in Theory A but  that's ok because gauge symmetries aren't real symmetries: they are merely redundancies in our description of the system. It's more important to match the global symmetries. We've seen that Theory A has a $U(1)$ global symmetry. But there is also a less obvious global symmetry in Theory B, with the current given by
\be j^\mu = \frac{1}{2\pi}\ep \partial_\nu \alpha_\rho\label{u1j}\ee
This is the kind of current that we were playing with in our theories of the quantum Hall effect. The conserved charge is the magnetic flux associated to the $U(1)$ gauge symmetry. This is to be identified with the global $U(1)$ symmetry in Theory A. 

\para
Now let's look at the different phases exhibited by Theory B. Again, assuming that $b'>0$, there are three phases depending on the sign of $a'$, 
\begin{itemize}
\item $a'>0$: In this phase, the $\tilde{\phi}$ fields are massive and the  $U(1)$ gauge symmetry is unbroken. Correspondingly, there is a massless photon in the spectrum. This is usually referred to as the {\it Coulomb phase}. However, in $d=2+1$ dimensions, the photon carries only a single polarisation state and can be alternatively described by a scalar field, usually referred to as the {\it dual photon}, $\sigma$. We can implement the change of variables in the path integral if we ignore the coupling to the $\tilde{\phi}$ fields. We can then  replace the integration over $\alpha_\mu$ with an integration over the field strength $\tilde{f}_{\mu\nu}$ then, schematically (ignoring issues of gauge fixing) the partition function reads
\be Z &=& \int {\cal D}\alpha\ \exp\left(i\int d^3x\ - \frac{1}{4g^2}\tilde{f}_{\mu\nu}\tilde{f}^{\mu\nu}\right) 
 \nn\\ &=& \int {\cal D}\tilde{f}{\cal D}\sigma\ \exp\left(i\int d^3x\ -\frac{1}{4g^2}\tilde{f}_{\mu\nu}\tilde{f}^{\mu\nu}+ \frac{1}{2\pi}\sigma\ep \partial_\mu \tilde{f}_{\nu\rho}\right)\nn\ee
Here $\sigma$ is playing the role of a Lagrange multiplier whose role is to impose the Bianchi identity $\ep\partial_\mu \tilde{f}_{\nu\rho}=0$. If the field strength obeys the Dirac quantisation condition, then $\sigma$ has periodicity $2\pi$. Now we integrate out the field strength, leaving ourselves only with an effective action for $\sigma$,
\be Z = \exp\left(i\int d^3x\ \frac{g^2}{2\pi}\partial_\mu \sigma\partial^\mu\sigma\right)\nn\ee
This is the dual photon. It is related to the original field strength by the equation of motion
\be \tilde{f}^{\mu\nu} = \frac{g^2}{\pi}\ep \partial_\rho \sigma\nn\ee
Note that the current \eqn{u1j} can be easily written in terms of the dual photon: it is 
\be j_\mu = \frac{g^2}{\pi}\partial_\mu \sigma\nn\ee
Another way of saying this is that the global $U(1)$ symmetry  acts by shifting the value of the dual photon: $\sigma\rightarrow \sigma + {\rm const}. $ 

\para
The upshot of this is that the global $U(1)$ symmetry is spontaneously broken in this phase. This means that we should identify the Coulomb phase of Theory B with the $a<0$ phase of Theory A. The dual photon $\sigma$ can be viewed as the Goldstone mode of this broken symmetry. This is to be identified with the Goldstone mode of the $a<0$ phase of Theory A. (Indeed, we even took the liberty of giving the two Goldstone modes the same name.)

\para
The charged $\tilde{\phi}$ fields are massive in this phase. These are to be identified with the vortices of the $a<0$ phase of Theory A. As a check, note that the $\tilde{\phi}$ excitations interact  through the Coulomb force which, in $d=2+1$ dimensions, results in a logarithmically confining force between charges of opposite sign, just like the vortices of Theory A. 

\item $a'<0$: In this phase $\tilde{\phi}$ gets an expectation value and the $U(1)$ gauge symmetry is broken. Now the photon gets a mass by the Higgs mechanism and all excitations are gapped. This is the {\it Higgs} phase of the theory. 

\para
The global $U(1)$ symmetry is unbroken in this phase. This means that we should identify the Higgs phase of Theory B with the gapped $a>0$ phase of Theory A. 

\para
The breaking of the $U(1)$ gauge symmetry means that there are vortex solutions in the Higgs phase. These are defined by the asymptotic winding of the expectation value of $\tilde{\phi}$. The resulting  solutions exhibit some nice properties\footnote{For a more detailed discussion of these properties, see the \href{http://www.damtp.cam.ac.uk/user/tong/tasi.html}{\it TASI Lectures on Solitons}.}. First, unlike the global vortices of Theory A, vortices associated to a gauge symmetry have finite mass. Second, they also carry quantised magnetic flux
\be \oint dx^i \partial_i\tilde{\phi} = \frac{1}{2\pi}\int d^2x\ \tilde{f}_{12} = n'\nn\ee
where $n'\in {\bf Z}$ is the number of vortices. The fact that these vortices carry magnetic flux means that they are charged under the current \eqn{u1j}. These vortices are identified with the $\phi$ excitations of Theory A in the $a>0$ phase. 

\item $a'=0$: Lying between these two phases, there is again a quantum critical point.  Numerical simulations show that this is the same quantum critical point that exists in Theory A. 
 \end{itemize}

We can see that, viewed through a blurred lens, the theories share the same phase diagram. Roughly, the parameters of are related by
\be a \approx- a'\nn\ee
Note, however, that we're only described how qualitative features match. If you want to go beyond this, and see how the interactions match in detail then it's much harder and you have to worry about all the $\ldots$ interactions in the two theories that we didn't write down. (For what it's worth, you can go much further in supersymmetric theories where the analog of this duality is referred to as {\it mirror symmetry}). 

\para
The qualitative level of the discussion above will be more than adequate for our purposes. Our goal now is to apply these ideas to the effective field theories that we previously wrote down for the fractional quantum Hall effect.

\subsubsection{Duality and the Chern-Simons Ginzburg-Landau Theory}\label{csglsec}

So far, the duality that we've described has nothing to do with the quantum Hall effect. However, it's simple to tinker with this duality to get the kind of theory that we want.
We start with Theory A given in \eqn{theorya} . It's just a complex scalar field with a $U(1)$ global symmetry $\phi\rightarrow e^{i\theta}\phi$. We'll deform this theory in the following way: we gauge the global symmetry and add a Chern-Simons term at level $m$. We end up with
\be S_A[a,\phi] = \int d^3x\   |\partial_\mu\phi-ia_\mu\phi|^2 - V(\phi) - \frac{m}{4\pi} \ep a_\mu \partial_\nu a_\rho 
\label{theoryaa}\ee
But this is precisely our earlier effective action for the Laughlin state at filling fraction $\nu=1/m$. In this context, the  excitations of the field $\phi$ describe quasi-holes and quasi-particles in the theory, with fractional charge and statistics. The background gauge field of electromagnetism $A_\mu$ couples to the electron current which is
\be j^\mu = \frac{1}{2\pi} \ep \partial_\nu a_\rho\nn\ee
Now we can repeat this procedure for Theory B defined in \eqn{theoryb}. We again couple a $U(1)$ gauge field $a_\mu$ to the current which is now given by \eqn{u1j}. We find
\be S_B[a,\alpha,\phi] = \int d^3x\ |\partial_\mu\tilde{\phi} - i\alpha_\mu\tilde{\phi}|^2 - V(\tilde{\phi}) +   \frac{1}{2\pi} \ep a_\mu \partial_\nu \alpha_\rho -\frac{m}{4\pi} \ep a_\mu \partial_\nu a_\rho + \ldots\nn\ee
where  the Maxwell term in \eqn{theoryb} has been relegated to the $\ldots$ in the expression above as it won't play an important role in what follows. Next we simply integrate out the gauge field $a_\mu$ in this Lagrangian. Because $a_\mu$ appears quadratically in the action, we can naive just replace it by its equation of motion which is
\be f_{\mu\nu} = \frac{1}{m}\tilde{f}_{\mu\nu}
\nn\ee
Note, however, that we run into the same kind of issues that we saw in Section \ref{laughlinfoffsec}. This equation of motion is not consistent with the flux quantisation of both $f_{\mu\nu}$ and $\tilde{f}_{\mu\nu}$. This means that we should not take the resulting action too seriously when dealing with subtle topological issues, but hopefully it will capture the correct local physics. This action is:
\be S_B[\alpha,\phi] = \int d^3x\ |\partial_\mu\tilde{\phi} - i\alpha_\mu\tilde{\phi}|^2 - V(\tilde{\phi}) +   \frac{1}{4\pi m} \ep \alpha_\mu \partial_\nu \alpha_\rho  \ldots\label{theorybb}\ee
This is the theory dual to \eqn{theoryaa}. It is the dual description of the quantum Hall fluid. In the original theory \eqn{theoryaa}, the elementary quanta $\phi$ are the quasi-particles while the vortices are the electrons. In the new description \eqn{theorybb}, the elementary quanta of $\tilde{\phi}$ are the electrons while the vortices are the quasi-particles. 

\para
There is one last step that is usually taken before we get to the final Ginzburg-Landau theory. The field  $\tilde{\phi}$ in \eqn{theorybb} has second order kinetic terms, which means that, upon quantisation, it will give rise to both particles and anti-particles. The particles are  electrons (we will make this clearer below), while the anti-particles are holes. The existence of both particles and holes arises because both \eqn{theoryaa} and \eqn{theorybb} describe physics around the quantum Hall state which, of course, is built upon a sea of electrons.

\para
In contrast, in the Ginzburg-Landau approach to this problem it is more common to write down a field theory for electrons above the vacuum state. This is slightly odd because the resulting physics clearly requires a large number of electrons to be present but we can always insist upon this by including an appropriate chemical potential. We'll call the bosonic field that gives rise to electrons $\Phi$. This field now has first order kinetic terms, reflecting the fact that  there are no longer anti-particles. (Well, there are but they require around $10^{10}$ more energy than is available in quantum Hall system; this is condensed matter physics, not particle physics!). The resulting Lagrangian is
\be S= \int d^3x\  \, i\Phi^\dagger(\partial_0 - i \alpha_0-i\mu)\Phi  - \frac{1}{2m^\star}|\partial_i\Phi - i\alpha_i\Phi|^2
- V(\Phi)  +   \frac{1}{4\pi m} \ep \alpha_\mu \partial_\nu \alpha_\rho \ \ \ \  \ \ \ \ \ \label{zhk1}\ee
with $\mu$ the promised chemical potential and $m^\star$ is the effective mass of the electron (and is not to be confused with the integer $m$). This is the proposed Chern-Simons Ginzberg-Landau description of the fractional quantum Hall effect. This Lagrangian was first written down by Zhang, Hansson and Kivelson and is sometime referred to as the ZHK theory\footnote{The original paper is S. C. Zhang, T. Hansson and S. Kivelson, ``{\it Effective-Field-Theory Model for the Fractional Quantum Hall Effect}", Phys. Rev. Lett. {\bf 62}, 82 (1989) which can be \href{http://so5.stanford.edu/Research/Projects/Zhang1989.pdf}{downloaded here}.}.

\subsubsection*{Composite Bosons}

We know from our previous discussion that the excitations of $\tilde{\phi}$ in \eqn{theorybb}  (or $\Phi$ in \eqn{zhk1}) are supposed to describe the vortices of theory \eqn{theoryaa}. Yet those vortices should carry the same quantum numbers as the original electrons. Let's first check that this makes sense. 

\para
Recall that the $\tilde{\phi}$ (or $\Phi$) field is bosonic: it obeys commutation relations rather than anti-commutation relations. But, by the same arguments that we saw in Section \ref{laughlinfoffsec}, the presence of the Chern-Simons term will change the statistics of these excitations. In particular, if we work with the non-relativistic theory, the equation of motion for $\alpha_0$ reads
\be \frac{1}{2\pi}\tilde{f}_{12} = -m \Phi^\dagger\Phi \label{smokymorning}\ee
Here $\Phi^\dagger\Phi$ is simply the particle density $n({\bf x})$. 
This tells us that each particle in the quantum Hall fluid has $-m$ units of attached flux. 
By the usual Aharonov-Bohm arguments, these particles are bosonic if $m$ is even and fermionic if $m$ is odd. But that's exactly the right statistics for the ``electrons" underlying the quantum Hall states.

\para 
Let's briefly restrict to the case of $m$ odd, so that the ``electrons" are actual electrons. They can be thought of as bosons $\Phi$ attached to $-m$ flux units. Alternatively, the bosons $\Phi$ can be thought of as electrons attached to $+m$ units of flux. This object is referred to as a {\it composite boson}. Notice that it's very similar in spirit to the composite fermion that we met earlier. The difference is that we attach an odd number of fluxes to the electron to make a composite boson, while an even number of fluxes gives a composite fermion. In the next section, we'll see how to make a composite fermion in this language.

\subsubsection*{Off-Diagonal Long-Range Order}

We took a rather round-about route to get to Lagrangian \eqn{zhk1}: we first looked at the most general description of a fractional quantum Hall effect, and subsequently dualised. However, it's possible to motivate \eqn{zhk1} directly. In this short section, we briefly explain how.

\para
The usual construction of a Ginzburg-Landau effective theory involves first identifying a symmetry which is broken. The symmetry breaking is then described by an appropriate local order parameter, and the effective theory is written in terms of this order parameter. If we want to do this for the quantum Hall fluid, we first need to figure out what this order parameter could possibly be.

\para
We're going to take a hint from the theory of superfluidity where one works with an object called the {\it density matrix}. (Beware: this means something different than in previous courses on quantum mechanics and quantum information). There are two, equivalent, definitions of the density matrix. First, suppose that  we have some many-body system with particles created by the operator $\Psi^\dagger({\bf r})$. In a given state, we  define the {\it density matrix} to be 
\be \rho({\bf r},{\bf r}') = \langle \Psi^\dagger ({\bf r})\Psi({\bf r}')\rangle\nn\ee
Alternatively, there is also  simple definition in the first quantised framework. Suppose that our system of $N$ particles is described by the  the wavefunction $\psi({\bf x}_i)$. We focus on the position of just a single particle, say ${\bf x}_1\equiv {\bf r}$ and the density matrix is constructed as
\be \rho({\bf r},{\bf r}') = N\int \prod_{i=2}^N d{\bf x}_i\ \psi^\star({\bf r},{\bf x}_2,\ldots,{\bf x}_N)
\psi({\bf r}',{\bf x}_2,\ldots,{\bf x}_N)\nn\ee
The definition of a {\it superfluid} state is that the density matrix exhibits {\it off-diagonal long range order}. This means that
\be \rho({\bf r},{\bf r}') \rightarrow \rho_0\ \ \ \ {\rm as}\ \ \ |{\bf r}-{\bf r}'|\rightarrow \infty\nn\ee
Here $\rho_0$ is the density of the superfluid. 

\para
What does this have to do with our quantum Hall fluids? They certainly don't act like superfluids. And, indeed, you can check that quantum Hall fluids are {\it not} superfluids. If you compute the density matrix for the Laughlin wavefunction \eqn{laughlin}, you find
\be \rho(z,z') = N \int \prod_{i=2}^N d^2z_i\  \prod_i (z-z_i)^m (\bar{z}'-\bar{z}_i)^m\prod_{j<k}  |z_j-z_k|^{2m} e^{-\sum_j |z_j|^2/2l_B^2}\nn\ee
This does not exhibit off-diagonal long-range order. The first two terms ensure that the phase fluctuates wildly and this results in exponential decay of the density matrix: $\rho(z,z')\sim e^{-|z-z'|^2}$. 

\para
However, one can construct an object which does exhibit off-diagonal long-range order. This is not apparent in the electrons, but instead in the composite bosons $\Phi$. These operators are related to the electrons by the addition of $-m$ flux units,
\be \Phi^\dagger(z) = \Psi^\dagger(z) U^{-m}\label{sticktoit}\ee
where $U$ is the operator which inserts a single unit of flux of the gauge field $\alpha_\mu$. It can be shown that this is the operator which exhibits off-diagonal long-range order in the quantum Hall state\footnote{The first suggestion of long-range order in the Hall states was given by S. Girvin and A. H Macdonald, ``{\it Off-Diagonal Long-Range Order, Oblique Confinement, and the Fractional Quantum Hall Effect}, \href{http://journals.aps.org/prl/abstract/10.1103/PhysRevLett.58.1252}{Phys. Rev. Lett 58, 12 (1987)}. The refined, second-quantised arguments were given later by N. Read ``{\it Order Parameter and Ginzburg-Landau Theory for the Fractional Quantum Hall Effect}", \href{http://journals.aps.org/prl/abstract/10.1103/PhysRevLett.62.86}{Phys. Rev. Lett. 62, 1 (1989)}.} 
\be \langle \Phi^\dagger(z)\Phi(z')\rangle\rightarrow \rho_0\ \ \ \ {\rm as}\ \ \ |z-z'|\rightarrow \infty\nn\ee
Alternatively, if you're working with wavefunctions, you need to include a singular gauge transformation to implement the flux attachment.

\para
Note that,  usually in Ginzburg-Landau theories, one is interested in phases where the order parameter condensed. Indeed, if we follow through our duality transformations, the original theory \eqn{theoryaa} describes quantum Hall Hall physics when $\phi$ is a gapped excitation. (This is the phase $a>0$ of Theory A in the previous section). But the particle-vortex duality tells us that the dual theory \eqn{theorybb} should lie in the phase in which $\tilde{\phi}$ gets an expectation value. Equivalently, in the non-relativistic picture, $\Phi$ condenses. 

\para
This kind of thinking provided  the original motivation for writing down the Ginzburg-Landau theory and, ultimately, to finding the link to Chern-Simons theories. However, the presence of the flux attachment in \eqn{sticktoit} means that $\Phi$ is not a local operator. This is one of the reasons why this approach misses some of the more subtle effects such as topological order.

%
%
%
%

\subsubsection*{Adding Background Gauge Fields}

To explore more physics, we need to re-introduce the background gauge field $A_\mu$ into our effective Lagrangian. It's simple to re-do the integrating out with $A_\mu$ included; we find the effective Lagrangian
\be S&=& \int d^3x\ \Bigg\{ i\Phi^\dagger(\partial_0 - i (\alpha_0+A_0 + \mu))\Phi  - \frac{1}{2m^\star}|\partial_i\Phi - i(\alpha_i+A_i)\Phi|^2 \label{zhk}\\ && \ \ \ \ \ \ \ \ \ \ \ \ \ \ \ \ \ \ \ \ \ \ \ \ \ \ \ \ \ \ \ \ \ \ \ \ \ \ \ \ \ \ \ \ \ \ \ \ \ \ - V(\Phi)  +   \frac{1}{4\pi m} \ep \alpha_\mu \partial_\nu \alpha_\rho \Bigg\}  \nn\ee
%
%
Because we're working with the non-relativistic theory, the excitations of $\Phi$ in the ground state should include all electrons in our system. Correspondingly, the gauge field $A_\mu$ should now include the background magnetic field that we apply to the system.

\para
We've already seen that the Hall state is described when the $\Phi$ field condenses: $\langle \Phi^\dagger \Phi\rangle = n$, with $n$ the density of electrons. But we pay an energy cost if there is a non-vanishing magnetic field $B$ in the presence of such a condensate. This is the essence of the Meissner effect in a superconductor. However, our Hall fluid is not a superconductor. In this low-energy approach, it differs by the existence of the Chern-Simons gauge field $\alpha_\mu$ which can turn on to cancel the magnetic field,
\be \alpha_i + A_i =0 \ \ \ \Rightarrow\ \ \ \ \tilde{f}_{12} =- B\nn\ee
But we've already seen that the role of the Chern-Simons term is to bind the flux $\tilde{f}_{12}$ to the particle density $n({\bf x})$ \eqn{smokymorning}. We learn that
\be n({\bf x}) = \frac{1}{2\pi m}B({\bf x})\nn\ee
This is simply the statement that the theory is restricted to describe the lowest  Landau level with filling fraction $\nu = 1/m$ 

\para
We can also look at the vortices in this theory. These arise from the phase of $\Phi$ winding around the core of the vortex. The minimum vortex carries flux $\int d^2x \ \tilde{f}_{12} = \pm 2\pi$. From the flux attachment \eqn{smokymorning}, we see that they carry charge $e^\star =  \pm1/m$. This is as expected from our general arguments of particle-vortex duality: the vortices in the ZHK theory should correspond to the fundamental excitations of the original theory \eqn{theoryaa}: these are the quasi-holes and quasi-particles.

\para
So far, we've seen that this dual formalism can reproduce many of the  results that we saw earlier.  However, the theory \eqn{zhk} provides a framework to compute much more detailed response properties of the quantum Hall fluid. For most of these, it is not enough to consider just the classical theory as we've done above. One should take into account the quantum fluctuations of the Chern-Simons field, as well as the Coulomb interactions between electrons which we've buried in the potential. We won't describe any of this here\footnote{For a nice review article, see Shou Cheng Zhang, ``{\it The Chern-Simons-Landau-Ginzburg Theory of the Fractional Quantum Hall Effect}, Int. Jour. Mod. Phys. {\bf B6} (1992).}.

\subsubsection{Composite Fermions and the Half-Filled Landau Level}\label{hlrsec}

We can also use this Chern-Simons approach to make contact with the composite fermion picture that we met in Section \ref{fqhesec}. Recall that the basic idea was to attach an {\it even} number of vortices to each electron. In the language of Section \ref{fqhesec}, these vortices were simply zeros of the wavefunction, with holomorphicity ensuring that each zero is accompanied by a $2\pi$ winding of the phase. In the present language, we can think of the vortex attachment as flux attachment.
Adding an even number of fluxes to an electron doesn't change the statistics. The resulting object is the composite fermion. 

\para
As we saw in Section \ref{hfllsec}, one of the most interesting predictions of the composite fermion picture arises at $\nu=1/2$ where one finds a compressible fermi-liquid-type state. We can write down an effective action for the half-filled Landau level as follows,
\be S = \int d^3x\ &&  \Bigg\{ i\psi^\dagger(\partial_0 - i(\alpha_0 + A_0+ \mu)\psi - \frac{1}{2m^\star} |\partial_i\psi - i(\alpha_i+A_i)\psi|^2 \label{hlract}\\ 
  &&  \ \ \ \ \ +\  \frac{1}{2}\frac{1}{4\pi} \ep \alpha_\mu\partial_\nu\alpha_\rho +\frac{1}{2}\int d^2x'\ \psi^\dagger({\bf x})\psi(x) V({\bf x}-{\bf x}')\psi^\dagger({\bf x}') \psi({\bf x}') \Bigg\}
\nn\ee
Here $\psi$ is to be quantised as  a fermion, obeying anti-commutation relations. We have also explicitly written the potential between electrons, with $V({\bf x})$ usually taken to the be the Coulomb potential. Note that the Chern-Simons term has coefficient $1/2$, as befits a theory at half-filling.

\para
The action \eqn{hlract} is the starting point for the Halperin-Lee-Read theory of the half-filled Landau level. The basic idea is that an external  magnetic field $B$ can be screened by the emergent gauge field $\tilde{f}_{12}$, leaving the fermions free to fill up a Fermi sea. However, the fluctuations of the Chern-Simons gauge field mean that the resulting properties of this metal are different from the usual Fermi-liquid theory. It is, perhaps, the simplest example of a ``non-Fermi liquid". Many detailed calculations of properties of this state can be performed and successfully compared to experiment. We won't describe any of this here\footnote{Details can be found in the original paper by Halperin, Lee and Read,   ``{\it Theory of the half-filled Landau level}\ ", 
\href{http://journals.aps.org/prb/abstract/10.1103/PhysRevB.47.7312}{Phys. Rev. B 47, 7312 (1993)}, and in the nice review by Steve Simon, ``{\it The Chern-Simons Fermi Liquid Description of Fractional Quantum Hall States}\,", \href{http://arxiv.org/abs/cond-mat/9812186}{cond-mat/9812186}.}.

\subsubsection*{Half-Filled or Half-Empty?}

While the HLR theory \eqn{hlract} can claim many successes, there remains one issue that is poorly understood. When a Landau level is half full, it is also half empty. One would expect that the resulting theory would then exhibit a symmetry exchanging particles and holes. 
But the action \eqn{hlract} does not exhibit any such symmetry.

\para
There are a number of logical possibilities. The first is that, despite appearances, the theory \eqn{hlr} does secretly preserve particle-hole symmetry. The second possibility is that this symmetry is spontaneously broken at $\nu = 1/2$ and there are actually two possible states. (This turns out to be true at $\nu=5/2$ where the Pfaffian state we've already met has a brother, known as the anti-Pfaffian state). 

\para
Here we will focus on a third possibility: that the theory \eqn{hlract} is not quite correct. An alternative theory was suggested by Son who proposed that the composite fermion at $\nu=1/2$ should be rightly viewed as a two-component Dirac fermion\footnote{Son's original paper is ``{\it Is the Composite Fermion a Dirac Particle?}\,", Phys. Rev. {\bf X5}, 031027 (2015), \href{http://arxiv.org/abs/1502.03446}{arXiv:1502.03446}.}.

\para
The heart of Son's proposal is a new duality that can be thought of as a fermionic version of the particle-vortex duality that we met in Section \ref{xysec}. Here we first describe this duality. In the process of explaining how it works, we will see the connection to the half-filled Landau level.

\subsubsection*{\underline{Theory A: The Dirac Fermion}}

Our first theory consists of a single Dirac fermion $\psi$ in $d=2+1$ dimensions
\be S_A = \int d^3 x\   i\bar{\psi}(\!\delslash - i \slsh{A})\psi + \ldots\label{sona}\ee
In $d=2+1$ dimensions, the representation of the Clifford algebra $\{\gamma^\mu,\gamma^\nu\} = 2\eta^{\mu\nu}$ has  dimension 2. The gamma matrices can be written in terms of  the Pauli matrices, with a useful representation given by
\be \gamma^0 = i\sigma^2\ \ \ ,\ \ \ \gamma^1 = \sigma^1\ \ \ ,\ \ \ \ \gamma^2 = \sigma^3\nn\ee
Correspondingly, the Dirac spinor $\psi$ is a two-component object with complex components. As usual, $\bar{\psi} = \psi^\dagger \gamma^0$. (See the lectures on {\it Quantum Field Theory} for more information about the construction of spinors). Quantising the Dirac spinor in $d=2+1$ dimensions gives rise to spin-up particles and spin-down anti-particles. 

\para
Theory  A has a global $U(1)$ symmetry with current
\be J^\mu = \bar{\psi}\gamma^\mu\psi\label{sonnyjim}\ee
In the action \eqn{sona}, we've coupled this to a background electromagnetic gauge field $A_\mu$. 

\subsubsection*{\underline{Theory B: QED${}_3$}}

The second theory also consists of  a single Dirac fermion, $\tilde{\psi}$, this time coupled to a dynamical $U(1)$ gauge field $\alpha_\mu$. 
\be S_B =  \int d^3x\  i\bar{\tilde{\psi}}(\!\delslash - 2i\! \slsh{\,\alpha})\tilde{\psi} + \frac{1}{2\pi} \ep\alpha_\mu \partial_\nu A_\rho +\ldots\label{sonb}\ee
This is essentially QED in $d=2+1$ dimensions. However, there is one crucial subtlety: $\tilde{\psi}$ carries charge 2 under this gauge field, not charge 1. To avoid rescaling of the gauge field, we should accompany this charge with the statement that the fluxes of $\alpha$ remain canonically normalised
\be \frac{1}{2\pi} \int_{{\bf S}^2} \tilde{f}_{12} \in {\bf Z}\nn\ee
The charge $2$ is crucial for this theory to make sense. If the fermion $\tilde{\psi}$ had charge $1$ then the theory wouldn't make sense: it suffers from a discrete gauge anomaly, usually referred to as a {\it parity anomaly} in this context. However, with charge $2$ this is avoided\footnote{This is actually a bit too quick. A more careful analysis was given by T. Senthil, N. Seiberg, E. Witten and C. Wang in ``{\it A Duality Web in 2+1 Dimensions and Condensed Matter Physics}", \href{http://arxiv.org/abs/1606.01989}{ArXiv:1606.01989}.}
. 

\para
The theory \eqn{sonb} has a $U(1)$ symmetry with the kind of current that is, by now, familiar
\be J^\mu = \frac{1}{2\pi} \ep \partial_\nu \alpha_\rho\nn\ee
This is to be identified with the current \eqn{sonnyjim} of Theory A.

\subsubsection*{Half-Filling in the Two Theories}

\EPSFIGURE{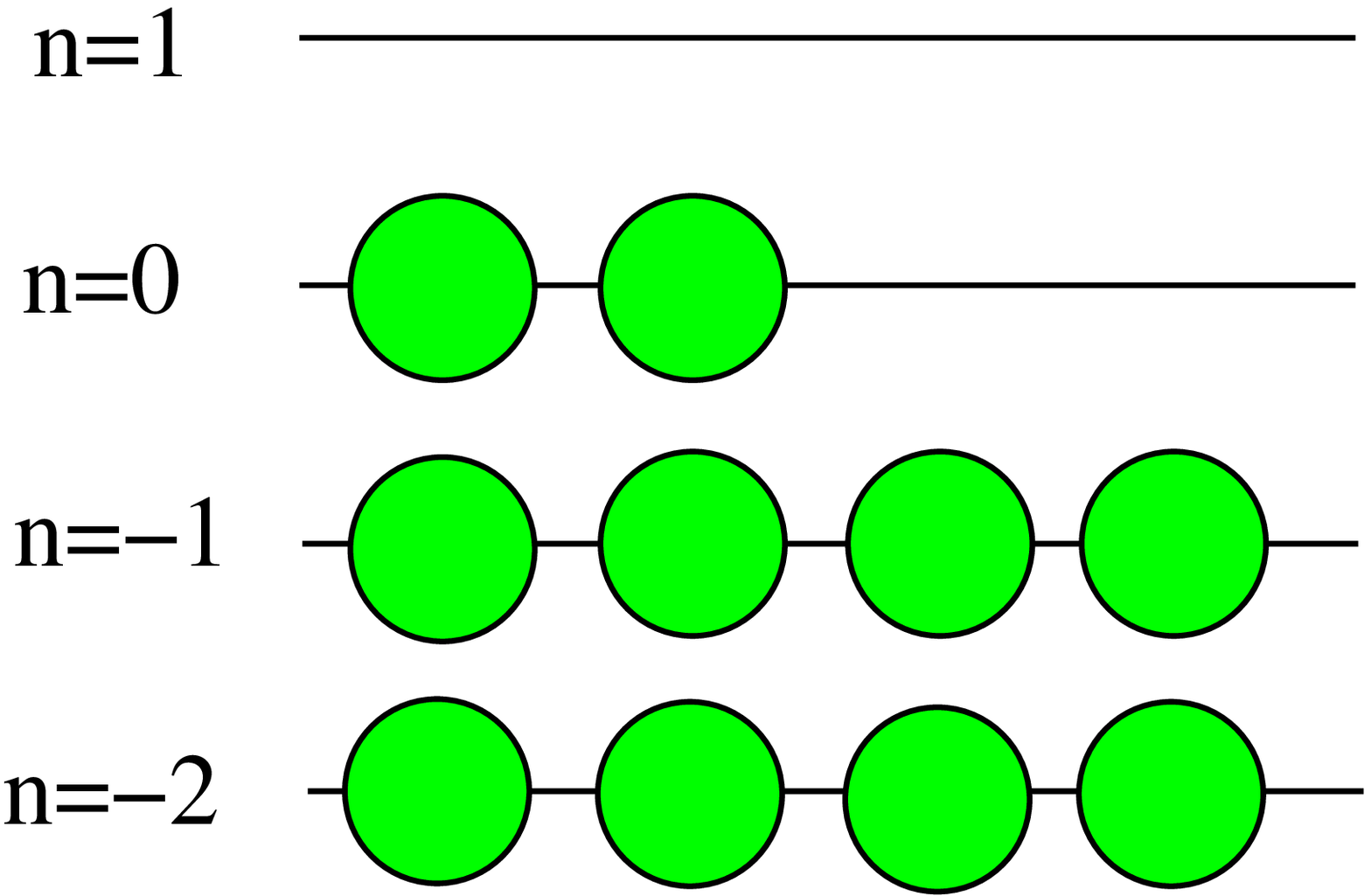,height=90pt}{The view from Theory A}
\noindent
Let's start with Theory A and turn on a background magnetic field $B$. The Dirac fermions form Landau levels. However, because of the relativistic dispersion relation, these Landau levels are somewhat different from those we met in Section \ref{basicsec}. A simple generalisation of these calculations shows that the Landau levels have energy
\be E^2_n =  2B |n|  \ \ \ n\in{\bf Z}\nn\ee
Note, in particular, that there is a zero energy $n=0$ Landau level. This arises because  the zero-point energy $\ft12 \hbar \omega_B$ seen in the non-relativistic Landau levels is exactly compensated by the Zeeman splitting. 

\EPSFIGURE{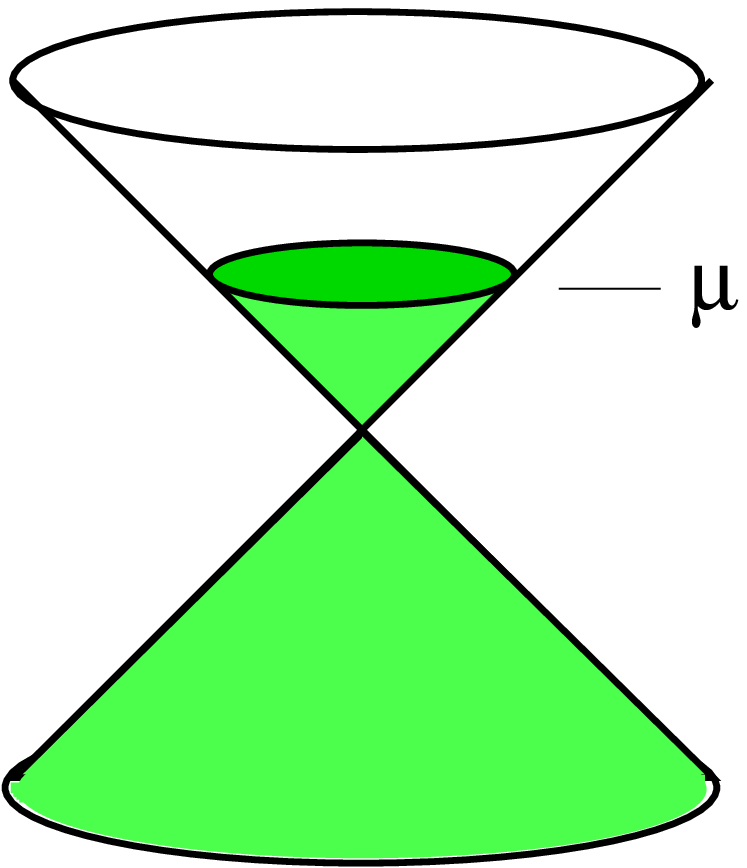,height=110pt}{and from Theory B}
\para
In the Dirac sea picture, we can think of filling the negative energy Landau levels, which we label with $n<0$. However, if we restrict to zero density then the $n=0$ Landau level is necessarily at half-filling. This is shown in the picture. In the absence of any interactions there is a large degeneracy. We rely on the interactions, captured by the $\ldots$ in \eqn{sona}, to resolve this degeneracy. In this way, the Dirac fermion in a magnetic field automatically sits at half filling.  Note that this picture is, by construction, symmetric under interchange  of particles and holes.

\para
Let's now see what this same picture looks like in Theory B. The background magnetic field contributes a term $\frac{1}{2\pi} B \alpha_0$ to the action \eqn{sonb}. This is a background charge density, $\tilde{n} = \frac{1}{2} (B/2\pi)$, where the factor of 1/2 can be traced to to the charge $2$ carried by the fermion. This means that the fermions in QED${}_3$ pile up to form a Fermi sea, with chemical potential $\mu$  set by the background magnetic field. This is shown in the figure to the right.

\para
This is the new proposed dual of the half-filled Landau level. We see that there is no hint of the magnetic field in the dual picture. Instead we get a Fermi surface which, just as in the HLR theory \eqn{hlract}, is coupled to a fluctuating gauge field. However, in this new proposal this gauge field no longer has a Chern-Simons coupling.

\para
It turns out that many, if not all, of the successful predictions  of the HLR theory \eqn{hlract} also hold for QED${}_3$ \eqn{sonb}. The difference between the theories two turns out to be rather subtle: the relativistic electrons in QED${}_3$ pick up an extra factor of Berry phase $\pi$ as they are transported around the Fermi surface. At the time of writing, there is an ongoing effort to determine whether this phase can be observed experimentally  see which of these two theories is correct.

\subsection{Non-Abelian Chern-Simons Theories}\label{nonabcssec}

So far we have discussed the effective theories only for Abelian quantum Hall states. As we have seen, these are described by Chern-Simons theories involving emergent $U(1)$ gauge fields. Given this, it seems plausible that the effective field theories for non-Abelian quantum Hall states involve emergent non-Abelian Chern-Simons gauge fields. This is indeed the case. Here we sketch some of the novel properties associated to non-Abelian Chern-Simons terms.

\subsubsection{Introducing Non-Abelian Chern-Simons Theories}

We start by describing the basics of non-Abelian Chern-Simons theories. Everything we say will hold for arbitrary gauge group $G$, but we will focus on $G=SU(N)$. For the most prominent applications to quantum Hall physics, $G=SU(2)$ will suffice. We work with Hermitian gauge connections $a_\mu$, valued in the Lie algebra. The associated field strength is
\be f_{\mu\nu} = \partial_\mu a_\nu - \partial_\nu a_\mu - i[a_\mu,a_\nu]\nn\ee
We take the basis generators in the fundamental representation with normalisation $\tr (T^a T^b) = \frac{1}{2} \delta^{ab}$. With this choice, the Yang-Mills action takes the familiar form
\be S_{YM} =  -\frac{1}{2g^2}\int d^3x\ \tr\,f^{\mu\nu} f_{\mu\nu}\nn\ee
However, just as we saw for the Abelian gauge fields, we are not interested in the Yang-Mills action. Instead, there is an alternative action that we can write down in $d=2+1$ dimensions. This is the non-Abelian Chern-Simons action
\be S_{CS} =  \frac{k}{4\pi}\int d^3x\ \ep \tr\,\left(a_\mu \partial_\nu a_\rho - \frac{2i}{3}a_\mu a_\nu a_\rho\right)\label{nonabcs}\ee
Chern-Simons theories with gauge group $G$ and level $k$ are sometimes denoted as $G_k$. 

\para
Our first goal is to understand some simple properties of this theory. The equation of motion is
\be f_{\mu\nu} = 0\nn\ee
This is deceptively simple! Yet, as we will see, many of the subtleties arise from the interesting solutions to this equation and its generalisations. Indeed, we've already seen our first hint that this equation has interesting solutions when we looked at Abelian Chern-Simons theories on the torus in Section \ref{cstorussec}.

\para
Let's start by seeing how the Chern-Simons action fares under a gauge transformation. The gauge potential transforms as
\be a_\mu \rightarrow g^{-1} a_\mu g +i g^{-1} \partial_\mu g \nn\ee
with  $g\in SU(N)$. The field strength transforms as $f_{\mu\nu}\rightarrow g^{-1}f_{\mu\nu}g$. A simple calculation shows that the Chern-Simons action changes as
\be S_{CS} \rightarrow S_{CS} + \frac{k}{4\pi}\int d^3x\  \Bigg\{\ep \partial_\nu \tr\,(\partial_\mu g\,g^{-1}a_\rho) + \frac{1}{3}\ep \tr\,(g^{-1}\partial_\mu g\, g^{-1}\partial_\nu g\,g^{-1}\partial_\rho g)\Bigg\}\nn\ee
The first term is a total derivative. The same kind of term arose in Abelian Chern-Simons theories. It will have an interesting role to play on manifolds with boundaries.

\para
For now, our interest lies in the second term. This is novel to non-Abelian gauge theories and has a beautiful interpretation. To see this, consider our theory on Euclidean ${\bf S}^3$ (or on  ${\bf R}^3$ with the requirement that gauge transformations asymptote to the same value at infinity). Then the gauge transformations can ``wind" around spacetime. This follows from the homotopy group $\Pi_3(SU(N)) \cong {\bf Z}$. The winding is counted by the function
\be w(g) = \frac{1}{24\pi^2} \int d^3x\ \ep \tr\,(g^{-1}\partial_\mu g \,g^{-1}\partial_\nu g \,g^{-1}\partial_\rho g ) \ \ \in {\bf Z}\label{winding}\ee
We recognise this as the final term that appears in the variation of the Chern-Simons action. This means that the Chern-Simons action is not invariant under these large gauge transformations; it changes as 
\be S_{CS} \rightarrow S_{CS} + \frac{k}{12\pi}\,24\pi^2 w(g)  = S_{CS}+ 2\pi kw(g)\nn\ee
However, just as we saw earlier, we need not insist that the Chern-Simons action is invariant. We need only insist  that the exponential that appears in the path integral, $e^{iS_{CS}}$ is invariant. We see that this holds providing
\be k\in {\bf Z}\nn\ee
This is the same quantisation that we saw for the Abelian theory, although this requirement arises in a more direct fashion for the non-Abelian theory. (Note that we're using the convention $e=\hbar=1$; if we put these back in, we find $\hbar k /e^2 \in {\bf Z}$).

\subsubsection*{Chern-Simons Term as a Boundary Term}

There is one other basic property of the Chern-Simons term that is useful to know. Consider a theory in $d=3+1$ dimensions. 
A natural quantity is the Pontryagin density $\epsilon^{\mu\nu\rho\sigma}\Tr(f_{\mu\nu}f_{\rho\sigma})$. It's not hard to show that this is a total derivative, 
\be \epsilon^{\mu\nu\rho\sigma}\tr(f_{\mu\nu}f_{\rho\sigma}) = 4\epsilon^{\mu\nu\rho\sigma}\partial_\mu \,\tr\,\left(a_\mu \partial_\rho a_\sigma - \frac{2i}{3}a_\nu a_\rho a_\sigma\right)\nn\ee
The object in brackets is precisely the Chern-Simons term.

\subsubsection{Canonical Quantisation and Topological Order}\label{csqsec}

Let's now quantise the Chern-Simons theory \eqn{nonabcs}. Here, and also in Section \ref{wilsonsec}, we explain how to do this. However, both sections will be rather schematic, often stating results rather than deriving them\footnote{There is a long and detailed literature on this material, starting with  Edward Witten's  Fields medal winning work, ``{\it Quantum Field Theory and the Jones Polynomial}", \href{http://projecteuclid.org/euclid.cmp/1104178138}{Comm. Math. Phys. Volume 121, Number 3, 351  (1989)}.}. We'll consider the theory on a manifold ${\bf R}\times \Sigma$ where ${\bf R}$ is time and $\Sigma$ is a spatial manifold which we'll take to be compact. Mostly in what follows we'll be interested in $\Sigma = {\bf S}^2$ and $\Sigma = {\bf T}^2$, but we'll also present results for more general manifolds. The action \eqn{nonabcs} can then be written as 
\be S_{CS} = \frac{k}{4\pi}\int dt\int_\Sigma d^2x\ \tr\,\left(\epsilon^{ij} a_i\ppp{}{t}a_j + a_0 f_{12}\right)\label{csab}\ee
This is crying out to be quantised in $a_0=0$ gauge. Here, the dynamical degrees of freedom $a_i$ obey the commutation relations
\be [a_i^a({\bf x}),a_j^b({\bf y})] = \frac{2\pi i}{k}\,\epsilon_{ij}\,\delta^{ab}\,\delta^2({\bf x}- {\bf y})\label{stanford}\ee
Subject to the constraint
\be f_{12} = 0\label{flat}\ee
As always with a gauge theory, there are two ways to proceed. We could either quantise and then impose the constraint. Or we could impose the constraint classically and quantise the resulting degrees of freedom.  Here, we start by describing the latter approach.

\para
We're looking for solutions to \eqn{flat} on the background $\Sigma$. This is the problem of finding {\it flat connections} on $\Sigma$ and has been well studied in the mathematical literature. We offer only a sketch of the solution. We already saw in Section \ref{cstorussec} how to do this for  Abelian Chern-Simons theories on a torus: the solutions are parameterised by the holonomies of $a_i$ around the cycles of the torus. The same is roughly true here. For gauge group $SU(N)$, there are $N^2-1$ such holonomies for each cycle, but we also need to identify connections that are related by gauge transformations. The upshot is that the moduli space ${\cal M}$ of flat connections has dimension $(2g-2)(N^2-1)$ where $g$ is the genus $\Sigma$.

\para
Usually in classical mechanics, we would view the space of solutions to the constraint -- such as ${\cal M}$ -- as the configuration space of the system. But that's not correct in the present context. Because we started with a first order action \eqn{csab}, the $a_i$ describe both positions and momenta of the system. This means that ${\cal M}$ is the {\it phase space}. Now,  importantly, it turns out that the moduli space ${\cal M}$ is compact (admittedly with some singularities that have to be dealt with). So we're in the slightly unusual situation of having a compact phase space. When you quantise you (very roughly) parcel the phase space up into chunks of area $\hbar$. Each of these chunks corresponds to a different state in the quantum Hilbert space. This means that  when you have a compact phase space, you will get a finite number of states. Of course, this is precisely what we saw for the $U(1)$ Chern-Simons theory on a torus in Section \ref{cstorussec}. What we're seeing here is just a fancy way of saying the same thing.

\para
So the question we need to answer is: what is the dimension of the Hilbert space ${\cal H}$ that you get from quantising $SU(N)$ Chern-Simons theory on a manifold $\Sigma$?

\para
When $\Sigma = {\bf S}^2$, the answer is easy. There are no flat connections on ${\bf S}^2$ and the quantisation is trivial. There is just a unique state: ${\rm dim}({\cal H})=1$. In Section \ref{wilsonsec}, we'll see how we can endow this situation with something a little more interesting. 

\para
When $\Sigma$ has more interesting topology, the quantisation of $G_k$ leads to a more interesting Hilbert space. 
When $G=SU(2)$, it turns out that the dimension of the Hilbert space for $g\geq 1$ is\footnote{This formula was first derived using a connection to conformal field theory. We will touch on this in Section \ref{edgesec}. The original paper is by Eric Verlinde, ``{\it  Fusion Rules and Modular Invariance in 2d Conformal Field Theories}", \href{https://inspirehep.net/record/23353/}{Nucl. Phys. {\bf B300}, 360 (1988)}. It is sometimes referred to the Verlinde formula.}
\be {\rm dim}({\cal H}) = \left(\frac{k+2}{2}\right)^{g-1}\sum_{j=0}^k\left(\sin\frac{(j+1)\pi}{k+2}\right)^{2(g-1)}\label{verlinde}\ee
Note that for $\Sigma = {\bf T}^2$, which has $g=1$, this is simply ${\rm dim}({\cal H}) = k+1$. It's not obvious, but nonetheless true, that the formula above gives an integer for all $g$. There is a generalisation of this formula for general gauge group which  involves various group theoretic factors such as sums over weights. 

\para
Finally, note that the dimension of the Hilbert space can be computed directly within the path integral. One simply needs to compute the partition function on the manifold ${\bf S}^1\times\Sigma$,
\be Z = \int {\cal D} a\ \exp\left[\frac{ik}{4\pi} \int_{{\bf S}^1\times \Sigma} d^3x\ \ep \tr\,\left(a_\mu \partial_\nu a_\rho - \frac{2i}{3}a_\mu a_\nu a_\rho\right)\right] = {\rm dim}({\cal H})\nn\ee
This provides a more direct way of computing the dimensions \eqn{verlinde} of the Hilbert spaces\footnote{ This calculation was described in M. Blau and G. Thompson, ``{\it Derivation of the Verlinde Formula from Chern-Simons Theory and the $G/G$ Model}", Nucl. Phys. {\bf 408}, 345 (1993),  \href{http://arxiv.org/abs/hep-th/9305010}{hep-th/9305010} where clear statements of the generalisation to other groups can be found.}.

\para
The discussion above has been rather brief. It turns out that the best way to derive these results is to map the problem into a $d=1+1$ conformal field theory known as the WZW model. Indeed, one of the most surprising results in this subject is that there is a deep connection between the states of the Chern-Simons theory and objects known as conformal blocks in the WZW model. We'll comment briefly on this in Section \ref{edgesec}.

\subsubsection{Wilson Lines}

So far we've only discussed the pure Chern-Simons. Now we want to introduce new degrees of freedom that are charged under the gauge field. These will play the role of non-Abelian anyons in the theory. 

\para
In the case of Abelian Chern-Simons theories, we could introduce quasi-holes by simply adding a background current to the Lagrangian. In the non-Abelian case, we need to be a little more careful. A current $J^\mu$ couples to the gauge field as, 
\be \int d^3x\  \tr\,(a_\mu J^\mu)\nn\ee
But now the current must transform under the gauge group. This means that we can't just stipulate some fixed background current because that wouldn't be gauge invariant. Instead, even if the charged particle is stationary, the current must include some dynamical degrees of freedom.  These describe the internal orientation of the particle within the gauge group. In the language of QCD, they are the ``colour" degrees of freedom of each quark and we'll retain this language here. In general, these colour degrees of freedom span some finite dimensional Hilbert space. For example, if we have an object transforming in the fundamental representation of $SU(N)$, then it will have an $N$-dimensional internal Hilbert space .

\para
In this section we'll see how to describe these colour degrees of freedom for each particle. Usually this is not done. Instead, one can work in a description where the colour degrees of freedom are integrated out in the path integral, leaving behind an object called a {\it Wilson line}. The purpose of this Section is really to explain  where these Wilson lines come from. In Section \ref{wilsonsec}, we will return to Chern-Simons theories and study their properties in the presence of these external sources. 

\para
Classically, we view each particle that is charged under the $SU(N)$ gauge field as carrying an internal  $N$-component complex vector with components $w_\gamma$, $\gamma=1,\ldots,N$. This vector has some special properties. First, it has a fixed length
\be w^\dagger w = \kappa\label{ww}\ee
Second, we  identify vectors which differ only by a phase: $w_\gamma\sim e^{i\theta} w_\gamma$. This means that the vectors parameterise the projective space ${\bf CP}^{N-1}$.

\para
Let's ignore the coupling of to the gauge field $a_\mu$ for now. The dynamics of these vectors is described by introducing an auxiliary $U(1)$ gauge field $\alpha$ which lives on the worldline of the particle. The action is
\be S_w  = \int dt\  \left(iw^\dagger{\cal D}_t  w - \kappa\alpha\right)\label{spinaction}\ee
where ${\cal D}_t = \partial_t w - i \alpha w$. The purpose of this gauge field is two-fold. Firstly, we have a gauge symmetry which identifies $w\rightarrow e^{i\theta(t)}w$. This means that two vectors which differ only by a phase are physically equivalent, just as we wanted. Second, the equation of motion for $\alpha$ is precisely the constraint equation \eqn{ww}. The net result is that $w_\gamma$ indeed parameterise ${\bf CP}^{N-1}$. 

\para
Note, however, that our action is first order, rather than second order. This means that ${\bf CP}^{N-1}$ is the phase space of the colour vector rather than the configuration space. But this too is  what we want: whenever we quantise a compact phase space, we end up with a finite dimensional Hilbert space. 

\para
Finally, we can couple the colour degree of freedom to the Chern-Simons gauge field. If the particle is stationary at some fixed position ${\bf x}= {\bf X}$, then the action is
\be
S_w = \int dt\  \left(iw^\dagger{\cal D}_t  w - \kappa\alpha - w^\dagger a_0(t)w\right)\nn\ee
where $a_0(t) = a_0(t,{\bf x}={\bf X})$ is the Chern-Simons gauge field at the location of the particle. The equation of motion for $w$ is then
\be i\frac{dw}{dt} = a_0(t) w\nn\ee
In other words, the Chern-Simons gauge field tells this colour vector how to precess. 

\subsubsection*{Quantising the Colour Degree of Freedom}

It's straightforward to quantise this system. Let's start with  the unconstrained variables $w_\gamma$ which obey the commutation relations,
\be [w_\gamma,w_{\gamma'}^\dagger] = \delta_{\gamma\gamma'}\label{wcommute}\ee
We define a ``ground state" $|0\rangle$ such that $w_\gamma|0\rangle=0$ for all $\gamma=1,\ldots,N$. A general state in the Hilbert space then takes the form
\be |\gamma_1\ldots \gamma_n\rangle = w_{\gamma_1}^\dagger\ldots w_{\gamma_n}^\dagger|0\rangle\nn\ee
However, we also need to take into account the constraint \eqn{ww} which, in this context, arises from the worldline gauge field $\alpha$. 
In the quantum theory, there is a normal ordering ambiguity in defining this constraint. The symmetric choice is to take the charge operator
\be Q =\frac{1}{2} (w_\gamma^\dagger w_\gamma + w_\gamma w_\gamma^\dagger)\label{qsym}\ee
and to impose the constraint
\be Q = \kappa\label{qk}\ee
The spectrum of $Q$ is quantised which means that the theory only makes sense if $\kappa$ is also quantised. In fact, the $\kappa \alpha$ term in \eqn{spinaction} is the one-dimensional analog of the 3d Chern-Simons term. (In particular, it is gauge invariant only up to a total derivative). The quantisation that we're seeing here is very similar to the kind of quantisations that we saw in the 3d case.

\para
However, the normal ordering implicit in the symmetric choice of  $Q$ in \eqn{qsym} gives rise to a shift in the spectrum. For $N$ even, $Q$ takes integer values; for $N$ odd, $Q$ takes half-integer values. It will prove useful to introduce the shifted Chern-Simons coefficient, 
\be \kappa_{\rm eff} = \kappa - \frac{N}{2}\label{kappashift}\ee
The quantisation condition then reads $\kappa_{\rm eff} \in {\bf Z}^+$.

\para
The  constraint \eqn{qk} now restricts the theory to a finite dimensional Hilbert space,  as expected from the quantisation of a compact phase space ${\bf CP}^{N-1}$. Moreover, for each value of $\kappa_{\rm eff}$, the Hilbert space inherits an action under the $SU(N)$ global symmetry. Let us look at some examples:
\begin{itemize}
\item $\kappa_{\rm eff} = 0$: The Hilbert space consists of a single state, $|0\rangle$. This is equivalent to putting a particle in the trivial representation of the gauge group. 
\item $\kappa_{\rm eff} = 1$: The Hilbert space consists of $N$ states, $w_\gamma^\dagger |0\rangle$. This describes a particle transforming in the fundamental representation of the $SU(N)$ gauge group.
\item $\kappa_{\rm eff} = 2$: The Hilbert space consists of $\frac{1}{2}N(N+1)$ states, $w_\gamma^\dagger w_{\gamma'}^\dagger |0\rangle$, transforming in the symmetric representation of the gauge group. 
\end{itemize}
By increasing the value of $\kappa_{\rm eff}$ in integer amounts, it is clear that we can build all symmetric representations of $SU(N)$ in this manner. If we were to replace the commutators in \eqn{wcommute} with anti-commutators, $\{w_\gamma,w_{\gamma'}^\dagger\} = \delta_{\gamma\gamma'}$, then it's easy to convince yourself that we will end up with particles in the anti-symmetric representations of $SU(N)$.

\subsubsection*{The Path Integral}

Let's now see what happens if we compute the path integral. For now, we will fix the Chern-Simons field $a_0(t)$ and consider only the integral over $w$ and the worldline gauge field $\alpha$. Subsequently, we'll also integrate over $a_\mu$. 

\para
The path integral is reasonably straightforward to compute. One has to be a little careful with the vacuum bubbles whose effect is to implement the shift \eqn{kappashift} from the path integral perspective. Let's suppose that we want to compute  in the theory with $\kappa_{\rm eff}=1$, so we're looking at objects in the ${\bf N}$ representation of $SU(N)$. It's not hard to see that the path integral over $\alpha$ causes the partition function to vanish unless we put in two insertions of $w$. We should therefore compute
\be W[a_0] = \int {\cal D}\alpha {\cal D}w {\cal D}w^\dagger \ e^{iS_{w}(w,\alpha;a_0)}w_\gamma(t=\infty) w_\gamma^\dagger(t=-\infty)\nn\ee
Note that we've called the partition function $W$ as opposed to its canonical name $Z$. We'll see the reason for this below. 
The insertion at $t=-\infty$ is simply placing the particle in some particular internal state and the partition function measures the amplitude that it remains in that state at $t=+\infty$

\para
Having taken this into account, we next perform the path integral over $w$ and $w^\dagger$. This is tantamount to summing a series of diagrams like this:
 \be   \raisebox{0.5ex}{\epsfxsize=5.4in\epsfbox{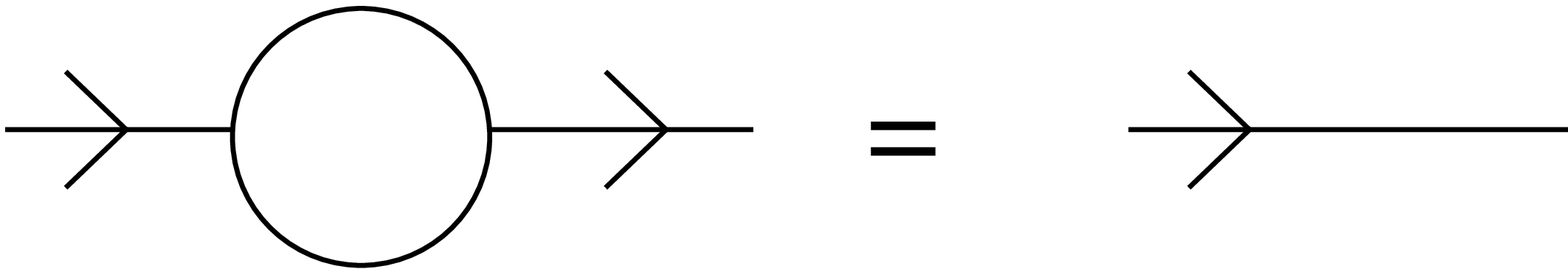}}\nn\ee
where the straight lines are propagators for $w_\gamma$ which are simply $\theta(t_1-t_2)\delta_{\gamma\gamma'}$, while the dotted lines represent insertions of the gauge fields. It's straightforward to sum these. The final result is something very simple: 
\be W[a_0] =  \Tr\, {\cal P} \exp \left(i \int dt\ a_0(t) \right)
\ee
Here ${\cal P}$ stands for path ordering which, since our particles are static, is the same thing as time ordering.   The trace is evaluated in the fundamental representation. This is the {\it Wilson line}. It is a classical function of the gauge field $a_0(t)$. However, as we've seen above, it should really be thought of as a quantum object, arising from integrating out the colour degrees of freedom of a particle. 

\para
We can also generalise this construction to other symmetric representations; you simply need to insert $\kappa_{\rm eff}$ factors of $w^\dagger$ at time $t=-\infty$ and a further  $\kappa_{\rm eff}$  factors of $w$ at $t=+\infty$. The end result is a Wilson line, with the trace evaluated in the ${\kappa}_{\rm eff}^{\rm th}$ symmetric representation.

\subsubsection{Chern-Simons Theory with Wilson Lines}\label{wilsonsec}

Let's now consider non-Abelian Chern-Simons theory with the insertion of some number of Wilson lines. Suppose that we insert $n$ Wilson lines, each in a representation $R_i$ and sitting  at position ${\bf X}_i$. For simplicity, we'll consider the theory on ${\bf R}\times {\bf S}^2$ where, previously, the theory had just a single state. Now we quantise in the presence of these Wilson lines. This will give a new Hilbert space that we'll denote ${\cal H}_{i_1\ldots i_n}$ with the labels denoting both position and representation of the Wilson lines. The first question that we want to ask is: what is the dimension of this new Hilbert space?

\para
The constraint equation in the presence of Wilson lines reads
\be \frac{k}{2\pi} f^a_{12}({\bf x}) = \sum_{i=1}^n \delta^2({\bf x}- {\bf X}_i)\, w^{(i)\dagger} T^a w^{(i)}\label{wilcons}\ee
with $w^{(i)}$ the colour degrees of freedom that we met in the previous section. These carry the information about the representation $R_i$  carried by the Wilson line

\para
Let's start by looking at the limit $k\rightarrow \infty$.  This is the weak coupling limit of the Chern-Simons theory (strictly, we need $k\gg N$) so we expect a classical analysis to be valid. However, we'll retain one element of the quantum theory: the Dirac quantisation of flux \eqn{fdirac}, now applied to each component $f_{12}^a$ individually. But, with $k$ very large, we see that it's impossible to reconcile Dirac quantisation with any non-trivial charge on the right-hand side. This means that the only way we can solve \eqn{wilcons} is if the charges on the right-hand side can somehow add up to zero. In the language of group theory, this means that we take need to decompose the tensor product of the representations $R_i$ into irreducible representations. We only get solutions to \eqn{wilcons} only if singlets appear in this decomposition. We write
\be \otimes_{i=1}^n R_i =  {\bf 1}^p \oplus\ldots\nn\ee
where $p$ is the number of singlets ${\bf 1}$ appearing in the decomposition and $\ldots$ are all the non-singlet representations. 
Each of these different decompositions gives rise to a different state in the Hilbert space ${\cal H}_{i_1\ldots i_n}$. In the weak coupling limit, we then have
\be \lim_{k\rightarrow \infty}\ {\rm dim}({\cal H}_{i_1\ldots i_n}) = p\nn\ee
Typically, when we have a large number $n$ of Wilson lines, there will be several different ways to make singlets so $p\geq 2$. 

\para
For finite $k$ when quantum effects become more important, one finds that
\be {\rm dim}({\cal H}_{i_1\ldots i_n})  \leq p\nn\ee
The possible reduction of the number of states  arises in an intuitive fashion through screening. At finite $k$, new solutions to \eqn{wilcons} exist in which the integrated flux is non-zero.  But we should sum over flux sectors in the path integral which means that these states become indistinguishable from the vacuum. This not only cuts down the dimension of the Hilbert space, but reduces the kinds of representations that we can insert to begin with. Let's illustrate this idea with some simple examples: 



\subsubsection*{An Example: $SU(2)_k$}

For $G=SU(2)$, representations are labelled by the spin $s$. Classically, of course, $s$ can take any half-integer value. There is no bound on how large the spin can be. However, at finite $k$ the spin is bounded by
\be 0 \leq s \leq \frac{k}{2}\label{su2k}\ee
The insertion of any Wilson line with spin $s>k/2$ can be screened by flux so that it is equivalent to spin $|s-k|$. 

\subsubsection*{Another Example: $SU(N)_k$.}

Let's first recall some $SU(N)$ group theory. Irreducible representations can be characterised by a Young tableau with rows of length $l_1\geq l_2\geq \ldots \geq l_{N-1}\geq 0$. In this notation, the fundamental representation ${\bf N}$ is simply a single box
 \be   \raisebox{0.5ex}{\epsfxsize=0.3in\epsfbox{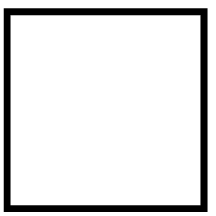}}\nn\ee
The $p^{\rm th}$ symmetric representation is a row of boxes
 \be   \raisebox{0.5ex}{\epsfxsize=1.2in\epsfbox{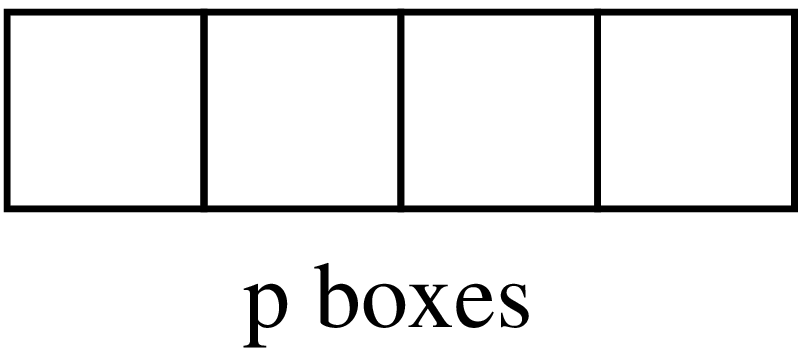}}\nn\ee
The anti-symmetric representation is a column of $p$ boxes, while the adjoint is a full column plus an extra guy stuck on the top, 
\be \raisebox{-7.3ex}{\epsfxsize=0.8in\epsfbox{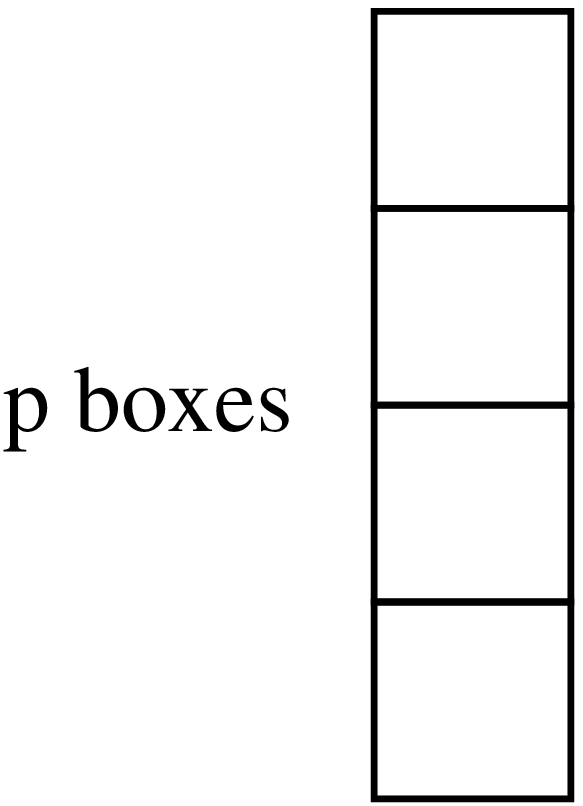}}\ \ \ \ \ \ {\rm and}\ \ \ \ \ \ \ \raisebox{-7.3ex}{\epsfxsize=1.2in\epsfbox{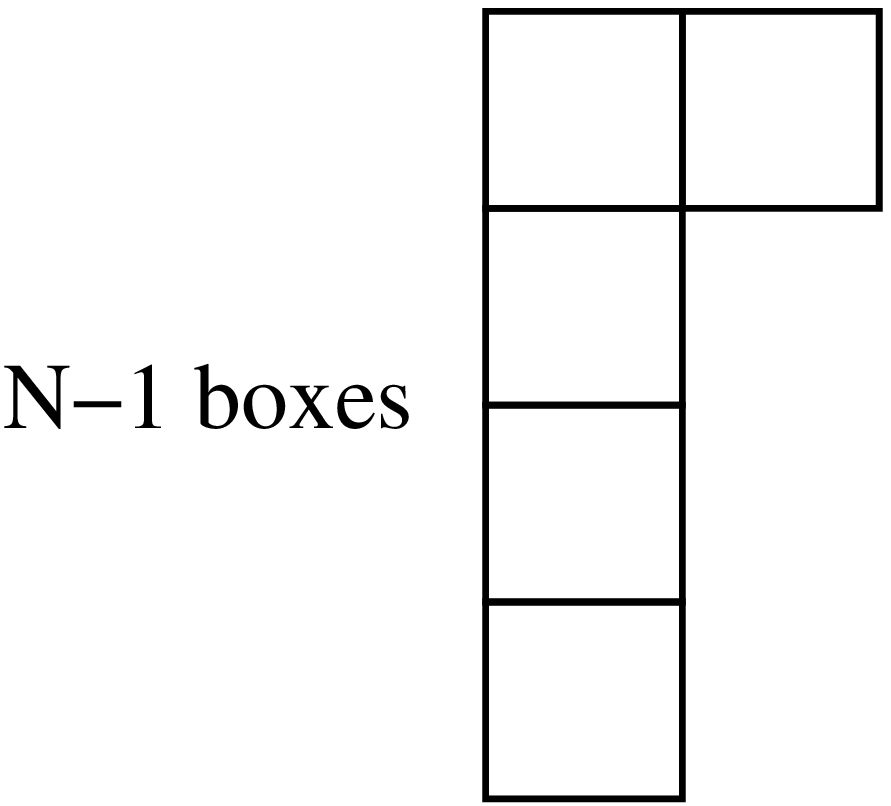}}\nn\ee
In particular, the anti-fundamental representation $\bar{\bf N}$ is the same as the $(N-1)^{\rm th}$ anti-symmetric representation. 

\para
The non-trivial Wilson lines at level $k$ are simply those with $l_1\leq k$. This means, in particular, that we can only have symmetric representations up to the $k^{\rm th}$ power of the fundamental. (This agrees with our result \eqn{su2k} for $SU(2)$). However, all anti-symmetric representations are allowed.  Most importantly, there are only a finite number of representations at any finite $k$.

\subsubsection*{Fusion Revisited}

Having specified the allowed representations, let's now return to the dimension of the Hilbert space  ${\cal H}_{i_1\ldots i_n}$. For two Wilson lines, the Hilbert space has dimension 1 if $R_1=\bar{R}_2$, so that their tensor product can form a singlet.  
The first non-trivial example arises with the insertion of three Wilson lines with representations as $R_i$, $R_j$ and $\bar{R}_k$. We'll denote the dimension of the Hilbert space as 
\be {\rm dim}({\cal H}_{ij\bar{k}}) = N_{ij}^k\nn\ee
As we described above, in the classical limit $N_{ij}^k$ is the number of times that $R_k$ appears in the tensor product of $R_i\otimes R_j$. However, it too can receive quantum corrections and, in general, $N_{ij}^k$ will be less than its classical value.

\para
There is a well-developed machinery to compute the numbers $N_{ij}^k$ in Chern-Simons theories. This involves replacing the tensor product of representations $\otimes$ with a modified operation called {\it fusion}. We will denote the fusion of two representations as $\star$. The number $N_{ij}^k$ is now the number of times that $R_k$ appears in the fusion product of $R_i\star R_j$.

\para
From knowledge of the $N_{ij}^k$, we can compute the dimension of the general Hilbert space ${\cal H}_{i_1\ldots i_n}$. It is given by
\be {\rm dim}({\cal H}_{i_1\ldots i_n}) = \sum_{j_1,\ldots,j_{n-2}} N_{i_1i_2}^{j_1}N_{j_1 i_3}^{j_2}\ldots N_{j_{n-2}i_n}^{j_{n-1}}\nn\ee
We've seen all of this before. This is the formal structure of {\it fusion} that underlies the theory of non-Abelian anyons that we described in Section \ref{nonabanyonsec}. The formula above is the same as \eqn{dimh}. In general, the Hilbert space of Wilson lines in Chern-Simons theory provides a concrete realisation of the somewhat abstract fusion rules.

\para
The fusion rules for Wilson lines in Chern-Simons theories are related to the representation theory of Kac-Moody algebras. We won't explain where these rules come from. Instead, we will just present the results\footnote{You can find all the details in the yellow  ``{\it Conformal Field Theory}" book by Di Francesco, Mathieu and S\'en\'echal.}.

\para
\underline{Fusion Rules for $SU(2)_k$}
\para

The representations of $SU(2)$ are labelled by the spin $s$ or the dimension ${\bf d} = 2s+1$. The tensor product between two representations follows from the familiar Clebsh-Gordon decomposition
\be r \otimes s = |r-s| \oplus |r-s|+1\oplus\ldots\oplus r+s\nn\ee
As we saw above, for a Chern-Simons theory $SU(2)_k$, the  spin $s$ of the representation must obey $s\leq k/2$. This means that we can't have any representations appearing on the right-hand side which are greater than $k/2$. You might think that we simply delete all representations in the tensor product that are too large. However, it turns out that the fusion rules are more subtle than that; sometimes we  need to delete some of the representations that appear to be allowed. The correct fusion rule is
\be r\star s  = |r-s|\oplus\ldots \oplus {\rm min}(k-r-s, r+s)\label{su2f}\ee
As an example, let's look at $SU(2)_2$. From \eqn{su2k}, we see that there are just three possible representations, with spin $j=0,1/2$ and $1$. We'll label these representations by their dimension, ${\bf 1},{\bf 2}$ and ${\bf 3}$. The fusion rules \eqn{su2f} in this case are

\be {\bf 2}\star {\bf 2} = {\bf 1} \oplus {\bf 3}  \ \ \ ,\ \ \ {\bf 2}\star {\bf 3} = {\bf 2} \ \ \ ,\ \ \ {\bf 3}\star {\bf 3} = {\bf 1}\label{su22}\ee
%
%
%
Note that the first two of these follow from standard Clebsh-Gordan coefficients, throwing out any spins greater than $1$. However, the final product does not have the representation ${\bf 3}$ on the right-hand side which one might expect.  We've seen the fusion rules \eqn{su22} before: they are identical to the fusion of Ising anyons \eqn{ising} with the identification
\be {\bf 2} \rightarrow \sigma\  \ \ {\rm and}\ \ \ {\bf 3}\rightarrow \psi\nn\ee
 Recall that these describe the anyonic excitations of the Moore-Read state. Similarly, one can check how many singlets you can form from $n$ spin-1/2 with the requirement that no group has spin greater than 1. The answer, for $n$ even, is $2^{n/2-1}$. We recognise this as the dimension of the Hilbert space of $n$ Ising anyons.  This leads us to suspect that the $SU(2)_2$ Chern-Simons theory plays some role in the description of the $\nu=5/2$ quantum Hall state. We'll look at this in more detail shortly. 

\para
\underline{Fusion Rules for $SU(N)_k$}
\para

For $SU(N)_k$, the fusion rules are simplest to explain using Young diagrams. However, like many aspects of Young diagrams, if you don't explain where the rules come from then they appear totally mysterious and arbitrary, like a weird  cross between sudoku and tetris. Here we're not going to explain. We're just going to have to put up with the mystery\footnote{A simple mathematica package to compute fusion rules, written by Carl Turner, can be found at \href{http://blog.suchideas.com/2016/03/computing-wzw-fusion-rules-in-mathematica/}{http://blog.suchideas.com/2016/03/computing-wzw-fusion-rules-in-mathematica/}}. 

\para
We start by writing down the usual tensor product of representations. For each representation on the right-hand side, we draw the corresponding Young diagram and define
\be t = l_1 - k-1\nn\ee
where, as before, $l_1$ is the length of the first row. Now we do one of three things, depending on the value of $t$.
\begin{itemize}
\item $t<0$: Keep this diagram.
\item $t=0$: Throw this diagram away.
\item $t>0$: Play. First, we remove a boundary strip of $t$ boxes, starting from the end of the first row and moving downwards and left. Next, we add a boundary strip of $t$ boxes, starting at the bottom of the first column and moving up and right. 

If the resulting Young diagram does not correspond to a representation of $SU(N)$, we throw it away. Otherwise, we repeat until the resulting diagram has $t\leq 0$. If $t=0$, we again throw it away. However, if $t<0$ then we keep it on the right-hand side, but with a sign given by
\be (-1)^{r_-+r_++1}\label{minussign}\ee
where $r_-$ is the number of columns from which boxes were removed,  while $r_+$ is the number of columns which had boxes added.
\end{itemize}

\subsubsection*{An Example: $SU(2)_2$ Again}

This probably sounds a little baffling. Let's first see how these rules reproduce what we saw for $SU(2)$. We'll consider $SU(2)_2$ which, as we saw, has representations ${\bf 1}$, ${\bf 2}$ and ${\bf 3}$. In terms of Young diagrams, these are $1$, $\Box$ and $\Box\!\Box$. Let's look at some tensor products. The first is 
\be {\bf 2}\otimes {\bf 2} = 1 \oplus {\bf 3} \ \ \ &\Rightarrow&\ \ \ \raisebox{-0.7ex}{\epsfxsize=0.2in\epsfbox{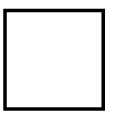}} \otimes \raisebox{-0.7ex}{\epsfxsize=0.2in\epsfbox{y1.eps}} = 1\oplus \raisebox{-0.7ex}{\epsfxsize=0.36in\epsfbox{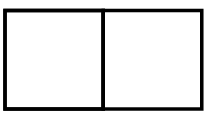}}\nn\ee
Both boxes on the right-hand side have $t<0$ so remain. In this case, the fusion rules are the same as the tensor product: ${\bf 2}\star{\bf 2} = 1\oplus{\bf 3}$. The next tensor product is
\be
{\bf 2}\otimes {\bf 3} = {\bf 2} \oplus {\bf 4} \ \ \ &\Rightarrow&\ \ \ \raisebox{-0.7ex}{\epsfxsize=0.2in\epsfbox{y1.eps}} \otimes \raisebox{-0.7ex}{\epsfxsize=0.36in\epsfbox{y2.eps}} = \raisebox{-0.7ex}{\epsfxsize=0.2in\epsfbox{y1.eps}} \oplus \raisebox{-0.7ex}{\epsfxsize=0.56in\epsfbox{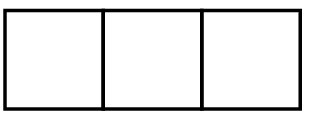}}\nn\ee
In this case, the final diagram is not an allowed representation of $SU(2)_2$. It has $t=3-2-1=0$ so we simply discard this diagram. We're left with the fusion rule ${\bf 2}\star{\bf 3} = {\bf 2}$. The final tensor product is 
\be
{\bf 3}\otimes {\bf 3} = 1 \oplus {\bf 3} \oplus {\bf 5} \ \ \ &\Rightarrow&\ \ \ \raisebox{-0.7ex}{\epsfxsize=0.36in\epsfbox{y2.eps}} \otimes \raisebox{-0.7ex}{\epsfxsize=0.36in\epsfbox{y2.eps}} = 1\oplus \raisebox{-0.7ex}{\epsfxsize=0.36in\epsfbox{y2.eps}}
\oplus \raisebox{-0.7ex}{\epsfxsize=0.69in\epsfbox{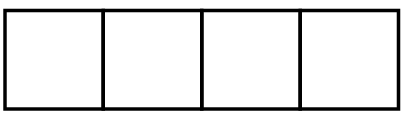}}
\label{wowee}\ee
The first two diagrams have $t<0$ and we leave them be. But the third has $l_1=4$ and so $t=1$. This means we can play. We remove a single box from the far right-hand end and replace it below the first box on the left: 
\be \raisebox{-0.7ex}{\epsfxsize=0.69in\epsfbox{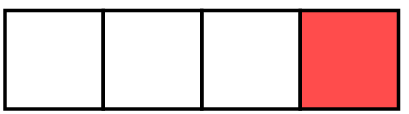}} \ \longrightarrow\ \raisebox{-2.7ex}{\epsfxsize=0.59in\epsfbox{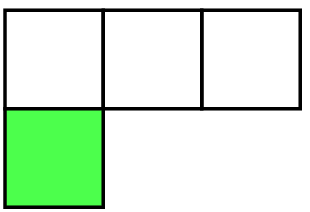}}\nn\ee
But a column of 2 boxes can be removed in $SU(2)$ Young diagrams. So the full result is
\be \raisebox{-0.7ex}{\epsfxsize=0.69in\epsfbox{y11.eps}} \ \longrightarrow\ \raisebox{-2.7ex}{\epsfxsize=0.59in\epsfbox{y12.eps}} \longrightarrow\raisebox{-0.7ex}{\epsfxsize=0.36in\epsfbox{y2.eps}} \nn\ee
This is another ${\bf 3}$ representation. But we should worry about the sign. The red box covers a single column, so $r_-=1$, while the green box also covers a single column so $r_+=1$. This means that this diagram comes with a sign $-1$. This cancels off the $\Box\!\Box$ that appeared on the right-hand side of \eqn{wowee}. This final result is  ${\bf 3}\star{\bf 3} = 1$. In this way, we see that our rules for manipulating Young diagram reproduce the $SU(2)_2$ fusion rules for Ising anyons \eqn{su22} that we introduced previously.

\subsubsection*{Another Example: $SU(3)_2$}

Let's now look another example. We choose $SU(3)_2$.  The allowed representations are ${\bf 3}= \raisebox{-0.2ex}{\epsfxsize=0.12in\epsfbox{y1.eps}}$, $\overline{\bf 3}= \raisebox{-0.7ex}{\epsfxsize=0.12in\epsfbox{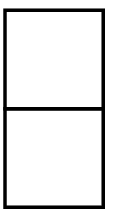}}$, ${\bf 6}= \raisebox{-0.2ex}{\epsfxsize=0.22in\epsfbox{y2.eps}}$, $\overline{\bf 6}= \raisebox{-0.7ex}{\epsfxsize=0.21in\epsfbox{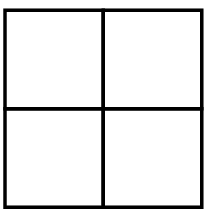}}$ and ${\bf 8} = \raisebox{-0.7ex}{\epsfxsize=0.21in\epsfbox{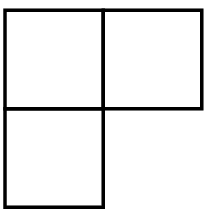}}$. Let's look at a simple example. The tensor product of two symmetric representations is
\be {\bf 6}\otimes {\bf 6} = \overline{\bf 6}\oplus {\bf 15} \oplus \overline{\bf 15} 
\ \ \ &\Rightarrow&\ \ \ \raisebox{-0.7ex}{\epsfxsize=0.36in\epsfbox{y2.eps}}\otimes \raisebox{-0.7ex}{\epsfxsize=0.36in\epsfbox{y2.eps}} =  \raisebox{-1.5ex}{\epsfxsize=0.36in\epsfbox{y6.eps}} \oplus  \raisebox{-0.7ex}{\epsfxsize=0.69in\epsfbox{y5.eps}} \oplus \raisebox{-1.5ex}{\epsfxsize=0.5in\epsfbox{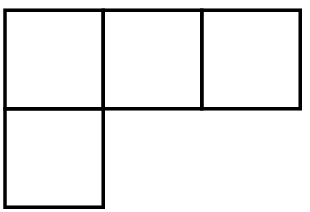}}
\nn\ee
The first of these diagrams has $t<0$. We keep it. The last of these diagrams has $t=0$. We discard it. More interesting is the middle diagram which has $t=1$. This we play with. We have the same manipulations that we saw in the $SU(2)_2$ case above, 
\be \raisebox{-0.7ex}{\epsfxsize=0.69in\epsfbox{y11.eps}} \ \longrightarrow\ \raisebox{-2.7ex}{\epsfxsize=0.59in\epsfbox{y12.eps}}\nn\ee
However, this time the two boxes in a single column don't cancel because we're dealing with $SU(3)$ rather than $SU(2)$. In fact, as we have seen, this diagram has $t=0$. We should just discard it. The upshot is that the fusion rules are simply
\be {\bf 6}\star{\bf 6} = \overline{\bf 6} \ \ \ &\Rightarrow&\ \ \ \raisebox{-0.7ex}{\epsfxsize=0.36in\epsfbox{y2.eps}}\otimes \raisebox{-0.7ex}{\epsfxsize=0.36in\epsfbox{y2.eps}} =  \raisebox{-1.5ex}{\epsfxsize=0.36in\epsfbox{y6.eps}}
\nn\ee
Let's look at another example. The tensor product for two adjoints is 
\be {\bf 8}\otimes {\bf 8} = \overline{\bf 1}\oplus {\bf 8} \oplus {\bf 8} \oplus{\bf 10} \oplus \overline{\bf 10} \oplus {\bf 27}\nn\ee
which, in diagrams, reads
\be
 \raisebox{-1.5ex}{\epsfxsize=0.36in\epsfbox{y8.eps}}\otimes \raisebox{-1.5ex}{\epsfxsize=0.36in\epsfbox{y8.eps}} = 1\oplus \raisebox{-1.5ex}{\epsfxsize=0.36in\epsfbox{y8.eps}} \oplus \raisebox{-1.5ex}{\epsfxsize=0.36in\epsfbox{y8.eps}}  \oplus \raisebox{-0.7ex}{\epsfxsize=0.56in\epsfbox{y3.eps}}  \oplus  \raisebox{-1.5ex}{\epsfxsize=0.50in\epsfbox{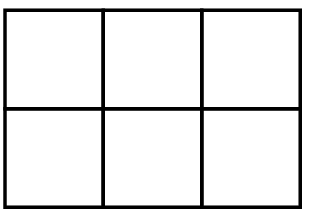}} \oplus \raisebox{-1.5ex}{\epsfxsize=0.7in\epsfbox{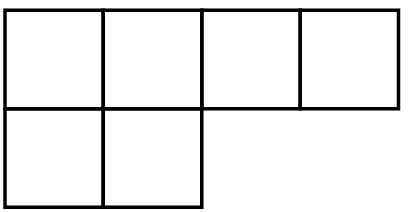}}
\nn\ee
The first three diagrams we keep. The ${\bf 10}$ and $\overline{\bf 10}$ diagrams have $t=0$ and we discard. This leaves us only with the final ${\bf 27}$ diagram. This we play with. Using the rules above, we have
\be \raisebox{-1.7ex}{\epsfxsize=0.69in\epsfbox{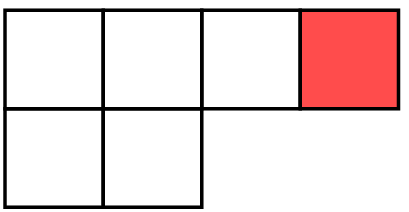}} \ \longrightarrow\  - \ \raisebox{-4.7ex}{\epsfxsize=0.59in\epsfbox{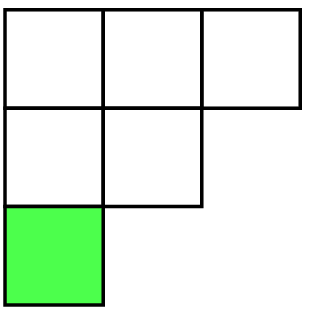}} \ \longrightarrow\  -\ \raisebox{-2.1ex}{\epsfxsize=0.4in\epsfbox{y8.eps}} \nn\ee
where we've now included the minus sign \eqn{minussign} in this expression, and the final step comes from removing the column of three boxes. The net result is that the ${\bf 27}$ diagram cancels one of the ${\bf 8}$ diagrams in the tensor product. We're left with the $SU(3)_2$ fusion rule
\be {\bf 8}\star{\bf 8} = {\bf 1} \oplus{\bf 8} \ \ \ \Rightarrow\ \ \  \raisebox{-1.5ex}{\epsfxsize=0.36in\epsfbox{y8.eps}}\otimes \raisebox{-1.5ex}{\epsfxsize=0.36in\epsfbox{y8.eps}} = 1\oplus \raisebox{-1.5ex}{\epsfxsize=0.36in\epsfbox{y8.eps}}\nn\ee
We recognise this as the fusion rule for Fibonacci anyons \eqn{fibnfuse}. This means that the adjoint Wilson lines in $SU(3)_2$ Chern-Simons theory acts like Fibonacci anyons.

\subsubsection*{Braiding Revisited}

We've seen above that Wilson lines in  non-Abelian Chern-Simons theories provide an arena to describe non-Abelian anyons. There is a finite dimensional Hilbert space arising from a process of fusion. The next step is obviously to understand braiding in this framework. The adiabatic motion of one Wilson line around another will give rise to a unitary operator on the Hilbert space. How can we calculate this?

\para
There is a long and beautiful story behind this which we will not describe here. The essence of this story is that the action of braiding on the Hilbert space can be translated into the computation of Wilson lines on ${\bf S}^3$, 
\be \langle W_R \rangle = \int {\cal D}a\ e^{iS_{CS}} W_R[a]\nn\ee
where $R$ describes the representation of the Wilson line which now traces out some closed, non-intersecting path $\gamma$ in ${\bf S}^3$. In general, such a path describes a tangled path known as a {\it knot}. Witten famously showed that the expectation value of the Wilson line provides an invariant to distinguish different knots. For $G=SU(2)$, with $R$ the fundamental representation, this invariant is the {\it Jones Polynomial}.

\subsubsection{Effective Theories of Non-Abelian Quantum Hall States}

It is clear that non-Abelian Chern-Simons theories give rise to non-Abelian anyons. Indeed, as we mentioned above, for $SU(2)_2$, the structure of anyons that arise is identical to the Ising anyons that describe the Moore-Read states. It's therefore very natural to think they provide effective field theories for the non-Abelian quantum Hall states. And this turns out to be correct. One can argue\footnote{The argument can be found in ``{\it A Chern-Simons effective field theory for the Pfaffian quantum Hall state}" by E. Fradkin, C. Nayak, A. Tsvelik and F. Wilczek, Nucl.Phys. {\bf B516} 3 704 (1998), \href{http://arxiv.org/abs/cond-mat/9711087}{cond-mat/9711087}.}  that the $SU(2)_2$ theory effectively  captures the braiding of anyons in the bosonic Moore-Read state at $\nu=1$. 

\para
However, the full description is somewhat involved. One very basic problem is as follows: to construct the full low-energy theory one should identify the electromagnetic current which couples to the background field $A_\mu$. And here gauge invariance works against us. The kind of trick that we used in the Abelian theory is not available for the non-Abelian theory since $\ep f_{\nu\rho}$ is not gauge invariant, while $\ep \tr f_{\nu\rho}=0$.

\para
The way to proceed is to look at $U(N) = U(1)\times SU(N)/{\bf Z}_N$ Chern-Simons theories. The background gauge field can easily couple to the $U(1)$ factor but we then need the $U(1)$ factor to couple to the rest of $SU(N)$ somehow. This is the part which is a little involved: it requires some discrete identifications of the allowed Wilson lines in a way which is compatible with gauge invariance\footnote{To my knowledge, this was first explained in Appendix C of the paper by Nati Seiberg and Edward Witten, ``{\it Gapped Boundary Phases of Topological Insulators at Weak Coupling}", \href{https://arxiv.org/abs/1602.04251}{arXiv:1602.0425}.}.

\para
However, the Chern-Simons theories also provide us with another way to look at quantum Hall states since these theories are intimately connected to $d=1+1$ dimensional conformal field theories. And it will turn out that these conformal field theories also capture many of the interesting aspects of quantum Hall physics. In our final section, we will look at this for some simple examples.

\newpage
\section{Edge Modes}\label{edgesec}

If a quantum Hall fluid is confined to a finite region, there will be gapless modes that live on the edge. We've already met these in Section \ref{llcondsec} for the integer quantum Hall states where we noticed that they are chiral: they propagate only in one direction. This is a key property shared by all edge modes.

\para
In this section we'll describe the edge modes for the fractional quantum Hall states. At first glance it may seem like this is quite an esoteric part of the  story. However, there's a surprise in store. The edge modes know much more about the quantum Hall states than you might naively imagine. Not only do they offer the best characterisation of these states, but they also provide a link between the Chern-Simons approach and the microscopic wavefunctions.

\subsection{Laughlin States}\label{ledgesec}

We start by looking at edge modes in the  $\nu=1/m$ Laughlin states. The basic idea is that the ground state forms an incompressible disc. The low-energy excitations of this state are deformations which change its shape, but not its area. These travel as waves around the edge of the disc, only in one direction.  In what follows, we will see this picture emerging from several different points of view. 

\subsubsection{The View from the Wavefunction}\label{chiralwfsec}

Let's first return to the description of the quantum Hall state in terms of the microscopic wavefunction. 
Recall that when we were discussing the toy Hamiltonians in Section \ref{toysec}, the Hamiltonian $H_{\rm toy}$ that we cooked up in \eqn{tinker} had the property that the zero energy ground states are
\be \psi(z_i) = s(z_i)\prod_{i<j} (z_i-z_j)^m\,e^{-\sum_i|z_i|^2/4l_B^2}\label{spoly}\ee
for any symmetric polynomial $s(z_i)$. The Laughlin wavefunction with $s(z_i)=1$ has the property that it is the most compact of these states. Equivalently, it is the state with the lowest angular momentum. We can pick this out as the unique ground state by adding a placing the system in a confining potential which we take to be the angular momentum operator $J$,
\be V_{\rm confining} =  \omega J \nn\ee
The Laughlin state, with $s(z_i)=1$, then has ground state energy
\be E_0 = \frac{\omega}{2}mN(N-1)\nn\ee
where $N$ is the number of electrons. What about the excited states? We can write down a basis of symmetric polynomials
\be s_n(z_i) = \sum_i z_i^n\nn\ee
The most general state \eqn{spoly} has polynomial
\be s(z_i) = \sum_{n=1}^\infty s_n(z_i)^{d_n}\nn\ee
which has energy
\be E = E_0 + \omega\sum_{n=1}^\infty n d_n\label{sspec}\ee
We see that each  polynomial $s_n$ contributes an energy 
\be E_n = \omega n\nn\ee
We're going to give an interpretation for this. Here we'll simply pull the interpretation out of thin air, but we'll spend the next couple of sections providing a more direct derivation. The idea is to interpret this as the Kaluza-Klein  spectrum as a gapless $d=1+1$ scalar field. We'll think of this scalar as living on the edge of the quantum Hall droplet. Recall that the Laughlin state has area $A= 2\pi m N l_B^2$ which means that the boundary is a circle of circumference  $L=  2\pi\sqrt{2mN}l_B$. The Fourier modes of such a scalar field have energies
\be E_n = \frac{2\pi n v}{L}\nn\ee
where $v$ is the speed of propagation the excitations. (Note: don't worry if this formula is unfamiliar: we'll derive it below). Comparing the two formulae, we see that the speed of propagation depends on the strength of the confining potential,
\be v = \frac{L\omega}{2\pi}\nn\ee
To see that this is a good interpretation of the spectrum \eqn{sspec}, we should also check that the degeneracies match. There's a nice formula for the number of quantum Hall states with energy $q\omega$ with $q\in {\bf Z}^+$. To see this, let's look at some examples. 
 There is, of course, a unique ground state. There is also a unique state with $\Delta E=\omega$ which has $d_1=1$ and $d_n=0$ for $n\geq 2$. However, for $\Delta E = 2\omega$ there are two states: $d_1=2$ or $d_2=1$. And for $\Delta E = 3\omega$ there are 3 states: $d_1=3$, or $d_1=1$ and $d_2=2$, or $d_3=1$. In general, the number of states at energy $\Delta E = q\omega$ is the number of {\it partitions} of the integer $q$. This is the number of ways of writing $q$ as a sum of positive integers. It is usually denoted as $P(q)$,
\be \left. \begin{array}{c} {\rm Degeneracy\ of\ states} \\ {\rm with}\ \Delta E = a\omega \end{array}\right\} \ =\  P(q)\label{partq}\ee
Now let's compare this to the Fourier modes of a scalar field. Suppose that we focus on the modes that only move one way around the circle, labelled by the momenta $n >0$. Then there's one way to create a state with energy $E=2\pi v/L$: we excite the first Fourier mode once. There are two ways to create a state with energies $E= 4\pi v/L$: we excite the first Fourier mode twice, or we excite the second Fourier mode once. And so on. What we're seeing is that the degeneracies match the quantum Hall result \eqn{partq} if we restrict the momenta to be positive. If we allowed the momenta to also be positive, we would not get the correct degeneracy of the spectrum. 
This is our first hint that the edge modes are described by a chiral scalar field, propagating only in one direction. 

\subsubsection{The View from Chern-Simons Theory}\label{csboundsec}

Let's see how this plays out in the effective Chern-Simons theory. We saw in Section \ref{fraccssec} that the low-energy effective action for the Laughlin state is
\be S_{CS}[a] = \frac{m}{4\pi} \int d^3x\ \epsilon^{\mu\nu\rho}a_\mu \partial_{\nu}a_{\rho}\label{csup}\ee
where we're working in units in which $e=\hbar=1$. 

\EPSFIGURE{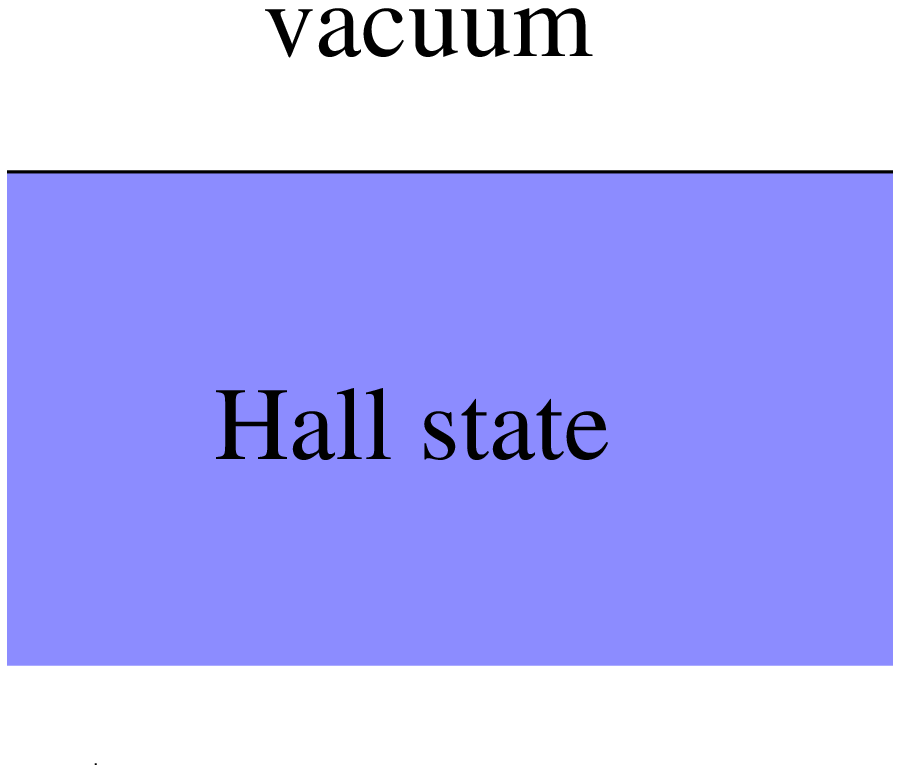,height=100pt}{}
\para
We'll now think about this action on a manifold with boundary. Ultimately we'll be interested in a disc-shaped quantum Hall droplet. But to get started it's simplest to think of the boundary as a straight line which we'll take to be at $y=0$. The quantum Hall droplet lies at $y<0$ while at $y>0$ there is only the vacuum. 

\para
There are a number of things to worry about in the presence of a boundary. The first is important for any field theory. When we derive the equations of motion from the action, we always integrate by parts and discard the boundary term. But now there's a boundary, we have to be more careful to make sure that this term vanishes. This is simply telling us that we should specify some boundary condition if we want to make the theory well defined. For our Chern-Simons theory, a variation of the fields gives
\be \delta S_{CS}  &=& \frac{m}{4\pi}\int d^3x\ \ep \left[ \delta a_\mu\partial_\nu a_\rho +  a_\mu \partial_\nu \delta a_\rho\right] \nn\\  &=& \frac{m}{4\pi}\int d^3x\ \ep \left[ \delta a_\mu f_{\nu\rho} +  \partial_\mu (a_\nu \delta a_\rho)\right]\nn\ee
Minimising the action gives the required equation of motion $f_{\mu\nu}=0$ only if we can set the last term to zero. We can do this if either by  setting  $a_t(y=0)=0$ on the boundary, or by setting $a_x(y=0)=0$. Alternatively, we can take a linear combination of these. We choose
\be (a_t - v a_x) \Big|_{y=0} = 0\label{abound}\ee
Here we've introduced a parameter $v$; this will turn out to be the velocity of excitations on the boundary. Note that the Chern-Simons theory alone has no knowledge of this speed. It's something that we have to put in by hand through the boundary condition. 

\para
The next issue is specific to Chern-Simons theory. 
As we've mentioned before, the action \eqn{csup} is only invariant up to a total derivative. Under a gauge transformation
\be a_\mu \rightarrow a_\mu + \partial_\mu \omega\nn\ee
we have
\be S_{CS}\rightarrow S_{CS} + \frac{m}{4\pi}\int_{y=0} dxdt\ \omega (\partial_t a_x - \partial_x a_t)\nn\ee
and the Chern-Simons action is no longer gauge invariant. We're going to have to deal with this. 
One obvious strategy is simply to insist that we only take gauge transformations that vanish on the boundary, so that $w(y=0)=0$. This has the happy corrolary that gauge transformations don't change our chosen boundary condition for the gauge fields. However, this approach has other consequences. Recall that the role of gauge transformations is to identify field configurations, ensuring that they are physically indistinguishable. Said another way, gauge transformations kill would-be degrees of freedom. This means that restricting the kinds of gauge transformations will resurrect some these degrees of freedom from the dead. 

\para
To derive an action for these degrees of freedom, we choose a gauge. The obvious one is to extend the boundary condition \eqn{abound} into the bulk, requiring that 
\be a_t - va_x = 0\label{gfv}\ee
everywhere. The easiest way to deal with this is to work in new coordinates
\be t' = t \ \ \ ,\ \ \ x' = x+vt \ \ \ ,\ \ \ y'=y\label{cprime}\ee
The Chern-Simons action is topological and so  invariant under such coordinate transformations if we also transform the gauge fields as
\be a'_{t'} = a_t-va_x\ \ \ ,\ \ \ a'_{x'} = a_x \ \ \ ,\ \ \ a'_{y'} = a_y\label{aprime}\ee
so the gauge fixing condition \eqn{gfv} becomes simply
\be a'_{t'}=0\label{tprime0}\ee
But now this is easy to deal with. The constraint imposed by the gauge fixing condition is simply $f'_{x'y'}=0$. Solutions to this are simply
\be a'_i = \partial_i\phi\nn\ee
with $ i=x',y'$. Of course, usually such solutions would be pure gauge. But that's what we wanted: a mode that was pure gauge which becomes dynamical. To see how this happens, we simply need to insert this solution back into the Chern-Simons action which, having set $a'_{t'}=0$, is 
\be S_{CS} &=& \frac{m}{4\pi}\int d^3x'\ \epsilon^{ij} a'_i\partial_{t'} a'_j \nn\\ &=&\frac{m}{4\pi}\int d^3x'\ \partial_{x'}\phi\, \partial_{t'}\partial_{y'}\phi - \partial_{y'}\phi\,\partial_{t'}\partial_{x'}\phi \nn\\ &=& \frac{m}{4\pi}\int_{y=0} d^2x'\ \partial_{t'}\phi\partial_{x'}\phi
\nn\ee
Writing this in terms of our original coordinates, we have
\be S = \frac{m}{4\pi}\int d^2x\ \partial_t\phi\partial_x\phi - v(\partial_x\phi)^2\label{jackyboy}\ee
This is sometimes called the Floreanini-Jackiw action. It looks slightly unusual, but it actually describes something very straightforward. The equations of motion are
\be \partial_t\partial_x\phi - v\partial_x^2\phi=0\label{funjacky}\ee
If we define a new field, 
\be \rho = \frac{1}{2\pi}\ppp{\phi}{x}\nn\ee
then the equation of motion is simply
\be \partial_t \rho(x,t) - v\partial_x \rho(x,t) =0\label{chiralwave1}\ee
This is the expression for a chiral wave propagating at speed $v$. The equation has solutions of the form $\rho(x+ vt)$. However, waves propagating in the other direction, described by $\rho(x-vt)$ are not solutions. 
The upshot of this analysis is that the $U(1)$ Chern-Simons theory has a chiral scalar field living on the boundary. This, of course, is the same conclusion that we came to by studying the excitations above the Laughlin state.

\subsubsection*{The Interpretation of $\rho$}

There's a nice physical interpretation of the chiral field $\rho$. To see this, recall that our Chern-Simons theory is coupled to a background gauge field $A_\mu$ through the coupling
\be S_J= \int d^3x\ A_\mu J^\mu = \frac{1}{2\pi}\int d^3x\ \ep A_\mu \partial_\nu a_\rho 
\nn\ee
This is invariant under gauge transformations of $a_\mu$ but, in the presence of a boundary, is not gauge invariant under transformations of $A_\mu$. That's not acceptable. While $a_\mu$ is an emergent gauge field, which only exists within the sample, $A_\mu$ is electromagnetism. It doesn't stop just because the sample stops and there's no reason that we should only consider electromagnetic gauge transformations that vanish on the boundary. However, there's a simple fix to this. We integrate the expression by parts and throw away the boundary term. We then get the subtly different coupling
\be S_J= \frac{1}{2\pi}\int d^3x\ \ep a_\mu \partial_\nu A_\rho \nn\ee
This is now invariant under electromagnetic gauge transformations and, as we saw above, under the restricted gauge transformations of $a_\mu$.  This is the correct way to couple electromagnetism in the presence of a boundary.

\para
 We'll set $A_y=0$ and turn on background fields $A_t$ and $A_x$, both of which are independent of the $y$ direction. Then,  working in the coordinate system \eqn{cprime}, \eqn{aprime}, and the gauge \eqn{tprime0}, the coupling becomes
\be
S_J&=& \frac{1}{2\pi}\int d^3x\ a'_{y'} (\partial_{t'}A'_{x'} - \partial_{x'}A'_{t'})\nn\\ &=& \frac{1}{2\pi}\int d^3x\ \partial_{y'}\phi (\partial_{t'}A'_{x'} - \partial_{x'}A'_{t'}) \nn\\ &=& \frac{1}{2\pi}\int_{y=0} d^2x\ \phi (\partial_{t'}A'_{x'} - \partial_{x'}A'_{t'})  \nn\ee
Integrating the first term by parts gives $\partial_{t'}\phi = \partial_t\phi-v\partial_x\phi$. (Recall that $\partial_{t'}$ transforms like $a'_{t'}$ and so is not the same thing as $\partial_t$). But this vanishes or, at least, is a constant  by the equation of motion \eqn{funjacky}. We'll set this term to zero. We're left with
\be S_J = \frac{1}{2\pi} \int_{y=0} dtdx \ (A_t-vA_x)\partial_x \phi \nn\ee
The coupling to $A_t$ tells us that the field 
\be \rho = \frac{1}{2\pi}\ppp{\phi}{x}\nn\ee
 is the charge density along the boundary. The coupling to $A_x$ tells us that $-v\rho$ also has the interpretation as the current. The same object is both charge density and current reflects the fact that the waves propagate in a chiral manner with speed $v$. The current is conserved by virtue of the chiral wave equation \eqn{chiralwave1}

\EPSFIGURE{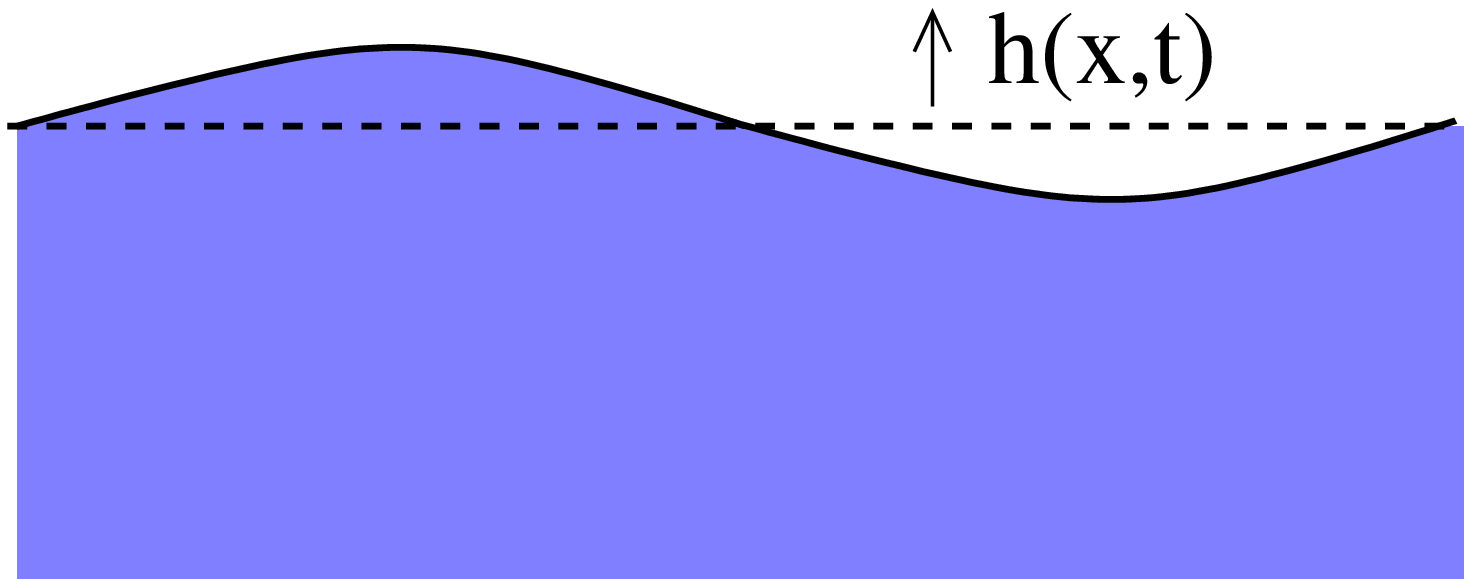,height=60pt}{}
\para
There is a simple intuitive way to think about $\rho$. Consider the edge of the boundary as shown in the figure. The excitations that we're considering  are waves in which the boundary deviates from a straight line. If the height of these waves is $h(x,t)$, then the charge density is $\rho(x,t) = nh(x,t)$ where $n=1/2\pi m l_B^2$ is the density of the Laughlin state at filling fraction $\nu=1/m$. 

\subsubsection*{Towards an Interpretation of $\phi$}

There's one important property  of $\phi$ that we haven't mentioned until now: it's periodic. This follows because the emergent gauge $U(1)$ gauge group is compact. When we write the flat connection $a_\mu = \partial_\mu \phi$, what we really mean is 
\be a_\mu = ig^{-1}\partial_\mu g\ \ \ {\rm with}\ \ \ g=e^{-i\phi}\nn\ee
This tells us that $\phi$ should be thought of as a scalar with period $2\pi$. It is sometimes called a {\it compact} boson. 

\para
As an aside: sometimes in the literature, people work with the rescaled field $\phi \rightarrow  \sqrt{m}\phi$. This choice is made so that the normalisation of the action \eqn{jackyboy} becomes $1/2\pi$ for all filling fractions. The price that's paid is that the periodicity of the boson becomes $2\pi\sqrt{m}$. In these lectures, we'll work with the normalisation \eqn{jackyboy} in which $\phi$ has period $2\pi$. 

\para
This possibility allows us to capture some new physics. Consider the more realistic situation where the quantum Hall fluid forms a disc and the boundary is a circle ${\bf S}^1$ of circumference $L$. We'll denote the coordinate around the boundary as $\sigma \in [0,L)$. The total charge on the boundary is
\be Q = \int_0^L d\sigma\ \rho =  \frac{1}{2\pi}\int_0^L d\sigma\ \ppp{\phi}{\sigma}\label{qwind}\ee
It's tempting to say that this vanishes because it's the integral of a total derivative. But if $\phi$ is compact, that's no longer true. We have the possibility that $\phi$ winds some number of times as we go around the circle. For example, the configuration $\phi =  2\pi p\sigma/L$ is single valued for any integer $p$. Evaluated on this configuration, the charge on the boundary is $Q=p$. Happily, the charge is quantised even though we haven't needed to invoke quantum mechanics anywhere: it's quantised for topological reasons.

\para
Although we've introduced $Q$ as the charge on the boundary, it's really capturing the charge in the bulk. This is simply because the quantum Hall fluid is incompressible. If you add $p$ electrons to the system, the boundary has to swell a little bit. That's what $Q$ is measuring. This is our first hint that the boundary knows about things that happen in the bulk.

\para
There's one other lesson to take from the compact nature of $\phi$. Observables should be single valued. This means that $\phi$ itself is not something we can measure. One way around this is to look at $\partial_x\phi$ which, as we have seen, gives the charge density. However, one could also consider the exponential operators $e^{i\phi}$. What is the interpretation of these? We will answer this in Section \ref{compactqhsec} where we will see that $e^{i\phi}$ describes quasi-holes in the boundary theory.

\subsubsection{The Chiral Boson}

We've seen that the edge modes of the quantum Hall fluid are described by a chiral wave. From now on, we'll think of the quantum Hall droplet as forming a disc, with the boundary a circle of circumference $L = 2\pi \sqrt{2m Nl_B}$. We'll parameterise the circle by $\sigma\in [0,L)$. The chiral wave equation obeyed by the density is 
\be \partial_t \rho(\sigma,t) - v\partial_\sigma \rho(\sigma,t) =0\label{chiralwave}\ee
which, as we've seen, arises from the action for a field 
\be S = \frac{m}{4\pi} \int_{{\bf R}\times {\bf S}^1} dtd\sigma\ \partial_t\phi\,\partial_\sigma\phi - v(\partial_\sigma\phi)^2\label{fjak}\ee
The original charge density is related to $\phi$ by  $\rho = \partial_\sigma\phi/2\pi$. 

\para
In this section, our goal is to quantise this theory.  It's clear from \eqn{fjak} that the momentum conjugate to $\phi$ is proportional to $\partial_\sigma\phi$. If you just naively go ahead and write down canonical commutation relations then there's an annoying factor of 2 that you'll get wrong, arising from the fact that there is a constraint on phase space. To avoid this, the simplest thing to do is to work 
with Fourier modes in what follows. Because these modes live on a circle of circumference $L$, we can write
\be \phi(\sigma,t) = \frac{1}{\sqrt{L}} \sum_{n=-\infty}^\infty \phi_n(t)\, e^{2\pi i n\sigma/L}\nn\ee
and
\be \rho(\sigma,t) = \frac{1}{\sqrt{L}} \sum_{n=-\infty}^\infty \rho_n(t)\, e^{2\pi i n\sigma/L}\nn\ee
The Fourier modes are related by 
\be \rho_n = \frac{ik_n}{2\pi} \phi_n\nn\ee
with $k_n$ the momentum carried by the $n^{\rm th}$ Fourier mode given by
\be k_n= \frac{2\pi n}{L}\nn\ee
The  condition on $\phi$ and $\rho$ means that $\phi_n^\star = \phi_{-n}$ and $\rho_n^\star = \rho_{-n}$.  Note that the zero mode $\rho_0$ vanishes according to this formula. This reflects the fact that the corresponding zero mode $\phi_0$  decouples from the dynamics since the action is written using $\partial_\sigma \phi$. The correct treatment of this zero mode is rather subtle. In what follows, we will simply ignore it and set $\phi_0=0$. Using these Fourier modes, the action \eqn{fjak} becomes
\be S &=& \frac{m}{4\pi}\int dt \sum_{n=-\infty}^\infty \left( i k_{-n} \dot{\phi}_n\phi_{-n} + v k_n k_{-n} \phi_n\phi_{-n} \right) \nn\\ &=& -\frac{m}{2\pi} \int dt\sum_{n=0}^\infty \left( ik_n\dot{\phi}_n\phi_{-n} + vk_n^2 \phi_n\phi_{-n}\right)\nn\ee
This final expression suggests that we treat the Fourier modes $\phi_n$ with $n>0$ as the ``coordinates" of the problem. The momenta conjugate  to $\phi_n$ is then proportional to $\phi_{-n}$. This gives us the Poisson bracket structure for the theory or, passing to quantum mechanics, the commutators
\be [\phi_n,\phi_{n'}] &=& \frac{2\pi}{m} \frac{1}{k_n} \delta_{n+n'}\nn\\ 
{[}\rho_n,\phi_{n'}] &=& \frac{i}{m} \delta_{n+n'}\nn\\
 {[}\rho_n,\rho_{n'}] &=& \frac{k_n}{2\pi m}  \delta_{n+n'}\nn\ee
This final equation is an example of a $U(1)$ {\it Kac-Moody algebra}. It's a provides a powerful constraint on the dynamics of conformal field theories. We won't have much use for this algebra in the present context, but its non-Abelian extension plays a much more important role in WZW conformal field theories. These commutation relations  can be translated back to equal-time commutation relations for the continuum fields. They read
\be 
[\phi(\sigma),\phi(\sigma')] &=& \frac{\pi i}{m}\,\sign(\sigma-\sigma')\label{phicom}\\
{[}\rho(\sigma),\phi(\sigma') ] &=& \frac{i}{m} \delta(\sigma-\sigma')\label{pscom}\\ 
{[}\rho(\sigma),\rho(\sigma')] &=& -\frac{i}{2\pi m}\partial_\sigma \delta(\sigma-\sigma')\label{rhocom}\ee

\subsubsection*{The Hamiltonian}

We can easily construct the Hamiltonian from the action \eqn{chiralwave}. It is
\be H = \frac{mv}{2\pi}\sum_{n=0}^\infty k_n^2\phi_n\phi_{-n} = 2\pi mv \sum_{n=0}^\infty \rho_n\rho_{-n}\nn\ee
where, in the quantum theory, we've chosen to normal order the operators. The time dependence of the operators is given by
\be \dot{\rho}_n =i[H,\rho_n] =  i v k_n \rho_n\nn\ee
One can  check that this is indeed the time dependence of the Fourier modes that follows from the equation of motion \eqn{chiralwave}.

\para
 Our final Hamiltonian is simply that of a bunch of harmonic oscillators. The ground state $|0\rangle$ satisfies $\rho_{-n}|0\rangle =0$ for $n>0$. The excited states can then be constructed by acting with
\be |\psi\rangle = \sum_{n=1}^\infty \rho_n^{d_n} |0\rangle\ \ \ \Rightarrow\ \ \ H|\psi\rangle = \frac{2\pi v}{L} \sum_{n=1}^\infty n d_n |\psi\rangle\nn\ee
We've recovered exactly the spectrum and degeneracy of the excited modes of the Laughlin wavefunction that we saw in Section \ref{chiralwfsec}.


\subsubsection{Electrons and Quasi-Holes}\label{compactqhsec}

All of the excitations that we saw above describe ripples of the edge. They do not change the total charge of the system. In this section, we'll see how we can build new operators in the theory that carry charge. As a hint, recall that we saw in  \eqn{qwind} that any object that changes the charge has to involve $\phi$ winding around the boundary. This suggests that it has something to do with the compact nature of the scalar field $\phi$.  

\para
We claim that the operator describing  an electron in the boundary is
\be \Psi = \ :e^{im\phi}:\label{electronop}\ee
where the dots denote normal ordering, which means that all $\phi_{-n}$, with $n$ positive, are moved to the right. 
In the language of conformal field theory, exponentials of this type are called {\it vertex operators}. 
To see that this operator carries the right charge, we can use the commutation relation \eqn{pscom} to show that
\be [\rho(\sigma),\Psi^\dagger(\sigma')] = \Psi^\dagger(\sigma')\,\delta(\sigma-\sigma')\ \ \ {\rm and}\ \ \  [\rho(\sigma),\Psi(\sigma')] = -\Psi(\sigma')\,\delta(\sigma-\sigma')\nn\ee
which tells us that $\Psi^\dagger$ inserts an object of unit charge while $\Psi$ removes an object of unit charge. This looks good. However, there's something rather surprising about the formula \eqn{electronop}. The field $\phi$ is a boson, but if $\Psi$ is really the electron operator then it should be a fermion. To see that this is indeed the case, we use  the Baker-Campbell-Hausdorff formula to get
\be \Psi(\sigma) \Psi(\sigma') =  e^{-m^2[\phi(\sigma),\phi(\sigma')]}\Psi(\sigma')\Psi(\sigma)\nn\ee
The commutator of $\phi$ was given in \eqn{phicom}. We find that when $\sigma\neq \sigma'$, 
\be [\Psi(\sigma), \Psi(\sigma')] = 0\ \ \ \ m\ {\rm even}\nn\\ \{\Psi(\sigma), \Psi(\sigma')\} = 0\ \ \ \ m\ {\rm odd}\nn\ee
We see that the field $\Phi$ acts like a boson if $m$ is even and a fermion if $m$ is odd. But we know from the Laughlin wavefunction that the objects underlying the quantum Hall state are bosons when $m$ is even and fermions when $m$ is odd. Miraculously,  our edge theory knows about the nature of the underlying constituents in the bulk. The formula \eqn{electronop} is one of the key formulas in the subject of {\it bosonisation}, in which fermions in $d=1+1$ dimensions can be written in terms of bosons and vice versa. 

\para
It should be clear that the electron operator \eqn{electronop} is not the simplest operator that we can construct in our theory. Since $\phi$ has periodicity $2\pi$, it also makes sense to look at the operator
\be \Psi_{qp} =\ : e^{i\phi}:\label{qhop}\ee
No prizes are awarded for guessing that this corresponds to the quasi-particle excitations in the quantum Hall fluid.  The commutator with $\rho$ 
\be [\rho(\sigma),\Psi_{qp}^\dagger(\sigma')] = \frac{1}{m}\Psi_{qp}^\dagger(\sigma')\,\delta(\sigma-\sigma')\ \ \ {\rm and}\ \ \  [\rho(\sigma),\Psi_{qp}(\sigma')] = -\frac{1}{m}\Psi_{qp}(\sigma')\,\delta(\sigma-\sigma')\nn\ee
tells us that these operators create particles with charge $\pm 1/m$. The  statistics of these operators can be seen by commuting
\be   \Psi_{qp}(\sigma) \Psi_{qp}(\sigma') =  e^{-[\phi(\sigma),\phi(\sigma')]}\Psi_{qp}(\sigma')\Psi_{qp}(\sigma) = e^{\pm {\pi i}/{m}}\Psi_{qp}(\sigma')\Psi_{qp}(\sigma) \nn\ee
We see that the particles are anyons, with statistical phase $e^{\pm \pi i/m}$ as expected. In this approach, the sign of the phase depends on the ${\rm sign}(\sigma-\sigma')$. This is analogous to the way the sign depends on whether to do a clockwise or anti-clockwise rotation in the bulk. 

\subsubsection*{Propagators}


Let's now turn to the propagators, starting with the compact boson $\phi$. Deriving the propagator directly from the action \eqn{jackyboy} involves a fiddly contour integral. However, the answer is straightforward and simple to understand intuitively: it is simply the left-moving part of the propagator for a normal boson. Let's start from action 
\be S = \frac{m}{8\pi}\int d^2x\ \partial_i\varphi\partial^i\varphi\nn\ee
The propagator for a free boson is simple to work out: it is
\be \langle \,\varphi(x,t)\varphi(0,0)\,\rangle = -\frac{1}{m}\log(v^2t^2-x^2) \nn\ee
where, as usual, there is an implicit time ordering in all correlation functions of this kind,  and there should really be a UV cut-off in the log which we've dropped. Of course, this action describes a scalar field which can propagate in both left-moving and right-moving directions. The equation of motion $(\partial_t^2-v^2\partial_x^2)\varphi=0$ ensures that all solutions decompose as $\varphi(x,t) = \varphi_L(x+vt) + \varphi_R(x-vt)$ (although there is, once again a subtlety with the zero mode which does not split into left- and right-moving pieces). 
The propagator above has a simple decomposition into left- and right- moving parts, with
\be \langle\,  \varphi_L(x+vt)\varphi_L(0) \,\rangle = -\frac{1}{m}\log(x+vt) +{\rm const.}\nn\ee
Our chiral boson $\phi$ is precisely this left-moving boson $\varphi_L$, albeit  without the accompanying right-moving partner. The propagator. Indeed, one can show the correct propagator derived from \eqn{jackyboy} is equal to that found above
\be \langle\,  \phi(x+vt)\phi(0,0) \,\rangle = -\frac{1}{m}\log(x+vt) +{\rm const.}\label{phiprop}\ee
(An aside: there is a seeming factor of 2 discrepancy between the normalisation of the boson action above and the normalisation of \eqn{jackyboy}. This can be traced to the Jacobian in going between Euclidean coordinates and the light-cone coordinates $X^\pm = \sigma\pm vt$ which are appropriate for the chiral boson). 

\para
The logarithmic dependence seen in \eqn{phiprop} reflects the fact that there are infra-red divergences if we work with massless scalar fields in $d=1+1$. It's telling us that the physical information is carried by other fields. The propagator for the charge density follows immediately from differentiating \eqn{phiprop},
\be\langle\, \rho(x+vt)\rho(0)\,\rangle =- \frac{1}{(2\pi)^2m}\frac{1}{(x+vt)^2}\nn\ee
However, more interesting for us is the electron propagator. 
\be G_F(x,t) = \langle \,\Psi^\dagger(x,t)\Psi(0,0)\,\rangle\label{eprop1}\ee
To compute this, we need to learn how to compute expectation values of normal ordered exponentials \eqn{electronop}. 
Since the field $\phi$ is free, this must ultimately reduce to a problem in terms of harmonic oscillators. Because this is a calculation that we'll need to use again later,  we pause briefly to explain how this works for the harmonic oscillator. . We'll then pick up our thread and compute the electron propagator \eqn{eprop1}.

\subsubsection*{An Aside: Coherent States in the Harmonic Oscillator}

Consider a harmonic oscillator with the usual creation and annihilation operators satisfying $[a,a^\dagger]=1$ and a vacuum $|0\rangle$ obeying $a|0\rangle=0$. A {\it coherent state} is defined as the exponential
\be |z\rangle = e^{za^\dagger}|0\rangle\nn\ee
with $z\in {\bf C}$. 
It's simple to show that $[a, e^{za^\dagger}] = ze^{za^\dagger}$ from which we see that $|z\rangle$ is the eigenstate of the annihilation operator: $a|z\rangle = z|z\rangle$.

\para
Now consider some linear combination of creation and annihilation operators, 
\be A_i = \alpha_i a + \beta_i a^\dagger\nn\ee
The analog of the electron vertex operator \eqn{electronop} is the normal ordered exponential
\be :e^{A_i}\!: \ = e^{\beta_i a^\dagger}\, e^{\alpha_i a}\nn\ee
Our goal is to compute the vacuum expectation value of a string of these vertex operators,
\be \langle 0 | :e^{A_1}\!: :e^{A_2}\!:\,\ldots\, :e^{A_N}\!:|0\rangle\label{vertexstring}\ee
To do this, we need to move all the  $e^{i\alpha_i a}$ to the right, commuting them past the $e^{i\beta_j a^\dagger}$ with $j>i$ as they go.  
By the Baker-Campbell-Hausdorff formula, this is achieved by
\be e^{\alpha a} e^{\beta a^\dagger} = e^{\beta a^\dagger}e^{\alpha a}e^{\alpha\beta[a,a^\dagger]} = e^{\beta a^\dagger}e^{\alpha a}e^{\alpha\beta}
\nn\ee
Applying this to the whole string of operators in \eqn{vertexstring}, we have
\be  :e^{A_1}\!: :e^{A_2}\!:\,\ldots\, :e^{A_N}\!: \ &=& e^{(\beta_1+\ldots +\beta_N)a^\dagger} e^{(\alpha_1+\ldots +\alpha_N) a} \,e^{\sum_{i<j} \alpha_i\beta_j}\nn\\ &=& \ :e^{A_1+\ldots + A_N}\!: \, e^{\sum_{i<j} \langle 0|A_iA_j|0\rangle}\ee
Taking the expectation value of both sides, we have our final result
\be \langle 0|:e^{A_1}\!: :e^{A_2}\!:\,\ldots\, :e^{A_N}\!:|0\rangle = 
\exp\left({\sum_{i<j} \langle 0|A_iA_j|0\rangle}\right)\label{coherent}\ee
This is the result that we want. Let's now see what it means for our electrons on the edge. 

\subsubsection*{The Electron Propagator}

Because the free field $\phi$ is simply a collection of harmonic oscillators, we can apply the formula \eqn{coherent} to vertex operators like \eqn{electronop}. We have
\be G_F(x+vt) =\langle \,\Psi^\dagger(x,t)\Psi(0,0)\,\rangle   = \exp\Big(m^2 \langle\, \phi(x,t)\phi(0,0)\,\rangle\Big)
\nn\ee
Using \eqn{phiprop}, we find that the electron Green's function is given by
\be G_F(x,t)  \sim \frac{1}{(x+vt)^m}\label{eprop}\ee
This is interesting because it's not the usual expression for an electron Green's function in $d=1+1$. 

\para
To explain this, let's first  review some condensed matter field theory. There's a simple theory that describes Fermi surfaces in $d=1+1$ dimensions (where they are really just Fermi points). Unlike in higher dimensions, these electrons are typically interacting, but in a way that is under control. The resulting theory is known as the {\it Luttinger liquid}. One of its key results is that the electron propagator for left-moving modes scales as $G_{\rm Luttinger} \sim 1/(x+vt)$. 

\para
Comparing to our propagator \eqn{eprop}, we see that it coincides with the Luttinger liquid result when $m=1$. This should not be surprising: $m=1$ describes a fully-filled Landau level which does not exhibit topological order. In contrast, in the fractional quantum Hall states with $m\neq 1$, the electrons on the edge of the sample do not follow the standard lore. This reflects the fact that they are strongly coupled. What we are calling an ``electron" in not the same thing as an electron in the Standard Model. Instead, it is some collective excitation that carries the same quantum numbers as the electron in the Standard model.  The resulting theory usually goes by the name of the {\it chiral Luttinger liquid}\,\footnote{These ideas were pioneered by Xiao-Gang Wen in a series of papers, starting with ``{\it Chiral Luttinger Liquid and the Edge Excitations in the Fractional Quantum Hall State}", Phys. Rev. {\bf B41} 12838 (1990) which can be \href{http://dao.mit.edu/~wen/pub/cll.pdf}{downloaded here}. A review can be found in ``{\it Chiral Luttinger Liquids at the Fractional Quantum Hall Edge}" by A. M. Chang, Rev. Mod. Phys. {\bf 75}, 1449 (2003) which can be \href{http://www.phy.duke.edu/~yingshe/RMP-Preprint.pdf}{found here}.}.

\para
The most important information to take from the propagator \eqn{eprop} comes from some simple dimensional analysis. Comparing both sides, we learn that the electron operator $\Psi$ has dimension $m/2$. This should be contrasted with the usual value of $1/2$
The fact that electrons are  fermions means that  $m$ has to be odd. But this means that the exponent in the propagator can't change continuously as the Hamiltonian underlying the quantum Hall state varies. For this reason, the dimension of the edge operator can be viewed as  a characterisation of the bulk state. It can only change if the bulk goes through a phase transition.

\subsubsection{Tunnelling}

The electron propagator \eqn{eprop} has some surprisingly physical consequences. There is a long and detailed literature on this subject. Here we provide only a baby version to explain the basic physics. 

\para
Suppose we connect the edge of the quantum Hall fluid to a wire, but put a small insulating material in between. This kind of set-up goes by the name of a {\it tunnel junction}. It means that if electrons want to get from the one side to the other, they have to tunnel. 
The way to model this in our theory is to add the interaction 
\be S_{\rm tunnel} = \tau \int dt \ e^{im\phi(0,t)}\,\Psi^\dagger_e (0,t) + {\rm h.c.}\nn\ee
where $\Psi^\dagger_e$ is the creation operator for the electron in the wire. Here we've inserted the junction at the point $\sigma=0$ on the edge. 

\para
The strength of the tunnelling is governed by the coupling constant $\tau$. The action must be dimensionless (in units with $\hbar=1$). We learned above that  $e^{im\phi}$ has dimension $m/2$. Meanwhile $\Psi_e$ refers to a ``common or garden" electron in a wire and has dimension $1/2$. This means that the  dimension of $\tau$ must be
\be [\tau] = \frac{1-m}{2}   \nn\ee
We learn  that for $m>1$, the tunnelling is an irrelevant interaction in the language of the renormalisation group. The tunnelling will be suppressed at low energies or low temperature where we can work perturbatively.  We can use dimensional analysis to determine the way various quantities scale. In $d=1+1$, the conductivity has dimension $[\sigma] = -1$, but this means that the conductance $G$ is dimensionless: $[G]=0$. 

\para
Fermi's golden rule tells us that the lowest order contribution to the tunnelling conductance $G$ scales as $\tau^2$. The deficit in dimensions must be made up by temperature $T$, simply because there's no other scale in the game. We have
\be G  \sim \tau^2\, T^{m-1}\nn\ee
Alternatively, if we're at zero temperature then the current is driven by a voltage $V$. We have $[I] = 1$ and $[V]=1$, so we 
\be I \sim \tau^2\,V^m\label{vnotir}\ee

\EPSFIGURE{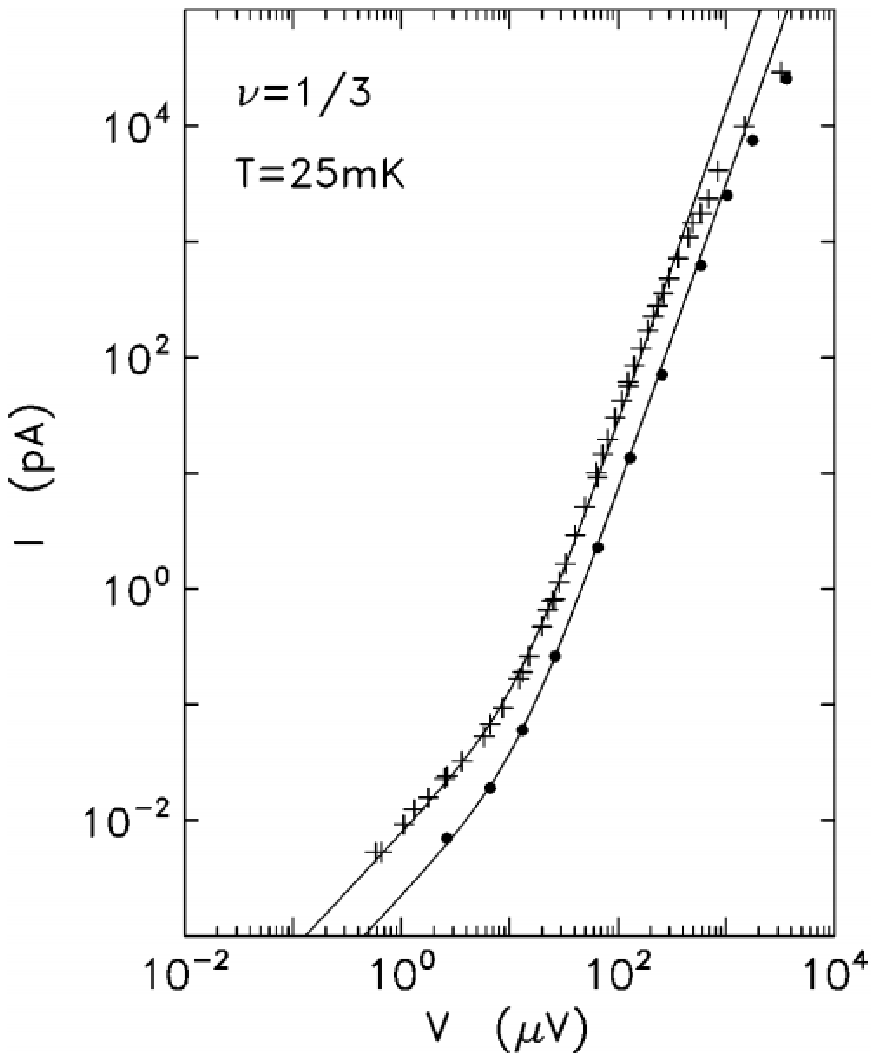,height=150pt}{}
\noindent
This final result is particularly striking as violates the form of Ohm's law, $V=IR$, that we all learned in high school. This prediction has been 
successfully tested for the $\nu=1/3$ quantum Hall state. The data shown in the figure\footnote{This plot is from A. M. Chang, L. N. Pfeiffer, and K. W. West, Ò{\it Observation of Chiral Luttinger Behavior in Electron Tunneling into Fractional Quantum Hall Edges},Ó \href{http://journals.aps.org/prl/abstract/10.1103/PhysRevLett.77.2538}{Phys. Rev. Lett. 77, 2538 (1996)}.} fits the solid line which matches \eqn{vnotir} with $m\approx 2.7$.

\para
We can also play variants on this game.  For example, suppose that we add a tunnel junction between two Hall fluids of the same type. Now the interaction is
\be
S_{\rm tunnel} = \tau \int dt \ e^{im\phi_1(0,t)}\, e^{-im\phi_2(0,t)} + {\rm h.c.}\nn\ee
This time we have $[\tau] = 1-m$ and, correspondingly, we have
\be G \sim \tau^2\,T^{2m-2}\ \ \ {\rm and}\ \ \ I \sim \tau^2\,V^{2m-1}\nn\ee

\subsubsection*{Quasi-Particle Tunnelling}

It's also possible to  set up a situation where the quasi-particles can tunnel. We do this by taking a single Hall fluid and putting in a {\it constriction} as shown in the figure. Because the bulk supports quasi-particles, these can tunnel from the top edge to the bottom. The tunnelling interaction is now
\be
S_{\rm tunnel} = \tau \int dt \ e^{i\phi_1(0,t)}\, e^{-i\phi_2(0,t)} + {\rm h.c.}\nn\ee
\EPSFIGURE{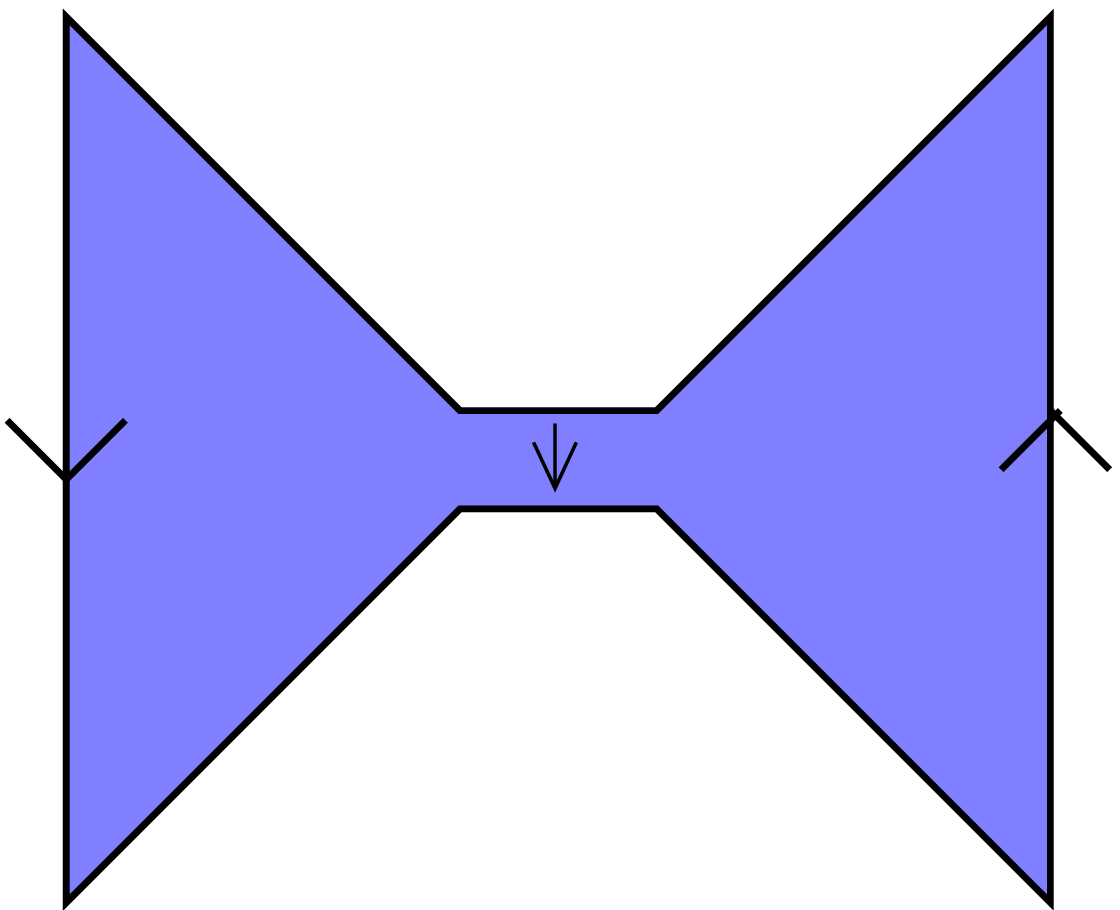,height=90pt}{A Constriction}
\noindent
To figure out the dimension of $\tau$ in this case, we first need the dimension of the quasi-particle operator. Repeating the calculation that led to \eqn{eprop} tells us that $[e^{i\phi} ] = 1/2m$, so now we have
\be [\tau] = 1-\frac{1}{m}\nn\ee
Now this is a relevant interaction. It becomes strong at low temperatures and our naive analysis does not work. (For example, the dimensions of operators at this point may be driven to something else at low temperatures).  Instead, the scaling is valid at high temperatures or high voltages, where ``high" means compared to the scale set by $\tau$ but, obviously not too high as to destroy the Hall state itself. When this scaling is valid, we get
\be G \sim \frac{\tau^2}{T^{2-2/m}}\ \ \ {\rm and}\ \ \ I \sim \frac{\tau^2}{V^{1-2/m}}\nn\ee
Again, we see a striking difference from the usual form of Ohm's law.

\subsection{The Bulk-Boundary Correspondence}\label{bulkboundsec}

We've seen that the theory of the edge modes know about the spectrum of quasi-holes in the bulk. However, it turns out that the edge knows somewhat more than this. Remarkably, it's possible to reconstruct the Laughlin wavefunction itself purely from knowledge about what's happening on the edge. In this section, we see how.

\subsubsection{Recovering the Laughlin Wavefunction}\label{wfmagicsec}

We'll work with the chiral boson theory that we introduced in the previous section. To make these arguments, we need to do some simply gymnastics. First, we set the speed of propagation $v=1$.  Next, we Wick rotate to Euclidean space, defining the complex variables
\be w= \frac{2\pi}{L}\sigma  + i t \ \ \ {\rm and}\  \ \ \ \bar{w} = \frac{2\pi}{L}\sigma - it\label{wickr}\ee
The complex coordinate $w$ parameterises the cylinder that lies at the edge of the Hall sample, with ${\rm Re}(w) \in [0,2\pi)$. The final step is to work with single-valued complex coordinates
\be z = e^{-iw}\ \ \ {\rm and}\ \ \ \bar{z} = e^{+i\bar{w}}\nn\ee
This can be thought of as a map from the cylinder to the plane as shown in the figure. 
If you know some conformal field theory, what we've done here the usual conformal transformation that implements the state-operator map. (You can learn more about this in the introduction to conformal field theory in the {\it String Theory}  lecture notes). 

\para
\EPSFIGURE{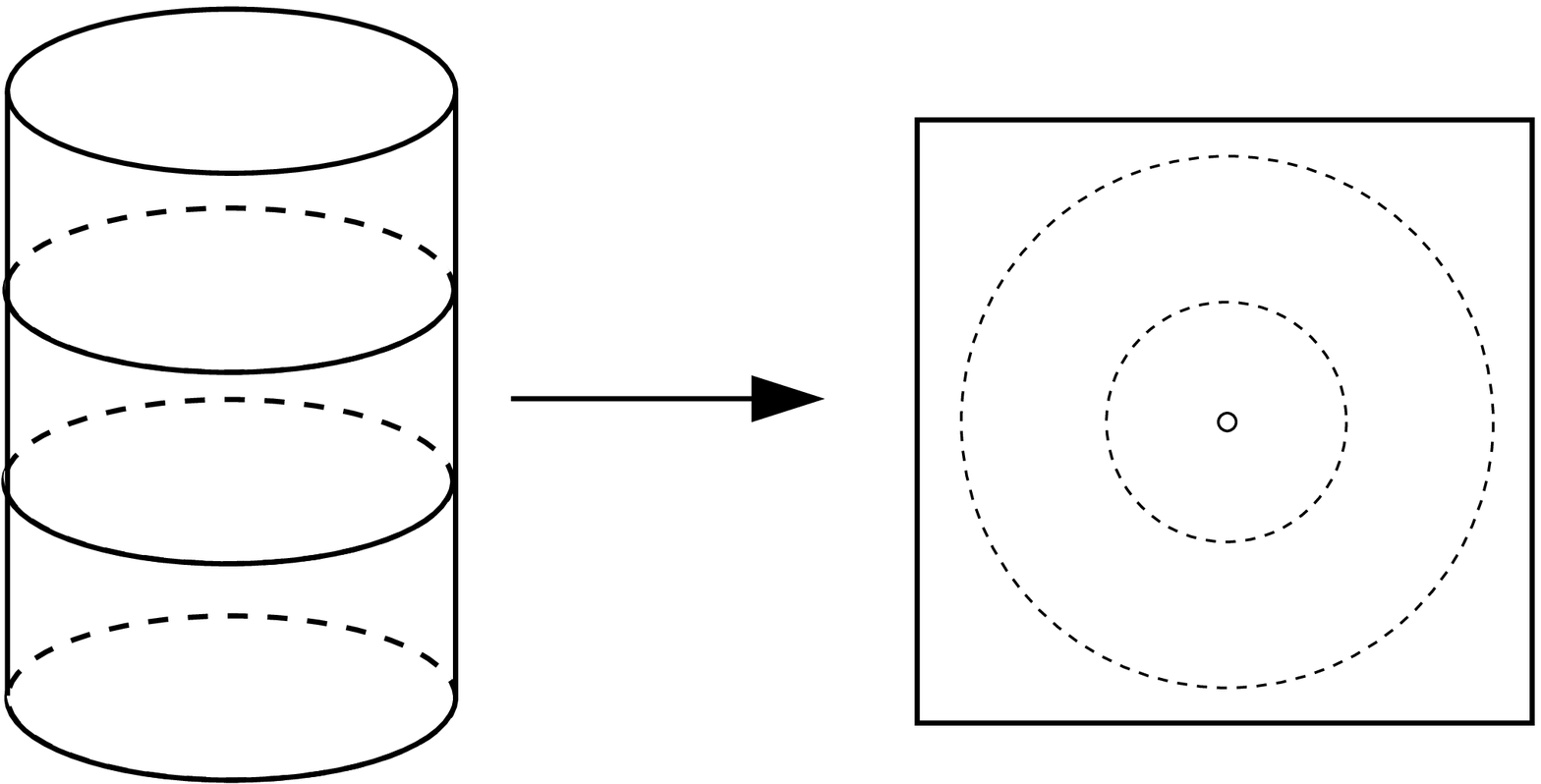,height=90pt}{}
\noindent
In this framework, the fact that the boson is chiral translates to the statement that $\phi$ is a holomorphic function of $z$, so $\phi=\phi(z)$. One can check that the propagator \eqn{phiprop} takes the same form, which now reads
\be \langle\,\phi(z)\phi(w)\,\rangle = -\frac{1}{m}\log(z-w) + {\rm const.}\nn\ee
The basis idea is to look at correlation functions involving insertions of electron operators of the form
\be \Psi = \ :e^{im\phi}\!:\nn\ee
Let's start by looking at something a little more general. We consider the correlation function involving a string of different vertex operators. Using \eqn{coherent}, it looks like we should have
\be \langle :e^{im_1\phi(z_1)}\!: :e^{im_2\phi(z_2)}\!: \,\ldots\,:e^{im_N\phi(z_N)}\!: \rangle &=& \exp \left(-\sum_{i<j} m_im_j \langle \phi(z_i)\phi(z_j)\rangle\right) \nn\\ &\sim& \prod_{i<j} (z_i-z_j)^{m_im_j/m}
\label{almost}\ee
For a bunch of electron operators, with $m_i=m$, this looks very close to the pre-factor of the Laughlin wavefunction. However, the result \eqn{almost} is not quite right. What we missed was a subtle issue to do with the zero mode $\phi_0$ which we were hoping that we could ignore. Rather than deal with this zero mode, let's just see why the calculation above must be wrong\footnote{A correct treatment of the zero mode can be found in the lecture notes on \href{http://www.damtp.cam.ac.uk/user/tong/string.html}{\it String Theory} where this same issue arises when computing scattering amplitudes and is ultimately responsible for momentum conservation in spacetime.
}. Our original theory was invariant under the shift $\phi\rightarrow \phi +\alpha$ for any constant $\alpha$. This means that all correlation functions should also be invariant under this shift. But the left-hand side above transforms picks up a phase $e^{i\alpha(m_1+\ldots+m_N)}$. This means that the correlation function can only be non-zero if 
\be \sum_{i=1}^N m_i = 0\nn\ee
Previously we computed the electron propagator $\langle \Psi^\dagger \Psi\rangle$ which indeed satisfies this requirement. In general the 
 the correct result for the correlation function is 
\be \langle :e^{im_1\phi(z_1)}\!: :e^{im_2\phi(z_2)}\!: \,\ldots\,:e^{im_N\phi(z_N)}\!: \rangle \sim \prod_{i<j} (z_i-z)^{m_im_j/m}\,\delta(\sum_im_i)\nn\ee
The upshot of this argument is that a correlation function involving only electron operators does not give us the Laughlin wavefunction. Instead, it vanishes.

\para
To get something non-zero, we need to insert another operator into the correlation function. We will look at
%
%
%
\be G(z_i,\bar{z}_i) = \langle\,\Psi(z_1)\ldots \Psi(z_N)\,\exp\Big(-\rho_0 \int_\gamma d^2z'\,\phi(z')\Big)\,\rangle\label{lcorr}\ee
This is often said to be inserting a background charge into the correlation function. We take $\rho_0= 1/2\pi l_B^2$. Note that this is the same as the background charge density  \eqn{plasmaback} that we found when discussing the plasma analogy. Meanwhile,   $\gamma$ is a disc-shaped region of radius $R$, large enough to encompass all point $z_i$. Now the requirement that the correlation function is invariant under the shift $\phi\rightarrow \phi + \alpha$ tells us that it can be non-zero only if
\be mN  =\rho_0 \int_\gamma d^2z' = \pi R^2 \rho_0\nn\ee
Using $\rho_0 = 1/2\pi l_B^2$, we see that we should take $R=\sqrt{2mN}l_B$ which we recognise as  the radius of the droplet described by the quantum Hall wavefunction. 

\para
Using \eqn{coherent}, the correlation function \eqn{lcorr} can be written as
%
%
\be  G(z_i,\bar{z}_i) \sim \prod_{i<j} (z_i-z_j)^m
\exp\left(-\rho_0\sum_{i=1}^N \int_\gamma d^2z'\,\log(z_i-z')\right)\nn\ee
We're still left with an integral to do. The imaginary part of this integral is ill-defined because of the branch cuts inherent in the logarithm. However, as its only a phase, it can be undone by a (admittedly very singular) gauge transformation. Omitting terms the overall constant, and terms that are suppressed by  $|z_i|/R$, the final result for the correlation function is
\be  G(z_i,\bar{z}_i) \sim \prod_{i<j} (z_i-z_j)^m e^{-\sum_i |z_i|^2/4l_B^2}\label{lisc}\ee
This, of course, is the Laughlin wavefunction. 

\DOUBLEFIGURE{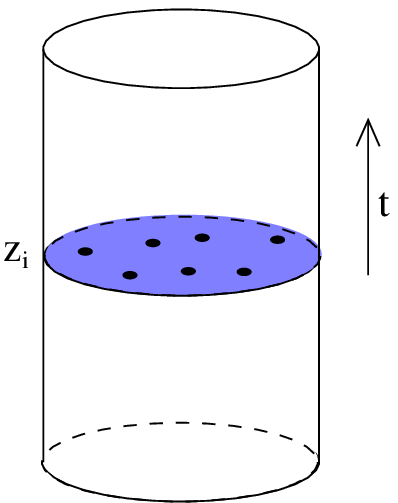,width=120pt}{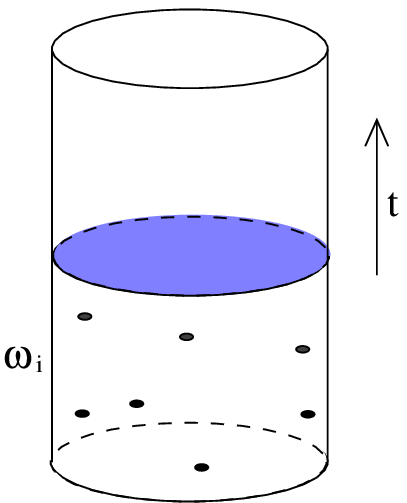,width=120pt}
{The wavefunction lives here}{The correlation function lives here}

\para
We can extend this to wavefunctions that involve quasi-holes. We simply need to insert some number of quasi-hole operators \eqn{qhop} into the correlation function
\be \tilde{G}(z_i,\bar{z}_i;\eta_a,\bar{\eta}_a)=  \langle\,\Psi_{qh}(\eta_1)\ldots \Psi_{qh}(\eta_p)\Psi(z_1)\ldots \Psi(z_N)\,\exp\Big(-\rho_0 \int_\gamma d^2z'\,\phi(z')\Big)\,\rangle\nn\ee
where the size of the disc $\gamma$ must now be extended so that the system remains charge neutral. 
The same calculations as above now yield
\be \tilde{G}(z_i,\bar{z}_i;\eta_a,\bar{\eta}_a)  = \prod_{a<b} (\eta_a-\eta_b)^{1/m}\prod_{a,i}(z_i-\eta_a)\prod_{k<l} (z_k-z_l)^m\,e^{-\sum_i |z_i|^2/4l_B^2 - \sum_a |\eta_a|^2/4 m l_B^2}\nn\ee
This is the Laughlin wavefunction for the quasi-hole excitations. Note that we've recovered the wavefunction in the form  \eqn{otherwf} where the Berry phase vanishes. Instead the correlation function is not single valued and all the statistical phases that arise from braiding  the quasi-hole positions are explicit.

\subsubsection*{What the Hell Just Happened?}

It's been a long journey. But finally, after travelling through Chern-Simons theories and the theory of edge states, we've come right back to where we started: the Laughlin wavefunction\footnote{The connection between correlation functions and quantum Hall wavefunctions was first noticed by Greg Moore and Nick Read in the ``{\it Nonabelions in the Fractional Quantum Hall Effect}", Nucl. Phys. {\bf B360}, 362 (1991) which can be \href{http://www.physics.rutgers.edu/~gmoore/MooreReadNonabelions.pdf}{downloaded here}. This was also the paper where they first proposed the Moore-Read wavefunction. This is not coincidence: they arrived at the wavefunction by thinking about correlation functions in different conformal field theories.}. How did this happen? It seems like magic!

\para
The most glaring issue  in identifying the correlation function with the wavefunction is that the two live in different spaces. 
Our quantum Hall fluid lives on a disc, so spacetime is a cylinder as shown in the figures. The wavefunction is defined on a spatial slice at a fixed time; this is the blue disc in the figure. In the wavefunction, the positions $z_i$ lies within this disc as shown in the left-hand figure. Meanwhile, the conformal field theory lives on the boundary. The  operators inserted in the correlation function sit at positions $w_i = 2\pi\sigma/L + it$ which are subsequently mapped to the plane by $z=e^{-i\omega}$.
Why should we identify the positions in these two different spaces? 

\para
The answer is that there are actually two different ways in which the Chern-Simons theory is related to the CFT. This arises because the bulk Chern-Simons theory is topological, which means that you can cut it in different way and get the same answer. Above we've considered cutting the bulk along a timelike boundary to give a CFT in $d=1+1$ dimensions. This, of course, is what happens in a physical system. However, we could also consider an alternative way to slice the bulk along a spacelike section, as in the left-hand figure above. This gives the same CFT, but now Wick rotated to $d=2+0$ dimensions. We will discuss this imminently in Section \ref{magicsec}. This new perspective will make us slightly more comfortable about  why bulk wavefunctions are related to this boundary CFT.



\para
Before describing this new perspective, let me also mention a separate surprise about the relationship between correlation functions and the Laughlin wavefunction. Our original viewpoint in Section \ref{fqhesec} was that there was nothing particularly special about the Laughlin wavefunction; it is simply a wavefunction that is easy to write down which lives in the right universality class. Admittedly it has good overlap with the true ground state for low number of electrons, but it's only the genuine ground state for artificial toy Hamiltonians.  But now we learn that there is something special about this state: it is the correlation function of primary operators in the boundary theory. I don't understand what to make of this. 


\para
Practically speaking, the connection between  bulk wavefunctions and boundary correlation functions has proven to be a very  powerful tool. It is conjectured that this correspondence extends to all quantum Hall states.  First, this means that you don't need to guess quantum Hall wavefunctions anymore. Instead you can just guess a boundary CFT and compute its correlation functions. But there's a whole slew of CFTs that people have studied. We'll look at another example in Section \ref{fermisec}. Second, it turns out that the CFT framework is most useful for understanding the properties of quantum Hall states, especially those with non-Abelian anyons. The braiding properties of anyons are related to well-studied properties of CFTs. We'll give some flavour of this in Section \ref{forwardsec}.

%
%

\subsubsection{Wavefunction for  Chern-Simons Theory}\label{magicsec}

Above we saw how the boundary correlation functions of the CFT capture the bulk Laughlin wavefunctions. But the origins of this connection appear completely mysterious. Although we won't give a full explanation of this result, we will at least  try to motivate why one may expect the boundary theory to know about the bulk wavefunction. 

\para
As we described above, the key is to consider a different cut of the Chern-Simons theory. With this in mind, we will place Chern-Simons theory on ${\bf R}\times {\bf S}^2$ where ${\bf R}$ is time and ${\bf S}^2$ is a compact spatial manifold which no longer has a boundary. Instead, we will consider the system at some fixed time. But in any quantum system, the kind of object that sits at a fixed time is a wavefunction. We will see how the wavefunction of Chern-Simons theory is related to the boundary CFT.

\para
We're going to proceed by implementing a canonical quantisation of $U(1)_m$ Chern-Simons theory. 
We already did this for Abelian Chern-Simons theory in  Section \ref{cstorussec}. Working in $a_0=0$ gauge, the canonical commutation relations \eqn{stanford}
\be [a_i({\bf x}),a_j({\bf y})] = \frac{2\pi i}{m}\,\epsilon_{ij}\,\delta^2({\bf x}- {\bf y})\nn\ee
subject to the  constraint $f_{12}=0$. 

\para
At this stage, we differ slightly from what went before. We introduce complex coordinates  $z$ and $\bar{z}$ on the spatial ${\bf S}^2$. 
As an aside, I should mention that if we were working on a general spatial manifold $\Sigma$ then there is no canonical choice of complex structure, but the end result is independent of the complex structure you pick. This complex structure can also be used to complexify the gauge fields, so we have $a_z$ and $a_{\bar{z}}$ which obey the commutation relation

\be [a_z(z,\bar{z}), a_{\bar{z}}(w,\bar{w}) ] = \frac{4\pi}{m}\,\delta^2({\bf z}-{\bf w})\label{holoquant}\ee
The next step is somewhat novel. We're going to write down a Schr\"odinger equation for the theory. That's something very familiar in quantum mechanics, but not something that we tend to do in field theory. Of course, to write down a Schr\"odinger equation, we first need to introduce a wavefunction which depends only on the ``position" degrees of freedom and not on the momentum. This means that we need to make a choice on what is position and what is momentum.  The commutation relations \eqn{holoquant} suggest that it's sensible to choose $a_{\bar{z}}$ as ``position" and $a_z$ as ``momentum". This kind of choice goes by the name of {\it holomorphic quantisation}. This means that we describe the state of the theory by a wavefunction
\be \Psi(a_{\bar{z}}(z,\bar{z}))\nn\ee
Meanwhile, the $a_z$ act as a momentum type operator, 
\be a_z^a = \frac{4\pi}{k}\frac{\delta}{\delta a_{\bar{z}}^a}\nn\ee
The Hamiltonian for the Chern-Simons theory vanishes. Instead, what we're calling the  Schr\"odinger equation arises from imposing the constraint $f_{z\bar{z}} =0$ as an operator equation on $\Psi$. Replacing $a_z$ with the momentum operator, this reads
\be \left(\partial_{\bar{z}}\frac{\delta}{\delta a_{\bar{z}}} - \frac{m}{4\pi} \partial_z a_{\bar{z}}\right)\Psi(a_{\bar{z}})=0 \label{cschrodinger}\ee
This is our Schr\"odinger equation. 

\subsubsection*{The Partition Function of the Chiral Boson}

We'll now show that this same equation arises from the conformal field theory of a chiral boson. The key idea is to couple the current in the CFT to a background gauge field. We will call this background gauge field $a$.

\para
Recall from our discussion in Section \ref{csboundsec} that the charge density is given by $\rho \sim \partial\phi/\partial x$ 
 and, for the chiral action \eqn{jackyboy}, the associated current density is simply $-v\rho$, reflecting the fact that charge, like all excitations, precesses along the edge. 

\para
Here we want to think about the appropriate action in the Euclidean theory. It's simplest to look at the action for a massless boson and subsequently focus on the chiral part of it. This means we take
\be S[\phi] = \frac{m}{2\pi}\int d^2x\ \partial_{\bar{z}}\phi\,\partial_z\phi\nn\ee
Now the charge becomes 
\be \rho = \frac{1}{2\pi} \ppp{\phi}{z}\nn\ee
The chiral conservation law is simply $\partial_{\bar{z}}\rho \sim \partial_{\bar{z}}\partial_z\phi=0$ by virtue of the equation of motion. 

\para
 We want to couple this charge to a  background gauge field. We  achieve this by writing
\be S[\phi;a] = \frac{m}{2\pi}\int d^2x\ {\cal D}_{\bar{z}}\phi\, \partial_z\phi\label{oddact}\ee
where 
\be {\cal D}_{\bar{z}} \phi = \partial_{\bar{z}}\phi - a_{\bar{z}}\nn\ee
The extra term in this action takes the form $a_{\bar{z}}\rho$, which is what we wanted. Moreover, the form of the covariant derivative tells us that we've essentially gauged the shift symmetry $\phi \rightarrow \phi + {\rm constant}$ which was responsible for the existence of the charge in the first place. Note that, although we've given the gauge field the same name as in the Chern-Simons calculation above, they are (at this stage) rather different objects. The Chern-Simons gauge field is dynamical but, in the equation above, $a_{\bar{z}}(z,\bar{z})$ is some fixed function. We will see shortly why it's sensible to give them the same name.

\para
The action \eqn{oddact} looks rather odd. We've promoted $\partial_{\bar{z}}$ into a covariant derivative ${\cal D}_{\bar{z}}$ but not $\partial_z$.  This is because we're  dealing with a chiral boson rather than a normal boson. It has an important consequence. The equation of motion from \eqn{oddact} is
\be \partial_{\bar{z}}\partial_z \phi = \frac{1}{2}\partial_za_{\bar{z}}\label{lemonchick}\ee
This tells us that the charge $\rho$ is no longer conserved! That's quite a dramatic change. It is an example of an {\it anomaly} in quantum field theory.

\para
If you've heard of anomalies in the past, it is probably in the more familiar (and more subtle) context of chiral fermions. The classical chiral symmetry of fermions is not preserved at the quantum level, and the associated charge can change in the presence of a background field. The anomaly for the chiral boson above is much simpler: it appears already in the classical equations of motion. It is related to the chiral fermion anomaly through bosonization. 

\para
Now consider the partition function for the chiral boson. It is a function of the background field. 
\be Z[a_{\bar{z}}]  = \int{\cal D}\phi\, e^{-S[\phi,a]}\nn\ee
This, of course, is the generating function for the conformal field theory. The partition function in the present case obeys a rather nice equation,
\be \left(\partial_{\bar{z}}\frac{\delta}{\delta a_{\bar{z}}} - \frac{m}{4\pi} \partial_z a_{\bar{z}}\right)Z(a_{\bar{z}}) =0 \label{garlic}\ee
To see this, simply move the $\delta/\delta a_{\bar{z}}$ into the path integral where it brings down a factor of $\partial_z\phi$. The left-hand side of the above equation is then equivalent to computing the expectation value $\langle \partial_{\bar{z}}\partial_z \phi - \frac{1}{2}\partial_za_{\bar{z}}\rangle_a$, where the subscript $a$ is there to remind us that we evaluate this in the presence of  the background gauge field. But this is precisely the equation of motion \eqn{lemonchick} and so vanishes. 

\para
Finally, note that we've seen the equation \eqn{garlic} before; it is the Schr\"odinger equation 
\eqn{cschrodinger} for the Chern-Simons theory. Because they solve the same equation, we can equate
\be \Psi(a_{\bar{z}}) = Z[a_{\bar{z}}]\label{dscft}\ee
This is a lovely and surprising equation. It provides a quantitative relationship between the boundary correlation functions, which are generated by $Z[a]$, and the bulk Chern-Simons wavefunction.

\para
The relationship \eqn{dscft} says that the bulk vacuum wavefunction $a_{\bar{z}}$ is captured by correlation functions of $\rho \sim \partial\phi$. This smells like  what we want, but it isn't quite the same. Our previous calculation looked at correlation functions of  vertex operators $e^{im\phi}$. One might expect that these are related to bulk wavefunctions in the presence of Wilson lines. This is what we have seen coincides with our quantum Hall wavefunctions.

\para
The bulk-boundary correspondence that we've discussed here is reminiscent of what happens in gauge/gravity duality. The relationship \eqn{dscft} is very similar to what happens in the ds/CFT correspondence (as opposed to the AdS/CFT correspondence).  In spacetimes which are asymptotically de Sitter, the bulk Hartle-Hawking wavefunction at spacelike infinity is captured by a boundary Euclidean conformal field theory.

%
%
%

\subsubsection*{Wavefunction for Non-Abelian Chern-Simons Theories}

The discussion above generalises straightforwardly to non-Abelian Chern-Simons theories. Although we won't need this result for our quantum Hall discussion, it is important enough to warrant comment. The canonical commutation relations were given in \eqn{csab} and, in complex coordinates, read
\be [a_z^a(z,\bar{z}), a_{\bar{z}}^b(w,\bar{w}) ] = \frac{4\pi}{k}\,\delta^{ab}\,\delta^2({\bf z}-{\bf w})\nn\ee
with $a,b$ the group indices and $k$ the level. The constraint $f_{zz'}=0$ is once again interpreted as an operator equation acting on the wavefunction $\Psi(a_{\bar{z}})$. The only difference is that there is an extra commutator term in the non-Abelian $f_{zz'}$. The resulting Schr\"odinger equation is now
\be \left(\partial_{\bar{z}}\frac{\delta}{\delta a_{\bar{z}}} + [a_{\bar{z}},\frac{\delta}{\delta a_{\bar{z}}}] \right)\Psi(a_{\bar{z}}) = \frac{k}{4\pi} \partial_z a_{\bar{z}}\Psi(a_{\bar{z}})\nn\ee
As before, this same equation governs the partition function $Z[a_{\bar{z}}]$ boundary CFT, with the gauge field $a_{\bar{z}}$ coupled to the current. In this case, the boundary CFT is a WZW model about which we shall say (infinitesimally) more in Section \ref{forwardsec}.

%
%
%
%
%
%
%
%
%
%
%




\subsection{Fermions on the Boundary}\label{fermisec}

In this section we give another example of the bulk/boundary correspondence. However, we're not going to proceed systematically by figuring out the edge modes. Instead, we'll ask the question: what happens when you have fermions propagating on the edge? We will that this situation  corresponds to  the Moore-Read wavefunction. We'll later explain the relationship between this and the Chern-Simons effective theories that we described in Section \ref{cssec}.

\subsubsection{The Free Fermion}

In $d=1+1$ dimensions, a Dirac fermion $\psi$ is a two-component spinor. The action for a massless fermion is
\be S = \frac{1}{4\pi}\int d^2x\  i\psi^\dagger \gamma^0\gamma^\mu\partial_\mu\psi\nn\ee
In Minkowski space we take the gamma matrices to be $\gamma^0 = i\sigma^2$ and $\gamma^1 = \sigma^1$ with $\sigma^i$ the Pauli matrices. These  obey the Clifford algebra $\{\gamma^\mu,\gamma^\nu\} = 2\eta^{\mu\nu}$. We can decompose the  Dirac spinor into chiral spinors by constructing the other ``$\gamma^5$" gamma matrix. In our chosen  basis this  is simply $\sigma^3$ and the left-moving and right-moving spinors, which are eigenstates of $\sigma^3$, are simply
\be \psi= \left(\begin{array}{c} \chi_L\\ \chi_R\end{array}\right)\nn\ee
Written in the terms of these one-component Weyl spinors, the action is 
\be S = -\frac{1}{4\pi}\int d^2x\ i\chi_L^\dagger(\partial_t- \partial_x)\chi_L +  i\chi^\dagger_R(\partial_t+\partial_x)\chi_R\nn\ee
The solutions to the equations of motion are $\chi_L=\chi_L(x+t)$ and $\chi_R=\chi_R(x-t)$.

\para
There's something rather special about spinors in $d=1+1$ dimensions (and, indeed in $d=4k+2$ dimensions): they can be both Weyl and Majorana at the same time. We can see this already in our gamma matrices which are both real and in a Weyl basis. From now on, we will be interested in a single left-moving Majorana-Weyl spinor. We will denote this as $\chi$. The Majorana condition simply tells us that $\chi=\chi^\dagger$.

\subsubsection*{Fermions on a Circle}

The edge of our quantum Hall state is a cylinder. We'll take the spatial circle to be parameterised by $\sigma\in [0,L)$. If the fermion is periodic around the circle, so $\chi(\sigma+L) = \chi(\sigma)$, then it can be decomposed in Fourier modes as
\be \chi(\sigma) = \sqrt{\frac{2\pi}{L}}\sum_{n\in {\bf Z}}\chi_n\, e^{2\pi i n \sigma/L}\label{ramond}\ee
The Majorana condition is $\chi^\dagger_n = \chi_{-n}$. 
However, for fermions there is a different choice: we could instead ask that they are anti-periodic around the circle. In this case $\chi(\sigma+L) = -\chi(\sigma)$, and the modes $n$ get shifted by $1/2$, so the decomposition becomes
\be \chi(\sigma) = \sqrt{\frac{2\pi}{L}}\sum_{n\in {\bf Z}+\frac{1}{2}}\chi_n\, e^{2\pi i n \sigma/L}\label{ns}\ee
The periodic case is known as {\it Ramond} boundary conditions; the anti-periodic case as {\it Neveu-Schwarz} (NS) boundary conditions. In both cases, the modes have canonical anti-commutation relations
\be \{\chi_n,\chi_m\} = \delta_{n+m}\label{fermiacom}\ee

\subsubsection*{Fermions on the Plane}

At this stage, we play the same games that we saw at the beginning of Section \ref{wfmagicsec}; we Wick rotate, define complex coordinates $w=2\pi\sigma/L + it$ as in \eqn{ramond}, and then map to the complex plane by introducing $z=e^{-iw}$. However, something new happens for the fermion that didn't happen for the boson: it picks up an extra contribution in the map from the cylinder to the plane:
\be \chi(w) \rightarrow \sqrt{\frac{2\pi z}{L}}\chi(z)\nn\ee
In the language of conformal field theory, this arises because $\chi$ has dimension $1/2$. However, one can also see the reason behind this if we look at the mode expansion on the plane. With Ramond boundary conditions we get
\be \chi(z) = \sum_{n\in {\bf Z}} \chi_{n}\, z^{-n-1/2}\ \ \ \Rightarrow\ \ \ \chi(e^{2\pi i}z) = - \chi(z)\nn\ee
We see that the extra factor of $1/2$ in the mode expansion leads to the familiar fact that fermions pick up a minus sign when rotated by $2\pi$.  

\para
In contrast, for NS boundary conditions we have
\be \chi(z) = \sum_{n\in {\bf Z}+\frac{1}{2}} \chi_n\, z^{-n-1/2}\ \ \ \Rightarrow\ \ \ \chi(e^{2\pi i}z) = + \chi(z)\nn\ee
As we will see, various aspects of the physics depend on which of these boundary conditions we use.  This is clear already when compute the propagators. These are simplest for the NS boundary condition, where $\chi$ is single valued on the plane. The propagator can be computed from the anti-commutation relations \eqn{fermiacom},
\be \langle\, \chi(z)\chi(w)\,\rangle &=& \sum_{n,m\in {\bf Z}+\frac{1}{2}} z^{-n} w^{-m}\langle \chi_n\chi_m\rangle
\nn\\ &=& \sum_{n=0}^\infty \frac{1}{z}\left(\frac{w}{z}\right)^n\nn\\ &=& \frac{1}{z-w}
\label{nsprop}\ee
Meanwhile, in the Ramond sector, the result is more complicated as we get an extra contribution  from $\langle\chi_0^2\rangle$.  This time we find
\be\langle\, \chi(z)\chi(w)\,\rangle &=& \sum_{n,m\in{\bf Z}} z^{-n-1/2} w^{-m}\langle \chi_n\chi_m\rangle \nn\\ &=& 
\frac{1}{2\sqrt{zw}} + \sum_{n=1}^\infty z^{-n-1/2}w^{n-1/2}\nn\\ &=& 
\frac{1}{\sqrt{zw}}\left(\frac{1}{2} + \sum_{n=1}^\infty \left(\frac{w}{z}\right)^n\right) \nn\\
&=& \frac{1}{2}\frac{\sqrt{z/w} + \sqrt{w/z}}{z-w}\nn\ee
We see that there propagator inherits some global structure that differs from the Ramond case. 

\subsubsection*{This is the Ising Model in Disguise!}

The free fermion that we've described provides the solution to one of the classic problems in theoretical physics: it is the critical point of the 2d Ising model! We won't prove this here, but will sketch the extra ingredient that we need to make contact with the Ising model. It is called the {\it twist operator} $\sigma(z)$. It's role is to switch between the two boundary conditions that we defined above.  Specifically, if we insert a twist operator at the origin and at infinity then it relates the correlation functions with different boundary conditions, 
\be \langle {\rm NS}|\,\sigma(\infty)\chi(z)\chi(w)\sigma(0)\,|{\rm NS}\rangle = \langle  {\rm Ramond}|\,\chi(z)\chi(w)\,|{\rm Ramond}\rangle\nn\ee
With this definition, one can show that the dimension of the twist operator is $h_\sigma = 1/16$. This is identified with the spin field of the Ising model. Meanwhile, the fermion $\chi$ is related to the energy density. 

\para
One reason for mentioning this connection is that it finally explains the name ``{\it Ising anyons}" that we gave to the quasi-particles of the Moore-Read state. In particular, the ``fusion rules" that we met in Section \ref{nonabanyonsec} have a precise analog in conformal field theories. (What follows involves lots of conformal field theory talk that won't make much sense if you haven't studied the subject.) In this context, a basic tool  is the operator product expansion (OPE) between different operators. Every operator lives in a conformal family determined by a primary operator. The fusion rules are the answer to the question: if I know the family that two operators live in, what are the families of operators that can appear in the OPE? 

\para
For the Ising model, there are two primary operators other than the identity: these are $\chi$ and $\sigma$. The fusion rules for the associated families are 
\be \sigma \star \sigma = 1 \oplus \chi\ \ \ ,\ \ \ \sigma\star \chi = \sigma\ \ \ ,\ \ \ \chi\star \chi = 1\nn\ee
But we've seen these equations before: they are precisely the fusion rules for the Ising anyons \eqn{ising} that appear in the Moore-Read state (although we've renamed $\psi$ in \eqn{ising} as $\chi$). 

\para
Of course, none of this is coincidence. As we will now see, we can reconstruct the Moore-Read wavefunction from correlators in a $d=1+1$ field theory that includes the free fermion. 

\subsubsection{Recovering the Moore-Read Wavefunction}

Let's now see how to write the Moore-Read wavefunction
\be {\psi}_{MR}(z_i,\bar{z}_i) = {\rm Pf}\left(\frac{1}{z_i-z_j}\right)\prod_{i<j} (z_i-z_j)^m \,e^{-\sum |z_i|^2/4l_B^2}\nn\ee
as a correlation function of a $d=1+1$ dimensional field theory. The new ingredient is obviously the Pfaffian. But this is easily built from a free, chiral Majorana fermion. As we have seen, armed  with NS boundary conditions such a fermion has propagator
\be \langle\,\chi(z)\chi(w)\,\rangle = \frac{1}{z-w}\nn\ee
Using this, we can then employ Wick's theorem to compute the general correlation function. The result is 
\be \langle \, \chi(z_1)\ldots\chi(z_N)\,\rangle = {\rm Pf}\left(\frac{1}{z_i-z_j}\right)\nn\ee
which is just what we want. The piece that remains is simply a Laughlin wavefunction and we know how to  build this from  a chiral boson with propagator
\be \langle\,\phi(z)\phi(w)\,\rangle = -\frac{1}{m}\log(z-w)+{\rm const.}\label{cprop}\ee
The net result is that the Moore-Read wavefunction can be constructed from the product of  correlation functions
\be {\psi}_{MR}(z_i,\bar{z}_i) = \langle \, \chi(z_1)\ldots\chi(z_N)\,\rangle \,\langle\, :e^{im\phi(z_1)}\!:\,\ldots :e^{im\phi(z_N)}\!:\,e^{-\rho_0 \int_\gamma d^2z'\,\phi(z')}\,\rangle\nn\ee
From this expression, it's clear that we should identify the electron operator as the combination
\be \Psi(z) = \chi(z):e^{im\phi(z)}:\nn\ee
These are fermions for $m$ even and bosons for $m$ odd. 

\para
What about the quasi-holes of the theory? We won't give details but will instead just state the answer: the quasi-hole operator is related to the twist operator
\be \Psi_{qh} = \sigma(z):e^{i\phi(z)/2}:\nn\ee
Note that the bosonic vertex operator has a charge which would be illegal in the pure bosonic theory. However, the multi-valued issues are precisely compensated by similar properties of the twist, so their product is single valued.  This factor of $1/2$ explains how the quasi-holes have half the charge than in the Laughlin state. One can show that inserting $\Psi_{qh}$ results in an ambiguity. There are a number of different correlation functions. These are precisely the different quasi-hole wavefunctions \eqn{mrhole} that we met in Section \ref{mrsec}.

\para
Finally, the theory also has the elementary excitation that we started with: the fermion $\chi$. This corresponds to a fermionic, neutral excitation of the Moore-Read state.

\subsubsection*{Relationship to Chern-Simons Theory}

In this section, we just conjured the fermion theory out of thin air and showed that one can reconstruct the Moore-Read state. It would be nice to do better and show that it arises as the boundary theory of the corresponding Chern-Simons theory.
This is (fairly) straightforward for the case of the bosonic, $m=1$ Moore-Read state. Again, we won't be able to describe the details without getting into a lot more conformal field theory, but here's a  sketch of  the basics.

\para
When $m=1$, the propagator \eqn{cprop} for the chiral boson has no fractional  piece in its normalisation. Or, said another way, if we normalise the action canonically, so we rescale $\phi\rightarrow \sqrt{m}\phi$, then the radius of the chiral boson remains $R=1$.
However, a chiral boson at this radius has the nice property that it is equivalent to a  chiral Dirac fermion.  This, in turn,  is the same as  two Majorana fermions. The upshot is that the conformal field theory for $m=1$ is really three Majorana fermions: the $\chi$ that we started with and two more that come from $\phi$. There is an $SU(2)$ symmetry which rotates these three fermions among themselves. Indeed, it's known that this is the theory that arises on the edge of the $SU(2)$ Chern-Simons theory at level $k=2$. 

\para
As we discussed in Section \ref{nonabcssec}, for $m>1$ the corresponding Chern-Simons theories are less clear. Instead, it's better to think of the quantum Hall states as characterised by the conformal field theories on the edge. It is conjectured that, in general, the correct edge theory is precisely the one whose correlation functions reproduce bulk wavefunctions. 
Moreover, there are many powerful techniques that have been developed for conformal field theory which allow one to determine the properties of the wavefunctions, in particular the braiding of non-Abelian anyons. In the final section, we paint a cartoon picture of these techniques. 

\subsection{Looking Forwards: More Conformal Field Theory}\label{forwardsec}

In the last few sections, we've seen an increasing need to import results from conformal field theory. This doesn't improve moving forward! To make progress, we would really need to first pause and better understand the structure of conformal field theories. However, this is a large subject which we won't cover in these lectures. Instead,  we will just attempt to paint a picture with a broad brush while stating a few facts. At the very least, this will hopefully provide some vocabulary that will be useful if you want to pursue these ideas further.

\subsubsection*{Fusion, Braiding and Conformal Blocks}

The key idea is that the formal structure underlying non-Abelian anyons that we described in Section \ref{nonabanyonsec} also appears in conformal field theory (CFT). Indeed, it was first discovered in this context\footnote{See the paper ``{\it Classical and Quantum Conformal Field Theory}", \href{http://projecteuclid.org/euclid.cmp/1104178762}{Comm. Math. Phys 123, 177 (1989)} by Greg Moore and Nati Seiberg, or their subsequent ``{\it Lectures on RCFT}" which can be \href{http://physics.rutgers.edu/~gmoore/LecturesRCFT.pdf}{downloaded here}. }.

\para
The role of the different kinds of anyons is now played by the different representations of the conformal algebra (by which we mean either the Virasoro algebra, or something larger, such as a current algebra) that appear in a given conformal field theory. Each of these representations can be labelled by a highest-weight state called a {\it primary operator}, ${\cal O}_i$. A rational conformal field theory is one which has a finite number of these primary operators. 

\para
Next up, we need to define fusion. We already met this briefly in the previous section in the context of the Ising model. If  you have two operators which live within representations associated to the primary operators ${\cal O}_i$ and ${\cal O}_j$ respectively, then the operator product expansion can contain operators in other representations associated to ${\cal O}_k$. We write these fusion rules, following \eqn{fusion}, as 
\be {\cal O}_i \star {\cal O}_j = \sum_k N_{ij}^k \ {\cal O}_k\nn\ee
where $N_{ij}^k$ are integers.

\para
Similarly, we can define braiding matrices for a CFT. The general idea of the braiding is as follows. Consider  a CFT which has both left-moving and right-moving modes.   In general, correlation functions of primary operators can be decomposed as
\be \langle\,\prod_{i=1}^N{\cal O}_i (z_i,\bar{z}_i)\,\rangle = \sum_{p} |{\cal F}_p(z_i)|^2\nn\ee
Here the $F_p(z_i)$ are multi-branched analytic functions of the $z_i$ which depend on the set of list of operators  inserted on the left-hand-side. They are known as {\it conformal blocks}. In a rational conformal field theory (which is defined to have a finite number of primary operators) the sum over $p$ runs over a finite range.

\para
Now vary the $z_i$, which has the effect of exchanging the particles. (In the context of the quantum Hall wavefunctions, we would exchange the positions of the quasi-hole insertions.) The conformal blocks will be analytically continued onto different branches. However, the final answer can be written in terms of some linear combination of the original function. This linear map is analogous to the braiding of anyons. One of the main results of Moore and Seiberg is that there are consistency relations on the kinds of braiding that can arise. These are precisely the pentagon and hexagon relations that we described in Section \ref{nonabanyonsec}. 

\para
We've already seen two examples of this. For the Laughlin states  with quasi-holes, there is a single  conformal block but it is multi-valued due to the presence of the factor $\prod (\eta_i-\eta_j)^{1/m}$ involving the quasi-hole positions $\eta$. Meanwhile, for the Moore-Read state there are multiple conformal blocks corresponding to the different wavefunctions \eqn{mrhole}. In both these cases, the conformal field theory gave the wavefunction in a form in which all the monodromy properties are explicit and there is no further contribution from the Berry phase. (Recall the discussion at the end of Section \ref{fracstatsec}.) It is conjectured that this is always the case although, to my knowledge, there is no proof of this. 

\subsubsection*{WZW Models}

The most important conformal field theories for our purposes are known as WZW models. (The initials stand for Wess, Zumino and Witten. Sometimes Novikov's name breaks the symmetry and they are called WZNW models.) Their importance stems in large part from their relationship to non-Abelian Chern-Simons theories. These models describe the modes  which live at the edge of a non-Abelian Chern-Simons theory with boundary. Further, it turns out that the braiding of their conformal blocks coincides with the braiding of Wilson lines in the Chern-Simons theory that we briefly described in Section \ref{wilsonsec}.

\para
The WZW models are defined by the choice of gauge group $G$, which we will take to be $SU(N)$, and a level $k\in {\bf Z}$. These theories are denoted as $SU(N)_k$. The CFT for a compact boson that we met in Section \ref{wfmagicsec} is a particularly simple example of a WZW model model with $U(1)_m$. 

\para
Unusually for conformal field theories, WZW models have a Lagrangian description which can be derived using the basic method that we saw in Section \ref{csboundsec} for $U(1)$ Chern-Simons theories. The Lagrangian is
\be S = \frac{k}{4\pi}\int d^2x\ {\rm tr}\,\Big(g^{-1}\partial_t g\,g^{-1}\partial_x g - v(g^{-1}\partial_x g)^2\Big) + 2\pi k\, w(g)\nn\ee
Here $g\in G$ is a group valued field in $d=1+1$ dimensions. The first term describes a chiral sigma model whose target space is the group manifold $G$. If we're working with a quantum Hall fluid on a disc then this theory lives on the ${\bf R}\times {\bf S}^1$ boundary.

\para
The second term is more subtle. It is defined as the integral over the full three-dimensional manifold ${\cal M}$ on which the quantum Hall fluid lives,  
\be w(g) = \frac{1}{24\pi^2} \int_{{\cal M}} d^3x\ \ep \tr\,(g^{-1}\partial_\mu g \,g^{-1}\partial_\nu g \,g^{-1}\partial_\rho g ) \nn\ee
which we recognise as the winding \eqn{winding} that we saw earlier. Although the quantum Hall fluid provides us with a natural 3-manifold ${\cal M}$, taking the level $k\in {\bf Z}$ ensures that the two-dimensional theory on the boundary is actually independent of our choice of ${\cal M}$. In this way, the WZW model is, despite appearances, an intrinsically two-dimensional theory. 

\para
The central charge of the $SU(N)_k$ WZW model is
\be c = \frac{k(N^2-1)}{k+N}\nn\ee
The theories are weakly coupled as $k\rightarrow \infty$ where the central charge is equal to the dimension of the group $SU(N)$. Theories becomes strongly coupled as $k$ gets smaller. In particular, for $k=0$ we have $c=0$. This reflects that the fact that the sigma-model on the group manifold without any topological term flows to a gapped theory in the infra-red. 

\para
The WZW models have a large symmetry $G$ known as a current algebra. Usually in quantum field theory, a symmetry implies a current $J^\mu$ which obeys $\partial_\mu J^\mu=0$. The symmetry of the WZW model is much stronger as the left-moving and right-moving parts of the current are independently conserved. In terms of complex coordinates, this means that we have holomorphic and anti-holomorphic currents $J = \partial g\,g^{-1}$ and $\bar{J}=g^{-1}\bar{\partial}g$ obeying 
\be \bar{\partial} J(z) = 0\ \ \ {\rm and}\ \ \ \partial\bar{J}(\bar{z})=0\nn\ee
This is very similar to the conditions on the stress-tensor that you first meet in the study of CFT. In that case, one writes the stress tensor in a Laurent expansion and the resulting modes obey the Virasoro algebra. Here we do the same thing. This time the resulting modes obey 
\be
[J_n^a,J_m^b] = if^{ab}_{\ \ c}J^c_{n+m} + k n\delta^{ab}\delta_{n+m}\label{km}\ee
Here $a,b$ label the different generators of the Lie algebra associated to $G$ and $f^{ab}_{\ \ c}$ are the structure constants of the Lie algebra. Meanwhile, $n,m$ label the modes of the current algebra. Note that if we restrict to the $n,m=0$ sector then this is contains the Lie algebra. Including all the modes gives an infinite dimensional generalisation of the Lie algebra known as the {\it Kac-Moody algebra}. 

\para
Both the Kac-Moody algebra and the Virasoro algebra are infinite. But the Kac-Moody algebra should be thought of as bigger. Indeed, one can build the generators of the Virasoro algebra from bi-linears of the current using what's known as the {\it Sugawara construction}. We therefore work with representations of \eqn{km}, each of which splits into an infinite number of representations of the Virasoro algebra. 

\para
The representations of \eqn{km} are characterised by their highest weight state, a  primary operator. Each of these can be characterised by the way it transforms under the zero modes. In other words, the primary operators of the Kac-Moody algebra are labelled by representations of the underlying Lie algebra. 
The question that remains is: what are the primary  operators? 

\para
In fact, we've already seen the answer to this in Section \ref{wilsonsec}: the primary operators are the same as the non-trivial Wilson lines allowed in the bulk. For $G=SU(2)$, this means that the primary operators are labelled by their spin $j=0,\frac{1}{2},\ldots,\frac{k}{2}$. For $G=SU(N)$, the primary operators are labelled by Young diagrams whose upper row has no more than $k$ boxes. 

\para
Armed with this list of primary operators, we can start to compute correlation functions and their braiding. However, there are  a number of powerful tools that aid in this, not least the {\it Knizhnik-Zamolodchikov equations}, which are a set of partial differential equations which the conformal blocks must obey. In many cases, these tools allow one to determine completely the braiding properties of the conformal blocks. 

\para
To end, we will simply list some of the theories that have been useful in describing fractional quantum Hall states
\begin{itemize}
\item $SU(2)_1$: The WZW models at level $k=1$ have Abelian anyons. For  $SU(2)_1$, the central charge is $c=1$ which is just that of a free boson. It turns out that theory describes the Halperin $(2,2,1)$ spin-singlet state that we met in Section \ref{multiwfsec}
\item $SU(2)_2$: The central charge is $c=3/2$, which is the same as that of a free boson and a free Majorana fermion. But this is precisely the content that we needed to describe the Moore-Read states. The $SU(2)_2$ theory describes the physics of the state at filling fraction $\nu=1$. For filling fraction $\nu=1/2$, we should resort to the description of the CFT that we met in the last section as $U(1)_2\times {\rm Ising}$. 
\item $SU(2)_k/U(1)$: One can use the WZW models as the starting point to construct further conformal field theories known as coset models. Roughly, this means that you mod out by a $U(1)$ symmetry. These are sometimes referred to as ${\bf Z}_k$ parafermion theories. They are associated to the $p=k$-clustered Read Rezayi states that we met in Section \ref{rrsec}. In particular, the ${\bf Z}_3$ theory exhibits Fibonacci anyons.
\end{itemize}

\end{document}